\makeatletter
\def\input@path{{/Users/taqm/qt/single//}}
\makeatother
\documentclass{book}
\usepackage[T1]{fontenc}
\usepackage[latin1]{inputenc}
\usepackage{fancybox}
\usepackage{makeidx}
\def\hyperparamadd{pdfcreator=DBLaTeX-0.2.2}
\usepackage[hyperlink]{db2latex}
\title{Quantum Time}
\author{John Ashmead}

\renewcommand{\DBKindexation}{
\begin{DBKindtable}
\DBKinditem{\writtenby}{John Ashmead}

\end{DBKindtable}
}

\makeindex
\makeglossary

\begin{document}
\long\def\hyper@section@backref#1#2#3{%
\typeout{BACK REF #1 / #2 / #3}%
\hyperlink{#3}{#2}}

\maketitle
\tableofcontents
\listoffigures
\label{single}\hyperlabel{single}%


\chapter{Abstract}\label{abstract}\hyperlabel{abstract}%
\begin{quote}

Clearly, the Time Traveller proceeded, any real body must have extension in four directions: it must have Length, Breadth, Thickness, and\textendash{}Duration. But through a natural infirmity of
        the flesh, which I will explain to you in a moment, we incline to overlook this fact. There are really four dimensions, three which we call the three planes of Space, and a fourth, Time.
        There is, however, a tendency to draw an unreal distinction between the former three dimensions and the latter, because it happens that our consciousness moves intermittently in one direction
        along the latter from the beginning to the end of our lives.

\hspace*\fill---~H. G. Wells
 \cite{Wells-1935}\end{quote}
\begin{quote}

This is often the way it is in physics -{} our mistake is not that we take our theories too seriously, but that we do not take them seriously enough. It is always hard to realize that
        these numbers and equations we play with at our desks have something to do with the real world. Even worse, there often seems to be a general agreement that certain phenomena are just not fit
        subjects for respectable theoretical and experimental effort.

\hspace*\fill---~Steven Weinberg
 \cite{Weinberg-1977}\end{quote}

Normally we quantize along the space dimensions but treat time classically. But from relativity we expect a high level of symmetry between time and space. What happens if we quantize time
    using the same rules we use to quantize space?

To do this, we generalize the paths in the Feynman path integral to include paths that vary in time as well as in space. We use Morlet wavelet decomposition to ensure convergence and
    normalization of the path integrals. We derive the Schrödinger equation in four dimensions from the short time limit of the path integral expression. We verify that we recover standard quantum
    theory in the non-{}relativistic, semi-{}classical, and long time limits.

Quantum time is an experiment factory: most foundational experiments in quantum mechanics can be modified in a way that makes them tests of quantum time. We look at single and double slits
    in time, scattering by time-{}varying electric and magnetic fields, and the Aharonov-{}Bohm effect in time.


\chapter{Time and Quantum Mechanics}\label{int}\hyperlabel{int}%

\section{The Problem of Time}\label{int-intro}\hyperlabel{int-intro}%
\begin{quote}

In the world about us, the past is distinctly different from the future. More precisely, we say that the processes going on in the world about us are asymmetric in time, or display an
        arrow of time. Yet, this manifest fact of our experience is particularly difficult to explain in terms of the fundamental laws of physics. Newton's laws, quantum mechanics, electromagnetism,
        Einstein's theory of gravity, etc., make no distinction between the past and future -{} they are time-{}symmetric.

\hspace*\fill---~Halliwell, Pérez-{}Mercador, and Zurek
 \cite{Halliwell-1994}\end{quote}
\begin{quote}

Einstein's theory of general relativity goes further and says that time has no objective meaning. The world does not, in fact, change in time; it is a gigantic stopped clock. This
        freaky revelation is known as the problem of frozen time or simply the problem of time.

\hspace*\fill---~George Musser
 \cite{Musser-2009}\end{quote}

\subsection{Two Views of the River}
Time is a problem: it is not only that we never have enough of it, but we do not know what it is exactly that we do not have enough of. The two poles of the problem have been
        established for at least 2500 years, since the pre-{}Socratic philosophers of ancient Greece (\cite{Kirk-1957}, 
 \cite{Barbour-2000}): Parmenides viewed all time as existing at once, with change and movement being illusions; Heraclitus focused on the instant-{}by-{}instant passage of time: you
        cannot step in the same river twice.

If time is a river, some see it from the point of view of a white-{}water rafter, caught up in the moment; others from the perspective of a surveyor, mapping the river as a whole.

The debate has sharpened considerably in the last century, since our two strongest theories of physics \textendash{} relativity and quantum mechanics \textendash{} take almost opposite views.

\subsection{Relativity}
Time and space are treated symmetrically in relativity: they are formally indistinguishable, except that they enter the metric with opposite signs. Even this breaks down crossing the
        Schwarzschild radius of a black hole. Consider the line element for such:

\begin{equation}
\label{equation-int-intro-int-intro-0}\hyperlabel{equation-int-intro-int-intro-0}%
d{s}^{2}=\left(1-\frac{2m}{r}\right)d{t}^{2}-\frac{d{r}^{2}}{1-\frac{2m}{r}}-{r}^{2}\left(d{\theta }^{2}+{\mathrm{sin}}^{2}\left(\theta \right)d{\phi }^{2}\right)
\end{equation}

Here the time and the radius elements swap sign and therefore roles when 
 $ r=2m$~ (\cite{Adler-1965}, 
 \cite{Novikov-1998}). The problem was resolved by Georges LeMaitre in 1932 (per 
 \cite{Kaku-2005}) but it is curious that it arose in the first place.

Further, in relativity, it takes (significant) work to recover the traditional forward-{}travelling time. We have to construct the initial spacelike hypersurface and subsequent steps,
        they do not appear naturally, see for instance 
 \cite{Barbour-2000}.

\subsection{Quantum Mechanics}
Problems with respect to the role of time in quantum mechanics include:
\begin{enumerate}

\item{}Time and space enter asymmetrically in quantum mechanics.

\item{}Treatments of quantum mechanics typically rely on the notion that we can define a series of presents, marching forwards in time. It is difficult to define what one means by
                this.

\item{}The uncertainty principle for time/energy has a different character than the uncertainty principle for space/momentum.

\end{enumerate}

\paragraph*{Time a Parameter, Not an Operator}

\noindent

In quantum mechanics we have the mantra: time is a parameter, not an operator. Time functions like a butler, escorting wave functions from one room to another, but not itself
        interacting with them.

This is alien to the spirit of quantum mechanics. Why should time, alone among coordinates, escape being quantized?

\paragraph*{Spacelike Hypersurface}

\noindent

In quantum mechanics, defining the spacelike foliations across which time marches is problematic.
\begin{enumerate}

\item{}These foliations are not well-{}defined, given that uncertainty in time precludes exact knowledge of which hypersurface you are on at any one time.

\item{}They are difficult to reconcile with relativity. If Alice and Bob are traveling at relativistic velocities with respect to each other, they will foliate the planes of the
                present in different ways; each "present moment" for one will be partly past, partly future for the other. The quantum fluctuations purely in space for one, will be partly in time for
                the other.

\end{enumerate}

There is a nice analysis of the difficulties in a series of papers by Suarez: 
 \cite{Suarez-1997}  \cite{Suarez-1997b}  \cite{Suarez-1998}  \cite{Suarez-1998b}  \cite{Suarez-1998c}  \cite{Suarez-1998d}  \cite{Suarez-1998e}  \cite{Suarez-2000}  \cite{Suarez-2003}. He points out that standard quantum theory implies a "preferred frame". Not only is this troubling in its own right, but it may imply the possibility of
        superluminal communication. Suarez's specific response, Multisimultaneity, was not confirmed experimentally (\cite{Stefanov-2001}  \cite{Stefanov-2002}) but his objections remain.

\paragraph*{Uncertainty Relations}

\noindent

The existence of an uncertainty principle between time and energy was assumed by Heisenberg (\cite{Heisenberg-1927}) as a matter of course. Much work has been done since then and matters are no longer simple. References include: 
 \cite{Hilgevoord-1996}, 
 \cite{Hilgevoord-1998}, 
 \cite{Oppenheim-1997}, 
 \cite{Oppenheim-1998b}, 
 \cite{Oppenheim-1998c},
 \cite{Oppenheim-1999},
 \cite{Oppenheim-2000},
 \cite{Busch-2001}, 
 \cite{Hilgevoord-2001},
 \cite{Hilgevoord-2005},
 \cite{Hilgevoord-2005b},
 \cite{Hilgevoord-2007}. To over-{}summarize some fairly subtle discussions:
\begin{enumerate}

\item{}There is an uncertainty relationship between time and energy, but it does not stand on quite the same basis as the uncertainty relation between space and momentum.

The `not quite the same basis' is troubling. As Feynman has noted, if any experiment can break down the uncertainty principle, the whole structure of quantum mechanics will
                fail.

\item{}Great precision in the definition of terms is essential, if the disputants are not to be merely talking past one another.

In this connection, Oppenheim uses a particularly effective approach in his 1999 thesis (\cite{Oppenheim-1999}, like this work titled "Quantum Time"): he analyzes the effects of quantum mechanics along the time dimension using model experiments, which
                ensures that words are given operational meaning.

\end{enumerate}

\paragraph*{Feynman Path Integrals Not Full Solution}

\noindent
\begin{quote}

Although the path-{}integral formalism provides us with manifestly Lorentz-{}invariant rules, it does not make clear why the S-{}matrix calculated in this way is unitary. As far as I
            know, the only way to show that the path-{}integral formalism yields a unitary S-{}matrix is to use it to reconstruct the canonical formalism, in which unitarity is obvious.

\hspace*\fill---~Steven Weinberg
 \cite{Weinberg-1995}\end{quote}

One can argue that one does not expect covariance in non-{}relativistic quantum mechanics. But the problem does not go away in quantum electrodynamics.

In canonical quantization we have manifest unitarity, but not manifest covariance; in Feynman path integrals, we have manifest covariance, but not manifest unitarity.

If no single perspective has both manifest unitarity and manifest covariance, then it is possible that the underlying theory is incomplete.

We are in the position of a nervous accountant whose client never lets him see all the books at once, but only one set at a time. We can not be entirely sure that there is not some
        small but significant discrepancy, perhaps disguised in an off-{}book entry or hidden in an off-{}shore account.

\subsection{Bridging the Gap}
As relativity and quantum mechanics are arguably the two best confirmed theories we have, the dichotomy is troubling.

We are going to attack the problem from the quantum mechanics side. We will quantize time using the same rules we use to quantize space then see what breaks.

This does not mean cutting time up into small bits or quanta \textendash{} we do not normally do that to space after all \textendash{} it means applying the rules used to quantize space along the time axis as
        well.

Our objective is to create a version of standard quantum theory which satisfies the requirements of being (\cite{Langmuir-1953},
 \cite{Schick-1995}):
\begin{enumerate}

\item{}Well-{}defined,

\item{}Manifestly covariant,

\item{}Consistent with known experimental results,

\item{}Testable,

\item{}And reasonably simple.

\end{enumerate}

We will do this using path integrals, generalizing the usual single particle path integrals by allowing the paths to vary in time as well as in space. We will need to make no other
        changes to the path integrals themselves, but we will need to manage some of the associated mathematics a bit differently (see 
 \hyperlink{tq-fpi}{Feynman Path Integrals}). The defining assumption of complete covariance between time and space means we have no free parameters and no "wiggle room": quantum
        time as developed here is immediately falsifiable.

Our "work product" will be a well-{}defined set of rules \textendash{} manifestly symmetric between time and space \textendash{} which will let us, subject to the limits of our ingenuity and computing resources,
        predict the result of any experiment involving a single particle interacting with slits or electromagnetic fields.

As you might expect intuitively, the main effect expected is additional fuzziness in time. A particle going through a chopper might show up on the far side a bit earlier or later than
        expected. If it is going through a time-{}varying electromagnetic field, it will sample the future behavior of the field a bit too early, remember the previous behavior of the field a bit too
        long. These are the sorts of effects that might easily be discarded as experimental noise if they are not being specifically looked for.

In general, to see an effect from quantum time we need both beam and target to be varying in time. If either is steady, the effects of quantum time will be averaged out. Therefore a
        typical experimental setup will have a prep stage, presumably a chopper of some kind, to force the particle to have a known width in time, followed by the experiment proper.

We may classify the possible experimental outcomes as:
\begin{enumerate}

\item{}The behavior of time in quantum mechanics is fully covariant; all quantum effects seen along the space dimensions are seen along the time dimension.

\item{}We see quantum mechanical effects along the time direction, but they are not fully covariant: the effects along the time direction are less (or greater) than those seen in
                space.

Presumably there would be a frame in which the quantum mechanics effects in time were least (or greatest); such a frame would be a candidate "preferred frame of the universe".
                The rest frame of the center of mass of the universe might define such a frame (see for instance a re-{}analysis of the Michelson-{}Morley data by Cahill 
 \cite{Cahill-2003}).

\item{}We see no quantum mechanical effects along the time dimension. In this case (and the previous) we might look for associated failures of Lorentz invariance
 \footnote{For recent reviews of the experimental/observational state of Lorentz invariance see 
 \cite{Jacobson-2004}, 
 \cite{Mattingly-2005}, 
 \cite{MattinglyD-2007}, 
 \cite{Maccione-2009}, 
 \cite{Stecker-2009}. At this point, the assumption of Lorentz invariance appears reasonably safe, but for an opposite point of view see the recent work by Horara (\cite{Horara-2009}).}.

\end{enumerate}

Any of these results would be interesting in its own right
 \footnote{We are therefore in the position of a bookie who so carefully balanced the incoming wagers and the odds as to be indifferent as to which horse wins.}.

\section{Laboratory and Quantum Time}\label{int-key}\hyperlabel{int-key}%
\begin{quote}

Wheeler's often unconventional vision of nature was grounded in reality through the principle of radical conservatism, which he acquired from Niels Bohr: Be conservative by sticking to
        well-{}established physical principles, but probe them by exposing their most radical conclusions.

\hspace*\fill---~Kip S. Thorne
 \cite{Thorne-2009}\end{quote}

\subsection{Laboratory Time}\begin{figure}[H]

\begin{center}
\imgexists{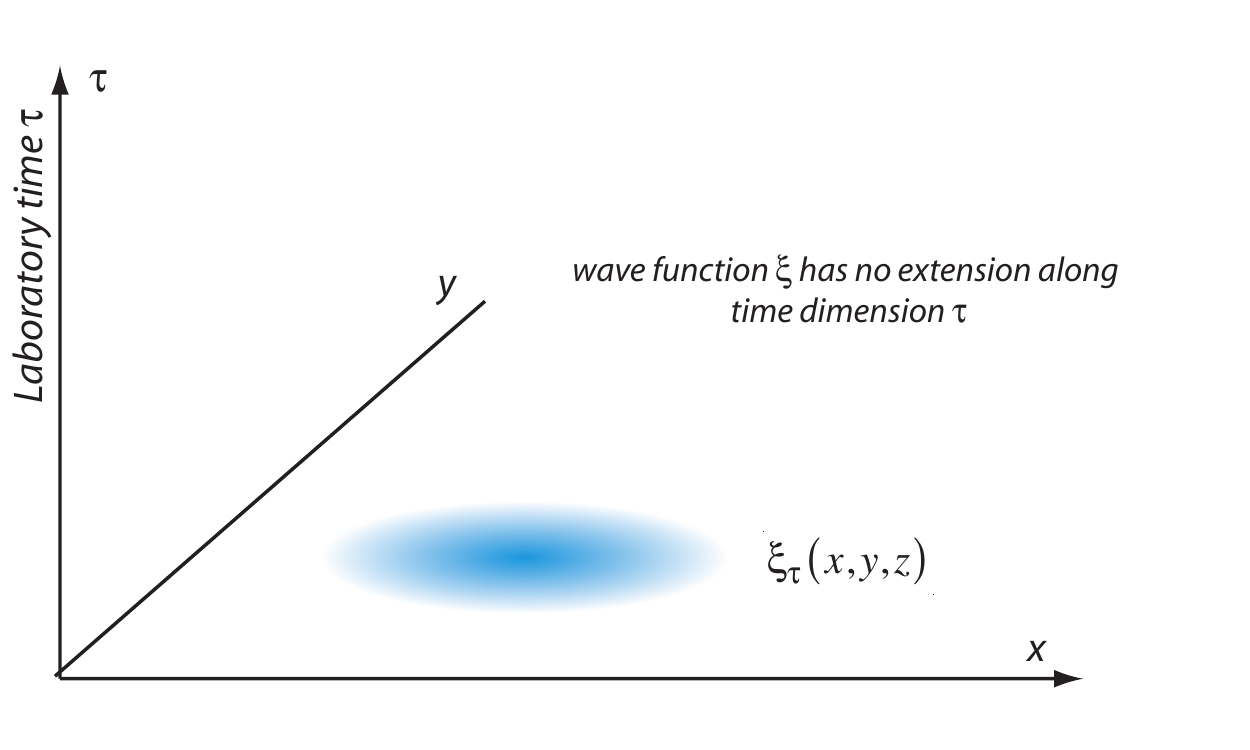}{{\imgevalsize{images/int-key-0.pdf}{\includegraphics[width=\imgwidth,height=\imgheight,keepaspectratio=true]{images/int-key-0.pdf}}\quad
}
}{}
\end{center}
\caption{Laboratory Time}
\label{figure-images-int-key-0}\hyperlabel{figure-images-int-key-0}%
\end{figure}

We start with the laboratory time or clock time 
 $ \tau $, measured by Alice using clocks, laser beams, and graduate students. Laboratory time is defined operationally; in terms of seconds, clock ticks, cycles of a cesium atom. The
        term is used by Busch (\cite{Busch-2001}) and others. We will take laboratory time as understood "well enough" for our purposes. (For a deeper examination see, for instance, 
 \cite{Helling-2008}, 
 \cite{Prati-2008}.)

The usual wave function 
 $ \xi $~ is "flat" in time: it represents a well-{}defined measure of our uncertainty about the particle's position in space, but shows no evidence of any uncertainty in time.
        This seems "unquantum-{}mechanical". Given that any observer, Bob say, going at high velocity with respect to Alice will mix time and space, what to Alice looks like uncertainty only in space
        will to Bob look like uncertainty in a blend of time and space.

\subsection{Quantum Time}\begin{figure}[H]

\begin{center}
\imgexists{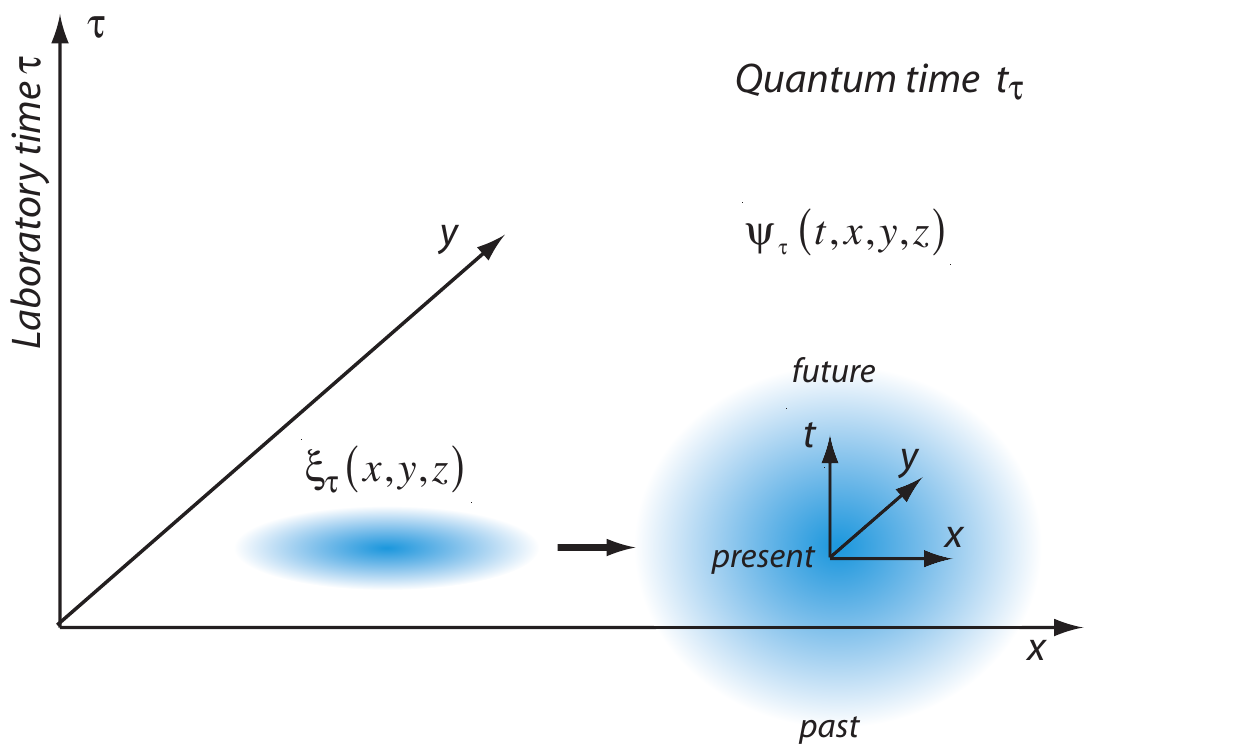}{{\imgevalsize{images/int-key-1.pdf}{\includegraphics[width=\imgwidth,height=\imgheight,keepaspectratio=true]{images/int-key-1.pdf}}\quad
}
}{}
\end{center}
\caption{Quantum Time}
\label{figure-images-int-key-1}\hyperlabel{figure-images-int-key-1}%
\end{figure}

If we are to treat time and space symmetrically \textendash{} our basic assumption \textendash{} there can be no justification for treating time as flat but space as fuzzy.

We will therefore extrude Alice's wave function into the time dimension, positing that the wave function, at any given instant, is a function of time as well. Alice will now have to add
        uncertainty about the particle's position in time to her existing uncertainty about the particle's position in space:

\begin{equation}
\label{equation-int-key-int-key-1}\hyperlabel{equation-int-key-int-key-1}%
{\xi }_{\tau }\left(x,y,z\right)\rightarrow {\psi }_{\tau }\left(t,x,y,z\right)
\end{equation}

This extruded wave function represents uncertainty in time and space, just as the wave function normally does in just space. How the extruded wave function depends on quantum time,
        Latin 
 $ t$, is strongly constrained by covariance. Of this much much more below (\hyperlink{tq}{Formal Development}).

To see the effects of the extrusion of the wave function into quantum time 
 $ t$, we can treat quantum time like any other unmeasured quantum variable, computing its indirect effects by taking expectations against reduced density matrices and the like
 \footnote{
Implicit in this use of quantum time is the assumption of the block universe, that all time exists at once (\cite{Price-1995}, 
 \cite{Nahin-1999},
 \cite{Barbour-2000}, 
 \cite{Stenger-2000}). While there is no question that this is counter-{}intuitive, it is difficult to reconcile the more intuitive concept of a fleeting and momentary present
            with special relativity and its implications for simultaneity. See Petkov for a vigorous defense of this point: 
 \cite{Petkov-2005},
 \cite{Petkov-2007b}.

There is evidence for the block universe view within quantum mechanics as well, in the delayed choice quantum eraser (\cite{Scully-1982},
 \cite{Kim-2000b}). The most straightforward way to make sense of this experiment is to see all time as existing at once.

Asymmetry between time and space is customary in quantum mechanics, but not mandatory. Aharonov, Bergmann, and Lebowitz have given a time-{}symmetric approach to measurement 
 \cite{Aharonov-1964}. Cramer has given a time-{}symmetric interpretation of quantum mechanics 
 \cite{Cramer-1986},
 \cite{Cramer-1988}. There is no quantum arrow of time per Maccone 
 \cite{Maccone-2009}~ amended in 
 \cite{Maccone-2009b}.
}.

\subsection{Relationship of Quantum and Laboratory Time}
Hilgevoord cautions us to distinguish between the use of coordinates as parameter and as operator (\cite{Hilgevoord-1996}  \cite{Hilgevoord-1998}). For instance, we have 
 \emph{x}~ the coordinate and 
 \emph{x}~ the operator, with different roles in a typical construction:

\begin{equation}
\label{equation-int-key-int-key-4}\hyperlabel{equation-int-key-int-key-4}%
\langle {x}^{\left(op\right)}\rangle \equiv {\displaystyle \int \text{d}{x}^{\left(coord\right)}{\xi }^{\ast }\left({x}^{\left(coord\right)}\right)}{x}^{\left(op\right)}\xi \left({x}^{\left(coord\right)}\right)
\end{equation}

He argues (correctly in our view) that in standard quantum theory there is no time operator:
\begin{quote}

If 
 $ t$~ is not the relativistic partner of 
 \emph{q}~ [the space operator], what is the true partner of the latter? The answer is simply that such a partner does not exist; the position variable of a point
            particle is a non-{}covariant concept.

\hspace*\fill---~Jan Hilgevoord and David Atkinson
 \cite{Hilgevoord-2007}\end{quote}

While time is not an operator in standard quantum theory, in this work \textendash{} by assumption \textendash{} it is. We can therefore write:

\begin{equation}
\label{equation-int-key-int-key-5}\hyperlabel{equation-int-key-int-key-5}%
\langle {t}^{\left(op\right)}\rangle \equiv {\displaystyle \int \text{d}{t}^{\left(coord\right)}\text{d}{x}^{\left(coord\right)}{\psi }^{\ast }\left({t}^{\left(coord\right)},{x}^{\left(coord\right)}\right){t}^{\left(op\right)}\psi \left({t}^{\left(coord\right)},{x}^{\left(coord\right)}\right)}
\end{equation}

The usual wave function changes shape as laboratory time advances; if it did not it would not be interesting. The quantum time wave function must evolve with laboratory time as well. At
        each tick of the laboratory clock we expect that 
 $ \psi $~ will have in general a slightly different shape with respect to quantum time.

It will be (extremely) convenient to define the relative quantum time 
 $ {t}_{\tau }$~ as the offset in quantum time from the current value of Alice's laboratory time:

\begin{equation}
\label{equation-int-key-conv-symbols-reltime}\hyperlabel{equation-int-key-conv-symbols-reltime}%
{t}_{\tau }\equiv t-\tau 
\end{equation}

If the lab clock says 10 seconds past the hour, the relative quantum time might be 10 attoseconds before or after that. In most cases, we expect that the expectation of the quantum time
        will be approximately equal to the laboratory time and therefore that the expectation of the relative time will be approximately zero:

\begin{equation}
\label{equation-int-key-int-key-2b}\hyperlabel{equation-int-key-int-key-2b}%
\begin{array}{l}\langle t\rangle \sim \tau \\ \langle {t}_{\tau }\rangle \sim 0\end{array}
\end{equation}

The situation is analogous to the use of "center of mass" coordinates. We use center of mass coordinates to subtract off the average value of the space coordinates, letting us focus on
        the interesting part. And we can use "center of time" coordinates the same way, to focus on what is essential.

As an example, suppose Alice is travelling by train from Berne to Zurich. She decides to while away the time by doing quantum mechanics experiments (we are not explaining, merely
        reporting). If she is doing, say, a standard double slit experiment, then she will compute 
 \emph{x}, 
 \emph{y}, and 
 \emph{z}~ relative to her current location on the train. An outside observer, say Bob, might compute his 
 \emph{x}~ as the sum of the train's 
 \emph{x}~ and Alice's 
 \emph{x}. Alice's 
 \emph{x}~ may be thought of as a relative space coordinate. The same with time. Alice may find it convenient to compute her experimental times in terms of attoseconds; Bob
        may compute the times as the clock time in the train plus the attoseconds. Alice is then using relative time; Bob is using block time.

With quantum time we are 
 \emph{not}~ inventing a new time dimension or assigning new properties to the existing time dimension. We are merely treating, for purposes of quantum mechanics, time the
        same as the three space dimensions.

\subsection{Evolution of the Wave Function}\begin{figure}[H]

\begin{center}
\imgexists{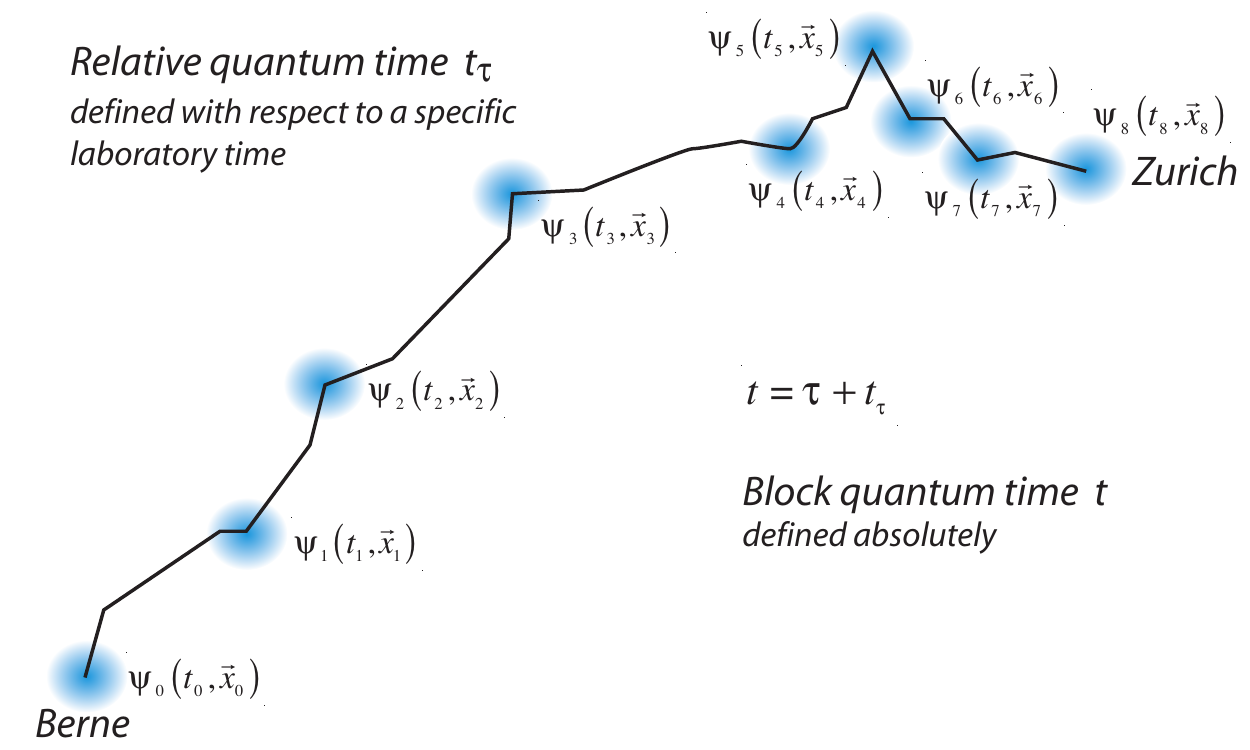}{{\imgevalsize{images/int-key-2.pdf}{\includegraphics[width=\imgwidth,height=\imgheight,keepaspectratio=true]{images/int-key-2.pdf}}\quad
}
}{}
\end{center}
\caption{Evolution of Wave Function}
\label{figure-images-int-key-2}\hyperlabel{figure-images-int-key-2}%
\end{figure}

How are we to compute the wave function at the next tick of the laboratory clock when we know it at the current clock tick? We need dynamics.

We will use Feynman path integrals as our defining methodology; we will derive the Schrödinger equation, operator mechanics, and canonical path integrals from them
 \footnote{Particularly readable introductions to Feynman path integrals are found in 
 \cite{Feynman-1965}~ and 
 \cite{Schulman-1981}.}.

A path in the usual three dimensional Feynman path integrals is defined as a series of coordinate locations; to specify the path we specify a specific location in three space at each
        tick of the laboratory clock. To do the path integral we sum over all such paths using an appropriate weighting factor.

In temporal quantization we specify the paths as a specific location in four space \textendash{} time plus the three space dimensions \textendash{} at each tick of the laboratory clock. To do the path integral
        we sum over all such paths using an appropriate weighting factor. Curiously enough, we can use the same 
 \hyperlink{tq-fpi-lag}{weighting factor}~ in temporal quantization as in standard quantum theory.

The paths in temporal quantization can be a bit ahead or behind the laboratory time; they can \textendash{} and typically will \textendash{} have a non-{}zero relative time. We will have to first show that these
        effects normally average out (or else someone would have seen them); we will then show that with a bit of ingenuity they should be detectable.

In Feynman path integrals we do not normally use a continuous laboratory time; we break it up into slices and then let the number of slices go to infinity. We can see each slice as
        corresponding to a frame in a movie. The laboratory time functions as an index, like the frame count in a movie. It is not part of the dynamics. Laboratory time is time as parameter.

If Alice is walking her dog, her path corresponds to laboratory time, a smooth steadily increasing progression. Her dog's path corresponds to quantum time, frisking ahead or behind at
        any moment, but still centered on the laboratory time
 \footnote{Of course, Alice is herself a quantum mechanical system, made of atoms and their bonds. Alice's own wave function is a product of the many many wave functions of her particles,
        amino acids, sugars, water molecules, and so on. Her average quantum time will be almost exactly her laboratory time.}.

\section{Literature}\label{int-lit}\hyperlabel{int-lit}%

With 2500 years to work up a running start, the literature on time is enormous. Popular discussions include: 
 \cite{Carroll-1865}, 
 \cite{Carroll-1871}, 
 \cite{Gold-1958}, 
 \cite{Davies-1974}, 
 \cite{Horwich-1987}, 
 \cite{Earman-1989}, 
 \cite{Thorne-1994}, 
 \cite{Davies-1995}, 
 \cite{Hawking-1996}, 
 \cite{Hawking-1996b}, 
 \cite{Hawking-1996c}, 
 \cite{Price-1996}, 
 \cite{Nahin-1999}, 
 \cite{Toomey-2007}, 
 \cite{Barbour-2000}, 
 \cite{Stenger-2000}, 
 \cite{Galison-2003}, 
 \cite{Halpern-2008}, 
 \cite{Kaku-2008}, 
 \cite{Carroll-2010}; more technical include: 
 \cite{Reichenbach-1956}, 
 \cite{Penrose-1962}, 
 \cite{Cramer-1983}, 
 \cite{Penrose-1986}, 
 \cite{Hawking-1993}, 
 \cite{Halliwell-1994}, 
 \cite{Macey-1994}, 
 \cite{Oaklander-1994}, 
 \cite{Omnes-1994b}, 
 \cite{Savitt-1997}, 
 \cite{Schulman-1997}, 
 \cite{Atmanspacher-1997}, 
 \cite{Krasnikov-1996b}, 
 \cite{Zeh-1999}, 
 \cite{Muga-2001b}, 
 \cite{Zeh-2001}, 
 \cite{Zeh-2003c}, 
 \cite{Greenberger-2005}, 
 \cite{Kiefer-2005b}, 
 \cite{Gross-2007}, 
 \cite{Barbour-2008}.

The approach we are taking here is most similar to some work by Feynman (\cite{Feynman-1950}  \cite{Feynman-1951}). Note particularly his variation on the Klein-{}Gordon equation:

\begin{equation}
\label{equation-int-lit-int-lit-0}\hyperlabel{equation-int-lit-int-lit-0}%
i\frac{\partial {\psi }_{u}\left(x\right)}{\partial u}=-\frac{1}{2}\left(i\frac{\partial }{\partial {x}^{\mu }}-e{A}_{\mu }\right)\left(i\frac{\partial }{\partial {x}_{\mu }}-e{A}^{\mu }\right){\psi }_{u}\left(x\right)
\end{equation}

Where 
 \emph{u}~ is a formal time parameter "somewhat analogous to proper time".

Using proper time makes it difficult to handle multiple particles \textendash{} whose proper time should we use? \textendash{} hence our preference for using laboratory time as a starting point.

We see some resemblances of our propagators and Schrödinger equation to the Stuckelberg propagator and Schrödinger equation used by Land and by Horwitz (\cite{Land-1996}  \cite{Horwitz-1998}). They add a fifth parameter, treated dynamically, so that it takes part in gauge transformations and the like. Another fifth parameter formalism is found in 
 \cite{Seidewitz-2005}. There is an ongoing series of conferences on such: 
 \cite{IARD-2010}.

The principal difference between fifth parameter formalisms in general and quantum time here is that here we have only four parameters: laboratory time and quantum time have to share: they
    are really only different views of a single time dimension. Neither is formal; both are real.

Among the many other variations on the theme of time are: stochastic time 
 \cite{Bonifacio-1999}, random time 
 \cite{Chung-2003}, complex time 
 \cite{Asaro-1996}, discrete times 
 \cite{Bender-1985}  \cite{Jaroszkiewicz-1998}  \cite{Qi-2000}  \cite{Date-2002}, labyrinthean time 
 \cite{Halpern-2008}, multiple time dimensions 
 \cite{Chen-2005},
 \cite{Sparling-2007},
 \cite{Weinstein-2008}, time generated from within the observer \textendash{} internal time 
 \cite{Svozil-1996}, and most recently crystallizing time 
 \cite{Ellis-2009}.

There is an excellent summary of possible times in a Scientific American article by Max Tegmark 
 \cite{Tegmark-2003}. He describes massively parallel time, forking time, distant times, and more.

There is no end to alternate times: in his novel Einstein's Dreams 
 \cite{Lightman-1993}~ A. Lightman imagines A. Einstein imagining thirty or more different kinds of times, before settling on relativity.

Temporal quantization \textendash{} as we will refer to the process of quantizing along the time dimensions \textendash{} plays nicely with many of these variations on the theme of time. For instance we assume
    time is smooth. But suppose time is quantized at the scale of the Planck time:

\begin{equation}
\label{equation-int-lit-int-lit-1}\hyperlabel{equation-int-lit-int-lit-1}%
{t}^{\left(Planck\right)}\equiv \sqrt{\frac{\hslash G}{{c}^{5}}}\approx 5.39x{10}^{-44}s
\end{equation}

We would only insist that space be quantized in the same way, at the scale of the Planck length:

\begin{equation}
\label{equation-int-lit-int-lit-2}\hyperlabel{equation-int-lit-int-lit-2}%
{l}^{\left(Planck\right)}\equiv c{t}^{\left(Planck\right)}=\sqrt{\frac{\hslash G}{{c}^{3}}}\approx 1.62x{10}^{-35}m
\end{equation}

\section{Plan of Attack}\label{int-plan}\hyperlabel{int-plan}%
\begin{figure}[H]

\begin{center}
\imgexists{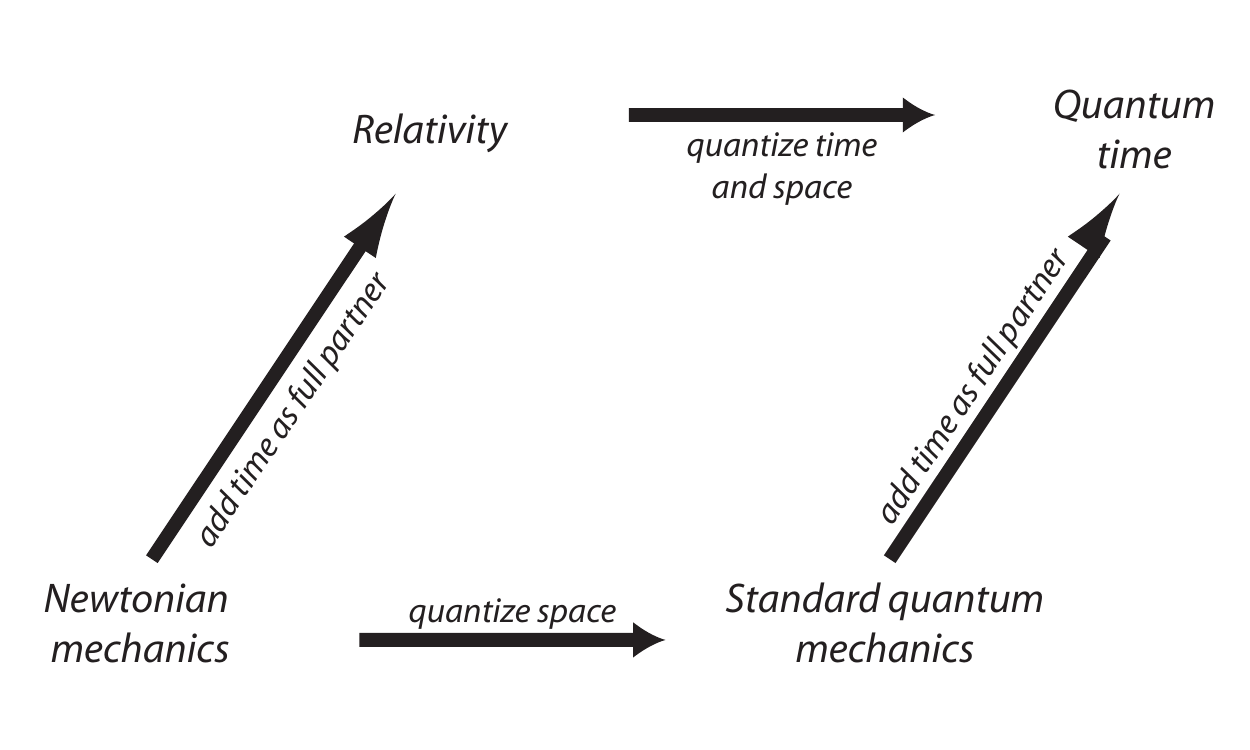}{{\imgevalsize{images/int-plan-0.pdf}{\includegraphics[width=\imgwidth,height=\imgheight,keepaspectratio=true]{images/int-plan-0.pdf}}\quad
}
}{}
\end{center}
\caption{Objective: a Manifestly Covariant Quantum Mechanics}
\label{figure-images-int-plan-0}\hyperlabel{figure-images-int-plan-0}%
\end{figure}

\subsection{Organization}
In the interests of biting off a "testable chunk", we will only look at the single particle case here. We will do so in a way that does not exclude extending the ideas to multiple
        particles.

We primarily interested in "proof-{}of-{}concept" here, so we will only look at the lowest nontrivial corrections resulting from quantum time.

We have organized the rest of this paper in roughly the order of the 
 \hyperlink{int-intro}{five requirements}, that temporal quantization be:
\begin{enumerate}

\item{}Well-{}defined,

\item{}Manifestly covariant,

\item{}Consistent with known experimental results,

\item{}Testable,

\item{}And reasonably simple.

\end{enumerate}

By chapters:
\begin{enumerate}

\item{}In 
 \hyperlink{tq}{Formal Development}, we work out the formalism, using path integrals and the requirement of manifest covariance (\hyperlink{tq-fpi}{Feynman Path Integrals}).

We use the path integral result to derive the Schrödinger equation in four dimensions (\hyperlink{tq-seqn-deriv}{Derivation of the Schrödinger Equation}). We use the Schrödinger equation to prove unitarity (\hyperlink{tq-seqn-unitarity}{Unitarity}) and to analyze the effect of gauge transformations (\hyperlink{tq-seqn-gauge}{Gauge Transformations for the Schrödinger Equation}).

We derive an operator formalism from the Schrödinger equation (\hyperlink{tq-op}{Operators in Time}), then derive the canonical path integral from the operator formalism (\hyperlink{tq-pq-deriv}{Derivation of the Canonical Path Integral}).

From the canonical path integral we derive the Feynman path integral (\hyperlink{tq-pq-circle}{Closing the Circle}), closing the circle.

All four formalisms make use of the laboratory time; in a final section, 
 \hyperlink{covar}{Covariant Definition of Laboratory Time}, we show we can define the laboratory time in a way which is covariant, thereby establishing full covariance of
                the formalisms.

This establishes that temporal quantization is well-{}defined (in a formal sense) and covariant (by construction).

\item{}In 
 \hyperlink{comp}{Comparison of Temporal Quantization To Standard Quantum Theory}, we look at various limits in which we recover standard quantum theory from temporal
                quantization:
\begin{enumerate}

\item{}The 
 \hyperlink{nonrel}{Non-{}relativistic Limit},

\item{}The 
 \hyperlink{semi}{Semi-{}classical Limit},

\item{}And the 
 \hyperlink{stat}{Long Time Limit}.

\end{enumerate}

\item{}In 
 \hyperlink{xt}{Experimental Tests}, we look at a starter set of experimental tests.

In general, to see an effect of temporal quantization 
 \emph{both}~ beam and apparatus should vary in time. With this condition met, temporal quantization will tend to produce increased dispersion in time and \textendash{}
                occasionally \textendash{} slightly different interference patterns.

We look at the cases of:
\begin{enumerate}

\item{}  \hyperlink{xt-gates}{Slits in Time},

\item{}  \hyperlink{xt-fields}{Time-{}varying Magnetic and Electric Fields},

\item{}And the 
 \hyperlink{xt-ab}{Aharonov-{}Bohm Experiment}.

\end{enumerate}

These experiments establish that:
\begin{itemize}

\item{}Temporal quantization is well-{}defined in an operational sense; its formalism can be mapped into well-{}defined counts of clicks in a detector.

\item{}Temporal quantization is testable.

\end{itemize}

\item{}Finally, in the 
 \hyperlink{disc}{Discussion}~ we summarize the analysis, argue that temporal quantization has met the requirements, and look at further areas for investigation.

\end{enumerate}

Post-{}finally, we summarize some useful facts about free wave functions in an appendix 
 \hyperlink{free}{Free Particles}.

\subsection{Notations and Conventions}\begin{enumerate}

\item{}We use Latin 
 $ t$~ for quantum time; Greek 
 $ \tau $~ for laboratory time.

\item{}We use an over-{}dot for the derivative with respect to laboratory time, e.g:

\begin{equation}
\label{equation-int-plan-conv-symbols-overdot}\hyperlabel{equation-int-plan-conv-symbols-overdot}%
{\dot{\chi }}_{\tau }\equiv \frac{d}{d\tau }{\chi }_{\tau }
\end{equation}

\item{}We use an over-{}bar to indicate averaging. We also use an over-{}bar to indicate the standard quantum theory/space part of an object.

We use a 'frown' character ($ \frown $) to indicate the quantum time part.

We use the absence of a mark to mark a fully four dimensional ~ object.

For example, we will see that the free kernel can be written as a product of time and space parts:

\begin{equation}
\label{equation-int-plan-conv-symbols-3}\hyperlabel{equation-int-plan-conv-symbols-3}%
{K}_{\tau }^{\left(free\right)}\left({t}^{{''}},{\overrightarrow{x}}^{{''}};{t}^{\prime },{\overrightarrow{x}}^{\prime }\right)={\stackrel{\frown }{K}}_{\tau }^{\left(free\right)}\left({t}^{{''}};{t}^{\prime }\right){\overline{K}}_{\tau }^{\left(free\right)}\left({\overrightarrow{x}}^{{''}};{\overrightarrow{x}}^{\prime }\right)
\end{equation}

\item{}When we can, we will use 
 $ \chi $~ for the time part of wave functions (from the initial letter of 
 $ \chi \rho \acute{o}\nu o\varsigma $, the Greek word for time), and 
 $ \xi $~ (Greek 
 \emph{x}) for the space part, e.g.:

\begin{equation}
\label{equation-int-plan-conv-symbols-2}\hyperlabel{equation-int-plan-conv-symbols-2}%
{\psi }_{\tau }\left(t,\overrightarrow{x}\right)={\chi }_{\tau }\left(t\right){\xi }_{\tau }\left(\overrightarrow{x}\right)
\end{equation}

\end{enumerate}

We will be using natural units, 
 \emph{c}~ and 
 $ \hslash $~ set to one, except as explicitly noted.

In an effort to reduce notational clutter we will:
\begin{enumerate}

\item{}Use two indexes together to mean the difference of the first indexed variable and the second:

\begin{equation}
\label{equation-int-plan-conv-symbols-short1}\hyperlabel{equation-int-plan-conv-symbols-short1}%
{\tau }_{BA}\equiv {\tau }_{B}-{\tau }_{A}
\end{equation}

\item{}Replace an indexed laboratory time, e.g. 
 $ {\tau }_{\text{A}}$, with its index, e.g. 
 \emph{A}, when we can do so without loss of clarity:

\begin{equation}
\label{equation-int-plan-conv-symbols-short2}\hyperlabel{equation-int-plan-conv-symbols-short2}%
{\psi }_{A}\equiv {\psi }_{{\tau }_{A}}
\end{equation}

\end{enumerate}

For the square root of complex numbers we use a branch cut from zero to negative infinity.

We use the Einstein summation convention with Greek indices being summed from 0 to 3, Latin from 1 to 3.


\chapter{Formal Development}\label{tq}\hyperlabel{tq}%

\section{Overview}\label{tq-intro}\hyperlabel{tq-intro}%
\begin{quote}

The rules of quantum mechanics and special relativity are so strict and powerful that it's very hard to build theories that obey both.

\hspace*\fill---~Frank Wilczek
 \cite{Wilczek-2008}\end{quote}
\begin{quote}

Now we could travel anywhere we wanted to go. All a man had to do was to think of what he says and to look where he was going.

\hspace*\fill---~The Legend of the Flying Canoe
 \cite{Doucet-1996}\end{quote}
\begin{figure}[H]

\begin{center}
\imgexists{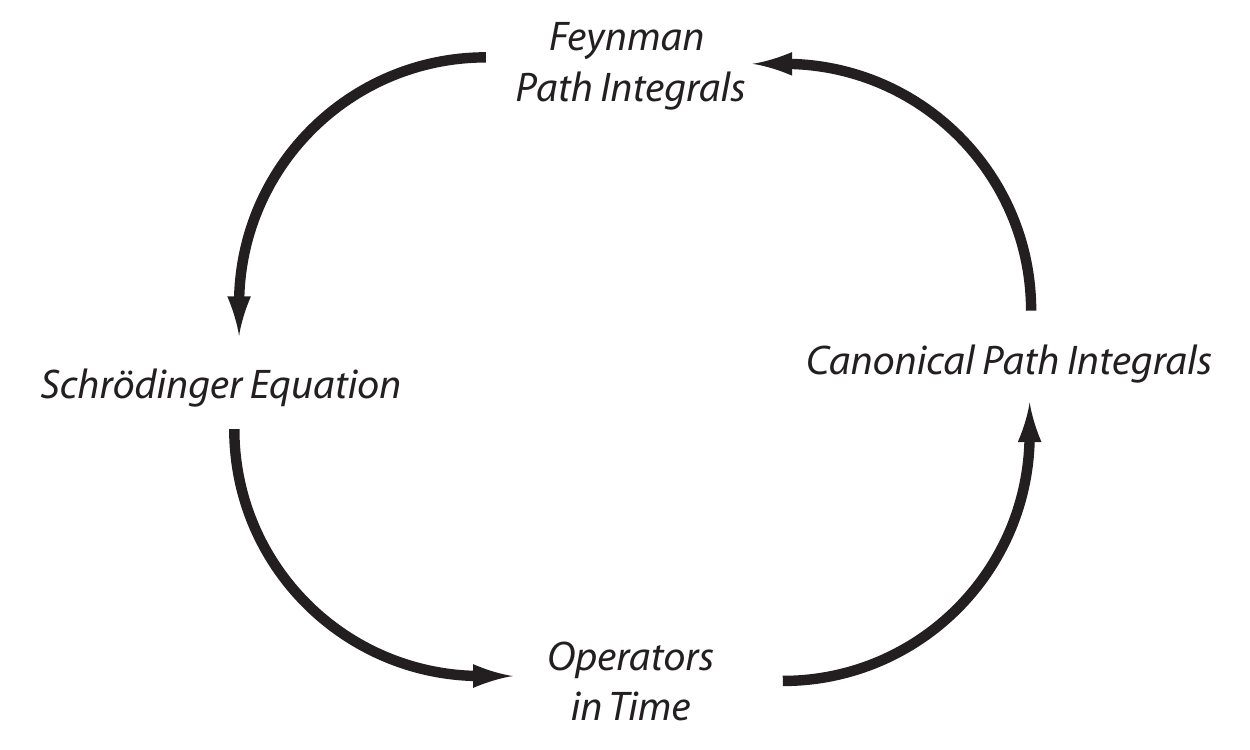}{{\imgevalsize{images/tq-disc-0.pdf}{\includegraphics[width=\imgwidth,height=\imgheight,keepaspectratio=true]{images/tq-disc-0.pdf}}\quad
}
}{}
\end{center}
\caption{Four Formalisms}
\label{figure-images-tq-disc-0}\hyperlabel{figure-images-tq-disc-0}%
\end{figure}

In this chapter we develop the formal rules for temporal quantization. Like a snake headed out for its morning rat, we will need to take some twists and turns to get to our objective. We
    use 
 \hyperlink{tq-fpi}{Feynman path integrals}~ as the defining formalism. Comprehensive treatments of path integrals are provided in: 
 \cite{Feynman-1965d}, 
 \cite{Schulman-1981}, 
 \cite{Swanson-1992}, 
 \cite{Khandekar-1993}, 
 \cite{Kleinert-2004}, 
 \cite{Zinn-2005}.

We will derive three other formalisms, each in turn, from the Feynman path integrals:
\begin{enumerate}

\item{}  \hyperlink{tq-seqn}{Schrödinger Equation},

\item{}  \hyperlink{tq-op}{Operators in Time},

\item{}And 
 \hyperlink{tq-pq}{Canonical Path Integrals}.

\end{enumerate}

We will close the circle by deriving the Feynman path integral from the canonical path integral.

All four approaches make use of the laboratory time. To get a completely covariant treatment we need to define the laboratory time in a covariant way as well. The proper time of the
    particle will not do; what if we have many particles? What of virtual particles, those evanescent dolphins of the Bose and Fermi seas? What of the massless and therefore timeless photons?

Instead in the last section of this chapter, 
 \hyperlink{covar}{Covariant Definition of Laboratory Time}, we break the initial wave function down using Morlet wavelet decomposition (\cite{Morlet-1982},
 \cite{Chui-1992},
 \cite{Meyer-1992},
 \cite{Kaiser-1994},
 \cite{vandenBerg-1999},
 \cite{Addison-2002},
 \cite{Bratteli-2002},
 \cite{Antoine-2004},
 \cite{Ashmead-2009w}), then evolve each part along its own personal classical path to its destined detector. These classical paths each have a well-{}defined proper time which will
    serve as the laboratory time for the part. At the detector, we assemble the parts back into one self-{}consistent whole.

With this done, we will have satisfied the first two requirements (\hyperlink{int-intro}{The Problem of Time}), that temporal quantization be:
\begin{enumerate}

\item{}Well-{}defined

\item{}And manifestly covariant.

\end{enumerate}

\section{Feynman Path Integrals}\label{tq-fpi}\hyperlabel{tq-fpi}%
\begin{quote}

The path integral method is perhaps the most elegant and powerful of all quantization programs.

\hspace*\fill---~Michio Kaku
 \cite{Kaku-1993}\end{quote}
\begin{quote}

Furthermore, we wish to emphasize that in future in all cities, markets and in the country, the only ingredients used for the brewing of beer must be Barley, Hops and Water. Whosoever
        knowingly disregards or transgresses upon this ordinance, shall be punished by the Court authorities' confiscating such barrels of beer, without fail.

\hspace*\fill---~Duke of Bavaria
 \cite{Reinheitsgebot-1516}\end{quote}
\begin{figure}[H]

\begin{center}
\imgexists{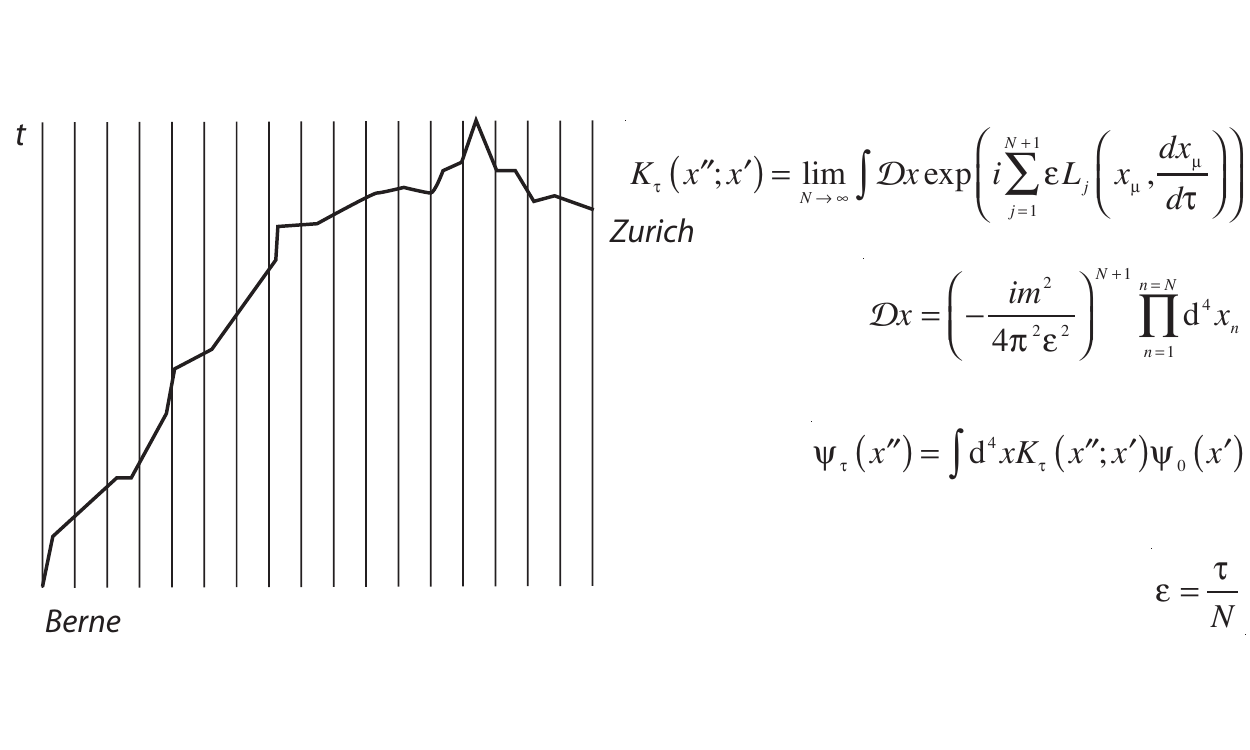}{{\imgevalsize{images/tq-fpi-0.pdf}{\includegraphics[width=\imgwidth,height=\imgheight,keepaspectratio=true]{images/tq-fpi-0.pdf}}\quad
}
}{}
\end{center}
\caption{Path Integrals}
\label{figure-images-tq-fpi-0}\hyperlabel{figure-images-tq-fpi-0}%
\end{figure}

With path integrals as with beer, all we need to get good results are a few simple ingredients, a recipe for combining them, and a bit of time. For path integrals the ingredients are the
    initial wave functions, the paths, and a Lagrangian; the recipe is the procedure for summing over the paths
 \footnote{Usually we would start an analysis with the Schrödinger equation and then derive the kernel as its inverse. Here it is more natural to start with the path integral expression for the
    kernel then derive the Schrödinger equation from that.}.

Our guiding principle is manifest covariance. We look at:
\begin{enumerate}

\item{}The 
 \hyperlink{tq-fpi-waves}{Wave Functions},

\item{}The 
 \hyperlink{tq-fpi-paths}{Paths},

\item{}The 
 \hyperlink{tq-fpi-lag}{Lagrangian},

\item{}The 
 \hyperlink{tq-fpi-sum}{Sum over Paths},

\item{}How to ensure 
 \hyperlink{tq-fpi-conv}{Convergence}~ of the sum over paths,

\item{}How to ensure correct 
 \hyperlink{tq-fpi-norm}{Normalization},

\item{}And the 
 \hyperlink{tq-fpi-final}{Formal Expression for the Path Integral}.

\end{enumerate}

\subsection{Wave Functions}\label{tq-fpi-waves}\hyperlabel{tq-fpi-waves}%

\paragraph*{Requirements}

\noindent

To define our wave functions we need a basis which is:
\begin{enumerate}

\item{}Non-{}singular,

\item{}General,

\item{}And reasonably simple.

\end{enumerate}

\paragraph*{Plane Waves}

\noindent

Using a Fourier decomposition in time is general but problematic. The use of singular functions or non-{}normalizable functions, e.g. plane waves or 
 $ \delta $~ functions, may introduce artifacts 
 \footnote{A further disadvantage of using plane waves as the basis functions is that demonstrating the convergence of our path integrals then becomes tricky, as will be discussed below 
 \hyperlink{tq-fpi-conv}{Convergence}.}. It is safer to use more physical wave functions, i.e. wave packets.

There is a good example of the benefits of using more realistic wave functions in Gondran and Gondran, 
 \cite{Gondran-2005}. When they analyze the Stern-{}Gerlach experiment using a Gaussian initial wave function (as opposed to a plane wave) they see the usual split of the beam \textendash{}
    without any need to invoke the notorious and troubling "collapse of the wave function".

An implication of Gondran and Gondran's work is that some of the difficulties in the analysis of the measurement problem may be the result not of problems with quantum mechanics per se but
    rather of using unphysical approximations.

This is an implication of the program of decoherence as well.

The assumption that we can isolate a quantum system from its environment is unphysical. The program of decoherence (see 
 \cite{Zeh-1996},
 \cite{Giulini-1996},
 \cite{Schlosshauer-2005}~ and references therein) has been able to explain much of the "problem of measurement" by relaxing this assumption, by explicitly including
    interactions with the environment in the analysis.

Avoiding unphysical assumptions is as important in the analysis of time as of measurement. Time is already a subtle and difficult subject; we do not want to introduce any unnecessary
    complexities, especially not at the start.

\paragraph*{Wave Packets}

\noindent

Wave packets are more physical but are not general.

Consider an incoming beam, say of electrons. A typical wave packet might look like:

\begin{equation}
\label{equation-tq-fpi-waves-tq-fpi-waves-0}\hyperlabel{equation-tq-fpi-waves-tq-fpi-waves-0}%
{\psi }_{\tau }\left(x\right)=\sqrt[4]{\frac{1}{\pi {\sigma }_{1}^{2}}}\mathrm{exp}\left(-i\omega \tau +ikx-\frac{{\left(x-{x}_{0}\right)}^{2}}{2{\sigma }_{1}^{2}}\right)
\end{equation}

We could generalize the time part by adding a bit of dispersion along our hypothesized quantum time axis:

\begin{equation}
\label{equation-tq-fpi-waves-tq-fpi-waves-1}\hyperlabel{equation-tq-fpi-waves-tq-fpi-waves-1}%
{\psi }_{\tau }\left(t,x\right)\sim \sqrt[4]{\frac{1}{{\pi }^{2}{\sigma }_{0}^{2}{\sigma }_{1}^{2}}}\mathrm{exp}\left(-i\omega t-\frac{{\left(t-\tau \right)}^{2}}{2{\sigma }_{0}^{2}}+ikx-\frac{{\left(x-{x}_{0}\right)}^{2}}{2{\sigma }_{1}^{2}}\right)
\end{equation}

If the 
 $ \sigma $'s are large, we may think of this as a gently rounded plane wave.

This is physically reasonable, but not general. Not every wave function is a Gaussian.

\paragraph*{Morlet Wavelet Decomposition}

\noindent
\begin{figure}[H]

\begin{center}
\imgexists{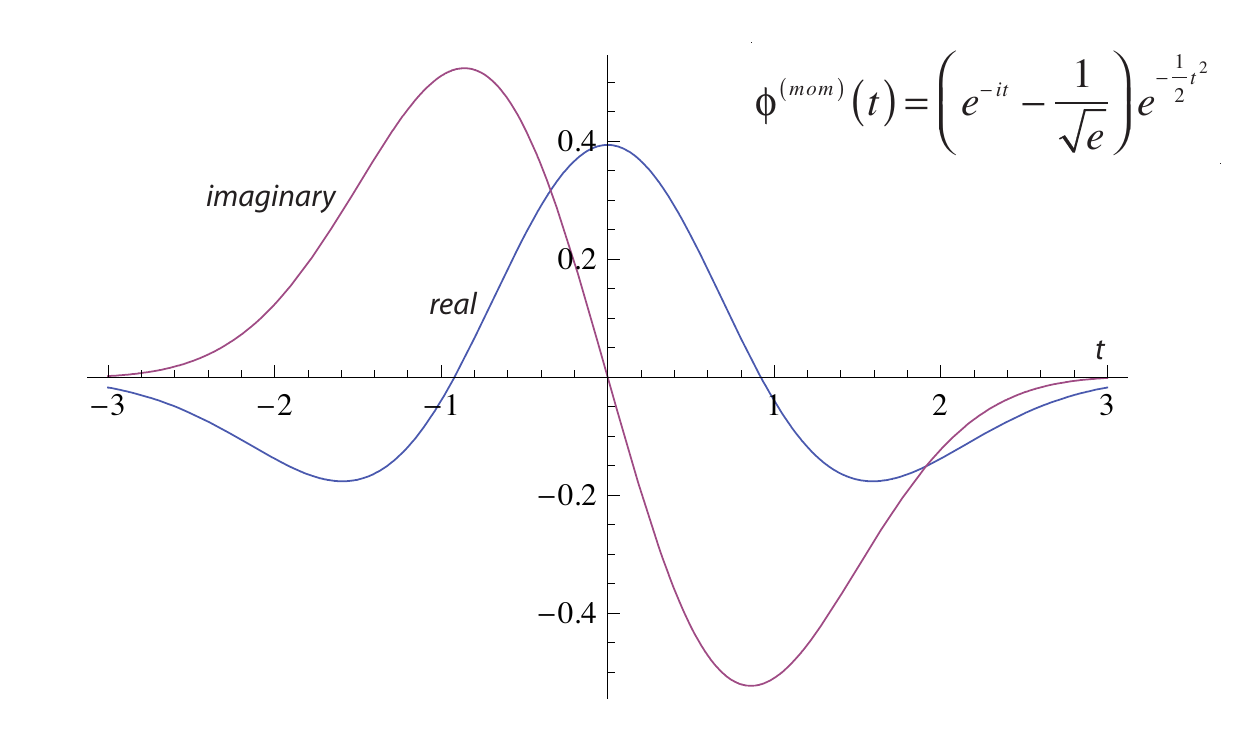}{{\imgevalsize{images/mor-intro-0.pdf}{\includegraphics[width=\imgwidth,height=\imgheight,keepaspectratio=true]{images/mor-intro-0.pdf}}\quad
}
}{}
\end{center}
\caption{Typical Morlet Wavelet}
\label{figure-images-mor-intro-0}\hyperlabel{figure-images-mor-intro-0}%
\end{figure}

To make this general, we recall that any square-{}integrable wave function may be written as a sum over Morlet wavelets
 \footnote{Morlet's initial reference is: 
 \cite{Morlet-1982}. Good discussions of Morlet wavelets and wavelets in general are found in 
 \cite{Chui-1992},
 \cite{Meyer-1992},
 \cite{Kaiser-1994},
 \cite{vandenBerg-1999},
 \cite{Addison-2002},
 \cite{Bratteli-2002},
 \cite{Antoine-2004}.}. We can then break an arbitrary square-{}integrable wave function up into its component wavelets, propagate each wavelet individually, then sum over
    the wavelets at the end.

A Morlet wavelet in one dimension has the form:

\begin{equation}
\label{equation-tq-fpi-waves-conv-trans-mw}\hyperlabel{equation-tq-fpi-waves-conv-trans-mw}%
{\phi }_{sd}\left(t\right)=\frac{1}{\sqrt{\left\vert s\right\vert }}\left({e}^{-i\left(\frac{t-d}{s}\right)}-\frac{1}{\sqrt{e}}\right){e}^{-\frac{1}{2}{\left(\frac{t-d}{s}\right)}^{2}}
\end{equation}

Here 
 \emph{s}~ is the scale and 
 \emph{d}~ is the displacement. The wavelet with scale one and displacement zero is the mother wavelet:

\begin{equation}
\label{equation-tq-fpi-waves-conv-trans-mother}\hyperlabel{equation-tq-fpi-waves-conv-trans-mother}%
{\phi }^{\left(mom\right)}\left(t\right)=\left({e}^{-it}-\frac{1}{\sqrt{e}}\right){e}^{-\frac{1}{2}{t}^{2}}
\end{equation}

All wavelets are derived from her by changing the values of 
 \emph{s}~ and 
 \emph{d}:

\begin{equation}
\label{equation-tq-fpi-waves-conv-trans-scale}\hyperlabel{equation-tq-fpi-waves-conv-trans-scale}%
{\phi }_{sd}\left(t\right)=\frac{1}{\sqrt{\left\vert s\right\vert }}{\phi }^{\left(mom\right)}\left(\frac{t-d}{s}\right)
\end{equation}

The Morlet wavelet components of a square-{}integrable wave function 
 \emph{f}~ are:

\begin{equation}
\label{equation-tq-fpi-waves-conv-trans-mw_analysis}\hyperlabel{equation-tq-fpi-waves-conv-trans-mw_analysis}%
{\tilde{f}}_{sd}={\displaystyle \underset{-\infty }{\overset{\infty }{\int }}\text{d}t{\phi }_{sd}^{\ast }\left(t\right)f\left(t\right)}
\end{equation}

To recover the original wave function 
 \emph{f}~ from the Morlet wavelet components 
 $ {\tilde{f}}_{sd}$:

\begin{equation}
\label{equation-tq-fpi-waves-conv-trans-mw_synthesis}\hyperlabel{equation-tq-fpi-waves-conv-trans-mw_synthesis}%
f\left(t\right)=\frac{1}{C}{\displaystyle \underset{-\infty }{\overset{\infty }{\int }}\frac{\text{d}s\text{d}d}{{s}^{2}}{\phi }_{sd}\left(t\right){\tilde{f}}_{sd}}
\end{equation}

  \emph{C}, the "admissibility constant", is computed in 
 \cite{Ashmead-2009w}.

As each Morlet wavelet is a sum over two Gaussians, any normalizable function may be decomposed into a sum over Gaussians.

Therefore, to compute the path integral results for an arbitrary normalizable wave function, we use Morlet wavelet analysis to decompose it into Gaussian test functions, compute the result
    for each Gaussian test function, then sum to get the full result.

\paragraph*{Gaussian Test Functions}

\noindent

A Gaussian test function (a squeezed state) in 
 \emph{x}~ is defined by its values for the average 
 \emph{x}, average 
 \emph{p}, and dispersion in 
 \emph{x}. For a four dimensional function, 
 \emph{x}~ and 
 \emph{p}~ are vectors and the dispersion is a four-{}by-{}four matrix. Therefore there are potentially four plus four plus sixteen or twenty-{}four of these values. For test
    functions we will always use a diagonal dispersion matrix, letting us write our test functions as direct products of functions in 
 $ t$, 
 \emph{x}, 
 \emph{y}, and 
 \emph{z}:

\begin{equation}
\label{equation-tq-fpi-waves-free-fourt-phi0}\hyperlabel{equation-tq-fpi-waves-free-fourt-phi0}%
{\psi }_{0}\left(x\right)=\sqrt[4]{\frac{1}{{\pi }^{4}\mathrm{det}\left({\Sigma }_{0}\right)}}\mathrm{exp}\left(-i{p}_{\mu }^{\left(0\right)}{x}^{\mu }-\frac{1}{2{\Sigma }_{0}^{\mu \nu }}\left({x}^{\mu }-{\overline{x}}_{0}^{\mu }\right)\left({x}^{n}-{\overline{x}}_{0}^{\nu }\right)\right)
\end{equation}

With the definition of the dispersion matrix:

\begin{equation}
\label{equation-tq-fpi-waves-free-fourt-sigma0}\hyperlabel{equation-tq-fpi-waves-free-fourt-sigma0}%
{\Sigma }_{0}^{\mu \nu }=\left(\begin{array}{cccc}{\sigma }_{0}^{2}& 0& 0& 0\\ 0& {\sigma }_{1}^{2}& 0& 0\\ 0& 0& {\sigma }_{2}^{2}& 0\\ 0& 0& 0& {\sigma }_{3}^{2}\end{array}\right)
\end{equation}

And only twelve free parameters, three per dimension.

We can break this out into time and space parts. We use 
 $ \chi $~ for the time part, 
 $ \xi $~ for the space part:

\begin{equation}
\label{equation-tq-fpi-waves-free-fourt-breakout}\hyperlabel{equation-tq-fpi-waves-free-fourt-breakout}%
{\psi }_{0}\left(t,x\right)={\chi }_{0}\left(t\right){\xi }_{0}\left(\overrightarrow{x}\right)
\end{equation}

Time part:

\begin{equation}
\label{equation-tq-fpi-waves-free-onet-chi0}\hyperlabel{equation-tq-fpi-waves-free-onet-chi0}%
{\chi }_{0}\left(t\right)=\sqrt[4]{\frac{1}{\pi {\sigma }_{0}^{2}}}\mathrm{exp}\left(-i{E}_{0}t-\frac{1}{2{\sigma }_{0}^{2}}{\left(t-{\overline{t}}_{0}\right)}^{2}\right)
\end{equation}

Space part:

\begin{equation}
\label{equation-tq-fpi-waves-free-fourt-xi0}\hyperlabel{equation-tq-fpi-waves-free-fourt-xi0}%
{\xi }_{0}\left(\overrightarrow{x}\right)=\sqrt[4]{\frac{1}{{\pi }^{3}\mathrm{det}\left({\Sigma }_{ij}^{\left(0\right)}\right)}}\mathrm{exp}\left(i{\overrightarrow{p}}_{0}\cdot \overrightarrow{x}-{\left(\overrightarrow{x}-{\overrightarrow{x}}_{0}\right)}_{i}\frac{1}{2{\Sigma }_{ij}^{\left(0\right)}}{\left(\overrightarrow{x}-{\overrightarrow{x}}_{0}\right)}_{j}\right)
\end{equation}

Expectations of coordinates:

\begin{equation}
\label{equation-tq-fpi-waves-tq-fpi-waves-gtf-6}\hyperlabel{equation-tq-fpi-waves-tq-fpi-waves-gtf-6}%
\langle {\psi }_{0}\vert {x}^{\mu }\vert {\psi }_{0}\rangle ={\overline{x}}_{0}^{\mu }
\end{equation}

And uncertainty:

\begin{equation}
\label{equation-tq-fpi-waves-tq-fpi-waves-gtf-7}\hyperlabel{equation-tq-fpi-waves-tq-fpi-waves-gtf-7}%
\langle {\psi }_{0}\vert \left({x}^{\mu }-{\overline{x}}_{0}^{\mu }\right)\left({x}^{\nu }-{\overline{x}}_{0}^{\nu }\right)\vert {\psi }_{0}\rangle =\frac{{\Sigma }_{0}^{\mu \nu }}{2}
\end{equation}

We use the Gaussian test functions as "typical" wave forms, to see what the system is likely to do, and as the components (in a Morlet wavelet decomposition) of a completely general
    solution
 \footnote{The Gaussian test functions here are covariant, but the Morlet wavelet decomposition is not. We provide a covariant form for the Morlet wavelet decomposition in 
 \cite{Ashmead-2009w}.}.

\subsection{Paths}\label{tq-fpi-paths}\hyperlabel{tq-fpi-paths}%

A path is a series of wave functions indexed by 
 $ \tau $. For a particle in coordinate representation, we can model a path as a series of 
 $ \delta $~ functions ~ indexed by 
 $ \tau $, i.e.: 
 $ \delta \left(x-{x}_{\tau }\right)$.

More formally, we consider the laboratory time from start to finish, sliced into 
 \emph{N}~ pieces. A path is then given by the value of the coordinates at each slice. At the end, we let the number of slices go to infinity.

Like all paths in path integrals, our paths are going to be jagged, darting around forward and back in time, like a frisky dog being walked by its much slower and more sedate owner.

\subsection{Lagrangian}\label{tq-fpi-lag}\hyperlabel{tq-fpi-lag}%

To sum over our paths we have to weight each path by the exponential of 
 \emph{i}~ times the action 
 \emph{S}, the action being the integral of the Lagrangian over laboratory time:

\begin{equation}
\label{equation-tq-fpi-lag-tq-fpi-lag-exp}\hyperlabel{equation-tq-fpi-lag-tq-fpi-lag-exp}%
\text{exp}\left(\text{i}{S}_{BA}\right)=\text{exp}\left(\text{i}{\displaystyle \underset{A}{\overset{B}{\int }}\text{d}\tau L\left({x}^{\mu },{\dot{x}}^{\mu }\right)}\right)
\end{equation}

\paragraph*{Requirements for Lagrangian}

\noindent

Our requirements for the Lagrangian are that it:
\begin{enumerate}

\item{}Produce the correct classical equations of motion,

\item{}Be manifestly covariant,

\item{}Be reasonably simple,

\item{}Give the correct Schrödinger equation,

\item{}And give the correct non-{}relativistic limit.

\end{enumerate}

\paragraph*{Selection of Lagrangian}

\noindent

A Lagrangian of the form:

\begin{equation}
\label{equation-tq-fpi-lag-tq-fpi-lag-lag}\hyperlabel{equation-tq-fpi-lag-tq-fpi-lag-lag}%
L\left({x}^{\mu },{\dot{x}}^{\mu }\right)=-\frac{1}{2}m{\dot{x}}^{\mu }{\dot{x}}_{\mu }-e{\dot{x}}^{\mu }{A}_{\mu }\left(x\right)
\end{equation}

With definition:

\begin{equation}
\label{equation-tq-fpi-lag-tq-fpi-lag-0c}\hyperlabel{equation-tq-fpi-lag-tq-fpi-lag-0c}%
{A}^{\mu }\left(x\right)\equiv \left(\Phi \left(x\right),\overrightarrow{A}\left(x\right)\right)
\end{equation}

Will satisfy the first three requirements.

These requirements do not fully constrain our choice of Lagrangian. We may think of the classical trajectory as being like the river running through the center of a valley; the quantum
    fluctuations as corresponding to the topography of the surrounding valley. Many different topologies of the valley are consistent with the same course for the river.

In particular, Goldstein (\cite{Goldstein-1980}) notes that we could also look at Lagrangians where the velocity squared term is replaced by a general function of the velocity squared:

\begin{equation}
\label{equation-tq-fpi-lag-tq-fpi-lag-1a}\hyperlabel{equation-tq-fpi-lag-tq-fpi-lag-1a}%
-mf\left({\dot{x}}^{\mu }{\dot{x}}_{\mu }\right)
\end{equation}

Subject to the condition:

\begin{equation}
\label{equation-tq-fpi-lag-tq-fpi-lag-1b}\hyperlabel{equation-tq-fpi-lag-tq-fpi-lag-1b}%
\frac{\partial f\left(a\right)}{\partial a}=\frac{1}{2}
\end{equation}

However our choice is the only Lagrangian which is no worse than quadratic in the velocities. It is therefore simplest.

We are still free to select an overall scale and an additive constant. The scale is constrained by the space part, see 
 \hyperlink{tq-fpi-norm}{Normalization}. The additive part is constrained by the Schrödinger equation; with the additive constant 
 $ -m/2$~ we get the Klein-{}Gordon equation back as our Schrödinger equation, see below in 
 \hyperlink{tq-seqn-deriv}{Derivation of the Schrödinger Equation}.

Our candidate Lagrangian is therefore:

\begin{equation}
\label{equation-tq-fpi-lag-tq-fpi-lag-final}\hyperlabel{equation-tq-fpi-lag-tq-fpi-lag-final}%
L\left({x}^{\mu },{\dot{x}}^{\mu }\right)=-\frac{1}{2}m{\dot{x}}^{\mu }{\dot{x}}_{\mu }-e{\dot{x}}^{\mu }{A}_{\mu }\left(x\right)-\frac{m}{2}
\end{equation}

We break out the Lagrangian into time and space parts:

\begin{equation}
\label{equation-tq-fpi-lag-tq-fpi-lag-2}\hyperlabel{equation-tq-fpi-lag-tq-fpi-lag-2}%
L\left(t,\overrightarrow{x},\dot{t},\dot{\overrightarrow{x}}\right)=-\frac{1}{2}m{\dot{t}}^{2}+\frac{1}{2}m\dot{\overrightarrow{x}}\cdot \dot{\overrightarrow{x}}-e\dot{t}\Phi \left(t,\overrightarrow{x}\right)+e{\dot{x}}_{j}{A}_{j}\left(t,\overrightarrow{x}\right)-\frac{1}{2}m
\end{equation}

This Lagrangian gives the correct Euler-{}Lagrange ~ equations of motion:

\begin{equation}
\label{equation-tq-fpi-lag-tq-fpi-lag-el}\hyperlabel{equation-tq-fpi-lag-tq-fpi-lag-el}%
\begin{array}{l}m\ddot{t}=-e\dot{\Phi }+e\dot{t}{\Phi }_{,0}-e{\dot{x}}_{j}{A}_{j,0}=-e{\dot{x}}_{j}\left({\Phi }_{,j}+{A}_{j,0}\right)\\ m{\ddot{x}}_{i}=-e{\dot{A}}_{i}-e\dot{t}{\Phi }_{,i}+e{\dot{x}}_{j}{A}_{j,i}=-e\dot{t}{A}_{i,0}-e{\dot{x}}_{j}{A}_{i,j}-e{\Phi }_{,i}\dot{t}+e{\dot{x}}_{j}{A}_{j,i}\end{array}
\end{equation}

In terms of electric and magnetic fields:

\begin{equation}
\label{equation-tq-fpi-lag-tq-fpi-lag-force}\hyperlabel{equation-tq-fpi-lag-tq-fpi-lag-force}%
\begin{array}{l}m\ddot{t}=e\overrightarrow{E}\cdot \dot{\overrightarrow{x}}\\ m\ddot{\overrightarrow{x}}=e\dot{t}\overrightarrow{E}+e\dot{\overrightarrow{x}}\times \overrightarrow{B}\end{array}
\end{equation}

In manifestly covariant form:

\begin{equation}
\label{equation-tq-fpi-lag-tq-fpi-lag-4}\hyperlabel{equation-tq-fpi-lag-tq-fpi-lag-4}%
m{\ddot{x}}_{\mu }=e{F}_{\mu \nu }{\dot{x}}^{\nu }
\end{equation}

With:

\begin{equation}
\label{equation-tq-fpi-lag-tq-fpi-lag-5}\hyperlabel{equation-tq-fpi-lag-tq-fpi-lag-5}%
{F}_{\mu \nu }\equiv \frac{\partial {A}_{\nu }}{\partial {x}^{\mu }}-\frac{\partial {A}_{\mu }}{\partial {x}^{\nu }}=\left(\begin{array}{cccc}0& {E}_{x}& {E}_{y}& {E}_{z}\\ -{E}_{x}& 0& -{B}_{z}& {B}_{y}\\ -{E}_{y}& {B}_{z}& 0& -{B}_{x}\\ -{E}_{z}& -{B}_{y}& {B}_{x}& 0\end{array}\right)
\end{equation}

The effect of enforcing complete symmetry between time and space is to replace the electric potential with three new terms:

\begin{equation}
\label{equation-tq-fpi-lag-tq-fpi-lag-6}\hyperlabel{equation-tq-fpi-lag-tq-fpi-lag-6}%
-e\Phi \Rightarrow -e\dot{t}\Phi -\frac{m{\dot{t}}^{2}}{2}-\frac{m}{2}
\end{equation}

The action is defined using the Lagrangian:

\begin{equation}
\label{equation-tq-fpi-lag-tq-fpi-lag-7}\hyperlabel{equation-tq-fpi-lag-tq-fpi-lag-7}%
{S}_{BA}\equiv {\displaystyle \underset{A}{\overset{B}{\int }}\text{d}sL\left(\frac{dx}{ds},x\right)}
\end{equation}

The choice of the parameter 
 \emph{s}~ involves some subtleties. Per 
 \cite{Goldstein-1980}, it can be any Lorentz invariant parameter.

Perhaps the most popular choice is the particle's own proper time:

\begin{equation}
\label{equation-tq-fpi-lag-tq-fpi-lag-proper_time}\hyperlabel{equation-tq-fpi-lag-tq-fpi-lag-proper_time}%
s={\displaystyle \int \text{d}t\sqrt{1-\frac{dx}{dt}\frac{dx}{dt}-\frac{dy}{dt}\frac{dy}{dt}-\frac{dz}{dt}\frac{dz}{dt}}}
\end{equation}

However this is unacceptable as it would make generalization to the multi-{}particle case impossible. The problem is not only that each real particle would have its own, different time, but
    also each virtual particle would as well. No coherent theory can result from this.

We will use Alice's proper time for the time being:

\begin{equation}
\label{equation-tq-fpi-lag-tq-fpi-lag-8}\hyperlabel{equation-tq-fpi-lag-tq-fpi-lag-8}%
{S}_{BA}\equiv {\displaystyle \underset{{\tau }_{A}}{\overset{{\tau }_{B}}{\int }}\text{d}{\tau }^{\left(Alice\right)}L\left(\frac{dx}{d\tau },x\right)}
\end{equation}

However once we have worked out the rules using this, we will (re)define the laboratory time in a manifestly covariant (if slightly fractured) way in 
 \hyperlink{covar}{Covariant Definition of Laboratory Time}.

\paragraph*{Hamiltonian Form}

\noindent

We derive the Hamiltonian from the Lagrangian. The Hamiltonian gives insight as to the Schrödinger equation (\hyperlink{tq-seqn-deriv}{Derivation of the Schrödinger Equation}) and the evolution of operators (\hyperlink{tq-op}{Operators in Time}), and it provides the starting point for the derivation of the canonical form of path integrals (\hyperlink{tq-pq}{Canonical Path Integrals}).

The conjugate momentum for quantum time is given by:

\begin{equation}
\label{equation-tq-fpi-lag-tq-fpi-ham-1}\hyperlabel{equation-tq-fpi-lag-tq-fpi-ham-1}%
{\pi }_{0}\equiv \frac{\delta L}{\delta \dot{t}}=-m\dot{t}-e\Phi 
\end{equation}

So:

\begin{equation}
\label{equation-tq-fpi-lag-tq-fpi-ham-4}\hyperlabel{equation-tq-fpi-lag-tq-fpi-ham-4}%
\dot{t}=-\frac{{\pi }_{0}+e\Phi }{m}
\end{equation}

The conjugate momentum to space with respect to laboratory time is given by:

\begin{equation}
\label{equation-tq-fpi-lag-tq-fpi-ham-2}\hyperlabel{equation-tq-fpi-lag-tq-fpi-ham-2}%
\overrightarrow{\pi }\equiv \frac{\delta L}{\delta \dot{\overrightarrow{x}}}=m\dot{\overrightarrow{x}}+e\overrightarrow{A}
\end{equation}

So:

\begin{equation}
\label{equation-tq-fpi-lag-tq-fpi-ham-5}\hyperlabel{equation-tq-fpi-lag-tq-fpi-ham-5}%
\dot{\overrightarrow{x}}=\frac{\overrightarrow{\pi }-e\overrightarrow{A}}{m}
\end{equation}

The Hamiltonian is given by:

\begin{equation}
\label{equation-tq-fpi-lag-tq-fpi-ham-3}\hyperlabel{equation-tq-fpi-lag-tq-fpi-ham-3}%
H={\pi }_{0}\dot{t}+\overrightarrow{\pi }\cdot \dot{\overrightarrow{x}}-L
\end{equation}

Or:

\begin{equation}
\label{equation-tq-fpi-lag-tq-fpi-ham-ham}\hyperlabel{equation-tq-fpi-lag-tq-fpi-ham-ham}%
H=-\frac{1}{2m}{\left({\pi }_{0}+e\Phi \right)}^{2}+\frac{1}{2m}{\left(\overrightarrow{\pi }-e\overrightarrow{A}\right)}^{2}+\frac{m}{2}
\end{equation}

The Hamiltonian equations for the coordinates are:

\begin{equation}
\label{equation-tq-fpi-lag-tq-fpi-ham-eomq}\hyperlabel{equation-tq-fpi-lag-tq-fpi-ham-eomq}%
\begin{array}{l}\dot{t}=\frac{\partial H}{\partial {\pi }_{0}}=-\frac{{\pi }_{0}+e\Phi }{m}\\ {\dot{x}}_{i}=\frac{\partial H}{\partial {\pi }_{i}}=\frac{{\pi }_{i}-e{A}_{i}}{m}\end{array}
\end{equation}

And for the momenta are:

\begin{equation}
\label{equation-tq-fpi-lag-tq-fpi-ham-eomp}\hyperlabel{equation-tq-fpi-lag-tq-fpi-ham-eomp}%
\begin{array}{l}{\dot{\pi }}_{0}=-\frac{\partial H}{\partial t}=\frac{{\pi }_{0}+e\Phi }{m}e{\Phi }_{,0}+\frac{{\pi }_{j}-e{A}_{j}}{m}e{A}_{j,0}\\ {\dot{\pi }}_{i}=-\frac{\partial H}{\partial {x}_{i}}=\frac{{\pi }_{0}+e\Phi }{m}e{\Phi }_{,i}+\frac{{\pi }_{j}-e{A}_{j}}{m}e{A}_{j,i}\end{array}
\end{equation}

\subsection{Sum over Paths}\label{tq-fpi-sum}\hyperlabel{tq-fpi-sum}%

Our path integral measure has to include fluctuations in time as well as the more familiar fluctuations in space. By analogy with the usual three-{}space kernel, he kernel is:

\begin{equation}
\label{equation-tq-fpi-sum-tq-fpi-sum-5}\hyperlabel{equation-tq-fpi-sum-tq-fpi-sum-5}%
{K}_{BA}={\displaystyle \underset{A}{\overset{B}{\int }}\mathcal{D}x\mathrm{exp}\left(i{S}_{BA}\right)}
\end{equation}

With measure:

\begin{equation}
\label{equation-tq-fpi-sum-tq-fpi-sum-4}\hyperlabel{equation-tq-fpi-sum-tq-fpi-sum-4}%
\mathcal{D}x\equiv \underset{N\rightarrow \infty }{\mathrm{lim}}{C}_{N}{\displaystyle \int {\displaystyle \prod _{n=1}^{n=N}\text{d}{t}_{n}\text{d}{\overrightarrow{x}}_{n}}}
\end{equation}

And with 
 $ {C}_{N}$~ ~ a normalization constant to be determined below in 
 \hyperlink{tq-fpi-norm}{Normalization}.

We break out the path integral into time slices:

\begin{equation}
\label{equation-tq-fpi-sum-tq-fpi-sum-6}\hyperlabel{equation-tq-fpi-sum-tq-fpi-sum-6}%
{K}_{BA}=\underset{N\rightarrow \infty }{\mathrm{lim}}{C}_{N}{\displaystyle \int {\displaystyle \prod _{n=1}^{n=N}\text{d}{t}_{n}\text{d}{\overrightarrow{x}}_{n}\mathrm{exp}\left(i\epsilon {\displaystyle \sum _{j=1}^{N+1}{L}_{j}}\right)}}
\end{equation}

A single time step has duration:

\begin{equation}
\label{equation-tq-fpi-sum-tq-fpi-sum-7}\hyperlabel{equation-tq-fpi-sum-tq-fpi-sum-7}%
\epsilon \equiv \frac{{\tau }_{B}-{\tau }_{A}}{N}=\frac{{\tau }_{BA}}{N}
\end{equation}

We use the discrete form of the Lagrangian:

\begin{equation}
\label{equation-tq-fpi-sum-tq-fpi-sum-lag_step}\hyperlabel{equation-tq-fpi-sum-tq-fpi-sum-lag_step}%
{L}_{j}\equiv {L}_{j}^{t}+{L}_{j}^{\overrightarrow{x}}+{L}_{j}^{m}
\end{equation}

\begin{equation}
\label{equation-tq-fpi-sum-tq-fpi-sum-9}\hyperlabel{equation-tq-fpi-sum-tq-fpi-sum-9}%
{L}_{j}^{t}\equiv -\frac{m}{2}{\left(\frac{{t}_{j}-{t}_{j-1}}{\epsilon }\right)}^{2}-e\frac{{t}_{j}-{t}_{j-1}}{\epsilon }\frac{\Phi \left({x}_{j}\right)+\Phi \left({x}_{j-1}\right)}{2}
\end{equation}

\begin{equation}
\label{equation-tq-fpi-sum-tq-fpi-sum-a}\hyperlabel{equation-tq-fpi-sum-tq-fpi-sum-a}%
{L}_{j}^{\overrightarrow{x}}\equiv \frac{m}{2}{\left(\frac{{\overrightarrow{x}}_{j}-{\overrightarrow{x}}_{j-1}}{\epsilon }\right)}^{2}+e\frac{{\overrightarrow{x}}_{j}-{\overrightarrow{x}}_{j-1}}{\epsilon }\cdot \frac{\overrightarrow{A}\left({x}_{j}\right)+\overrightarrow{A}\left({x}_{j-1}\right)}{2}
\end{equation}

\begin{equation}
\label{equation-tq-fpi-sum-tq-fpi-sum-lag_mass}\hyperlabel{equation-tq-fpi-sum-tq-fpi-sum-lag_mass}%
{L}_{j}^{m}\equiv -\frac{m}{2}
\end{equation}

The laboratory time functions as a kind of stepper. It is as if we were making a stop motion film, i.e. a Wallace and Gromit feature, with each click forward of 
 $ \tau $~ by 
 $ \epsilon $~ corresponding to an advance by a single frame.

Compare this Lagrangian to the discrete standard quantum theory Lagrangian:

\begin{equation}
\label{equation-tq-fpi-sum-tq-fpi-sum-b}\hyperlabel{equation-tq-fpi-sum-tq-fpi-sum-b}%
{\overline{L}}_{j}=\frac{m}{2}{\left(\frac{{\overrightarrow{x}}_{j}-{\overrightarrow{x}}_{j-1}}{\epsilon }\right)}^{2}-e\Phi \left({x}_{j}\right)+e\frac{{\overrightarrow{x}}_{j}-{\overrightarrow{x}}_{j-1}}{\epsilon }\cdot \frac{\overrightarrow{A}\left({x}_{j}\right)+\overrightarrow{A}\left({x}_{j-1}\right)}{2}
\end{equation}

There are three changes:
\begin{enumerate}

\item{}With temporal quantization, the usual potential term becomes more complex:

\begin{equation}
\label{equation-tq-fpi-sum-tq-fpi-sum-c}\hyperlabel{equation-tq-fpi-sum-tq-fpi-sum-c}%
e\Phi \left({x}_{j}\right)\rightarrow e\frac{{t}_{j}-{t}_{j-1}}{\epsilon }\frac{\Phi \left({x}_{j}\right)+\Phi \left({x}_{j-1}\right)}{2}
\end{equation}

We now multiply the potential by the velocity of 
 $ t$~ with respect to 
 $ \tau $. In the non-{}relativistic case (see below in 
 \hyperlink{nonrel}{Non-{}relativistic Limit}), this factor will turn into approximately one, giving the usual result back.

\item{}The time velocity squared term is completely new. We may think of this term as a kinetic energy in time: identical in form to the usual space term, but opposite in sign.

\item{}The mass term, our additive constant, is less interesting. It is needed to make sure we get the right Schrödinger equation, see below in 
 \hyperlink{tq-seqn-deriv}{Derivation of the Schrödinger Equation}. It can be gauged out of existence, see further below in 
 \hyperlink{tq-seqn-gauge}{Gauge Transformations for the Schrödinger Equation}.

\end{enumerate}

\paragraph*{Midpoint Rule}

\noindent

We are using the midpoint rule: we evaluate the potential at the midpoint between the end times for a slice:

\begin{equation}
\label{equation-tq-fpi-sum-tq-fpi-midpoint-0}\hyperlabel{equation-tq-fpi-sum-tq-fpi-midpoint-0}%
\Phi \left({x}_{j}\right)\rightarrow \frac{\Phi \left({x}_{j}\right)+\Phi \left({x}_{j-1}\right)}{2}
\end{equation}

This is already required for evaluations of the vector potential:

\begin{equation}
\label{equation-tq-fpi-sum-tq-fpi-midpoint-1}\hyperlabel{equation-tq-fpi-sum-tq-fpi-midpoint-1}%
\overrightarrow{A}\left({x}_{j}\right)\rightarrow \frac{\overrightarrow{A}\left({x}_{j}\right)+\overrightarrow{A}\left({x}_{j-1}\right)}{2}
\end{equation}

Schulman points out that failure to use the midpoint rule for the vector potential causes spurious terms to appear in the Schrödinger equation. Our principle of the most complete symmetry
    between space and time therefore mandates use of the midpoint rule for the electric potential as well.

\subsection{Convergence}\label{tq-fpi-conv}\hyperlabel{tq-fpi-conv}%

\paragraph*{Convergence Problems With Path Integrals}

\noindent

Path integrals usually involve long series of Gaussian integrals, of the general form
 \footnote{See any of the path integral references cited earlier, or any quantum electrodynamics text that deals with path integrals (e.g. 
 \cite{Kaku-1993}, 
 \cite{Peskin-1995}, 
 \cite{Huang-1998},
 \cite{Zee-2003}).}:

\begin{equation}
\label{equation-tq-fpi-conv-tq-fpi-conv-0}\hyperlabel{equation-tq-fpi-conv-tq-fpi-conv-0}%
{\displaystyle \underset{-\infty }{\overset{\infty }{\int }}\text{d}t\mathrm{exp}\left(-a{t}^{2}+bt\right)}=\sqrt{\frac{\pi }{a}}\mathrm{exp}\left(\frac{{b}^{2}}{4a}\right)
\end{equation}

For these to converge, the real part of 
 \emph{a}~ should be greater than zero. However, if we are using a plane wave decomposition of the initial wave function, it is exactly zero. Unfortunate.

The traditional response to this problem is to add a small positive real part to 
 \emph{a}, then let the small real part go to zero. In our path integrals we have integrals like:

\begin{equation}
\label{equation-tq-fpi-conv-tq-fpi-conv-1}\hyperlabel{equation-tq-fpi-conv-tq-fpi-conv-1}%
{\displaystyle \underset{-\infty }{\overset{\infty }{\int }}\text{d}{t}_{j}\mathrm{exp}\left(-im\frac{{\left({t}_{j}-{t}_{j-1}\right)}^{2}}{2\Delta \tau }\right)}
\end{equation}

And we may add this small real part to either the mass:

\begin{equation}
\label{equation-tq-fpi-conv-tq-fpi-conv-1b}\hyperlabel{equation-tq-fpi-conv-tq-fpi-conv-1b}%
m\rightarrow m-i\epsilon 
\end{equation}

Or the time:

\begin{equation}
\label{equation-tq-fpi-conv-tq-fpi-conv-1c}\hyperlabel{equation-tq-fpi-conv-tq-fpi-conv-1c}%
\Delta \tau \rightarrow \Delta \tau +i\epsilon 
\end{equation}

There is the obvious question: where did that come from? The candid answer is that it is a magic convergence factor added to make things come out.

Unfortunately, with temporal quantization the magic fails. Look at a single step in the path integral:

\begin{equation}
\label{equation-tq-fpi-conv-tq-fpi-conv-2}\hyperlabel{equation-tq-fpi-conv-tq-fpi-conv-2}%
{\displaystyle \underset{-\infty ,-\infty }{\overset{\infty ,\infty }{\int }}\text{d}{t}_{j}\text{d}{x}_{j}\mathrm{exp}\left(-im\frac{{\left({t}_{j}-{t}_{j-1}\right)}^{2}}{2\Delta \tau }+im\frac{{\left({x}_{j}-{x}_{j-1}\right)}^{2}}{2\Delta \tau }\right)}
\end{equation}

No matter which sign we choose for the small real part, it cannot be the same sign for both time and space.

And if we choose different signs for time and space, then we lose manifest symmetry between time and space.

An alternative response is to Wick rotate in time, shifting to an imaginary time coordinate. This has the same problem: no matter in which sense we Wick rotate, either the past or the
    future integral will be infinite.

\paragraph*{Resolution Via Wavelets}

\noindent

However while the magic fails, if we use Morlet wavelet decomposition we see the magic is not needed in the first place. We consider several points in turn:
\begin{enumerate}

\item{}If we limit our wave functions to square integrable functions \textendash{} which includes all physically meaningful wave functions \textendash{} then by using Morlet wavelet decomposition we may write any
            allowed wave function as a sum of Gaussian test functions.

\item{}If we then integrate from our starting slice forward, each integral will in turn be well-{}defined. The convergence comes naturally from the wave function, not from the kernel.

\item{}We have then no need of artificial means to ensure convergence: it is a natural consequence of restricting our examination to physically meaningful wave functions.

\item{}To be sure, the integrals we have to do are more complex than usual: we have to do each integral with respect to a specific incoming wave function.

\item{}And, we will need to pay careful attention to how we normalize the path integrals: the normalization could depend on the specific initial wave function.

\end{enumerate}

As we will see, the dispersion of the wave function gets larger with time on a per Gaussian test function basis:

\begin{equation}
\label{equation-tq-fpi-conv-tq-fpi-conv-3}\hyperlabel{equation-tq-fpi-conv-tq-fpi-conv-3}%
\langle {\sigma }_{\tau }^{2}\rangle \sim {\sigma }_{0}^{2}\frac{1}{1+\frac{{\tau }^{2}}{{m}^{2}{\sigma }_{0}^{4}}}
\end{equation}

So the rate of convergence is different for each Gaussian test function. At each step it is a function of the initial $\sigma$ and the laboratory time.

However, once the Gaussian test function is picked, convergence is assured.

This means different parts of a general wave function will converge to the final result at different rates.

So long as they do converge, this does not matter.

By relying on Morlet wavelet decomposition, we have avoided magic at the cost of trading unconditional convergence for conditional convergence.

\subsection{Normalization}\label{tq-fpi-norm}\hyperlabel{tq-fpi-norm}%

\paragraph*{Definition of Normalization}

\noindent

In the usual development of path integrals, the normalization is inherited from the Schrödinger equation. Here it has to be supplied by the path integrals themselves.

We start with a Gaussian test function:

\begin{equation}
\label{equation-tq-fpi-norm-free-fourt-phi0}\hyperlabel{equation-tq-fpi-norm-free-fourt-phi0}%
{\psi }_{0}\left(x\right)=\sqrt[4]{\frac{1}{{\pi }^{4}\mathrm{det}\left({\Sigma }_{0}\right)}}\mathrm{exp}\left(-i{p}_{\mu }^{\left(0\right)}{x}^{\mu }-\frac{1}{2{\Sigma }_{0}^{\mu \nu }}\left({x}^{\mu }-{\overline{x}}_{0}^{\mu }\right)\left({x}^{n}-{\overline{x}}_{0}^{\nu }\right)\right)
\end{equation}

The wave function at the end is given by an integral over the kernel and the initial wave function:

\begin{equation}
\label{equation-tq-fpi-norm-free-fourt-phipush}\hyperlabel{equation-tq-fpi-norm-free-fourt-phipush}%
{\psi }_{\tau }\left({x}^{{''}}\right)={\displaystyle \int {\text{d}}^{\text{4}}{x}^{\prime }{K}_{\tau }\left({x}^{{''}};{x}^{\prime }\right){\psi }_{0}\left({x}^{\prime }\right)}
\end{equation}

The kernel is correctly normalized if we have:

\begin{equation}
\label{equation-tq-fpi-norm-tq-fpi-norm-2}\hyperlabel{equation-tq-fpi-norm-tq-fpi-norm-2}%
1={\displaystyle \int \text{d}x{\left\vert {\psi }_{\tau }\left(x\right)\right\vert }^{2}}
\end{equation}

We define the unnormalized or raw kernel as the kernel we get from a straight computation of the path integral, with no normalization. The full kernel is given by the raw kernel times a
    normalization factor 
 $ {C}_{\tau }$:

\begin{equation}
\label{equation-tq-fpi-norm-tq-fpi-norm-3}\hyperlabel{equation-tq-fpi-norm-tq-fpi-norm-3}%
{K}_{\tau }\left({x}^{{''}};{x}^{\prime }\right)={C}_{\tau }{K}_{\tau }^{\left(raw\right)}\left({x}^{{''}};{x}^{\prime }\right)
\end{equation}

The normalization factor is:

\begin{equation}
\label{equation-tq-fpi-norm-tq-fpi-norm-4}\hyperlabel{equation-tq-fpi-norm-tq-fpi-norm-4}%
{C}_{\tau }=\frac{1}{\sqrt{{{\displaystyle \int \text{d}{x}^{{''}}\left\vert {\displaystyle \int \text{d}{x}^{\prime }}{K}_{\tau }^{\left(raw\right)}\left({x}^{{''}};{x}^{\prime }\right){\psi }_{0}\left({x}^{\prime }\right)\right\vert }}^{2}}}
\end{equation}

Obviously we are free to add an overall phase at each step; thereby creating a gauge degree of freedom, see below in 
 \hyperlink{tq-seqn-gauge}{Gauge Transformations for the Schrödinger Equation}~ and further below in 
 \hyperlink{semi}{Semi-{}classical Limit}  \footnote{A similar gauge degree of freedom shows up in discussions of fifth parameter formalisms, as cited in 
 \hyperlink{int-lit}{Literature}~ above.}.

\paragraph*{Normalization in Time}

\noindent

Here we look at the free case only. Later we will complete the analysis by using the Schrödinger equation to demonstrate unitarity (in 
 \hyperlink{tq-seqn-unitarity}{Unitarity}), implying the normalization is correct in the general case.

We separate variables in time and space. We work first with the time part, then generalize to all four dimensions.

We start with a Gaussian test function:

\begin{equation}
\label{equation-tq-fpi-norm-free-onet-chi0}\hyperlabel{equation-tq-fpi-norm-free-onet-chi0}%
{\chi }_{0}\left(t\right)=\sqrt[4]{\frac{1}{\pi {\sigma }_{0}^{2}}}\mathrm{exp}\left(-i{E}_{0}t-\frac{1}{2{\sigma }_{0}^{2}}{\left(t-{\overline{t}}_{0}\right)}^{2}\right)
\end{equation}

And write the kernel as:

\begin{equation}
\label{equation-tq-fpi-norm-tq-fpi-norm-t-kernel_raw_def}\hyperlabel{equation-tq-fpi-norm-tq-fpi-norm-t-kernel_raw_def}%
{\stackrel{\frown }{K}}_{\tau }^{\left(raw\right)}\left({t}_{N+1};{t}_{0}\right)={\displaystyle \int \text{d}{t}_{1}\text{d}{t}_{2}\dots \text{d}{t}_{N}}\mathrm{exp}\left(-i{\displaystyle \sum _{j=1}^{N+1}\left(\frac{m}{2\epsilon }{\left({t}_{j}-{t}_{j-1}\right)}^{2}+\frac{m\epsilon }{2}\right)}\right)
\end{equation}

The wave function after the first step is:

\begin{equation}
\label{equation-tq-fpi-norm-tq-fpi-norm-t-2}\hyperlabel{equation-tq-fpi-norm-tq-fpi-norm-t-2}%
{\chi }_{\tau }\left({t}_{1}\right)={\displaystyle \int \text{d}{t}_{0}}\mathrm{exp}\left(-i\frac{m}{2\epsilon }{\left({t}_{1}-{t}_{0}\right)}^{2}-i\frac{m}{2}\epsilon \right)\chi \left({t}_{0}\right)
\end{equation}

Which gives:

\begin{equation}
\label{equation-tq-fpi-norm-tq-fpi-norm-t-3}\hyperlabel{equation-tq-fpi-norm-tq-fpi-norm-t-3}%
{\chi }_{\epsilon }\left({t}_{1}\right)=\sqrt{\frac{2\pi \epsilon }{im}}\sqrt[4]{\frac{1}{\pi {\sigma }_{0}^{2}}}\sqrt{\frac{1}{{f}_{\epsilon }}}\mathrm{exp}\left(-i{E}_{0}{t}_{1}+i\frac{{E}_{0}^{2}}{2m}\epsilon -\frac{1}{2{\sigma }_{0}^{2}{f}_{\epsilon }}{\left({t}_{1}-{\overline{t}}_{0}-\frac{{E}_{0}}{m}\epsilon \right)}^{2}-i\frac{m}{2}\epsilon \right)
\end{equation}

With the definition of 
 $ {f}_{\tau }^{\left(0\right)}$:

\begin{equation}
\label{equation-tq-fpi-norm-free-onet-ftau}\hyperlabel{equation-tq-fpi-norm-free-onet-ftau}%
{f}_{\tau }^{\left(0\right)}\equiv 1-i\frac{\tau }{m{\sigma }_{0}^{2}}
\end{equation}

The normalization requirement is:

\begin{equation}
\label{equation-tq-fpi-norm-tq-fpi-norm-t-norm}\hyperlabel{equation-tq-fpi-norm-tq-fpi-norm-t-norm}%
1={\displaystyle \int \text{d}{t}_{1}}{\chi }_{\epsilon }^{\ast }\left({t}_{1}\right){\chi }_{\epsilon }\left({t}_{1}\right)
\end{equation}

The first step normalization is correct if we multiply the kernel by a factor of:

\begin{equation}
\label{equation-tq-fpi-norm-tq-fpi-norm-t-5}\hyperlabel{equation-tq-fpi-norm-tq-fpi-norm-t-5}%
\sqrt{\frac{im}{2\pi \epsilon }}
\end{equation}

Since this normalization factor does not depend on the laboratory time the overall normalization for 
 \emph{N}~ + 1 infinitesimal kernels is the product of 
 \emph{N}~ + 1 of these factors:

\begin{equation}
\label{equation-tq-fpi-norm-tq-fpi-norm-t-6}\hyperlabel{equation-tq-fpi-norm-tq-fpi-norm-t-6}%
{C}_{N}\equiv {\sqrt{\frac{im}{2\pi \epsilon }}}^{N+1}
\end{equation}

As noted, the phase is arbitrary. If we were working the other way, from Schrödinger equation to path integral, the phase would be determined by the Schrödinger equation itself. The
    specific phase choice we are making here has been chosen to ensure the four dimensional Schrödinger equation is manifestly covariant, see 
 \hyperlink{tq-seqn-deriv}{Derivation of the Schrödinger Equation}.

Therefore the expression for the free kernel in time is:

\begin{equation}
\label{equation-tq-fpi-norm-tq-fpi-norm-t-7}\hyperlabel{equation-tq-fpi-norm-tq-fpi-norm-t-7}%
{\stackrel{\frown }{K}}_{\tau }^{\left(0\right)}\left({t}^{{''}};{t}^{\prime }\right)={\displaystyle \int \text{d}{t}_{1}\text{d}{t}_{2}\dots \text{d}{t}_{N}}{\sqrt{\frac{im}{2\pi \epsilon }}}^{N+1}\mathrm{exp}\left(-i{\displaystyle \sum _{j=1}^{N+1}\left(\frac{m}{2\epsilon }{\left({t}_{j}-{t}_{j-1}\right)}^{2}+\frac{m}{2}\epsilon \right)}\right)
\end{equation}

Doing the integrals gives for the kernel:

\begin{equation}
\label{equation-tq-fpi-norm-free-onet-kernel}\hyperlabel{equation-tq-fpi-norm-free-onet-kernel}%
{\stackrel{\frown }{K}}_{\tau }\left({t}^{{''}};{t}^{\prime }\right)=\sqrt{\frac{im}{2\pi \tau }}\mathrm{exp}\left(-im\frac{{\left({t}^{{''}}-{t}^{\prime }\right)}^{2}}{2\tau }-im\frac{\tau }{2}\right)
\end{equation}

Applying the kernel to the initial wave function gives:

\begin{equation}
\label{equation-tq-fpi-norm-free-onet-chitq}\hyperlabel{equation-tq-fpi-norm-free-onet-chitq}%
{\chi }_{\tau }\left(t\right)=\sqrt[4]{\frac{1}{\pi {\sigma }_{0}^{2}}}\sqrt{\frac{1}{{f}_{\tau }^{\left(0\right)}}}\mathrm{exp}\left(-i{E}_{0}t-\frac{1}{2{\sigma }_{0}^{2}{f}_{\tau }^{\left(0\right)}}{\left(t-{\overline{t}}_{\tau }\right)}^{2}+i\frac{{E}_{0}^{2}-{m}^{2}}{2m}\tau \right)
\end{equation}

With this normalization we have for the probability distribution in time:

\begin{equation}
\label{equation-tq-fpi-norm-free-onet-probtau}\hyperlabel{equation-tq-fpi-norm-free-onet-probtau}%
{\left\vert {\chi }_{\tau }\left(t\right)\right\vert }^{2}=\sqrt{\frac{1}{\pi {\sigma }_{0}^{2}{\left\vert {f}_{\tau }^{\left(0\right)}\right\vert }^{2}}}\mathrm{exp}\left(-\frac{1}{{\sigma }_{0}^{2}{\left\vert {f}_{\tau }^{\left(0\right)}\right\vert }^{2}}{\left(t-{\overline{t}}_{0}-\frac{{E}_{0}}{m}\tau \right)}^{2}\right)
\end{equation}

With the expectation for 
 $ t$:

\begin{equation}
\label{equation-tq-fpi-norm-free-onet-average}\hyperlabel{equation-tq-fpi-norm-free-onet-average}%
{\overline{t}}_{\tau }={\overline{t}}_{0}+\frac{{E}_{0}}{m}\tau 
\end{equation}

Implying a velocity for quantum time with respect to laboratory time:

\begin{equation}
\label{equation-tq-fpi-norm-tq-fpi-norm-t-c}\hyperlabel{equation-tq-fpi-norm-tq-fpi-norm-t-c}%
{v}_{0}=\frac{{E}_{0}}{m}=\gamma \equiv \frac{1}{\sqrt{1-{\overrightarrow{v}}^{2}}}
\end{equation}

The uncertainty is given by:

\begin{equation}
\label{equation-tq-fpi-norm-free-onet-dispersion}\hyperlabel{equation-tq-fpi-norm-free-onet-dispersion}%
\langle {\left(t-{\overline{t}}_{\tau }\right)}^{2}\rangle =\frac{{\sigma }_{0}^{2}{\left\vert {f}_{\tau }^{\left(0\right)}\right\vert }^{2}}{2}=\frac{{\sigma }_{0}^{2}}{2}\left(1+\frac{{\tau }^{2}}{{m}^{2}{\sigma }_{0}^{4}}\right)
\end{equation}

We have done the analysis for an arbitrary Gaussian test function; we recall that any square-{}integrable function will be a sum over these.

The most important thing about the normalization is what we do not see in it: it is not a function of the frequency, dispersion, or offset of the Gaussian test function. Since any
    square-{}integrable wave function may be built up of sums of Gaussian test functions, the normalization \textendash{} at least for the free case \textendash{} is independent of the wave function.

\paragraph*{Normalization in Space}

\noindent

We recapitulate the analysis for time in space. We use the correspondences:

\begin{equation}
\label{equation-tq-fpi-norm-tq-fpi-norm-x-2}\hyperlabel{equation-tq-fpi-norm-tq-fpi-norm-x-2}%
t\rightarrow x,m\rightarrow -m,{\overline{t}}_{0}\rightarrow {\overline{x}}_{0},{E}_{0}\rightarrow -{p}_{0}^{\left(x\right)},{\sigma }_{0}^{2}\rightarrow {\sigma }_{1}^{2}
\end{equation}

With these we can write down the equivalent set of results by inspection.

Initial Gaussian test function:

\begin{equation}
\label{equation-tq-fpi-norm-tq-fpi-norm-x-0}\hyperlabel{equation-tq-fpi-norm-tq-fpi-norm-x-0}%
\xi \left({x}_{0}\right)=\sqrt[4]{\frac{1}{\pi {\sigma }_{1}^{2}}}\mathrm{exp}\left(i{p}_{0}^{\left(x\right)}{x}_{0}-\frac{{\left({x}_{0}-{\overline{x}}_{0}\right)}^{2}}{2{\sigma }_{1}^{2}}\right)
\end{equation}

Free kernel:

\begin{equation}
\label{equation-tq-fpi-norm-tq-fpi-norm-x-1}\hyperlabel{equation-tq-fpi-norm-tq-fpi-norm-x-1}%
{\overline{K}}_{\tau }^{\left(0\right)}\left({x}^{{''}};{x}^{\prime }\right)={\displaystyle \int \text{d}{x}_{1}\text{d}{x}_{2}\dots \text{d}{x}_{N}}\mathrm{exp}\left(i{\displaystyle \sum _{j=1}^{N+1}\frac{m}{2\epsilon }{\left({x}_{j}-{x}_{j-1}\right)}^{2}}\right)
\end{equation}

We choose the phase:

\begin{equation}
\label{equation-tq-fpi-norm-tq-fpi-norm-x-3}\hyperlabel{equation-tq-fpi-norm-tq-fpi-norm-x-3}%
\sqrt{\frac{2i\pi \epsilon }{m}}\rightarrow \sqrt{\frac{2\pi \epsilon }{im}}
\end{equation}

So the kernel matches the usual standard quantum theory kernel, the familiar (\cite{Feynman-1965d},
 \cite{Schulman-1981}):

\begin{equation}
\label{equation-tq-fpi-norm-tq-fpi-norm-x-4}\hyperlabel{equation-tq-fpi-norm-tq-fpi-norm-x-4}%
{\overline{K}}_{\tau }^{\left(0\right)}\left({x}^{{''}};{x}^{\prime }\right)=\sqrt{-\frac{im}{2\pi \tau }}\mathrm{exp}\left(im\frac{{\left({x}^{{''}}-{x}^{\prime }\right)}^{2}}{2\tau }\right)
\end{equation}

The wave function as a function of laboratory time is therefore:

\begin{equation}
\label{equation-tq-fpi-norm-tq-fpi-norm-x-5}\hyperlabel{equation-tq-fpi-norm-tq-fpi-norm-x-5}%
{\xi }_{\tau }\left(x\right)=\sqrt[4]{\frac{1}{\pi {\sigma }_{1}^{2}}}\sqrt{\frac{1}{{f}_{\tau }^{\left(1\right)}}}\mathrm{exp}\left(i{p}_{0}^{\left(x\right)}x+i\frac{{\left({p}_{0}^{\left(x\right)}\right)}^{2}}{2m}\tau -\frac{1}{2{\sigma }_{1}^{2}{f}_{\tau }^{\left(1\right)}}{\left(x-{\overline{x}}_{0}-\frac{{p}_{0}^{\left(x\right)}}{m}\tau \right)}^{2}\right)
\end{equation}

With definition of 
 $ {f}_{\tau }^{\left(1\right)}$:

\begin{equation}
\label{equation-tq-fpi-norm-tq-fpi-norm-x-6}\hyperlabel{equation-tq-fpi-norm-tq-fpi-norm-x-6}%
{f}_{\tau }^{\left(1\right)}\equiv 1+i\frac{\tau }{m{\sigma }_{1}^{2}}
\end{equation}

Probability distribution:

\begin{equation}
\label{equation-tq-fpi-norm-tq-fpi-norm-x-7}\hyperlabel{equation-tq-fpi-norm-tq-fpi-norm-x-7}%
{\left\vert {\xi }_{\tau }\left(x\right)\right\vert }^{2}=\sqrt{\frac{1}{\pi {\sigma }_{1}^{2}{\left\vert {f}_{\tau }^{\left(1\right)}\right\vert }^{2}}}\mathrm{exp}\left(-\frac{1}{{\sigma }_{1}^{2}{\left\vert {f}_{\tau }^{\left(1\right)}\right\vert }^{2}}{\left(x-{\overline{x}}_{0}-\frac{{p}_{0}^{\left(x\right)}}{m}\tau \right)}^{2}\right)
\end{equation}

With expectation of position:

\begin{equation}
\label{equation-tq-fpi-norm-tq-fpi-norm-x-8}\hyperlabel{equation-tq-fpi-norm-tq-fpi-norm-x-8}%
{\overline{x}}_{\tau }={\overline{x}}_{0}+\frac{{p}_{0}^{\left(x\right)}}{m}\tau 
\end{equation}

Implying a velocity with respect to laboratory time:

\begin{equation}
\label{equation-tq-fpi-norm-tq-fpi-norm-x-9}\hyperlabel{equation-tq-fpi-norm-tq-fpi-norm-x-9}%
{v}_{0}^{\left(x\right)}=\frac{{p}_{0}^{\left(x\right)}}{m}
\end{equation}

And uncertainty in position:

\begin{equation}
\label{equation-tq-fpi-norm-tq-fpi-norm-x-dispersionx}\hyperlabel{equation-tq-fpi-norm-tq-fpi-norm-x-dispersionx}%
\langle {\left(x-{\overline{x}}_{\tau }\right)}^{2}\rangle =\frac{{\sigma }_{1}^{2}}{2}\left\vert 1+\frac{{\tau }^{2}}{{m}^{2}{\sigma }_{1}^{4}}\right\vert 
\end{equation}

The three-{}space kernel is the product of 
 \emph{x}, 
 \emph{y}, and 
 \emph{z}~ parts; it is in fact the usual non-{}relativistic free kernel:

\begin{equation}
\label{equation-tq-fpi-norm-free-fourt-kernel3}\hyperlabel{equation-tq-fpi-norm-free-fourt-kernel3}%
{\overline{K}}_{\tau }\left(\overrightarrow{{x}^{{''}}};\overrightarrow{{x}^{\prime }}\right)={\sqrt{\frac{m}{2\pi i\tau }}}^{3}\mathrm{exp}\left(\frac{im}{2\tau }{\left(\overrightarrow{{x}^{{''}}}-\overrightarrow{{x}^{\prime }}\right)}^{2}\right)
\end{equation}

This confirms our choice of scale for the Lagrangian. If we had multiplied the Lagrangian by a scale 
 \emph{s}, we would have gotten a different kernel
 \footnote{Another way of saying this is that the scale is fixed by the value of 
 $ \hslash $.}:

\begin{equation}
\label{equation-tq-fpi-norm-tq-fpi-norm-x-scale}\hyperlabel{equation-tq-fpi-norm-tq-fpi-norm-x-scale}%
L\rightarrow sL\Rightarrow {\overline{K}}_{\tau }\left(\overrightarrow{{x}^{{''}}};\overrightarrow{{x}^{\prime }}\right)\rightarrow {\sqrt{\frac{ms}{2\pi i\tau }}}^{3}\mathrm{exp}\left(\frac{im}{2\tau }{\left(\overrightarrow{{x}^{{''}}}-\overrightarrow{{x}^{\prime }}\right)}^{2}s\right)
\end{equation}

\paragraph*{Normalization in Time and Space}

\noindent

The full kernel is the product of the time and the three-{}space kernels:

\begin{equation}
\label{equation-tq-fpi-norm-free-fourt-kernel}\hyperlabel{equation-tq-fpi-norm-free-fourt-kernel}%
{K}_{\tau }\left({x}^{{''}};{x}^{\prime }\right)=-\frac{i{m}^{2}}{4{\pi }^{2}{\tau }^{2}}\mathrm{exp}\left(-\frac{im}{2\tau }{\left({x}^{{''}}-{x}^{\prime }\right)}^{2}-i\frac{m}{2}\tau \right)
\end{equation}

Full wave function as a function of laboratory time:

\begin{equation}
\label{equation-tq-fpi-norm-free-fourt-phitau}\hyperlabel{equation-tq-fpi-norm-free-fourt-phitau}%
{\psi }_{\tau }\left(x\right)=\sqrt[4]{\frac{\mathrm{det}\left({\Sigma }_{0}^{\mu \nu }\right)}{{\pi }^{4}}}\sqrt{\frac{1}{\mathrm{det}\left({\Sigma }_{\tau }^{\mu \nu }\right)}}\mathrm{exp}\left(-i{p}_{0}^{\mu }{x}_{\mu }-\frac{1}{2{\Sigma }_{\tau }^{\mu \nu }}\left({x}^{\mu }-{\overline{x}}_{\tau }^{\mu }\right)\left({x}^{\nu }-{\overline{x}}_{\tau }^{\nu }\right)+i\frac{{p}_{0}^{2}-{m}^{2}}{2m}\tau \right)
\end{equation}

With expectation of coordinates:

\begin{equation}
\label{equation-tq-fpi-norm-free-fourt-average}\hyperlabel{equation-tq-fpi-norm-free-fourt-average}%
{\overline{x}}_{\tau }^{\mu }\equiv {\langle {x}^{\mu }\rangle }_{\tau }={\overline{x}}_{0}^{\mu }+\frac{{p}_{0}^{\mu }}{m}\tau 
\end{equation}

And dispersion matrix:

\begin{equation}
\label{equation-tq-fpi-norm-free-fourt-sigmatau}\hyperlabel{equation-tq-fpi-norm-free-fourt-sigmatau}%
{\Sigma }_{\tau }^{\mu \nu }=\left(\begin{array}{cccc}{\sigma }_{0}^{2}-i\frac{\tau }{m}& 0& 0& 0\\ 0& {\sigma }_{1}^{2}+i\frac{\tau }{m}& 0& 0\\ 0& 0& {\sigma }_{2}^{2}+i\frac{\tau }{m}& 0\\ 0& 0& 0& {\sigma }_{3}^{2}+i\frac{\tau }{m}\end{array}\right)=\left(\begin{array}{cccc}{\sigma }_{0}^{2}{f}_{\tau }^{\left(0\right)}& 0& 0& 0\\ 0& {\sigma }_{1}^{2}{f}_{\tau }^{\left(1\right)}& 0& 0\\ 0& 0& {\sigma }_{2}^{2}{f}_{\tau }^{\left(2\right)}& 0\\ 0& 0& 0& {\sigma }_{3}^{2}{f}_{\tau }^{\left(3\right)}\end{array}\right)
\end{equation}

We give a summary of the free Gaussian test functions and kernels in coordinate and momentum space and in block time and relative time in the appendix 
 \hyperlink{free}{Free Particles}.

\subsection{Formal Expression for the Path Integral}\label{tq-fpi-final}\hyperlabel{tq-fpi-final}%

The full kernel is therefore:

\begin{equation}
\label{equation-tq-fpi-final-tq-fpi-final-fullkernel}\hyperlabel{equation-tq-fpi-final-tq-fpi-final-fullkernel}%
{K}_{\tau }\left({x}^{{''}};{x}^{\prime }\right)={\displaystyle \int \mathcal{D}x\mathrm{exp}\left(-i{\displaystyle \sum _{j=1}^{N+1}m\frac{{\left({x}_{j}-{x}_{j-1}\right)}^{2}}{2\epsilon }}-ie\left({x}_{j}-{x}_{j-1}\right)\frac{A\left({x}_{j}\right)+A\left({x}_{j-1}\right)}{2}-i\frac{m}{2}\epsilon \right)}
\end{equation}

With the definition of the measure:

\begin{equation}
\label{equation-tq-fpi-final-tq-fpi-final-measure}\hyperlabel{equation-tq-fpi-final-tq-fpi-final-measure}%
\mathcal{D}x\equiv {\left(-\frac{i{m}^{2}}{4{\pi }^{2}{\epsilon }^{2}}\right)}^{N+1}{\displaystyle \prod _{n=1}^{n=N}{\text{d}}^{\text{4}}{x}_{n}}
\end{equation}

This was derived for an arbitrary Gaussian test function, but by Morlet wavelet decomposition is valid for an arbitrary square-{}integrable wave function.

As of this point in the argument, we have only verified the normalization for the free case; we will verify it more generally below, in 
 \hyperlink{tq-seqn-unitarity}{Unitarity}.

\section{Schrödinger Equation}\label{tq-seqn}\hyperlabel{tq-seqn}%

\subsection{Derivation of the Schrödinger Equation}\label{tq-seqn-deriv}\hyperlabel{tq-seqn-deriv}%
\begin{quote}

The next few steps involve a small nightmare of Taylor expansions and Gaussian integrals.

\hspace*\fill---~L. S. Schulman
 \cite{Schulman-1981}\end{quote}

Schulman (\cite{Schulman-1981}) has derived the Schrödinger equation from the path integral; we generalize his derivation to include quantum time.

We start with the sliced form of the path integral. We consider a single step.

Only terms first order in 
 $ \epsilon $~ appear in the limit as 
 \emph{N}~ goes to infinity. We define the coordinate difference:

\begin{equation}
\label{equation-tq-seqn-deriv-tq-seqn-expand-0}\hyperlabel{equation-tq-seqn-deriv-tq-seqn-expand-0}%
\xi \equiv {x}_{j}-{x}_{j+1}
\end{equation}

Giving:

\begin{equation}
\label{equation-tq-seqn-deriv-tq-seqn-expand-1}\hyperlabel{equation-tq-seqn-deriv-tq-seqn-expand-1}%
{A}_{\nu }\left({x}_{j}\right)={A}_{\nu }\left({x}_{j+1}\right)+\left({\xi }^{\mu }{\partial }_{\mu }\right){A}_{\nu }\left({x}_{j+1}\right)+\dots 
\end{equation}

\begin{equation}
\label{equation-tq-seqn-deriv-tq-seqn-expand-2}\hyperlabel{equation-tq-seqn-deriv-tq-seqn-expand-2}%
{\psi }_{\tau }\left({x}_{j}\right)={\psi }_{\tau }\left({x}_{j+1}\right)+\left({\xi }^{\mu }{\partial }_{\mu }\right){\psi }_{\tau }\left({x}_{j+1}\right)+\frac{1}{2}{\xi }^{\mu }{\xi }^{\nu }{\partial }_{\mu }{\partial }_{\nu }{\psi }_{\tau }\left({x}_{j+1}\right)+\dots 
\end{equation}

For one step we have:

\begin{equation}
\label{equation-tq-seqn-deriv-tq-seqn-deriv-1}\hyperlabel{equation-tq-seqn-deriv-tq-seqn-deriv-1}%
\begin{array}{c}{\psi }_{\tau +\epsilon }\left({x}_{j+1}\right)=\sqrt{\frac{im}{2\pi \epsilon }}{\sqrt{-\frac{im}{2\pi \epsilon }}}^{3}{\displaystyle \int {\text{d}}^{4}\xi \mathrm{exp}\left(-\frac{im{\xi }^{2}}{2\epsilon }-i\frac{m}{2}\epsilon \right)}\\ \times \mathrm{exp}\left(ie{\xi }^{\nu }\left({A}_{\nu }\left({x}_{j+1}\right)+\frac{1}{2}\left({\xi }^{\mu }{\partial }_{\mu }\right){A}_{\nu }\left({x}_{j+1}\right)+\dots \right)\right)\\ \times \left({\psi }_{\tau }\left({x}_{j+1}\right)+\left({\xi }^{\mu }{\partial }_{\mu }\right){\psi }_{\tau }\left({x}_{j+1}\right)+\frac{1}{2}{\xi }^{\mu }{\xi }^{\nu }{\partial }_{\mu }{\partial }_{\nu }{\psi }_{\tau }\left({x}_{j+1}\right)+\dots \right)\end{array}
\end{equation}

Or using 
 $ \xi \sim \sqrt{\epsilon }$:

\begin{equation}
\label{equation-tq-seqn-deriv-tq-seqn-deriv-2}\hyperlabel{equation-tq-seqn-deriv-tq-seqn-deriv-2}%
\begin{array}{c}{\psi }_{\tau +\epsilon }\left({x}_{j+1}\right)=\sqrt{\frac{im}{2\pi \epsilon }}{\sqrt{-\frac{im}{2\pi \epsilon }}}^{3}{\displaystyle \int {\text{d}}^{4}\xi \mathrm{exp}\left(-\frac{im{\xi }^{2}}{2\epsilon }\right)}\\ \times \left(1-i\frac{m\epsilon }{2}+ie{\xi }^{\nu }{A}_{\nu }\left({x}_{j+1}\right)+\frac{ie}{2}{\xi }^{\mu }{\xi }^{\nu }{\partial }_{\mu }{A}_{\nu }\left({x}_{j+1}\right)-\frac{{e}^{2}}{2}{\xi }^{\mu }{A}_{\mu }\left({x}_{j}\right){\xi }^{\nu }{A}_{\nu }\left({x}_{j+1}\right)\right)\\ \times \left({\psi }_{\tau }\left({x}_{j+1}\right)+\left({\xi }^{\mu }{\partial }_{\mu }\right){\psi }_{\tau }\left({x}_{j+1}\right)+\frac{1}{2}{\xi }^{\mu }{\xi }^{\nu }{\partial }_{\mu }{\partial }_{\nu }{\psi }_{\tau }\left({x}_{j+1}\right)+\dots \right)\end{array}
\end{equation}

The term zeroth order in 
 $ \epsilon $~ gives:

\begin{equation}
\label{equation-tq-seqn-deriv-tq-seqn-deriv-3}\hyperlabel{equation-tq-seqn-deriv-tq-seqn-deriv-3}%
\sqrt{\frac{im}{2\pi \epsilon }}{\sqrt{-\frac{im}{2\pi \epsilon }}}^{3}{\displaystyle \int {\text{d}}^{4}\xi \mathrm{exp}\left(-\frac{im{\xi }^{2}}{2\epsilon }\right)}=\sqrt{\frac{im}{2\pi \epsilon }}{\sqrt{-\frac{im}{2\pi \epsilon }}}^{3}\sqrt{\frac{2\pi \epsilon }{im}}{\sqrt{\frac{2\pi \epsilon }{-im}}}^{3}=1
\end{equation}

Odd powers of $\xi$ give zero, off diagonal powers of order $\xi$ squared give zero. Diagonal $\xi$ squared terms give:

\begin{equation}
\label{equation-tq-seqn-deriv-tq-seqn-deriv-4}\hyperlabel{equation-tq-seqn-deriv-tq-seqn-deriv-4}%
\begin{array}{c}\sqrt{\frac{im}{2\pi \epsilon }}{\displaystyle \int \text{d}{\xi }_{0}\mathrm{exp}\left(-\frac{im{\xi }_{0}^{2}}{2\epsilon }\right)}{\xi }_{0}^{2}=\frac{\epsilon }{im}\\ \sqrt{-\frac{im}{2\pi \epsilon }}{\displaystyle \int \text{d}{\xi }_{i}\mathrm{exp}\left(\frac{im{\xi }_{i}^{2}}{2\epsilon }\right)}{\xi }_{i}^{2}=-\frac{\epsilon }{im}\end{array}
\end{equation}

The expression for the wave function is therefore:

\begin{equation}
\label{equation-tq-seqn-deriv-tq-seqn-deriv-5}\hyperlabel{equation-tq-seqn-deriv-tq-seqn-deriv-5}%
{\psi }_{\tau +\epsilon }\left({x}_{j+1}\right)={\psi }_{\tau }+\frac{e}{2m}\epsilon \left(\partial A\right){\psi }_{\tau }+\frac{i{e}^{2}\epsilon }{2m}{A}^{2}{\psi }_{\tau }-\frac{i\epsilon }{2m}{\partial }^{2}{\psi }_{\tau }+\frac{e\epsilon }{m}\left(A\partial \right){\psi }_{\tau }-\frac{im\epsilon }{2}{\psi }_{\tau }
\end{equation}

Taking the limit as 
 $ \epsilon $~ goes to zero, we get the four dimensional ~ Schrödinger equation:

\begin{equation}
\label{equation-tq-seqn-deriv-tq-seqn-deriv-6}\hyperlabel{equation-tq-seqn-deriv-tq-seqn-deriv-6}%
i\frac{d{\psi }_{\tau }}{d\tau }=\frac{1}{2m}{\partial }^{\mu }{\partial }_{\mu }{\psi }_{\tau }+\frac{ie}{m}\left({A}^{\mu }{\partial }_{\mu }\right){\psi }_{\tau }+\frac{ie}{2m}{\psi }_{\tau }\left({\partial }^{\mu }{A}_{\mu }\right)-\frac{{e}^{2}}{2m}{A}^{\mu }{A}_{\mu }{\psi }_{\tau }+\frac{m}{2}{\psi }_{\tau }
\end{equation}

Or:

\begin{equation}
\label{equation-tq-seqn-deriv-tq-seqn-coord-seqn}\hyperlabel{equation-tq-seqn-deriv-tq-seqn-coord-seqn}%
i\frac{d{\psi }_{\tau }}{d\tau }\left(t,\overrightarrow{x}\right)=\frac{1}{2m}\left({\left({\partial }_{t}+ie\Phi \left(t,\overrightarrow{x}\right)\right)}^{2}-{\left(\overrightarrow{\nabla }-ie\overrightarrow{A}\left(t,\overrightarrow{x}\right)\right)}^{2}+{m}^{2}\right){\psi }_{\tau }\left(t,\overrightarrow{x}\right)
\end{equation}

Or, in manifestly covariant form:

\begin{equation}
\label{equation-tq-seqn-deriv-tq-seqn-coord-1}\hyperlabel{equation-tq-seqn-deriv-tq-seqn-coord-1}%
i\frac{d{\psi }_{\tau }}{d\tau }\left(t,\overrightarrow{x}\right)=-\frac{1}{2m}\left(\left(i{\partial }_{\mu }-e{A}_{\mu }\left(t,\overrightarrow{x}\right)\right)\left(i{\partial }^{\mu }-e{A}^{\mu }\left(t,\overrightarrow{x}\right)\right)-{m}^{2}\right){\psi }_{\tau }\left(t,\overrightarrow{x}\right)
\end{equation}

If we make the customary identifications:

\begin{equation}
\label{equation-tq-seqn-deriv-tq-seqn-deriv-7b}\hyperlabel{equation-tq-seqn-deriv-tq-seqn-deriv-7b}%
i\frac{\partial }{\partial t}\rightarrow E,-i\overrightarrow{\nabla }\rightarrow \overrightarrow{p}\Rightarrow i{\partial }_{\mu }\rightarrow {p}_{\mu }
\end{equation}

We have
 \footnote{
We recover Feynman's Klein-{}Gordon equation cited in 
 \hyperlink{int-lit}{Literature}~ with the substitution 
 $ u=\tau /m$.
}:

\begin{equation}
\label{equation-tq-seqn-deriv-tq-seqn-deriv-seqn}\hyperlabel{equation-tq-seqn-deriv-tq-seqn-deriv-seqn}%
i\frac{d{\psi }_{\tau }\left(x\right)}{d\tau }=-\frac{1}{2m}\left({\left(p-eA\right)}^{2}-{m}^{2}\right){\psi }_{\tau }\left(x\right)
\end{equation}

The stationary solutions of this:

\begin{equation}
\label{equation-tq-seqn-deriv-tq-seqn-kg-0}\hyperlabel{equation-tq-seqn-deriv-tq-seqn-kg-0}%
i\frac{d{\psi }_{\tau }\left(x\right)}{d\tau }=0
\end{equation}

Satisfy the Klein-{}Gordon equation:

\begin{equation}
\label{equation-tq-seqn-deriv-tq-seqn-kg-1}\hyperlabel{equation-tq-seqn-deriv-tq-seqn-kg-1}%
\left({\left(p-eA\right)}^{2}-{m}^{2}\right)\psi \left(x\right)=0
\end{equation}

With the minimal substitution 
 $ p\rightarrow p-eA$. We will argue below ~ that picking out the stationary solutions recovers the standard quantum theory in the 
 \hyperlink{stat}{Long Time Limit}. This is the motivation for the addition of the mass term 
 $ -1/2m$~ to the Lagrangian earlier (\hyperlink{tq-fpi-lag}{Lagrangian}).

If we make the identifications:

\begin{equation}
\label{equation-tq-seqn-deriv-tq-seqn-deriv-7c}\hyperlabel{equation-tq-seqn-deriv-tq-seqn-deriv-7c}%
\begin{array}{c}E=-{\pi }_{0}\\ \overrightarrow{p}\rightarrow \overrightarrow{\pi }\end{array}
\end{equation}

We recover the Hamiltonian from the 
 \hyperlink{tq-fpi-lag}{Lagrangian}:

\begin{equation}
\label{equation-tq-seqn-deriv-tq-seqn-expand-3}\hyperlabel{equation-tq-seqn-deriv-tq-seqn-expand-3}%
i\frac{d{\psi }_{\tau }\left(x\right)}{d\tau }=-\frac{1}{2m}\left({\left({\pi }_{0}+e\Phi \right)}^{2}-{\left(\overrightarrow{\pi }-e\overrightarrow{A}\right)}^{2}-{m}^{2}\right){\psi }_{\tau }\left(x\right)=H{\psi }_{\tau }\left(x\right)
\end{equation}

\paragraph*{New Terms}

\noindent

Splitting out the electric and vector potential we get:

\begin{equation}
\label{equation-tq-seqn-deriv-tq-seqn-deriv-a}\hyperlabel{equation-tq-seqn-deriv-tq-seqn-deriv-a}%
i\frac{d{\psi }_{\tau }\left(t,\overrightarrow{x}\right)}{d\tau }=\left(-\frac{{\left(E-e\Phi \left(t,\overrightarrow{x}\right)\right)}^{2}}{2m}+\frac{{\left(\overrightarrow{p}-e\overrightarrow{A}\left(t,\overrightarrow{x}\right)\right)}^{2}}{2m}+\frac{m}{2}\right){\psi }_{\tau }\left(t,\overrightarrow{x}\right)
\end{equation}

The magnetic potential contributes cross terms in 
 \emph{p}~ and 
 \emph{A}, where the momentum operator acts directly on the vector potential. We have to be mindful of the ordering of 
 $ \overrightarrow{p}$~ and 
 $ \overrightarrow{A}$~ in the cross terms:

\begin{equation}
\label{equation-tq-seqn-deriv-tq-seqn-deriv-a2}\hyperlabel{equation-tq-seqn-deriv-tq-seqn-deriv-a2}%
e\frac{\overrightarrow{p}\cdot \overrightarrow{A}\left(t,\overrightarrow{x}\right)+\overrightarrow{A}\left(t,\overrightarrow{x}\right)\cdot \overrightarrow{p}}{2m}
\end{equation}

When we have a non-{}zero electric potential we have similar cross terms from 
 $ E$~ and 
 $ \Phi $, where the energy operator acts directly on the electric potential. We have to be mindful of the ordering of 
 \emph{E}~ and 
 $ \Phi $~ when 
 $ \Phi $~ depends on 
 $ t$:

\begin{equation}
\label{equation-tq-seqn-deriv-tq-seqn-deriv-b}\hyperlabel{equation-tq-seqn-deriv-tq-seqn-deriv-b}%
e\frac{E\Phi \left(t,\overrightarrow{x}\right)+\Phi \left(t,\overrightarrow{x}\right)E}{2m}
\end{equation}

And just as we have an 
 $ \overrightarrow{A}$~ squared term:

\begin{equation}
\label{equation-tq-seqn-deriv-tq-seqn-deriv-b2}\hyperlabel{equation-tq-seqn-deriv-tq-seqn-deriv-b2}%
{e}^{2}\frac{{\overrightarrow{A}}^{2}\left(t,\overrightarrow{x}\right)}{2m}
\end{equation}

We also have a term which is the square of the electric potential:

\begin{equation}
\label{equation-tq-seqn-deriv-tq-seqn-deriv-c}\hyperlabel{equation-tq-seqn-deriv-tq-seqn-deriv-c}%
{e}^{2}\frac{{\Phi }^{2}\left(t,\overrightarrow{x}\right)}{2m}
\end{equation}

We shall worry further about this below, in 
 \hyperlink{scatter-elecx}{Time Independent Electric Field}.

\subsection{Unitarity}\label{tq-seqn-unitarity}\hyperlabel{tq-seqn-unitarity}%

We demonstrate unitarity using the same proof as for the Schrödinger equation in standard quantum theory. We form the probability:

\begin{equation}
\label{equation-tq-seqn-unitarity-tq-seqn-unitarity-1}\hyperlabel{equation-tq-seqn-unitarity-tq-seqn-unitarity-1}%
P\equiv {\displaystyle \int {\text{d}}^{4}x{\psi }^{\ast }\left(x\right)\psi \left(x\right)}
\end{equation}

We have for the rate of change of probability in time:

\begin{equation}
\label{equation-tq-seqn-unitarity-tq-seqn-unitarity-2}\hyperlabel{equation-tq-seqn-unitarity-tq-seqn-unitarity-2}%
\frac{dP}{d\tau }={\displaystyle \int {\text{d}}^{4}x}\left({\psi }^{\ast }\left(x\right)\frac{d\psi }{d\tau }\left(x\right)+\frac{d{\psi }^{\ast }}{d\tau }\left(x\right)\psi \left(x\right)\right)
\end{equation}

The Schrödinger equations for the wave function and its complex conjugate are:

\begin{equation}
\label{equation-tq-seqn-unitarity-tq-seqn-unitarity-0}\hyperlabel{equation-tq-seqn-unitarity-tq-seqn-unitarity-0}%
\begin{array}{l}\frac{d\psi }{d\tau }=\frac{-i}{2m}{\partial }^{\mu }{\partial }_{\mu }\psi +\frac{e}{m}\left(A\partial \right)\psi +\frac{e}{2m}\left(\partial A\right)\psi +i\frac{{e}^{2}}{2m}{A}^{\mu }{A}_{\mu }\psi -i\frac{m}{2}\psi \\ \frac{d{\psi }^{\ast }}{d\tau }=\frac{i}{2m}{\partial }^{\mu }{\partial }_{\mu }{\psi }^{\ast }+\frac{e}{m}\left(A\partial \right){\psi }^{\ast }+\frac{e}{2m}\left(\partial A\right){\psi }^{\ast }-i\frac{{e}^{2}}{2m}{A}^{\mu }{A}_{\mu }{\psi }^{\ast }+i\frac{m}{2}{\psi }^{\ast }\end{array}
\end{equation}

We rewrite 
 $ \frac{d\psi }{d\tau }$~ and 
 $ \frac{d{\psi }^{\ast }}{d\tau }$~ using these, throw out cancelling terms, and choose the Lorentz gauge to get:

\begin{equation}
\label{equation-tq-seqn-unitarity-tq-seqn-unitarity-3}\hyperlabel{equation-tq-seqn-unitarity-tq-seqn-unitarity-3}%
\frac{dP}{d\tau }={\displaystyle \int {\text{d}}^{4}x}\left({\psi }^{\ast }\left(-\frac{i}{2m}{\partial }^{\mu }{\partial }_{\mu }\psi +\frac{e}{m}\left(A\partial \right)\psi \right)+\left(\frac{i}{2m}{\partial }^{\mu }{\partial }_{\mu }{\psi }^{\ast }+\frac{e}{m}\left(A\partial \right){\psi }^{\ast }\right)\psi \right)
\end{equation}

We integrate by parts; we are left with zero on the right:

\begin{equation}
\label{equation-tq-seqn-unitarity-tq-seqn-unitarity-4}\hyperlabel{equation-tq-seqn-unitarity-tq-seqn-unitarity-4}%
\frac{dP}{d\tau }=0
\end{equation}

So the rate of change of probability is zero, as was to be shown.

The normalization is therefore correct. Probability is conserved. And unitarity is demonstrated, in four dimensions rather than three.

\subsection{Gauge Transformations for the Schrödinger Equation}\label{tq-seqn-gauge}\hyperlabel{tq-seqn-gauge}%

We can write the wave function as a product of a gauge function in quantum time, space, and laboratory time and a gauged wave function:

\begin{equation}
\label{equation-tq-seqn-gauge-tq-seqn-gauge-gauge2}\hyperlabel{equation-tq-seqn-gauge-tq-seqn-gauge-gauge2}%
{{\psi }^{\prime }}_{\tau }\left(t,\overrightarrow{x}\right)=\mathrm{exp}\left(ie{\Lambda }_{\tau }\left(t,\overrightarrow{x}\right)\right){\psi }_{\tau }\left(t,\overrightarrow{x}\right)
\end{equation}

If the original wave function satisfies a gauged Schrödinger equation:

\begin{equation}
\label{equation-tq-seqn-gauge-tq-seqn-gauge-e}\hyperlabel{equation-tq-seqn-gauge-tq-seqn-gauge-e}%
\left(i\frac{d}{d\tau }-e{\mathcal{A}}_{\tau }\left(x\right)\right){\psi }_{\tau }\left(x\right)=-\frac{1}{2m}\left({\left(p-eA\right)}^{2}-{m}^{2}\right){\psi }_{\tau }\left(x\right)
\end{equation}

The gauged wave function also satisfies a gauged Schrödinger equation:

\begin{equation}
\label{equation-tq-seqn-gauge-tq-seqn-gauge-f}\hyperlabel{equation-tq-seqn-gauge-tq-seqn-gauge-f}%
\left(i\frac{d}{d\tau }-e{{\mathcal{A}}^{\prime }}_{\tau }\left(x\right)\right){{\psi }^{\prime }}_{\tau }\left(x\right)=-\frac{1}{2m}\left({\left(p-e{A}^{\prime }\right)}^{2}-{m}^{2}\right){{\psi }^{\prime }}_{\tau }\left(x\right)
\end{equation}

Provided:

\begin{equation}
\label{equation-tq-seqn-gauge-tq-seqn-gauge-b}\hyperlabel{equation-tq-seqn-gauge-tq-seqn-gauge-b}%
{{\mathcal{A}}^{\prime }}_{\tau }\left(x\right)={\mathcal{A}}_{\tau }\left(x\right)-\frac{d{\Lambda }_{\tau }\left(x\right)}{d\tau }
\end{equation}

And we have the usual gauge transformations:

\begin{equation}
\label{equation-tq-seqn-gauge-tq-seqn-gauge-h}\hyperlabel{equation-tq-seqn-gauge-tq-seqn-gauge-h}%
{{A}^{\prime }}^{\mu }={A}^{\mu }-{\partial }^{\mu }{\Lambda }_{\tau }\left(x\right)
\end{equation}

Or:

\begin{equation}
\label{equation-tq-seqn-gauge-tq-seqn-gauge-4}\hyperlabel{equation-tq-seqn-gauge-tq-seqn-gauge-4}%
\begin{array}{l}\Phi \rightarrow \Phi -\frac{\partial \Lambda }{\partial t}\\ \overrightarrow{A}\rightarrow \overrightarrow{A}+\nabla \Lambda \end{array}
\end{equation}

If the gauge function is not a function of the laboratory time, then all we have is the usual gauge transformations for 
 $ \Phi $~ and 
 $ \overrightarrow{A}$.

We will call 
 $ {\mathcal{A}}_{\tau }\left(t,\overrightarrow{x}\right)$~ Alice's potential. The Schrödinger equation we derived above corresponds to an initial choice for 
 $ {\mathcal{A}}_{\tau }\left(t,\overrightarrow{x}\right)$~ of zero:

\begin{equation}
\label{equation-tq-seqn-gauge-tq-seqn-gauge-h2}\hyperlabel{equation-tq-seqn-gauge-tq-seqn-gauge-h2}%
{\mathcal{A}}_{\tau }\left(x\right)=0
\end{equation}

The gauge term is necessary. If we an integration by parts of the Lagrangian with respect to the laboratory time we will see gauge terms show up, e.g.:

\begin{equation}
\label{equation-tq-seqn-gauge-tq-seqn-gauge-ibyparts}\hyperlabel{equation-tq-seqn-gauge-tq-seqn-gauge-ibyparts}%
-{\displaystyle \underset{{\rm A}}{\overset{B}{\int }}\text{d}\tau \frac{1}{2}m{\dot{t}}^{2}}=-m{t}_{B}{\dot{t}}_{B}+m{t}_{A}{\dot{t}}_{A}+{\displaystyle \underset{{\rm A}}{\overset{B}{\int }}\text{d}\tau \frac{1}{2}mt\ddot{t}}
\end{equation}

The Lagrangian changes:

\begin{equation}
\label{equation-tq-seqn-gauge-tq-seqn-gauge-lag}\hyperlabel{equation-tq-seqn-gauge-tq-seqn-gauge-lag}%
\mathscr{L}=-\frac{1}{2}m{\dot{t}}^{2}\rightarrow {\mathscr{L}}^{\prime }=\frac{1}{2}mt\ddot{t}
\end{equation}

And therefore the kernel and wave function:

\begin{equation}
\label{equation-tq-seqn-gauge-tq-seqn-gauge-kerwf}\hyperlabel{equation-tq-seqn-gauge-tq-seqn-gauge-kerwf}%
\begin{array}{l}{K}_{BA}\rightarrow \mathrm{exp}\left(ie{\Lambda }_{B}\right){{K}^{\prime }}_{BA}\mathrm{exp}\left(-ie{\Lambda }_{A}\right)\\ {\psi }_{A}\rightarrow \mathrm{exp}\left(ie{\Lambda }_{A}\right){\psi }_{A}\end{array}
\end{equation}

With gauge:

\begin{equation}
\label{equation-tq-seqn-gauge-tq-seqn-gauge-gauge3}\hyperlabel{equation-tq-seqn-gauge-tq-seqn-gauge-gauge3}%
\Lambda =-m\frac{t\dot{t}}{e}
\end{equation}

Coordinate changes often induce gauge changes; for an example see 
 \hyperlink{semi-elec}{Constant Electric Field}.

We can use a gauge change 
 $ {\Lambda }_{\tau }=m\tau /2e$~ to eliminate the mass term; for an example see again in 
 \hyperlink{semi-elec}{Constant Electric Field}.

And we make use of a time gauge to establish the connection between temporal quantization and standard quantum theory in 
 \hyperlink{scatter-elecx}{Time Independent Electric Field}~ and 
 \hyperlink{scatter-elect}{Time Dependent Electric Field}.

\section{Operators in Time}\label{tq-op}\hyperlabel{tq-op}%

\paragraph*{Equation of Motion}

\noindent

The Hamiltonian acts as the generator of translations in laboratory time:

\begin{equation}
\label{equation-tq-op-tq-op-1}\hyperlabel{equation-tq-op-tq-op-1}%
i\frac{d}{d\tau }\psi =H\psi 
\end{equation}

We use this to compute the change of expectation values of an observable operator 
 \emph{O}~ with laboratory time. The proofs in the quantum mechanics textbooks (\cite{Messiah-1958},
 \cite{Baym-1969},
 \cite{Liboff-1997},
 \cite{Merzbacher-1998}) ~ apply. The derivative of the expectation is given by:

\begin{equation}
\label{equation-tq-op-tq-op-proof-0}\hyperlabel{equation-tq-op-tq-op-proof-0}%
\frac{d\langle O\rangle }{d\tau }=\underset{\epsilon \rightarrow 0}{\mathrm{lim}}\frac{{\langle O\rangle }_{\tau +\epsilon }-{\langle O\rangle }_{\tau }}{\epsilon }=\frac{{\displaystyle \int \text{d}q}{\psi }_{\tau +\epsilon }^{\ast }{O}_{\tau }{\psi }_{\tau +\epsilon }-{\displaystyle \int \text{d}q}{\psi }_{\tau }^{\ast }{O}_{\tau }{\psi }_{\tau }}{\epsilon }
\end{equation}

The Hamiltonian operator is responsible for evolving the wave function in laboratory time:

\begin{equation}
\label{equation-tq-op-tq-op-proof-1}\hyperlabel{equation-tq-op-tq-op-proof-1}%
{\psi }_{\tau +e}=\mathrm{exp}\left(-i\epsilon H\right){\psi }_{\tau }\approx \left(1-i\epsilon H\right){\psi }_{\tau }
\end{equation}

Taking advantage of the fact that the Hamiltonian is Hermitian:

\begin{equation}
\label{equation-tq-op-tq-op-proof-2}\hyperlabel{equation-tq-op-tq-op-proof-2}%
\frac{d\langle O\rangle }{d\tau }=\underset{\epsilon \rightarrow 0}{\mathrm{lim}}\frac{1}{\epsilon }\left({\displaystyle \int \text{d}q}{\psi }_{\tau }^{\ast }\left(1+i\epsilon H\right)\left(O+\epsilon \frac{\partial O}{\partial \tau }\right)\left(1-i\epsilon H\right){\psi }_{\tau }-{\psi }_{\tau }^{\ast }O{\psi }_{\tau }\right)
\end{equation}

And taking the limit as 
 $ \epsilon $~ goes to zero:

\begin{equation}
\label{equation-tq-op-tq-op-proof-first}\hyperlabel{equation-tq-op-tq-op-proof-first}%
\frac{d\langle O\rangle }{d\tau }=-i\left\lbrack O,H\right\rbrack +\frac{\partial O}{\partial \tau }
\end{equation}

This simplifies when the operators are not dependent on laboratory time:

\begin{equation}
\label{equation-tq-op-tq-op-3}\hyperlabel{equation-tq-op-tq-op-3}%
\langle \frac{\partial O}{\partial \tau }\rangle =0\Rightarrow i\langle \frac{dO}{d\tau }\rangle =\langle \left\lbrack O,H\right\rbrack \rangle 
\end{equation}

This always the case for the operators of interest here
 \footnote{If an observable were to depend on laboratory time, as opposed to quantum time, it would imply that Alice was changing the definition of an observable "on the fly", not the sort of
    conduct we have come to expect from the conscientious and reliable Alice.}.

If we apply this rule to the coordinate and momentum operators we get for the coordinates:

\begin{equation}
\label{equation-tq-op-tq-op-eom-0}\hyperlabel{equation-tq-op-tq-op-eom-0}%
\begin{array}{l}\dot{t}=-i\left\lbrack t,-\frac{{\left(p-eA\right)}^{2}-{m}^{2}}{2m}\right\rbrack =\frac{E-e\Phi }{m}\\ \dot{\overrightarrow{x}}=-i\left\lbrack \overrightarrow{x},-\frac{{\left(p-eA\right)}^{2}-{m}^{2}}{2m}\right\rbrack =\frac{\overrightarrow{p}-e\overrightarrow{A}}{m}\end{array}
\end{equation}

And for the momentum:

\begin{equation}
\label{equation-tq-op-tq-op-eom-1}\hyperlabel{equation-tq-op-tq-op-eom-1}%
\begin{array}{l}\dot{E}=-i\left\lbrack E,-\frac{{\left(p-eA\right)}^{2}-{m}^{2}}{2m}\right\rbrack =e\frac{E-e\Phi }{2m}{\Phi }_{,0}+e{\Phi }_{,0}\frac{E-e\Phi }{2m}-e\frac{\overrightarrow{p}-e\overrightarrow{A}}{2m}\cdot {\overrightarrow{A}}_{,0}-e{\overrightarrow{A}}_{,0}\cdot \frac{\overrightarrow{p}-e\overrightarrow{A}}{2m}\\ {\dot{p}}_{i}=-i\left\lbrack {p}_{i},-\frac{{\left(p-eA\right)}^{2}-{m}^{2}}{2m}\right\rbrack =-e\frac{E-e\Phi }{2m}{\Phi }_{,i}-e{\Phi }_{,i}\frac{E-e\Phi }{2m}+e\frac{\overrightarrow{p}-e\overrightarrow{A}}{2m}\cdot {\overrightarrow{A}}_{,i}+e{\overrightarrow{A}}_{,i}\cdot \frac{\overrightarrow{p}-e\overrightarrow{A}}{2m}\end{array}
\end{equation}

If we make the earlier substitutions:

\begin{equation}
\label{equation-tq-op-tq-op-eom-2}\hyperlabel{equation-tq-op-tq-op-eom-2}%
\begin{array}{l}E\rightarrow -{\pi }_{0}\\ \overrightarrow{p}\rightarrow \overrightarrow{\pi }\end{array}
\end{equation}

We get the earlier Hamiltonian equation of motion for coordinates and momentum.

The second derivative is defined by:

\begin{equation}
\label{equation-tq-op-tq-op-proof-4}\hyperlabel{equation-tq-op-tq-op-proof-4}%
\frac{{d}^{2}\langle O\rangle }{d{\tau }^{2}}=\underset{\epsilon \rightarrow 0}{\mathrm{lim}}\frac{1}{{\epsilon }^{2}}{\displaystyle \int \text{d}q}\left(\begin{array}{l}{\psi }_{\tau }^{\ast }\left(1+2i\epsilon H-2{\epsilon }^{2}{H}^{2}\right)\left(O+2\epsilon \frac{\partial O}{\partial \tau }+2{\epsilon }^{2}\frac{{\partial }^{2}O}{\partial {\tau }^{2}}\right)\left(1-2i\epsilon H-2{\epsilon }^{2}{H}^{2}\right){\psi }_{\tau }\\ -2{\psi }_{\tau }^{\ast }\left(1+i\epsilon H-\frac{{\epsilon }^{2}{H}^{2}}{2}\right)\left(O+\epsilon \frac{\partial O}{\partial \tau }+\frac{{\epsilon }^{2}}{2}\frac{{\partial }^{2}O}{\partial {\tau }^{2}}\right)\left(1-i\epsilon H-\frac{{\epsilon }^{2}{H}^{2}}{2}\right){\psi }_{\tau }\\ +{\psi }_{\tau }^{\ast }O{\psi }_{\tau }\end{array}\right)
\end{equation}

With the intuitive result:

\begin{equation}
\label{equation-tq-op-tq-op-proof-second}\hyperlabel{equation-tq-op-tq-op-proof-second}%
\frac{{d}^{2}\langle O\rangle }{d{\tau }^{2}}=-\left\lbrack \left\lbrack O,H\right\rbrack ,H\right\rbrack -2i\left\lbrack \frac{\partial O}{\partial \tau },H\right\rbrack +\frac{{\partial }^{2}O}{\partial {\tau }^{2}}
\end{equation}

\paragraph*{Uncertainty Principle in Time and Energy}

\noindent

By construction the quantum time 
 $ t$~ and quantum energy 
 \emph{E}~ are on same footing as the space 
 \emph{x}~ and the momentum 
 \emph{p}. Therefore the uncertainty principle between 
 $ t$~ and 
 \emph{E}~ is on the same footing as the one between 
 \emph{x}~ and 
 \emph{p}.

The laboratory time is still a parameter \textendash{} not an operator \textendash{} and therefore there is still no operator complimentary to the Hamiltonian, no operator corresponding to the laboratory time 
 $ \tau $,

We can define the quantum time operator for Alice as being such an operator, with the usual laboratory time being the expectation of this:

\begin{equation}
\label{equation-tq-op-tq-fpi-alice-1}\hyperlabel{equation-tq-op-tq-fpi-alice-1}%
{\langle \tau \rangle }^{\left(Alice\right)}\equiv {\displaystyle \int \text{d}t\text{d}\overrightarrow{x}{\psi }^{\left(Alice\right)\ast }\left(t,\overrightarrow{x}\right)t{\psi }^{\left(Alice\right)}\left(t,\overrightarrow{x}\right)}
\end{equation}

For that matter, there is nothing keeping us from extending this quantum time operator to apply to the rest of the universe:

\begin{equation}
\label{equation-tq-op-tq-fpi-alice-2}\hyperlabel{equation-tq-op-tq-fpi-alice-2}%
{\langle \tau \rangle }^{\left(rest-of-universe\right)}\equiv {\displaystyle \int \text{d}t\text{d}\overrightarrow{x}{\psi }^{\left(rest-of-universe\right)\ast }\left(t,\overrightarrow{x}\right)t{\psi }^{\left(rest-of-universe\right)}\left(t,\overrightarrow{x}\right)}
\end{equation}

So the laboratory time is becomes the expectation of the quantum time operator applied to the rest of the universe. Machian \textendash{} very Machian.

\section{Canonical Path Integrals}\label{tq-pq}\hyperlabel{tq-pq}%

\subsection{Derivation of the Canonical Path Integral}\label{tq-pq-deriv}\hyperlabel{tq-pq-deriv}%

We can now use the Hamiltonian as a starting point for developing path integrals. This adds another to our list of temporal quantization formalisms.

\paragraph*{Development}

\noindent

We follow 
 \cite{Schulman-1981}, 
 \cite{Kleinert-2004}. As before, things are simpler using Lorentz gauge, as 
 \emph{p}~ and 
 \emph{A}~ can be passed through each other:

\begin{equation}
\label{equation-tq-pq-deriv-tq-pq-deriv-00}\hyperlabel{equation-tq-pq-deriv-tq-pq-deriv-00}%
{\partial }_{\mu }{A}^{\mu }=0\Rightarrow \left\lbrack {p}_{\mu },{A}^{\mu }\right\rbrack =0
\end{equation}

Consider the Hamiltonian:

\begin{equation}
\label{equation-tq-pq-deriv-tq-pq-deriv-0}\hyperlabel{equation-tq-pq-deriv-tq-pq-deriv-0}%
H=-\frac{1}{2m}{p}^{2}+\frac{e}{2m}pA\left(x\right)+\frac{e}{2m}A\left(x\right)p-\frac{{e}^{2}}{2m}{A}^{2}\left(x\right)+\frac{m}{2}
\end{equation}

The kernel is (formally):

\begin{equation}
\label{equation-tq-pq-deriv-tq-pq-deriv-2}\hyperlabel{equation-tq-pq-deriv-tq-pq-deriv-2}%
{K}_{\tau }\left({x}^{{''}};{x}^{\prime }\right)=\mathrm{exp}\left(-i\tau H\right)=\mathrm{exp}{\left(-i\epsilon H\right)}^{N+1}
\end{equation}

We want to get every operator next to one of its eigenfunctions, whether on left or right. We break up the path from the start with 
 \emph{N}~ cuts into 
 \emph{N+1}~ pieces by inserting an expression for one between each:

\begin{equation}
\label{equation-tq-pq-deriv-tq-pq-deriv-3}\hyperlabel{equation-tq-pq-deriv-tq-pq-deriv-3}%
1={\displaystyle \int \text{d}{x}_{j}}\vert {x}_{j}\rangle \langle {x}_{j}\vert 
\end{equation}

Getting 
 \emph{N}~ 
 \emph{x}~ integrations:

\begin{equation}
\label{equation-tq-pq-deriv-tq-pq-deriv-4}\hyperlabel{equation-tq-pq-deriv-tq-pq-deriv-4}%
{K}_{\tau }\left({x}_{N+1};{x}_{0}\right)=\langle {x}_{N+1}\vert \mathrm{exp}\left(-i\epsilon H\right){\displaystyle \int \text{d}{x}_{N}}\vert {x}_{N}\rangle \langle {x}_{N}\vert \dots {\displaystyle \int \text{d}{x}_{2}}\vert {x}_{2}\rangle \langle {x}_{2}\vert \mathrm{exp}\left(-i\epsilon H\right){\displaystyle \int \text{d}{x}_{1}}\vert {x}_{1}\rangle \langle {x}_{1}\vert \mathrm{exp}\left(-i\epsilon H\right)\vert {x}_{0}\rangle 
\end{equation}

With the identifications 
 $ {x}_{0}={x}_{A},{x}_{N+1}={x}_{B}$.

Consider a single slice:

\begin{equation}
\label{equation-tq-pq-deriv-tq-pq-deriv-5}\hyperlabel{equation-tq-pq-deriv-tq-pq-deriv-5}%
{K}_{\tau }\left({x}_{j};{x}_{j-1}\right)=\langle {x}_{j}\vert \mathrm{exp}\left(-i\epsilon H\right)\vert {x}_{j-1}\rangle 
\end{equation}

We break the Hamiltonian into two pieces:

\begin{equation}
\label{equation-tq-pq-deriv-tq-pq-deriv-6}\hyperlabel{equation-tq-pq-deriv-tq-pq-deriv-6}%
\begin{array}{c}H={H}_{1}+{H}_{2}\\ {H}_{1}=-\frac{1}{4m}{p}^{2}+\frac{e}{2m}A\left(x\right)p-\frac{{e}^{2}}{4m}{A}^{2}\left(x\right)+\frac{m}{4}\\ {H}_{2}=-\frac{1}{4m}{p}^{2}+\frac{e}{2m}pA\left(x\right)-\frac{{e}^{2}}{4m}{A}^{2}\left(x\right)+\frac{m}{4}\end{array}
\end{equation}

We break the exponential into two pieces, ignoring terms of order 
 $ \epsilon $~ squared and higher:

\begin{equation}
\label{equation-tq-pq-deriv-tq-pq-deriv-7}\hyperlabel{equation-tq-pq-deriv-tq-pq-deriv-7}%
\mathrm{exp}\left(-i\epsilon H\right)=\mathrm{exp}\left(-i\epsilon {H}_{1}\right)\mathrm{exp}\left(-i\epsilon {H}_{2}\right)
\end{equation}

We insert another expression for one between the two pieces:

\begin{equation}
\label{equation-tq-pq-deriv-tq-pq-deriv-8}\hyperlabel{equation-tq-pq-deriv-tq-pq-deriv-8}%
1={\displaystyle \int \text{d}{p}_{j}}\vert {p}_{j}\rangle \langle {p}_{j}\vert 
\end{equation}

We get for the infinitesimal kernel:

\begin{equation}
\label{equation-tq-pq-deriv-tq-pq-deriv-9}\hyperlabel{equation-tq-pq-deriv-tq-pq-deriv-9}%
{K}_{\tau }\left({x}_{j};{x}_{j-1}\right)={\displaystyle \int \text{d}{p}_{j}}\langle {x}_{j}\vert \mathrm{exp}\left(-i\epsilon {H}_{1}\right)\vert {p}_{j}\rangle \langle {p}_{j}\vert \mathrm{exp}\left(-i\epsilon {H}_{2}\right)\vert {x}_{j-1}\rangle 
\end{equation}

This implies one integration between each pair of 
 \emph{x}'s. Counting the 
 \emph{x}'s at the ends, we have N+1 
 \emph{p}~ integrations, one more than the number of 
 \emph{x}~ integrations.

Since:

\begin{equation}
\label{equation-tq-pq-deriv-tq-pq-deriv-a}\hyperlabel{equation-tq-pq-deriv-tq-pq-deriv-a}%
\begin{array}{c}\langle {x}_{j}\vert {p}_{j}\rangle =\frac{1}{4{\pi }^{2}}\mathrm{exp}\left(-i{p}_{j}{x}_{j}\right)\\ \langle {p}_{j}\vert {x}_{j-1}\rangle =\frac{1}{4{\pi }^{2}}\mathrm{exp}\left(i{p}_{j}{x}_{j-1}\right)\end{array}
\end{equation}

We have for the integrand of the infinitesimal kernel:

\begin{equation}
\label{equation-tq-pq-deriv-tq-pq-deriv-b}\hyperlabel{equation-tq-pq-deriv-tq-pq-deriv-b}%
\frac{1}{16{\pi }^{4}}\mathrm{exp}\left(-i{p}_{j}\left({x}_{j}-{x}_{j-1}\right)\right)\mathrm{exp}\left(i\epsilon \frac{{p}_{j}^{2}}{2m}-i\epsilon \frac{e}{m}\left(\frac{A\left({x}_{j}\right){p}_{j}+{p}_{j}A\left({x}_{j-1}\right)}{2}\right)+i\epsilon \frac{{e}^{2}}{2m}\left(\frac{{A}^{2}\left({x}_{j}\right)+{A}^{2}\left({x}_{j-1}\right)}{2}\right)-i\epsilon \frac{m}{2}\right)
\end{equation}

We define the measure:

\begin{equation}
\label{equation-tq-pq-deriv-tq-pq-deriv-c}\hyperlabel{equation-tq-pq-deriv-tq-pq-deriv-c}%
{\displaystyle \int \mathcal{D}q}\equiv {\left(\frac{1}{4{\pi }^{2}}\right)}^{2N+2}{\displaystyle \int \text{d}{p}_{N+1}\text{d}{x}_{N}\text{d}{p}_{N}\dots \text{d}{x}_{2}\text{d}{p}_{2}\text{d}{x}_{1}\text{d}{p}_{1}}
\end{equation}

Giving the kernel:

\begin{equation}
\label{equation-tq-pq-deriv-tq-pq-deriv-d}\hyperlabel{equation-tq-pq-deriv-tq-pq-deriv-d}%
{K}_{\tau }\left({x}_{N+1};{x}_{0}\right)=\mathrm{exp}\left(-i\frac{m}{2}\tau \right){\displaystyle \int \mathcal{D}q}\left(\begin{array}{c}\mathrm{exp}\left(-i{\displaystyle \sum _{j=1}^{j=N+1}{p}_{j}\left({x}_{j}-{x}_{j-1}\right)}\right)\\ \times \mathrm{exp}\left(i\epsilon {\displaystyle \sum _{j=1}^{j=N+1}\left(\begin{array}{l}\frac{{p}_{j}^{2}}{2m}-\frac{e}{m}{p}_{j}\left(\frac{A\left({x}_{j}\right)+A\left({x}_{j-1}\right)}{2}\right)\\ +\frac{{e}^{2}}{2m}\left(\frac{{A}^{2}\left({x}_{j}\right)+{A}^{2}\left({x}_{j-1}\right)}{2}\right)\end{array}\right)}\right)\end{array}\right)
\end{equation}

We are observing the midpoint rule for the 
 \emph{pA}~ term, without having explicitly required that.

We rewrite the path integral in terms of the Hamiltonian:

\begin{equation}
\label{equation-tq-pq-deriv-tq-pq-deriv-e}\hyperlabel{equation-tq-pq-deriv-tq-pq-deriv-e}%
{K}_{\tau }\left({x}_{N+1};{x}_{0}\right)={\displaystyle \int \mathcal{D}q}\mathrm{exp}\left(-i{\displaystyle \sum _{j=1}^{j=N+1}{p}_{j}\left({x}_{j}-{x}_{j-1}\right)-i\epsilon {\displaystyle \sum _{j=1}^{j=N+1}{H}_{j}}}\right)
\end{equation}

With the per-{}slice Hamiltonian:

\begin{equation}
\label{equation-tq-pq-deriv-tq-pq-deriv-f}\hyperlabel{equation-tq-pq-deriv-tq-pq-deriv-f}%
{H}_{j}\equiv -\frac{{p}_{j}^{2}}{2m}+\frac{e}{m}{p}_{j}\left(\frac{A\left({x}_{j}\right)+A\left({x}_{j-1}\right)}{2}\right)-\frac{{e}^{2}}{2m}\left(\frac{{A}^{2}\left({x}_{j}\right)+{A}^{2}\left({x}_{j-1}\right)}{2}\right)-\frac{m}{2}
\end{equation}

\subsection{Closing the Circle}\label{tq-pq-circle}\hyperlabel{tq-pq-circle}%

The canonical path integral has integrals over 
 \emph{p}~ and 
 \emph{q}; the Feynman path integral over only 
 \emph{q}; we now do the 
 \emph{p}~ integrals. We thereby derive the Feynman path integral and close the circle. We start with the kernel just derived. A typical 
 \emph{p}~ integral is:

\begin{equation}
\label{equation-tq-pq-circle-tq-pq-circle-0b}\hyperlabel{equation-tq-pq-circle-tq-pq-circle-0b}%
{P}_{j}\equiv {\left(\frac{1}{4{\pi }^{2}}\right)}^{2}{\displaystyle \int \text{d}{p}_{j}}\mathrm{exp}\left(\frac{i\epsilon }{2m}{p}_{j}^{2}-i{p}_{j}\left({x}_{j}-{x}_{j-1}+\frac{e\epsilon }{m}\left(\frac{A\left({x}_{j}\right)+A\left({x}_{j-1}\right)}{2}\right)\right)\right)
\end{equation}

Or:

\begin{equation}
\label{equation-tq-pq-circle-tq-pq-circle-3}\hyperlabel{equation-tq-pq-circle-tq-pq-circle-3}%
{P}_{j}=-\frac{i{m}^{2}}{4{\pi }^{2}{\epsilon }^{2}}\mathrm{exp}\left(-im\frac{{\left({x}_{j}-{x}_{j-1}\right)}^{2}}{2\epsilon }-i\left({x}_{j}-{x}_{j-1}\right)e\left(\frac{A\left({x}_{j}\right)+A\left({x}_{j-1}\right)}{2}\right)-i\frac{{e}^{2}\epsilon }{2m}{\left(\frac{A\left({x}_{j}\right)+A\left({x}_{j-1}\right)}{2}\right)}^{2}\right)
\end{equation}

The 
 $ {A}^{2}$~ term almost cancels against the 
 $ {A}^{2}$~ term above:

\begin{equation}
\label{equation-tq-pq-circle-tq-pq-circle-5}\hyperlabel{equation-tq-pq-circle-tq-pq-circle-5}%
-{\left(\frac{A\left({x}_{j}\right)+A\left({x}_{j-1}\right)}{2}\right)}^{2}+\left(\frac{{A}^{2}\left({x}_{j}\right)+{A}^{2}\left({x}_{j-1}\right)}{2}\right)=\frac{{\left(A\left({x}_{j}\right)-A\left({x}_{j-1}\right)\right)}^{2}}{2}
\end{equation}

We are left with the difference between vector potentials at sequential slices:

\begin{equation}
\label{equation-tq-pq-circle-tq-pq-circle-6}\hyperlabel{equation-tq-pq-circle-tq-pq-circle-6}%
A\left({x}_{j}\right)\approx A\left({x}_{j-1}\right)+\left({x}_{j}^{\mu }-{x}_{j-1}^{\mu }\right)\frac{\partial A\left({x}_{j-1}\right)}{\partial {x}^{\mu }}
\end{equation}

We know the difference between sequential time/space coordinates is of order the square root of 
 $ \epsilon $:

\begin{equation}
\label{equation-tq-pq-circle-tq-pq-circle-7}\hyperlabel{equation-tq-pq-circle-tq-pq-circle-7}%
\left({x}_{j}^{\mu }-{x}_{j-1}^{\mu }\right)\sim \sqrt{\epsilon }
\end{equation}

(see the 
 \hyperlink{tq-seqn-deriv}{Derivation of the Schrödinger Equation}) so the difference in the vector potential terms is of order 
 $ \epsilon $:

\begin{equation}
\label{equation-tq-pq-circle-tq-pq-circle-8}\hyperlabel{equation-tq-pq-circle-tq-pq-circle-8}%
{\left(A\left({x}_{j}\right)-A\left({x}_{j-1}\right)\right)}^{2}\sim \epsilon 
\end{equation}

Since this is already being multiplied by a factor of 
 $ \epsilon $, it is of order 
 $ \epsilon $~ squared so drops out.

Therefore we get the previous coordinate space kernel back:

\begin{equation}
\label{equation-tq-pq-circle-tq-fpi-final-fullkernel}\hyperlabel{equation-tq-pq-circle-tq-fpi-final-fullkernel}%
{K}_{\tau }\left({x}^{{''}};{x}^{\prime }\right)={\displaystyle \int \mathcal{D}x\mathrm{exp}\left(-i{\displaystyle \sum _{j=1}^{N+1}m\frac{{\left({x}_{j}-{x}_{j-1}\right)}^{2}}{2\epsilon }}-ie\left({x}_{j}-{x}_{j-1}\right)\frac{A\left({x}_{j}\right)+A\left({x}_{j-1}\right)}{2}-i\frac{m}{2}\epsilon \right)}
\end{equation}

\section{Covariant Definition of Laboratory Time}\label{covar}\hyperlabel{covar}%
\begin{quote}

\ldots{}the same laws of electrodynamics and optics will be valid for all frames of reference for which the equations mechanics hold good. We will raise this conjecture to the status of a
        postulate\ldots{}

\hspace*\fill---~A. Einstein
 \cite{Einstein-1905}\end{quote}
\begin{quote}

I spark, I fizz, for the lady who knows what time it is.

\hspace*\fill---~Professor Harold Hill
 \cite{Willson-1957}\end{quote}

\paragraph*{Different Definitions of Laboratory Time}

\noindent

If Alice is in her lab, while Bob is jetting around like a fusion powered mosquito, they will have different notions of laboratory time. If Bob is going with velocity 
 \emph{v}~ with respect to Alice, his time and space coordinates are related by:

\begin{equation}
\label{equation-covar-covar-00}\hyperlabel{equation-covar-covar-00}%
\begin{array}{l}{t}^{\left(Bob\right)}=\gamma \left({t}^{\left(Alice\right)}-v{x}^{\left(Alice\right)}\right)\\ {x}^{\left(Bob\right)}=\gamma \left({x}^{\left(Alice\right)}-v{t}^{\left(Alice\right)}\right)\end{array}
\end{equation}

If we are to have a fully covariant formulation we have to find a definition of the laboratory time which is independent of the observer.

As noted, using the particle's proper time will work if we are dealing with one particle, but will break down if we have to include photons or more than one particle.

How are we to define a laboratory time that Alice and Bob can agree on?

We will assume Alice and Bob cross paths at one instant, which will be a zero of time for both. At this zero instant, we will use the same four dimensional ~ wave function for
    each:

\begin{equation}
\label{equation-covar-covar-0b}\hyperlabel{equation-covar-covar-0b}%
{\psi }_{0}^{Alice}\left({x}^{Alice}\right)={\psi }_{0}^{Bob}\left({x}^{Bob}\right)
\end{equation}

To get to the wave function at an arbitrary time we will need the kernel. However, Alice and Bob's definitions of the kernel differ:

\begin{equation}
\label{equation-covar-covar-1}\hyperlabel{equation-covar-covar-1}%
{K}_{\Delta {\tau }^{Alice}}^{Alice}\left({{x}^{{''}}}_{Alice};{{x}^{\prime }}_{Alice}\right)={\displaystyle \int \mathcal{D}{x}_{Alice}\mathrm{exp}\left(i{\displaystyle \underset{{{\tau }^{\prime }}_{Alice}}{\overset{{{\tau }^{{''}}}_{Alice}}{\int }}\text{d}{\tau }_{Alice}L\left({x}_{Alice},\frac{d{x}_{Alice}}{d{\tau }_{Alice}}\right)}\right)}
\end{equation}

\begin{equation}
\label{equation-covar-covar-2}\hyperlabel{equation-covar-covar-2}%
{K}_{\Delta {\tau }^{Bob}}^{Bob}\left({{x}^{{''}}}_{Bob};{{x}^{\prime }}_{Bob}\right)={\displaystyle \int \mathcal{D}{x}_{Bob}\mathrm{exp}\left(i{\displaystyle \underset{{{\tau }^{\prime }}_{Bob}}{\overset{{{\tau }^{{''}}}_{Bob}}{\int }}\text{d}{\tau }_{Bob}L\left({x}_{Bob},\frac{d{x}_{Bob}}{d{\tau }_{Bob}}\right)}\right)}
\end{equation}

The Lagrangian as a scalar is independent of Alice or Bob's coordinate system.

The coordinate systems at each point are related by a Lorentz transformation:

\begin{equation}
\label{equation-covar-covar-4b}\hyperlabel{equation-covar-covar-4b}%
{x}_{Bob}=\Lambda {x}_{Alice}
\end{equation}

This transformation has Jacobian one, so leaves the measures unaffected by the change of variables:

\begin{equation}
\label{equation-covar-covar-2b}\hyperlabel{equation-covar-covar-2b}%
\mathcal{D}{x}_{Alice}\rightarrow \mathcal{D}{x}_{Bob}
\end{equation}

Since there is no absolute way to define what is meant by simultaneous we do not know what Alice or Bob should use for the final laboratory time. And we do not therefore do not know how to
    relate the step sizes of Alice and Bob:

\begin{equation}
\label{equation-covar-covar-3}\hyperlabel{equation-covar-covar-3}%
{\epsilon }_{Alice}\equiv \frac{\Delta {\tau }_{Alice}}{N}\leftrightarrow {\epsilon }_{Bob}\equiv \frac{\Delta {\tau }_{Bob}}{N}
\end{equation}

As an example, the momentum space kernel for a free particle changes as:

\begin{equation}
\label{equation-covar-covar-5}\hyperlabel{equation-covar-covar-5}%
\mathrm{exp}\left(i\frac{{p}^{2}-{m}^{2}}{2m}\Delta {\tau }_{Alice}\right)\leftrightarrow \mathrm{exp}\left(i\frac{{p}^{2}-{m}^{2}}{2m}\Delta {\tau }_{Bob}\right)
\end{equation}

Clearly the effect on the offshell part of the wave functions will be different.

To be sure, realistic problems will normally be dominated by the onshell case:

\begin{equation}
\label{equation-covar-covar-5b}\hyperlabel{equation-covar-covar-5b}%
\frac{{p}^{2}-{m}^{2}}{2m}\approx 0
\end{equation}

But as it is actually the offshell bits that are most interesting to us, this is not helpful.

\paragraph*{Morlet Wavelets and Coordinate Patches}

\noindent
\begin{figure}[H]

\begin{center}
\imgexists{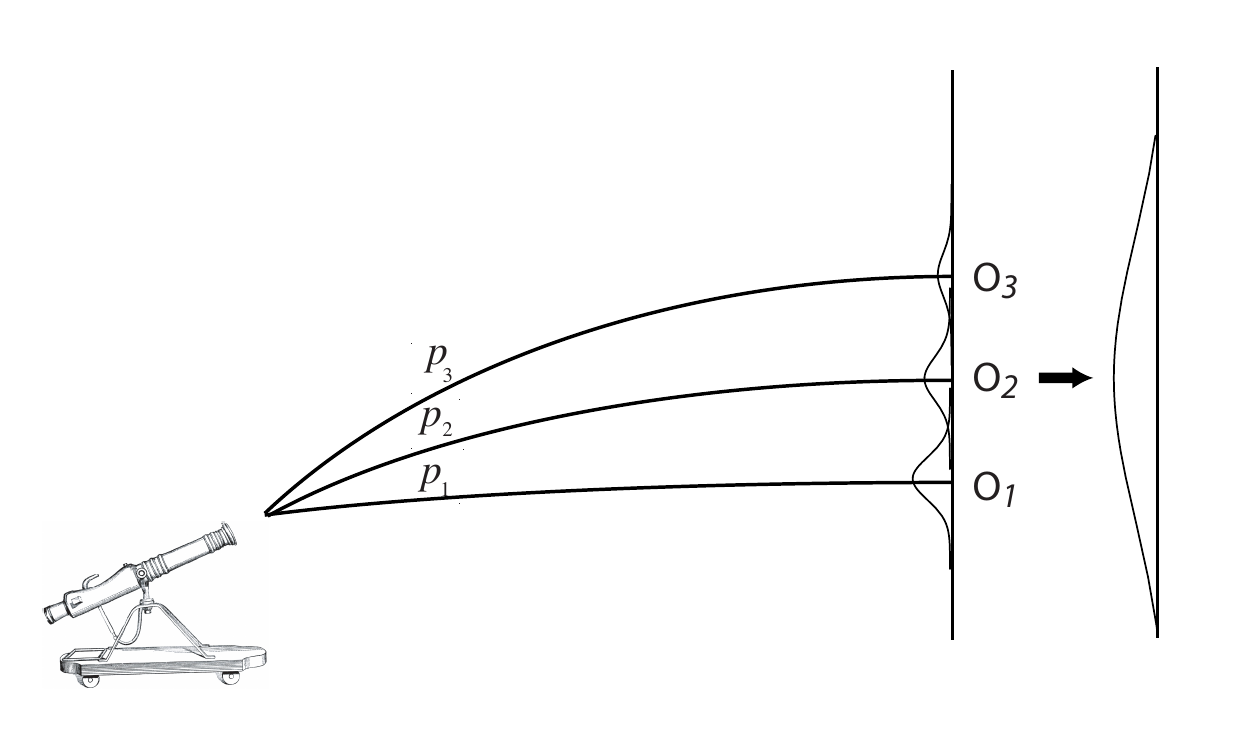}{{\imgevalsize{images/covar-1.pdf}{\includegraphics[width=\imgwidth,height=\imgheight,keepaspectratio=true]{images/covar-1.pdf}}\quad
}
}{}
\end{center}
\caption{Per Wavelet Paths}
\label{figure-images-covar-1}\hyperlabel{figure-images-covar-1}%
\end{figure}

We need a way to define the laboratory time which is independent of Alice or Bob.

We will do this not by finding a single laboratory time for the whole wave function, but by breaking the wave function into parts \textendash{} for each of which we can find a reasonable laboratory
    time \textendash{} propagating each part separately, and adding the parts back together at the end.

We work in the context of general relativity and think in terms of coordinate patches, small bits of space and time that are reasonably smooth. We will assume that our wave function is
    bounded within such a patch.

We need to have a way to deal with the extended character of the initial wave function and detector.

We break the wave function up into Morlet wavelets in four dimensions, so the full wave function is a sum over four dimensional Morlet wavelets, each of which is made of sixteen Gaussian
    test functions:

\begin{equation}
\label{equation-covar-covar-9}\hyperlabel{equation-covar-covar-9}%
{\tilde{\psi }}_{sd}={\displaystyle \int \mathcal{D}{x}_{A}{\phi }_{sd}^{\ast }\left({x}_{A}\right)\psi \left({x}_{A}\right)}
\end{equation}

For each wavelet, we have well-{}defined expectations for initial location:

\begin{equation}
\label{equation-covar-covar-9b}\hyperlabel{equation-covar-covar-9b}%
\langle {x}^{\mu }\rangle \equiv \frac{\langle {\phi }_{sd}\left\vert {x}^{\mu }\right\vert {\phi }_{sd}\rangle }{\langle {\phi }_{sd}\vert {\phi }_{sd}\rangle }
\end{equation}

And initial momentum:

\begin{equation}
\label{equation-covar-covar-9d}\hyperlabel{equation-covar-covar-9d}%
{p}^{\mu }\equiv i\frac{\langle {\phi }_{sd}\left\vert \frac{\partial }{\partial {x}_{\mu }}\right\vert {\phi }_{sd}\rangle }{\langle {\phi }_{sd}\vert {\phi }_{sd}\rangle }
\end{equation}

A well-{}defined initial position and momentum implies a well-{}defined classical trajectory. (We will assume that the detector is big enough that all such classical paths will hit it.)

The proper time along a classical trajectory is independent of the observer. This is our per wavelet laboratory time.

The kernel for each wavelet is now:

\begin{equation}
\label{equation-covar-covar-b}\hyperlabel{equation-covar-covar-b}%
{K}_{\Delta {\tau }^{wavelet}}^{wavelet}\left({x}_{B};{x}_{A}\right)={\displaystyle \int \mathcal{D}{x}_{wavelet}\mathrm{exp}\left(i{\displaystyle \underset{A}{\overset{B}{\int }}\text{d}{\tau }_{wavelet}L\left(x,\frac{dx}{d{\tau }_{wavelet}}\right)}\right)}
\end{equation}

With the laboratory time now defined on a per wavelet basis, we have at the detector:

\begin{equation}
\label{equation-covar-covar-c}\hyperlabel{equation-covar-covar-c}%
{\phi }_{sd}\left({x}_{B}\right)={\displaystyle \int \text{d}{x}_{A}{K}_{\Delta {\tau }^{wavelet}}^{wavelet}\left({x}_{B};{x}_{A}\right){\phi }_{sd}\left({x}_{A}\right)}
\end{equation}

To get the full wave, we add the components up
 \footnote{
Because the laboratory time is now defined on a per wavelet basis, we can see the wavelets as being, in a certain sense, more fundamental than the full wave.
}:

\begin{equation}
\label{equation-covar-covar-d}\hyperlabel{equation-covar-covar-d}%
\psi \left({x}_{B}\right)={\displaystyle \int \mathcal{D}s\mathcal{D}d{\phi }_{sd}^{\ast }\left({x}_{B}\right){\tilde{\psi }}_{sd}}
\end{equation}

We have implicitly assumed that Alice and Bob can use the same wavelet decomposition. We have shown elsewhere (\cite{Ashmead-2009w}) that we can define four dimensional Morlet wavelets in a covariant way; we will assume that Alice and Bob have read that paper and are willing to use that
    wavelet decomposition or an equivalent.

We will make use of this decomposition below in 
 \hyperlink{xt-gates}{Slits in Time}~ when we analyze a wave function going through a gate.

\section{Discussion}\label{tq-disc}\hyperlabel{tq-disc}%

We have therefore satisfied our first two requirements: we needed only one but we have found four formalisms for temporal quantization which are well-{}defined and manifestly
    covariant.

We have also made some progress with respect to reasonably simple:
\begin{enumerate}

\item{}We have no need for factors of 
 $ i\epsilon $, Wick rotation, or the like. We get well-{}behaved path integrals, if we are careful to apply them to well-{}behaved functions.

\item{}And we have reasonably simple expressions for the Schrödinger equation:

\begin{equation}
\label{equation-tq-disc-tq-seqn-deriv-seqn}\hyperlabel{equation-tq-disc-tq-seqn-deriv-seqn}%
i\frac{d{\psi }_{\tau }\left(x\right)}{d\tau }=-\frac{1}{2m}\left({\left(p-eA\right)}^{2}-{m}^{2}\right){\psi }_{\tau }\left(x\right)
\end{equation}

\item{}And its kernel:

\begin{equation}
\label{equation-tq-disc-tq-fpi-final-full}\hyperlabel{equation-tq-disc-tq-fpi-final-full}%
{K}_{\tau }\left(t,{\overrightarrow{x}}^{{''}};t,{\overrightarrow{x}}^{\prime }\right)={\left(-\frac{i{m}^{2}}{4{\pi }^{2}{\epsilon }^{2}}\right)}^{N+1}{\displaystyle \int {\displaystyle \prod _{n=1}^{n=N}{\text{d}}^{\text{4}}{x}_{n}}\mathrm{exp}\left(i{\displaystyle \underset{0}{\overset{\tau }{\int }}L\left\lbrack {t}_{\tau },{\overrightarrow{x}}_{\tau },\frac{d{t}_{\tau }}{d\tau },\frac{d{\overrightarrow{x}}_{\tau }}{d\tau }\right\rbrack \text{d}\tau }\right)}
\end{equation}

\end{enumerate}


\chapter{Comparison of Temporal Quantization To Standard Quantum Theory}\label{comp}\hyperlabel{comp}%

\section{Overview}\label{comp-intro}\hyperlabel{comp-intro}%

Quantum time functions as a fourth coordinate. It is like spin in being orthogonal to the other coordinates, unlike spin in being continuous and unbounded. Its properties are defined by
    covariance. Quantum time functions as a hidden coordinate \textendash{} like a small mammal in the age of the dinosaurs, unlikely to be seen unless looked for \textendash{} as spin did before the experiments of Stern
    and Gerlach (\cite{Gerlach-1922},
 \cite{Gerlach-1922b},
 \cite{Gerlach-1922c}).

If quantum time is real, why has it not been seen by chance? There are three reasons:
\begin{enumerate}

\item{}In terms of a classic beam/target experiment, unless both beam and target are varying in time, the effects of quantum time will tend to be averaged out.

\item{}Even if this is not the case, the principal effect of quantum time is increased dispersion in time. This is going to look like experimental noise, producing errors bars a bit wider
            in the time direction than might otherwise be the case. One seldom sees a headline in 
 \emph{Science}~ or 
 \emph{Nature}~ "Curiously large error bars seen in time measurement 
 \emph{X}".

\item{}In three limits, temporal quantization approaches standard quantum theory:
\begin{enumerate}

\item{}As the velocity goes to zero (\hyperlink{nonrel}{Non-{}relativistic Limit}).

\item{}As 
 $ \hslash $~ goes to zero (\hyperlink{semi}{Semi-{}classical Limit}).

\item{}As we average over longer stretches of laboratory time (\hyperlink{stat}{Long Time Limit}).

\end{enumerate}

\end{enumerate}

We look at each of these three limits in turn.

\section{Non-{}relativistic Limit}\label{nonrel}\hyperlabel{nonrel}%

We first look at the non-{}relativistic limit. We start with the path integral expression for the kernel:

\begin{equation}
\label{equation-nonrel-nonsuch-0}\hyperlabel{equation-nonrel-nonsuch-0}%
{K}_{\tau }\left({x}^{{''}};{x}^{\prime }\right)={\displaystyle \int \mathcal{D}x\mathrm{exp}\left(-i{\displaystyle \underset{0}{\overset{\tau }{\int }}\text{d}{\tau }^{\prime }\left(\frac{m}{2}\left({\dot{t}}^{2}-{\dot{\overrightarrow{x}}}^{2}\right)+e\dot{t}\Phi \left(t,\overrightarrow{x}\right)-e\dot{\overrightarrow{x}}\cdot \overrightarrow{A}\left(t,\overrightarrow{x}\right)+\frac{m}{2}\right)}\right)}
\end{equation}

We expand the potentials around the laboratory time:

\begin{equation}
\label{equation-nonrel-nonsuch-1}\hyperlabel{equation-nonrel-nonsuch-1}%
A\left(t,\overrightarrow{x}\right)=A\left(\tau ,\overrightarrow{x}\right)+{\frac{\partial A\left(t,\overrightarrow{x}\right)}{\partial t}\vert }_{t=\tau }{t}_{\tau }+\frac{1}{2}{\frac{{\partial }^{2}A\left(t,\overrightarrow{x}\right)}{\partial {t}^{2}}\vert }_{t=\tau }{t}_{\tau }^{2}+\dots 
\end{equation}

It is difficult to say much more in a general way about this without making some specific assumptions about the potentials. We now assume they are sufficiently slowly changing that the
    first and higher derivatives are negligible:

\begin{equation}
\label{equation-nonrel-nonsuch-2}\hyperlabel{equation-nonrel-nonsuch-2}%
A\left(t,\overrightarrow{x}\right)\approx A\left(\tau ,\overrightarrow{x}\right)
\end{equation}

The quantum time and standard quantum theory parts are entangled by the 
 $ e\dot{t}\Phi $~ term. In the non-{}relativistic case, if we do the integrals by steepest descents (see below 
 \hyperlink{semi}{Semi-{}classical Limit}), to lowest order the value of 
 $ \dot{t}$~ will be replaced by its average, approximately one in the non-{}relativistic case:

\begin{equation}
\label{equation-nonrel-nonrel-1}\hyperlabel{equation-nonrel-nonrel-1}%
\dot{t}\rightarrow \langle \dot{t}\rangle \approx 1
\end{equation}

This lets us separate the Lagrangian into a time part:

\begin{equation}
\label{equation-nonrel-nonrel-4b}\hyperlabel{equation-nonrel-nonrel-4b}%
\stackrel{\frown }{L}\equiv -\frac{m}{2}{\dot{t}}^{2}-\frac{m}{2}
\end{equation}

And a space part, the familiar (\cite{Feynman-1965d},
 \cite{Schulman-1981}):

\begin{equation}
\label{equation-nonrel-nonrel-5}\hyperlabel{equation-nonrel-nonrel-5}%
\overline{L}\equiv \frac{m}{2}{\dot{\overrightarrow{x}}}^{2}-e\Phi \left(\tau ,\overrightarrow{x}\right)+e\dot{\overrightarrow{x}}\cdot \overrightarrow{A}\left(\tau ,\overrightarrow{x}\right)
\end{equation}

Giving:

\begin{equation}
\label{equation-nonrel-nonrel-4}\hyperlabel{equation-nonrel-nonrel-4}%
L\approx \stackrel{\frown }{L}+\overline{L}
\end{equation}

We can factor the quantum time and space parts in the measure:

\begin{equation}
\label{equation-nonrel-nonrel-6}\hyperlabel{equation-nonrel-nonrel-6}%
\mathcal{D}x=\mathcal{D}t\mathcal{D}\overrightarrow{x}
\end{equation}

Using:

\begin{equation}
\label{equation-nonrel-nonrel-6b}\hyperlabel{equation-nonrel-nonrel-6b}%
\mathcal{D}t\equiv {\sqrt{\frac{im}{2\pi \epsilon }}}^{N}{\displaystyle \prod _{\text{j=1}}^{\text{N}}\text{d}{t}_{j}}
\end{equation}

\begin{equation}
\label{equation-nonrel-nonrel-6c}\hyperlabel{equation-nonrel-nonrel-6c}%
\mathcal{D}\overrightarrow{x}\equiv {\sqrt{\frac{m}{2\pi i\epsilon }}}^{3N}{\displaystyle \prod _{\text{j=1}}^{\text{N}}\text{d}{\overrightarrow{x}}_{j}}
\end{equation}

And therefore we can factor the kernel:

\begin{equation}
\label{equation-nonrel-nonrel-7}\hyperlabel{equation-nonrel-nonrel-7}%
{K}_{{\tau }^{{''}}{\tau }^{\prime }}\left({x}^{{''}};{x}^{\prime }\right)\approx {\stackrel{\frown }{K}}_{{\tau }^{{''}}{\tau }^{\prime }}\left({t}^{{''}};{t}^{\prime }\right){\overline{K}}_{{\tau }^{{''}}{\tau }^{\prime }}\left({\overrightarrow{x}}^{{''}};{\overrightarrow{x}}^{\prime }\right)
\end{equation}

Into a time part:

\begin{equation}
\label{equation-nonrel-nonrel-7b}\hyperlabel{equation-nonrel-nonrel-7b}%
{\stackrel{\frown }{K}}_{{\tau }^{{''}}{\tau }^{\prime }}\left({t}^{{''}};{t}^{\prime }\right)\equiv {\displaystyle \int \mathcal{D}t\mathrm{exp}\left(i{\displaystyle \underset{{\tau }^{\prime }}{\overset{{\tau }^{{''}}}{\int }}\text{d}\tau \stackrel{\frown }{L}}\right)}
\end{equation}

And the familiar space (standard quantum theory) part:

\begin{equation}
\label{equation-nonrel-nonrel-7c}\hyperlabel{equation-nonrel-nonrel-7c}%
{\overline{K}}_{{\tau }^{{''}}{\tau }^{\prime }}\left({\overrightarrow{x}}^{{''}};{\overrightarrow{x}}^{\prime }\right)={\displaystyle \int \mathcal{D}\overrightarrow{x}\mathrm{exp}\left(i{\displaystyle \underset{{\tau }^{\prime }}{\overset{{\tau }^{{''}}}{\int }}\text{d}\tau \overline{L}}\right)}
\end{equation}

Therefore we can separate the problem into a free quantum time part and the familiar standard quantum theory or space part if:
\begin{enumerate}

\item{}The velocities are non-{}relativistic,

\item{}And there is little or no time dependence in the potentials.

\end{enumerate}

And therefore the final wave functions will be products of the free wave function in quantum time and the usual standard quantum theory wave function in space.

The free wave function in quantum time will have the effect of increasing the dispersion in time. If the initial dispersion in quantum time is small and the distances traveled short, the
    effects are likely to be small, easily written off as experimental noise.

\section{Semi-{}classical Limit}\label{semi}\hyperlabel{semi}%
\begin{quote}

It is the outstanding feature of the path integral that the classical action of the system has appeared in a quantum mechanical expression, and it is this feature that is considered
        central to any extension of the path integral formalism.

\hspace*\fill---~Mark Swanson
 \cite{Swanson-1992}\end{quote}

\subsection{Overview}\label{semi-intro}\hyperlabel{semi-intro}%

In the limit as 
 $ \hslash $~ goes to zero we get the semi-{}classical approximation for standard quantum theory.

The semi-{}classical approximation for path integrals gives a clear connection to the classical picture: the classical trajectories mark the center of a quantum valley. The width of the
    quantum valley goes as 
 $ \hslash $, in the sense that as 
 $ \hslash $~ goes to zero the quantum fluctuations vanish.

From this perspective, temporal quantization is to standard quantum theory as standard quantum theory is to classical mechanics; temporal quantization posits quantum fluctuations in time
    just as standard quantum theory posits quantum fluctuations in space.

We now examine the semi-{}classical approximation in temporal quantization. We use block time. We look at:
\begin{enumerate}

\item{}The 
 \hyperlink{semi-calc}{Derivation of the Semi-{}classical Approximation};

\item{}  \hyperlink{semi-apps}{Applications of the Semi-{}classical Approximation}, specifically analyses of:
\begin{enumerate}

\item{}The 
 \hyperlink{semi-free}{Free Propagator},

\item{}  \hyperlink{semi-const}{Constant Potentials},

\item{}And the 
 \hyperlink{semi-elec}{Constant Electric Field}.

\end{enumerate}

\end{enumerate}

\subsection{Derivation of the Semi-{}classical Approximation}\label{semi-calc}\hyperlabel{semi-calc}%

To derive the semi-{}classical approximation, we rewrite the path integrals in terms of fluctuations around the classical path. We start with the path integral:

\begin{equation}
\label{equation-semi-calc-semi-calc-0}\hyperlabel{equation-semi-calc-semi-calc-0}%
{K}_{\tau }\left({x}^{{''}};{x}^{\prime }\right)={C}^{N+1}{\displaystyle \int \text{d}{x}_{1}\text{d}{x}_{2}\dots \text{d}{x}_{N}\mathrm{exp}\left(i\epsilon {\displaystyle \sum _{j=1}^{j=N+1}{L}_{j}}\right)}
\end{equation}

With:

\begin{equation}
\label{equation-semi-calc-semi-calc-1}\hyperlabel{equation-semi-calc-semi-calc-1}%
{L}_{j}\equiv -\frac{m}{2}{\left(\frac{{x}_{j}-{x}_{j-1}}{\epsilon }\right)}^{2}-e\frac{{x}_{j}-{x}_{j-1}}{\epsilon }\left(\frac{A\left({x}_{j}\right)+A\left({x}_{j-1}\right)}{2}\right)-\frac{m}{2}
\end{equation}

We rewrite the coordinates as the classical solution plus the quantum variation from that 
 $ {x}_{j}={\overline{x}}_{j}+\delta {x}_{j}$. We set the first variation with respect to 
 $ \delta {x}_{j}^{\mu }$~ to zero to get:

\begin{equation}
\label{equation-semi-calc-semi-calc-2}\hyperlabel{equation-semi-calc-semi-calc-2}%
-m\left(\frac{{\overline{x}}_{j}^{\mu }-{\overline{x}}_{j-1}^{\mu }}{{\epsilon }^{2}}\right)+m\left(\frac{{\overline{x}}_{j+1}^{\mu }-{\overline{x}}_{j}^{\mu }}{{\epsilon }^{2}}\right)-\frac{e}{2\epsilon }\left({A}^{\mu }\left({\overline{x}}_{j-1}\right)-{A}^{\mu }\left({\overline{x}}_{j+1}\right)\right)-\frac{e}{2\epsilon }\left({\overline{x}}_{j+1}^{\nu }-{\overline{x}}_{j-1}^{\nu }\right)\frac{\partial {A}_{\nu }\left({\overline{x}}_{j}\right)}{\partial {x}_{\mu }^{j}}=0
\end{equation}

In the continuum limit this is the classical equation of motion:

\begin{equation}
\label{equation-semi-calc-semi-calc-3}\hyperlabel{equation-semi-calc-semi-calc-3}%
m{\ddot{x}}_{j}^{\mu }=-e\frac{d{A}_{j}^{\mu }}{d\tau }+e\frac{d{x}_{j}^{\nu }}{d\tau }\frac{\partial {A}_{\nu }\left({x}_{j}\right)}{\partial {x}_{\mu }^{j}}=-e\left(\frac{\partial {A}^{\mu }\left({x}_{j}\right)}{\partial {x}_{\nu }^{j}}-\frac{\partial {A}_{\nu }\left({x}_{j}\right)}{\partial {x}_{\mu }^{j}}\right)\frac{d{x}_{j}^{\nu }}{d\tau }=e{F}_{\nu }^{\mu }\left({x}_{j}\right){\dot{x}}_{j}^{\nu }
\end{equation}

The full expression for the kernel is:

\begin{equation}
\label{equation-semi-calc-semi-calc-6}\hyperlabel{equation-semi-calc-semi-calc-6}%
{K}_{\tau }\left({x}^{{''}};{x}^{\prime }\right)={\displaystyle \int \text{d}{x}_{1}\text{d}{x}_{2}\dots \text{d}{x}_{N}\mathrm{exp}\left(i\epsilon {\displaystyle \sum _{j=1}^{j=N+1}{\overline{L}}_{j}+\frac{1}{2}\frac{{\partial }^{2}{\overline{L}}_{j}}{\partial \delta x\partial \delta x}\delta x\delta x+O{\left(\delta x\right)}^{3}}\right)}
\end{equation}

With:

\begin{equation}
\label{equation-semi-calc-semi-calc-7}\hyperlabel{equation-semi-calc-semi-calc-7}%
{\overline{L}}_{j}\equiv -\frac{m}{2}{\left(\frac{{\overline{x}}_{j}-{\overline{x}}_{j-1}}{\epsilon }\right)}^{2}-e\frac{{\overline{x}}_{j}-{\overline{x}}_{j-1}}{\epsilon }\left(\frac{A\left({\overline{x}}_{j}\right)+A\left({\overline{x}}_{j-1}\right)}{2}\right)-\frac{m}{2}\epsilon 
\end{equation}

This is exact when we include the cubic and higher terms. Including only terms through the quadratic we get:

\begin{equation}
\label{equation-semi-calc-semi-calc-8}\hyperlabel{equation-semi-calc-semi-calc-8}%
{K}_{\tau }\left({x}^{{''}};{x}^{\prime }\right)\approx {F}_{\tau }\left({x}^{{''}};{x}^{\prime }\right){\overline{K}}_{\tau }\left({x}^{{''}};{x}^{\prime }\right)
\end{equation}

With:

\begin{equation}
\label{equation-semi-calc-semi-calc-5}\hyperlabel{equation-semi-calc-semi-calc-5}%
{\overline{K}}_{\tau }\left({x}^{{''}};{x}^{\prime }\right)=\mathrm{exp}\left(i\epsilon {\displaystyle \sum _{j=1}^{j=N+1}{\overline{L}}_{j}}\right)
\end{equation}

And fluctuation factor 
 \emph{F}:

\begin{equation}
\label{equation-semi-calc-semi-calc-9}\hyperlabel{equation-semi-calc-semi-calc-9}%
{F}_{\tau }\left({x}^{{''}};{x}^{\prime }\right)\equiv {C}^{N+1}{\displaystyle \int \text{d}\delta {x}_{1}\text{d}\delta {x}_{2}\dots \text{d}\delta {x}_{N}\mathrm{exp}\left(i\epsilon {\displaystyle \sum _{j=1}^{j=N+1}\frac{1}{2}\frac{{\partial }^{2}{\overline{L}}_{j}}{\partial \delta {x}^{\mu }\partial \delta {x}^{\nu }}\delta {x}^{\mu }\delta {x}^{\nu }}\right)}
\end{equation}

The fluctuation factor can be computed in terms of the action and its derivatives. The derivation is nontrivial. It is treated in the one dimensional case in 
 \cite{Schulman-1981}~ and in one dimensional and higher dimensions in 
 \cite{Khandekar-1993}~ and 
 \cite{Kleinert-2004}.

To extend these derivations to apply to the four dimensional case we need to:
\begin{enumerate}

\item{}Go from three to four dimensions,

\item{}And rely on Morlet wavelets, not factors of 
 $ i\epsilon $~ or Wick rotation, for convergence.

\end{enumerate}

Neither of these changes has any material impact on the derivations. Therefore the result is the same in temporal quantization as in standard quantum theory:

\begin{equation}
\label{equation-semi-calc-semi-calc-ff-sca}\hyperlabel{equation-semi-calc-semi-calc-ff-sca}%
{K}_{\tau }\left({x}^{{''}};{x}^{\prime }\right)\approx \frac{1}{{\sqrt{2\pi i}}^{4}}\sqrt{\mathrm{det}\left(-\frac{{\partial }^{2}{S}_{\tau }\left({x}^{{''}};{x}^{\prime }\right)}{\partial {x}^{\prime }\partial {x}^{{''}}}\right)}\mathrm{exp}\left(i{S}_{\tau }\left({x}^{{''}};{x}^{\prime }\right)\right)
\end{equation}

We assume that the action is taken along the classical trajectory from 
 $ {x}^{\prime }$~ to 
 $ {x}^{{''}}$and that the determinant has not passed through a singularity on this trajectory. We will verify this for several cases.

\subsection{Applications of the Semi-{}classical Approximation}\label{semi-apps}\hyperlabel{semi-apps}%

The semi-{}classical approximation is exact for Lagrangians no worse than quadratic in the coordinates. We look at three cases:
\begin{enumerate}

\item{}The 
 \hyperlink{semi-free}{Free Propagator},

\item{}  \hyperlink{semi-const}{Constant Potentials},

\item{}And the 
 \hyperlink{semi-elec}{Constant Electric Field}.

\end{enumerate}

\subsubsection{Free Propagator}\label{semi-free}\hyperlabel{semi-free}%

\paragraph*{Free Propagator}

\noindent

The calculation of the free propagator in the semi-{}classical approximation provides a useful check. The Lagrangian is:

\begin{equation}
\label{equation-semi-free-semi-free-0}\hyperlabel{equation-semi-free-semi-free-0}%
\text{L}\left(x,\dot{x}\right)=-\frac{m}{2}{\dot{x}}^{2}-\frac{m}{2}
\end{equation}

The Euler-{}Lagrange ~ equations give:

\begin{equation}
\label{equation-semi-free-semi-free-1}\hyperlabel{equation-semi-free-semi-free-1}%
m\ddot{x}=0
\end{equation}

The classical trajectory is a straight line:

\begin{equation}
\label{equation-semi-free-semi-free-2}\hyperlabel{equation-semi-free-semi-free-2}%
{x}_{{\tau }^{\prime }}=\frac{{x}^{{''}}-{x}^{\prime }}{\tau }{\tau }^{\prime }+{x}^{\prime }
\end{equation}

The Lagrangian on the classical trajectory is constant:

\begin{equation}
\label{equation-semi-free-semi-free-3}\hyperlabel{equation-semi-free-semi-free-3}%
L=-\frac{m}{2}{\left(\frac{\Delta x}{\tau }\right)}^{2}-\frac{m}{2}
\end{equation}

The action is:

\begin{equation}
\label{equation-semi-free-semi-free-4}\hyperlabel{equation-semi-free-semi-free-4}%
S=-\frac{m}{2\tau }{\left(\Delta x\right)}^{2}-\frac{m}{2}\tau 
\end{equation}

The determinant of the action is:

\begin{equation}
\label{equation-semi-free-semi-free-5}\hyperlabel{equation-semi-free-semi-free-5}%
\mathrm{det}\left(-\frac{{\partial }^{2}S\left({x}^{{''}};{x}^{\prime }\right)}{\partial {x}^{{''}}\partial {x}^{\prime }}\right)=-\frac{{m}^{4}}{{\tau }^{4}}
\end{equation}

And the kernel is:

\begin{equation}
\label{equation-semi-free-free-fourt-kernel}\hyperlabel{equation-semi-free-free-fourt-kernel}%
{K}_{\tau }\left({x}^{{''}};{x}^{\prime }\right)=-\frac{i{m}^{2}}{4{\pi }^{2}{\tau }^{2}}\mathrm{exp}\left(-\frac{im}{2\tau }{\left({x}^{{''}}-{x}^{\prime }\right)}^{2}-i\frac{m}{2}\tau \right)
\end{equation}

This agrees with the free kernel derived in 
 \hyperlink{tq-fpi}{Feynman Path Integrals}.

\subsubsection{Constant Potentials}\label{semi-const}\hyperlabel{semi-const}%

Now we look at the case when the electric potential and vector potential are assumed constant:

\begin{equation}
\label{equation-semi-const-semi-const-3}\hyperlabel{equation-semi-const-semi-const-3}%
A=\left(\Phi ,\overrightarrow{A}\right)
\end{equation}

The Lagrangian is:

\begin{equation}
\label{equation-semi-const-semi-const-0}\hyperlabel{equation-semi-const-semi-const-0}%
L\left(x,\frac{dx}{d\tau }\right)=-\frac{m}{2}{\dot{t}}^{2}+\frac{m}{2}{\dot{\overrightarrow{x}}}^{2}-e\dot{t}\Phi +e\dot{\overrightarrow{x}}\cdot \overrightarrow{A}-\frac{m}{2}
\end{equation}

The equations of motion are unchanged.

The action is the free action plus 
 $ \Delta S$:

\begin{equation}
\label{equation-semi-const-semi-const-1}\hyperlabel{equation-semi-const-semi-const-1}%
\Delta S=-e{\displaystyle \underset{0}{\overset{\tau }{\int }}\text{d}{\tau }^{\prime }}\frac{\Delta t}{\tau }\Phi -\frac{\Delta \overrightarrow{x}}{\tau }\cdot \overrightarrow{A}=-e\Delta t\Phi +e\Delta \overrightarrow{x}\cdot \overrightarrow{A}
\end{equation}

The determinant of the action and therefore the fluctuation factor is unaffected. The kernel gets a change of phase:

\begin{equation}
\label{equation-semi-const-semi-const-2}\hyperlabel{equation-semi-const-semi-const-2}%
{K}_{\tau }^{\left(const\right)}\left({x}^{{''}};{x}^{\prime }\right)={K}_{\tau }^{\left(free\right)}\left({x}^{{''}};{x}^{\prime }\right)\mathrm{exp}\left(-ie\Delta t\Phi +ie\Delta \overrightarrow{x}\cdot \overrightarrow{A}\right)
\end{equation}

This is essentially a gauge change, with the choice of gauge:

\begin{equation}
\label{equation-semi-const-semi-const-4}\hyperlabel{equation-semi-const-semi-const-4}%
{\Lambda }_{\tau }\left(t,\overrightarrow{x}\right)=-{x}_{\mu }{A}^{\mu }
\end{equation}

For a single trajectory, this has no more impact on probabilities than any other gauge change would have.

However, if there are two (or more) classical trajectories from source to detector, different phase changes along the different paths may result in interference patterns at the detector.
    See the 
 \hyperlink{xt-ab}{Aharonov-{}Bohm Experiment}~ below.

\subsubsection{Constant Electric Field}\label{semi-elec}\hyperlabel{semi-elec}%

Now we look at the case of a constant electric field.

This is formally similar to that of the constant magnetic field, under interchange of 
 \emph{x}~ and 
 $ t$:

\begin{equation}
\label{equation-semi-elec-semi-elec-flip-0}\hyperlabel{equation-semi-elec-semi-elec-flip-0}%
\begin{array}{l}x\rightarrow y\\ t\rightarrow x\\ \overrightarrow{E}\rightarrow \overrightarrow{B}\end{array}
\end{equation}

The treatment here is modeled on the treatment of the constant magnetic field in 
 \cite{Kleinert-2004}. To emphasize the similarities, we assume we have gauged away the 
 $ -m/2$~ term in the Lagrangian (\hyperlink{tq-seqn-gauge}{Gauge Transformations for the Schrödinger Equation}).

\paragraph*{Derivation of Kernel}

\noindent

We focus on the 
 $ t$~ and 
 \emph{x}~ dimensions. We take the electric field along the 
 \emph{x}~ direction:

\begin{equation}
\label{equation-semi-elec-semi-elec-0}\hyperlabel{equation-semi-elec-semi-elec-0}%
\overrightarrow{E}=E\widehat{x}
\end{equation}

We choose the four-{}potential to be in Lorentz gauge and symmetric between 
 $ t$~ and 
 \emph{x}:

\begin{equation}
\label{equation-semi-elec-semi-elec-1}\hyperlabel{equation-semi-elec-semi-elec-1}%
A=\frac{E}{2}\left(-x,-t,0,0\right)
\end{equation}

This gives the Lagrangian:

\begin{equation}
\label{equation-semi-elec-semi-elec-2}\hyperlabel{equation-semi-elec-semi-elec-2}%
L=-\frac{m}{2}{\dot{t}}^{2}+\frac{m}{2}{\dot{\overrightarrow{x}}}^{2}+\frac{eE}{2}x\dot{t}-\frac{eE}{2}t\dot{x}
\end{equation}

And the Euler-{}Lagrange ~ equations:

\begin{equation}
\label{equation-semi-elec-semi-elec-3}\hyperlabel{equation-semi-elec-semi-elec-3}%
\begin{array}{l}\ddot{t}=\alpha \dot{x}\\ \ddot{x}=\alpha \dot{t}\end{array}
\end{equation}

Defining 
 $ \alpha \equiv eE/m$, the action for 
 \emph{x}~ and 
 $ t$~ is:

\begin{equation}
\label{equation-semi-elec-semi-elec-5}\hyperlabel{equation-semi-elec-semi-elec-5}%
{S}_{{\tau }^{{''}}{\tau }^{\prime }}^{\left(elec\right)}=\frac{m}{2}{\displaystyle \underset{0}{\overset{\tau }{\int }}\text{d}{\tau }^{\prime }\left(-{\dot{t}}^{2}+{\dot{x}}^{2}+\alpha \dot{t}x-\alpha t\dot{x}\right)}
\end{equation}

Integrating by parts we get:

\begin{equation}
\label{equation-semi-elec-semi-elec-6}\hyperlabel{equation-semi-elec-semi-elec-6}%
{S}_{{\tau }^{{''}}{\tau }^{\prime }}^{\left(elec\right)}=\frac{m}{2}{\left(-t\dot{t}+x\dot{x}\right)\vert }_{{\tau }^{\prime }}^{{\tau }^{{''}}}+\frac{m}{2}{\displaystyle \underset{{\tau }^{\prime }}{\overset{{\tau }^{{''}}}{\int }}\text{d}\tau \left(t\ddot{t}-x\ddot{x}+\alpha \dot{t}x-\alpha t\dot{x}\right)}
\end{equation}

From the equations of motion the integral is zero. Therefore the action is:

\begin{equation}
\label{equation-semi-elec-semi-elec-7}\hyperlabel{equation-semi-elec-semi-elec-7}%
{S}_{{\tau }^{{''}}{\tau }^{\prime }}^{\left(elec\right)}=\frac{m}{2}{\left(-t\dot{t}+x\dot{x}\right)\vert }_{{\tau }^{\prime }}^{{\tau }^{{''}}}=\frac{m}{2}\left(-{t}^{{''}}{\dot{t}}^{{''}}+{t}^{\prime }{\dot{t}}^{\prime }+{x}^{{''}}{\dot{x}}^{{''}}-{x}^{\prime }{\dot{x}}^{\prime }\right)
\end{equation}

We take the derivative of both sides of the Euler-{}Lagrange ~ equations with respect to laboratory time:

\begin{equation}
\label{equation-semi-elec-semi-elec-8}\hyperlabel{equation-semi-elec-semi-elec-8}%
\begin{array}{l}\dot{\ddot{t}}={\alpha }^{2}\dot{t}\\ \dot{\ddot{x}}={\alpha }^{2}\dot{x}\end{array}
\end{equation}

The solutions are:

\begin{equation}
\label{equation-semi-elec-semi-elec-9}\hyperlabel{equation-semi-elec-semi-elec-9}%
\begin{array}{l}t=\frac{1}{\mathrm{sinh}\left(\alpha \Delta \tau \right)}\left(\left({t}^{{''}}-{t}_{0}\right)\mathrm{sinh}\left(\alpha \left(\tau -{\tau }^{\prime }\right)\right)-\left({t}^{\prime }-{t}_{0}\right)\mathrm{sinh}\left(\alpha \left(\tau -{\tau }^{{''}}\right)\right)\right)+{t}_{0}\\ x=\frac{1}{\mathrm{sinh}\left(\alpha \Delta \tau \right)}\left(\left({x}^{{''}}-{x}_{0}\right)\mathrm{sinh}\left(\alpha \left(\tau -{\tau }^{\prime }\right)\right)-\left({x}^{\prime }-{x}_{0}\right)\mathrm{sinh}\left(\alpha \left(\tau -{\tau }^{{''}}\right)\right)\right)+{x}_{0}\end{array}
\end{equation}

The constants of integration can be gotten by requiring we satisfy the Euler-{}Lagrange ~ equations at the endpoints:

\begin{equation}
\label{equation-semi-elec-semi-elec-9b}\hyperlabel{equation-semi-elec-semi-elec-9b}%
\begin{array}{l}{\ddot{t}}^{\prime }=\alpha {\dot{x}}^{\prime },{\ddot{x}}^{\prime }=\alpha {\dot{t}}^{\prime }\\ {\ddot{t}}^{{''}}=\alpha {\dot{x}}^{{''}},{\ddot{x}}^{{''}}=\alpha {\dot{t}}^{{''}}\end{array}
\end{equation}

Giving:

\begin{equation}
\label{equation-semi-elec-semi-elec-a}\hyperlabel{equation-semi-elec-semi-elec-a}%
\begin{array}{l}{t}_{0}=\frac{1}{2}\left(\left({t}^{{''}}+{t}^{\prime }\right)-\left({x}^{{''}}-{x}^{\prime }\right)\mathrm{coth}\left(\frac{\alpha \Delta \tau }{2}\right)\right)\\ {x}_{0}=\frac{1}{2}\left(\left({x}^{{''}}+{x}^{\prime }\right)-\left({t}^{{''}}-{t}^{\prime }\right)\mathrm{coth}\left(\frac{\alpha \Delta \tau }{2}\right)\right)\end{array}
\end{equation}

The action is:

\begin{equation}
\label{equation-semi-elec-semi-elec-c}\hyperlabel{equation-semi-elec-semi-elec-c}%
{S}_{{\tau }^{{''}}{\tau }^{\prime }}^{\left(elec\right)}\left({t}^{{''}},{x}^{{''}};{t}^{\prime },{x}^{\prime }\right)=\frac{m}{2}\left(\frac{\alpha }{2}\mathrm{coth}\left(\frac{\alpha \Delta \tau }{2}\right)\left(-{\left({t}^{{''}}-{t}^{\prime }\right)}^{2}+{\left({x}^{{''}}-{x}^{\prime }\right)}^{2}\right)+\alpha \left({x}^{\prime }{t}^{{''}}-{x}^{{''}}{t}^{\prime }\right)\right)
\end{equation}

The first term is the square of the Minkowski distance times a coefficient. The second term is a gauge term. The determinant of the action is:

\begin{equation}
\label{equation-semi-elec-semi-elec-d}\hyperlabel{equation-semi-elec-semi-elec-d}%
\mathrm{det}\left(-\frac{{\partial }^{{''}}{\partial }^{\prime }{S}_{{\tau }^{{''}}{\tau }^{\prime }}^{\left(elec\right)}}{\partial {x}^{{''}}\partial {x}^{\prime }}\right)=\frac{{m}^{2}{\alpha }^{2}}{4}\left(-{\mathrm{coth}}^{2}\left(\frac{\alpha \Delta \tau }{2}\right)+1\right)=-\frac{{m}^{2}{\alpha }^{2}}{4{\mathrm{sinh}}^{2}\left(\frac{\alpha \Delta \tau }{2}\right)}
\end{equation}

The full kernel is therefore:

\begin{equation}
\label{equation-semi-elec-semi-elec-kernel}\hyperlabel{equation-semi-elec-semi-elec-kernel}%
{K}_{{\tau }^{{''}}{\tau }^{\prime }}^{\left(elec\right)}\left({t}^{{''}},{x}^{{''}};{t}^{\prime },{x}^{\prime }\right)=\frac{m\alpha }{4\pi \mathrm{sinh}\left(\frac{\alpha \Delta \tau }{2}\right)}\mathrm{exp}\left(i{S}_{{\tau }^{{''}}{\tau }^{\prime }}^{\left(elec\right)}\left({t}^{{''}},{x}^{{''}};{t}^{\prime },{x}^{\prime }\right)\right)
\end{equation}

As a quick check, we see that in the limit as the electric field goes to zero we recover the free kernel (in two dimensions):

\begin{equation}
\label{equation-semi-elec-semi-elec-tofree}\hyperlabel{equation-semi-elec-semi-elec-tofree}%
\underset{\alpha \rightarrow 0}{\mathrm{lim}}{K}_{{\tau }^{{''}}{\tau }^{\prime }}^{\left(elec\right)}\left({t}^{{''}},{x}^{{''}};{t}^{\prime },{x}^{\prime }\right)=\frac{m}{2\pi \Delta \tau }\mathrm{exp}\left(i\frac{m}{2\Delta \tau }\left(-{\left({t}^{{''}}-{t}^{\prime }\right)}^{2}+{\left({x}^{{''}}-{x}^{\prime }\right)}^{2}\right)\right)={K}_{{\tau }^{{''}}{\tau }^{\prime }}^{\left(free\right)}\left({t}^{{''}},{x}^{{''}};{t}^{\prime },{x}^{\prime }\right)
\end{equation}

\paragraph*{Meaning of Gauge}

\noindent

Consider a coordinate change:

\begin{equation}
\label{equation-semi-elec-semi-elec-f}\hyperlabel{equation-semi-elec-semi-elec-f}%
t,x\rightarrow t+{d}_{t},x+{d}_{x}
\end{equation}

This induces a change in the action:

\begin{equation}
\label{equation-semi-elec-semi-elec-g}\hyperlabel{equation-semi-elec-semi-elec-g}%
\Delta S=\frac{m\alpha }{2}\left({d}_{x}\left({t}^{{''}}-{t}^{\prime }\right)-{d}_{t}\left({x}^{{''}}-{x}^{\prime }\right)\right)
\end{equation}

This is a gauge term:

\begin{equation}
\label{equation-semi-elec-semi-elec-h}\hyperlabel{equation-semi-elec-semi-elec-h}%
\mathrm{exp}\left(ie{\Lambda }^{{''}}-ie{\Lambda }^{\prime }\right)
\end{equation}

With the gauge function:

\begin{equation}
\label{equation-semi-elec-semi-elec-i}\hyperlabel{equation-semi-elec-semi-elec-i}%
\Lambda =-\frac{m\alpha }{2e}\left({d}_{x}t-{d}_{t}x\right)
\end{equation}

So a coordinate change induces a gauge change in the kernel.

\paragraph*{Comparison To Magnetic Kernel}

\noindent

Consider a magnetic field in the 
 \emph{z}~ direction:

\begin{equation}
\label{equation-semi-elec-semi-elec-mag-0}\hyperlabel{equation-semi-elec-semi-elec-mag-0}%
\overrightarrow{B}=\left(0,0,B\right)
\end{equation}

With vector potential:

\begin{equation}
\label{equation-semi-elec-semi-elec-mag-1}\hyperlabel{equation-semi-elec-semi-elec-mag-1}%
\overrightarrow{A}=\frac{B}{2}\left(-y,x,0\right)
\end{equation}

The magnetic Lagrangian for temporal quantization is:

\begin{equation}
\label{equation-semi-elec-semi-elec-mag-2}\hyperlabel{equation-semi-elec-semi-elec-mag-2}%
L=-\frac{m}{2}{\dot{t}}^{2}+\frac{m}{2}{\dot{\overrightarrow{x}}}^{2}-\frac{eB}{2}\dot{x}y+\frac{eB}{2}\dot{y}x
\end{equation}

If we drop the 
 $ \dot{t}$~ squared ~ term we have the standard quantum theory magnetic Lagrangian.

Euler-{}Lagrange equations:

\begin{equation}
\label{equation-semi-elec-semi-elec-mag-3}\hyperlabel{equation-semi-elec-semi-elec-mag-3}%
\begin{array}{l}\ddot{x}=\omega \dot{y}\\ \ddot{y}=-\omega \dot{x}\end{array}
\end{equation}

With the definition of the Larmor frequency:

\begin{equation}
\label{equation-semi-elec-semi-elec-mag-larmor}\hyperlabel{equation-semi-elec-semi-elec-mag-larmor}%
\omega \equiv \frac{eB}{m}
\end{equation}

The trajectories are:

\begin{equation}
\label{equation-semi-elec-semi-elec-mag-5}\hyperlabel{equation-semi-elec-semi-elec-mag-5}%
\begin{array}{l}x=\frac{1}{\mathrm{sin}\left(\omega \left({\tau }^{{''}}-{\tau }^{\prime }\right)\right)}\left(\left({x}^{{''}}-{x}_{0}\right)\mathrm{sin}\left(\omega \left(\tau -{\tau }^{\prime }\right)\right)-\left({x}^{\prime }-{x}_{0}\right)\mathrm{sin}\left(\omega \left(\tau -{\tau }^{{''}}\right)\right)\right)\\ y=\frac{1}{\mathrm{sin}\left(\omega \left({\tau }^{{''}}-{\tau }^{\prime }\right)\right)}\left(\left({y}^{{''}}-{y}_{0}\right)\mathrm{sin}\left(\omega \left(\tau -{\tau }^{\prime }\right)\right)-\left({y}^{\prime }-{y}_{0}\right)\mathrm{sin}\left(\omega \left(\tau -{\tau }^{{''}}\right)\right)\right)\end{array}
\end{equation}

With constants of the motion:

\begin{equation}
\label{equation-semi-elec-semi-elec-mag-6}\hyperlabel{equation-semi-elec-semi-elec-mag-6}%
\begin{array}{l}{x}_{0}=\frac{1}{2}\left(\left({x}^{{''}}+{x}^{\prime }\right)+\left({y}^{{''}}-{y}^{\prime }\right)\mathrm{cot}\left(\frac{\omega }{2}\left({\tau }^{{''}}-{\tau }^{\prime }\right)\right)\right)\\ {y}_{0}=\frac{1}{2}\left(\left({y}^{{''}}+{y}^{\prime }\right)-\left({x}^{{''}}-{x}^{\prime }\right)\mathrm{cot}\left(\frac{\omega }{2}\left({\tau }^{{''}}-{\tau }^{\prime }\right)\right)\right)\end{array}
\end{equation}

Integrating the Lagrangian as a function of 
 $ \tau $~ along the trajectories we get the magnetic action:

\begin{equation}
\label{equation-semi-elec-semi-elec-mag-c}\hyperlabel{equation-semi-elec-semi-elec-mag-c}%
{S}_{{\tau }^{{''}}{\tau }^{\prime }}^{\left(mag\right)}\left({x}^{{''}},{y}^{{''}};{x}^{\prime },{y}^{\prime }\right)=\frac{m}{2}\left(\frac{\omega }{2}\mathrm{cot}\left(\frac{\omega \Delta \tau }{2}\right)\left({\left({x}^{{''}}-{x}^{\prime }\right)}^{2}+{\left({y}^{{''}}-{y}^{\prime }\right)}^{2}\right)+\omega \left({x}^{\prime }{y}^{{''}}-{x}^{{''}}{y}^{\prime }\right)\right)
\end{equation}

The magnetic kernel is:

\begin{equation}
\label{equation-semi-elec-semi-elec-mag-e}\hyperlabel{equation-semi-elec-semi-elec-mag-e}%
{K}_{{\tau }^{{''}}{\tau }^{\prime }}^{\left(mag\right)}\left({x}^{{''}},{y}^{{''}};{x}^{\prime },{y}^{\prime }\right)=-\frac{1}{4\pi }\frac{m\omega }{\mathrm{sin}\left(\frac{\omega \Delta \tau }{2}\right)}\mathrm{exp}\left(i{S}_{{\tau }^{{''}}{\tau }^{\prime }}^{\left(mag\right)}\left({x}^{{''}},{y}^{{''}};{x}^{\prime },{y}^{\prime }\right)\right)
\end{equation}

In the limit as 
 $ \omega $~ goes to zero we get the free kernel:

\begin{equation}
\label{equation-semi-elec-semi-elec-mag-g}\hyperlabel{equation-semi-elec-semi-elec-mag-g}%
\underset{\omega \rightarrow 0}{\mathrm{lim}}{K}_{{\tau }^{{''}}{\tau }^{\prime }}^{\left(mag\right)}\left({x}^{{''}},{y}^{{''}};{x}^{\prime },{y}^{\prime }\right)=-\frac{1}{2\pi }\frac{m}{\Delta \tau }\mathrm{exp}\left(i\frac{m}{2\Delta \tau }\left({\left({x}^{{''}}-{x}^{\prime }\right)}^{2}+{\left({y}^{{''}}-{y}^{\prime }\right)}^{2}\right)\right)
\end{equation}

We can go from the electric to the magnetic kernel with the substitutions
 \footnote{Because of the different conventions we chose for the time and space parts of the kernel, there is an overall factor of -{}1 difference between the magnetic and electric kernels as
    well.}:

\begin{equation}
\label{equation-semi-elec-semi-elec-mag-f}\hyperlabel{equation-semi-elec-semi-elec-mag-f}%
t,x,\alpha \rightarrow ix,y,i\omega 
\end{equation}

\paragraph*{Verification}

\noindent

As the electric Lagrangian is no worse than quadratic in the coordinates we expect our kernel will represent an exact solution. We verify this.

The Schrödinger equation is:

\begin{equation}
\label{equation-semi-elec-semi-elec-seqn-seqn}\hyperlabel{equation-semi-elec-semi-elec-seqn-seqn}%
i\frac{d}{d\tau }{\psi }_{\tau }\left(t,x\right)={H}^{\left(elec\right)}{\psi }_{\tau }\left(x\right)=-\frac{1}{2m}\left({\left(i{\partial }_{t}+\frac{eE}{2}x\right)}^{2}-{\left(-i{\partial }_{x}+\frac{eE}{2}t\right)}^{2}\right){\psi }_{\tau }\left(x\right)
\end{equation}

When 
 $ \tau \text{>0}$~ we verify by explicit calculation that the kernel satisfies the Schrödinger equation:

\begin{equation}
\label{equation-semi-elec-semi-elec-seqn-2}\hyperlabel{equation-semi-elec-semi-elec-seqn-2}%
\left(i\frac{d}{d\tau }-{H}^{\left(elec\right)}\right){K}_{\tau }^{\left(elec\right)}=0
\end{equation}

When 
 $ \tau $~ goes to zero, we require:

\begin{equation}
\label{equation-semi-elec-semi-elec-seqn-3}\hyperlabel{equation-semi-elec-semi-elec-seqn-3}%
\tau \rightarrow 0\Rightarrow {K}_{\tau }^{\left(elec\right)}\left(t,x;{t}^{\prime },{x}^{\prime }\right)\rightarrow \delta \left(t-{t}^{\prime }\right)\delta \left(x-{x}^{\prime }\right)
\end{equation}

For short laboratory times, we have:

\begin{equation}
\label{equation-semi-elec-semi-elec-short-1}\hyperlabel{equation-semi-elec-semi-elec-short-1}%
{K}_{\tau }^{\left(elec\right)}\left({x}^{{''}};{x}^{\prime }\right)\rightarrow \mathrm{exp}\left(i\frac{m\alpha }{2}\left({x}^{\prime }{t}^{{''}}-{x}^{{''}}{t}^{\prime }\right)\right){K}_{\tau }^{\left(free\right)}\left({x}^{{''}};{x}^{\prime }\right)
\end{equation}

For short times the free kernel becomes a 
 $ \delta $~ function:

\begin{equation}
\label{equation-semi-elec-semi-elec-short-2}\hyperlabel{equation-semi-elec-semi-elec-short-2}%
\underset{\tau \rightarrow 0}{\mathrm{lim}}{K}_{\tau }^{\left(free\right)}\left({x}^{{''}};{x}^{\prime }\right)\rightarrow \delta \left({t}^{{''}}-{t}^{\prime }\right)\delta \left({x}^{{''}}-{x}^{\prime }\right)
\end{equation}

The 
 $ \delta $~ functions in turn force the gauge factor to one:

\begin{equation}
\label{equation-semi-elec-semi-elec-short-3}\hyperlabel{equation-semi-elec-semi-elec-short-3}%
{x}^{{''}}={x}^{\prime },{t}^{{''}}={t}^{\prime }\Rightarrow \mathrm{exp}\left(i\frac{m\alpha }{2}\left({x}^{\prime }{t}^{{''}}-{x}^{{''}}{t}^{\prime }\right)\right)\rightarrow 1
\end{equation}

\section{Long Time Limit}\label{stat}\hyperlabel{stat}%

\subsection{Overview}\label{stat-intro}\hyperlabel{stat-intro}%

Over longer times we expect that the effects of temporal quantization will average out. If the fluctuations in time are small and high frequency, then slowly changing visitors from outside
    \textendash{} photons in the visible light range and the like \textendash{} simply will not see them. For most practical purposes, interactions will be dominated by the slowly changing or stationary parts of the
    temporal quantization wave function.

Consider the Schrödinger equation in temporal quantization:

\begin{equation}
\label{equation-stat-intro-stat-3}\hyperlabel{equation-stat-intro-stat-3}%
i\frac{\partial }{\partial \tau }{\psi }_{\tau }\left(t,\overrightarrow{x}\right)=H\left(t,\overrightarrow{x}\right){\psi }_{\tau }\left(t,\overrightarrow{x}\right)
\end{equation}

With the Hamiltonian:

\begin{equation}
\label{equation-stat-intro-stat-4}\hyperlabel{equation-stat-intro-stat-4}%
H\left(t,\overrightarrow{x}\right)=-\frac{{\left(p-eA\right)}^{2}-{m}^{2}}{2m}
\end{equation}

The stationary solutions are given by:

\begin{equation}
\label{equation-stat-intro-stat-5}\hyperlabel{equation-stat-intro-stat-5}%
H\left(t,\overrightarrow{x}\right){\psi }_{\tau }\left(t,\overrightarrow{x}\right)=0
\end{equation}

The stationary solutions are solutions of the Klein-{}Gordon equation, a promising development and the motivation for making the choice of additive constant that we did in 
 \hyperlink{tq-fpi-lag}{Lagrangian}.

Naively, we expect deviations from compliance with the Klein-{}Gordon equation will be of order 
 $ 1/2m$. This is (one-{}half of) the natural time scale (the Compton time) associated with a particle, the Compton wave length divided by speed of light. For an electron this is:

\begin{equation}
\label{equation-stat-intro-stat-7}\hyperlabel{equation-stat-intro-stat-7}%
{\tau }_{e}\sim \frac{\hslash }{{m}_{e}{c}^{2}}=\frac{6.58\cdot {10}^{-16}\text{eV-s}}{.511\cdot {10}^{6}eV}=1.29\cdot {10}^{-21}\text{s}=1.29\text{zs}
\end{equation}

Where 
 \emph{zs}~ is zeptoseconds, not a large unit of time, even by modern standards
 \footnote{It is apparently possible to look for effects at this scale, see Hestenes's and Catillion's papers 
 \cite{Hestenes-1990},
 \cite{Hestenes-2008},
 \cite{Hestenes-2008b},
 \cite{Hestenes-2008c},
 \cite{Catillon-2008}. Amazing. Hestenes and Catillion 
 \emph{et al}~ are looking for effects associated with Zitterbewegung. It is possible that variations on their experiments could be developed to look for quantum fluctuations in
    time.}.

With respect to interactions on a scale much slower than this, physical systems will be dominated by the solutions of the Klein-{}Gordon equation.

Now consider the case where we are looking at stationary solutions and can separate the time and space parts, i.e. we are dealing with static or slowly changing potentials.

The variation of the space part with laboratory time will go as:

\begin{equation}
\label{equation-stat-intro-stat-0}\hyperlabel{equation-stat-intro-stat-0}%
\xi \sim \mathrm{exp}\left(-i\mathscr{E}\tau \right)
\end{equation}

To keep the state as a whole stationary the variation of the time part with laboratory time must go as:

\begin{equation}
\label{equation-stat-intro-stat-1}\hyperlabel{equation-stat-intro-stat-1}%
\chi \sim \mathrm{exp}\left(i\mathscr{E}\tau \right)
\end{equation}

Thereby leaving the overall state stationary:

\begin{equation}
\label{equation-stat-intro-stat-2}\hyperlabel{equation-stat-intro-stat-2}%
\psi \sim 1
\end{equation}

This is an amusing picture: for stationary states the variations in time and space cancel each other out, like contra-{}rotating propellers where the two propellers spin in opposite senses
 \footnote{An arrangement which reduces yaw and improves efficiency.}.

We think the wave functions are moving in time because we have only been considering the space part; in reality or at least in temporal quantization the total wave function is stationary.
    Parmenides is pleased; Heraclitus is tapping his fingers waiting for us to explain how then standard quantum theory works so well.

Stationary states give a natural connection from temporal quantization to standard quantum theory: the standard quantum theory states are the stationary subset of all temporal quantization
    states.

To complete our understanding of the connection we will look at two cases:
\begin{enumerate}

\item{}  \hyperlink{scatter}{Non-{}singular Potentials},

\item{}And 
 \hyperlink{bound}{Bound States}.

\end{enumerate}

\subsection{Non-{}singular Potentials}\label{scatter}\hyperlabel{scatter}%

\subsubsection{Schrödinger Equation in Relative Time}\label{scatter-intro}\hyperlabel{scatter-intro}%

Consider the Schrödinger equation with the time dependence broken out:

\begin{equation}
\label{equation-scatter-intro-scatter-def-2}\hyperlabel{equation-scatter-intro-scatter-def-2}%
i\frac{d{\psi }_{\tau }\left(t,\overrightarrow{x}\right)}{d\tau }=\frac{1}{2m}{\left({\partial }_{t}+ie\Phi \left(t,\overrightarrow{x}\right)\right)}^{2}{\psi }_{\tau }\left(t,\overrightarrow{x}\right)-\frac{1}{2m}{\left(\nabla -ie\overrightarrow{A}\left(t,\overrightarrow{x}\right)\right)}^{2}{\psi }_{\tau }\left(t,\overrightarrow{x}\right)+\frac{m}{2}{\psi }_{\tau }\left(t,\overrightarrow{x}\right)
\end{equation}

We rewrite the Schrödinger equation in relative time 
 $ \langle {t}_{\tau }\rangle $~ getting:

\begin{equation}
\label{equation-scatter-intro-scatter-def-3}\hyperlabel{equation-scatter-intro-scatter-def-3}%
i\frac{d{\psi }_{\tau }^{\left(rel\right)}\left({t}_{\tau },\overrightarrow{x}\right)}{d\tau }=\left(i{\partial }_{{t}_{\tau }}+\frac{1}{2m}{\left({\partial }_{{t}_{\tau }}+ie{\Phi }_{\tau }\left({t}_{\tau },\overrightarrow{x}\right)\right)}^{2}-\frac{1}{2m}{\left(\nabla -ie{\overrightarrow{A}}_{\tau }\left({t}_{\tau },\overrightarrow{x}\right)\right)}^{2}+\frac{m}{2}\right){\psi }_{\tau }^{\left(rel\right)}\left({t}_{\tau },\overrightarrow{x}\right)
\end{equation}

And in energy momentum space:

\begin{equation}
\label{equation-scatter-intro-scatter-def-4}\hyperlabel{equation-scatter-intro-scatter-def-4}%
i\frac{d{\widehat{\psi }}_{\tau }^{\left(rel\right)}\left(E,\overrightarrow{p}\right)}{d\tau }=\left(E-\frac{1}{2m}{\left(E-e{\Phi }_{\tau }\left({t}_{\tau },\overrightarrow{x}\right)\right)}^{2}+\frac{1}{2m}{\left(\overrightarrow{p}-e{\overrightarrow{A}}_{\tau }\left({t}_{\tau },\overrightarrow{x}\right)\right)}^{2}+\frac{m}{2}\right){\widehat{\psi }}_{\tau }^{\left(rel\right)}\left(E,\overrightarrow{p}\right)
\end{equation}

With the notation:

\begin{equation}
\label{equation-scatter-intro-scatter-def-5}\hyperlabel{equation-scatter-intro-scatter-def-5}%
\begin{array}{l}{\Phi }_{\tau }\left({t}_{\tau },\overrightarrow{x}\right)\equiv \Phi \left(\tau +{t}_{\tau },\overrightarrow{x}\right)\\ {\overrightarrow{A}}_{\tau }\left({t}_{\tau },\overrightarrow{x}\right)\equiv \overrightarrow{A}\left(\tau +{t}_{\tau },\overrightarrow{x}\right)\end{array}
\end{equation}

We will look here at
 \footnote{The solutions to the free Schrödinger equation in relative time are given in an appendix 
 \hyperlink{free}{Free Particles}.}:
\begin{enumerate}

\item{}The 
 \hyperlink{scatter-magx}{Time Independent Magnetic Field},

\item{}The 
 \hyperlink{scatter-magt}{Time Dependent Magnetic Field},

\item{}The 
 \hyperlink{scatter-elecx}{Time Independent Electric Field},

\item{}The 
 \hyperlink{scatter-elect}{Time Dependent Electric Field},

\item{}And 
 \hyperlink{scatter-gen}{General Fields}.

\end{enumerate}

\subsubsection{Time Independent Magnetic Field}\label{scatter-magx}\hyperlabel{scatter-magx}%

For a time independent magnetic field the Schrödinger equation factors into a free quantum time part and the usual standard quantum theory part:

\begin{equation}
\label{equation-scatter-magx-scatter-magx-0}\hyperlabel{equation-scatter-magx-scatter-magx-0}%
i{\dot{\psi }}_{\tau }^{\left(rel\right)}\left({t}_{\tau },\overrightarrow{x}\right)={\stackrel{\frown }{H}}^{\left(free\right)}\left({t}_{\tau }\right){\psi }_{\tau }^{\left(rel\right)}\left({t}_{\tau },\overrightarrow{x}\right)+{\overline{H}}^{\left(mag\right)}\left(\overrightarrow{x}\right){\psi }_{\tau }^{\left(rel\right)}\left({t}_{\tau },\overrightarrow{x}\right)
\end{equation}

With free quantum time part:

\begin{equation}
\label{equation-scatter-magx-scatter-magx-hamfreettau}\hyperlabel{equation-scatter-magx-scatter-magx-hamfreettau}%
{\stackrel{\frown }{H}}^{\left(free\right)}\left({t}_{\tau }\right)=i{\partial }_{{t}_{\tau }}+\frac{1}{2m}{\partial }_{{t}_{\tau }}^{2}+\frac{m}{2}
\end{equation}

And space part the usual:

\begin{equation}
\label{equation-scatter-magx-scatter-magx-3}\hyperlabel{equation-scatter-magx-scatter-magx-3}%
{\overline{H}}^{\left(mag\right)}\left(\overrightarrow{x}\right)\equiv \frac{1}{2m}{\left(\overrightarrow{p}-e\overrightarrow{A}\left(\overrightarrow{x}\right)\right)}^{2}
\end{equation}

Therefore the full solutions are given by sums over the direct product of the plane wave solutions to the free quantum time part and the standard quantum theory solutions to the magnetic
    standard quantum theory part.

We expect additional dispersion in time, along the lines in the free case (\hyperlink{free}{Free Particles}). We do not expect anything qualitatively new.

\subsubsection{Time Dependent Magnetic Field}\label{scatter-magt}\hyperlabel{scatter-magt}%

For a magnetic field varying in time the Schrödinger equation in relative time is:

\begin{equation}
\label{equation-scatter-magt-scatter-magt-0}\hyperlabel{equation-scatter-magt-scatter-magt-0}%
i{\dot{\psi }}_{\tau }^{\left(rel\right)}\left({t}_{\tau },x\right)={\stackrel{\frown }{H}}^{\left(free\right)}\left({t}_{\tau }\right){\psi }_{\tau }^{\left(rel\right)}\left({t}_{\tau },x\right)+\frac{1}{2m}{\left(\overrightarrow{p}-e{\overrightarrow{A}}_{\tau }\left({t}_{\tau },\overrightarrow{x}\right)\right)}^{2}{\psi }_{\tau }^{\left(rel\right)}\left({t}_{\tau },x\right)
\end{equation}

The standard quantum theory Hamiltonian is ($ {\overrightarrow{A}}_{\tau }\left(\overrightarrow{x}\right)\equiv {\overrightarrow{A}}_{\tau }\left(0,\overrightarrow{x}\right)$):

\begin{equation}
\label{equation-scatter-magt-scatter-magt-hambarmag}\hyperlabel{equation-scatter-magt-scatter-magt-hambarmag}%
{\overline{H}}_{\tau }^{\left(mag\right)}\left(\overrightarrow{x}\right)\equiv \frac{1}{2m}{\left(\overrightarrow{p}-e{\overrightarrow{A}}_{\tau }\left(\overrightarrow{x}\right)\right)}^{2}
\end{equation}

We write the temporal quantization vector potential as:

\begin{equation}
\label{equation-scatter-magt-scatter-magt-3}\hyperlabel{equation-scatter-magt-scatter-magt-3}%
{\overrightarrow{A}}_{\tau }\left({t}_{\tau },\overrightarrow{x}\right)={\overrightarrow{A}}_{\tau }\left({t}_{\tau },\overrightarrow{x}\right)-{\overrightarrow{A}}_{\tau }\left(0,\overrightarrow{x}\right)+{\overrightarrow{A}}_{\tau }\left(0,\overrightarrow{x}\right)={\overrightarrow{A}}_{\tau }\left(\overrightarrow{x}\right)+\langle \frac{\partial {\overrightarrow{A}}_{\tau }}{\partial {t}_{\tau }}\left({t}_{\tau },\overrightarrow{x}\right)\rangle {t}_{\tau }
\end{equation}

With the average derivative of the vector potential with respect to the relative time defined as:

\begin{equation}
\label{equation-scatter-magt-scatter-magt-3b}\hyperlabel{equation-scatter-magt-scatter-magt-3b}%
\langle \frac{\partial {\overrightarrow{A}}_{\tau }\left({t}_{\tau },\overrightarrow{x}\right)}{\partial {t}_{\tau }}\rangle \equiv \frac{1}{{t}_{\tau }}{\displaystyle \underset{0}{\overset{{t}_{\tau }}{\int }}\text{d}{{t}^{\prime }}_{\tau }\frac{\partial {\overrightarrow{A}}_{\tau }\left({{t}^{\prime }}_{\tau },\overrightarrow{x}\right)}{\partial {{t}^{\prime }}_{\tau }}}
\end{equation}

The Schrödinger equation is now:

\begin{equation}
\label{equation-scatter-magt-scatter-magt-seqn}\hyperlabel{equation-scatter-magt-scatter-magt-seqn}%
i\frac{d{\psi }_{\tau }^{\left(rel\right)}\left({t}_{\tau },x\right)}{d\tau }=\left({\stackrel{\frown }{H}}^{\left(free\right)}\left({t}_{\tau }\right)+{\overline{H}}_{\tau }^{\left(mag\right)}\left(\overrightarrow{x}\right)+{V}_{\tau }^{\left(mag\right)\left(1\right)}\left({t}_{\tau },\overrightarrow{x}\right){t}_{\tau }+{V}_{\tau }^{\left(mag\right)\left(2\right)}\left({t}_{\tau },\overrightarrow{x}\right){t}_{\tau }^{2}\right){\psi }_{\tau }^{\left(rel\right)}\left({t}_{\tau },x\right)
\end{equation}

With the linear relative time correction:

\begin{equation}
\label{equation-scatter-magt-scatter-magt-first}\hyperlabel{equation-scatter-magt-scatter-magt-first}%
{V}_{\tau }^{\left(mag\right)\left(1\right)}\left({t}_{\tau },\overrightarrow{x}\right)=-e\frac{\left(\overrightarrow{p}-e{\overrightarrow{A}}_{\tau }\left(\overrightarrow{x}\right)\right).\langle \frac{\partial {\overrightarrow{A}}_{\tau }}{\partial {t}_{\tau }}\left({t}_{\tau },\overrightarrow{x}\right)\rangle +\langle \frac{\partial {\overrightarrow{A}}_{\tau }}{\partial {t}_{\tau }}\left({t}_{\tau },\overrightarrow{x}\right)\rangle \cdot \left(\overrightarrow{p}-e{\overrightarrow{A}}_{\tau }\left(\overrightarrow{x}\right)\right)}{2m}
\end{equation}

And the quadratic relative time correction:

\begin{equation}
\label{equation-scatter-magt-scatter-magt-second}\hyperlabel{equation-scatter-magt-scatter-magt-second}%
{V}_{\tau }^{\left(mag\right)\left(2\right)}\left({t}_{\tau },\overrightarrow{x}\right)={e}^{2}\frac{{\langle \frac{\partial {\overrightarrow{A}}_{\tau }}{\partial {t}_{\tau }}\left({t}_{\tau },\overrightarrow{x}\right)\rangle }^{2}}{2m}
\end{equation}

This is exact.

If the vector potential is constant with respect to time, there is no correction.

The corrections are proportional to linear and higher terms in the relative time. If we know the standard quantum theory solutions we can get the temporal quantization corrections via
    perturbation theory, by assuming that 
 $ {t}_{\tau }$~ and 
 $ {t}_{\tau }^{2}$~ are small. We do this below in 
 \hyperlink{xt-magt}{Experimental Tests/Electric and Magnetic Fields/Time Dependent Magnetic Fields}.

Time-{}varying magnetic fields will be associated with time-{}varying vector potentials. These will in turn induce an electric field, the displacement current:

\begin{equation}
\label{equation-scatter-magt-scatter-magt-dispcur}\hyperlabel{equation-scatter-magt-scatter-magt-dispcur}%
\overrightarrow{E}=-\frac{\partial \overrightarrow{A}}{\partial t}
\end{equation}

This is already accounted for in the Hamiltonian, via the vector potential, so requires no further attention.

\subsubsection{Time Independent Electric Field}\label{scatter-elecx}\hyperlabel{scatter-elecx}%

Recalling the derivation of the Schrödinger equation (\hyperlink{tq-seqn-deriv}{Derivation of the Schrödinger Equation}), the 
 $ e\dot{t}\Phi $~ coupling leads to 
 $ \Phi $~ and 
 $ {\Phi }^{2}$~ terms in the Hamiltonian:

\begin{equation}
\label{equation-scatter-elecx-scatter-electric-ham_static}\hyperlabel{equation-scatter-elecx-scatter-electric-ham_static}%
i\frac{d}{d\tau }{\psi }_{\tau }^{\left(rel\right)}\left({t}_{\tau },\overrightarrow{x}\right)=i{\partial }_{{t}_{\tau }}{\psi }_{\tau }^{\left(rel\right)}\left({t}_{\tau },\overrightarrow{x}\right)+\left(-\frac{1}{2m}{\left(i\frac{\partial }{\partial {t}_{\tau }}-e\Phi \left(t,\overrightarrow{x}\right)\right)}^{2}+\frac{{\overrightarrow{p}}^{2}}{2m}+\frac{m}{2}\right){\psi }_{\tau }^{\left(rel\right)}\left({t}_{\tau },\overrightarrow{x}\right)
\end{equation}

This is not at all like the Hamiltonian in standard quantum theory, where 
 $ \Phi $~ enters linearly:

\begin{equation}
\label{equation-scatter-elecx-scatter-electric-ham_sqt}\hyperlabel{equation-scatter-elecx-scatter-electric-ham_sqt}%
\overline{H}\left(\overrightarrow{x}\right)=\frac{{\overrightarrow{p}}^{2}}{2m}+e\Phi \left(\overrightarrow{x}\right)
\end{equation}

Our goal is to write the Hamiltonian in the form:

\begin{equation}
\label{equation-scatter-elecx-scatter-electric-ham}\hyperlabel{equation-scatter-elecx-scatter-electric-ham}%
H={\stackrel{\frown }{H}}^{\left(free\right)}\left({t}_{\tau }\right)+\overline{H}\left(\overrightarrow{x}\right)+{V}_{\tau }\left({t}_{\tau },\overrightarrow{x}\right)
\end{equation}

With 
 $ {V}_{\tau }$~ (the cross time and space potential or more simply the cross potential) defined as whatever is left over from the standard quantum theory and the free temporal
    quantization Hamiltonians.

We will do this by introducing a time gauge:

\begin{equation}
\label{equation-scatter-elecx-scatter-elecx-time_gauge}\hyperlabel{equation-scatter-elecx-scatter-elecx-time_gauge}%
{\Lambda }_{\tau }\left({t}_{\tau },\overrightarrow{x}\right)=\Phi \left(\overrightarrow{x}\right){t}_{\tau }
\end{equation}

This cancels the 
 $ \Phi $~ that accompanies the 
 $ i\frac{\partial }{\partial t}$~ in the second term. And it creates a term linear in 
 $ \Phi $. However it also induces some cross terms, all of order 
 $ {t}_{\tau }$~ to first and higher powers. Using:

\begin{equation}
\label{equation-scatter-elecx-scatter-elecx-7b}\hyperlabel{equation-scatter-elecx-scatter-elecx-7b}%
\overrightarrow{p}\rightarrow \overrightarrow{p}-e\nabla {\Lambda }_{\tau }=\overrightarrow{p}-e\nabla \Phi \left(\overrightarrow{x}\right){t}_{\tau }=\overrightarrow{p}+e\overrightarrow{E}\left(\overrightarrow{x}\right){t}_{\tau }
\end{equation}

We get:

\begin{equation}
\label{equation-scatter-elecx-scatter-elecx-vspace}\hyperlabel{equation-scatter-elecx-scatter-elecx-vspace}%
{V}_{\tau }\left({t}_{\tau },\overrightarrow{x}\right)=e\frac{\overrightarrow{E}\left(\overrightarrow{x}\right)\cdot \overrightarrow{p}+\overrightarrow{p}\cdot \overrightarrow{E}\left(\overrightarrow{x}\right)}{2m}{t}_{\tau }+\frac{{e}^{2}{\overrightarrow{E}}^{2}\left(\overrightarrow{x}\right)}{2m}{t}_{\tau }^{2}
\end{equation}

The full Schrödinger equation is now:

\begin{equation}
\label{equation-scatter-elecx-scatter-electric-result}\hyperlabel{equation-scatter-elecx-scatter-electric-result}%
H=\left(i{\partial }_{{t}_{\tau }}+\frac{1}{2m}\frac{{\partial }^{2}}{\partial {t}_{\tau }^{2}}+\frac{m}{2}\right)+\left(\frac{{\overrightarrow{p}}^{2}}{2m}+e\Phi \left(\overrightarrow{x}\right)\right)+\left(e\frac{\overrightarrow{E}\left(\overrightarrow{x}\right)\cdot \overrightarrow{p}+\overrightarrow{p}\cdot \overrightarrow{E}\left(\overrightarrow{x}\right)}{2m}{t}_{\tau }+\frac{{e}^{2}{\overrightarrow{E}}^{2}\left(\overrightarrow{x}\right)}{2m}{t}_{\tau }^{2}\right)
\end{equation}

This is exact.

Therefore even when the electric field is constant we can still see effects of temporal quantization.

\subsubsection{Time Dependent Electric Field}\label{scatter-elect}\hyperlabel{scatter-elect}%

Now we consider a time dependent electric field:

\begin{equation}
\label{equation-scatter-elect-scatter-elect-2}\hyperlabel{equation-scatter-elect-scatter-elect-2}%
\overrightarrow{E}\left(t,\overrightarrow{x}\right)=-\nabla \Phi \left(t,\overrightarrow{x}\right)
\end{equation}

If we are to get rid of the term quadratic in 
 $ \Phi $~ we again need:

\begin{equation}
\label{equation-scatter-elect-scatter-elect-4}\hyperlabel{equation-scatter-elect-scatter-elect-4}%
{\partial }_{{t}_{\tau }}{\Lambda }_{\tau }\left({t}_{\tau },\overrightarrow{x}\right)={\Phi }_{\tau }\left({t}_{\tau },\overrightarrow{x}\right)
\end{equation}

We generalize the time gauge
 \footnote{The time gauge here is similar to the gauge used by Aharonov and Rohrlich in Quantum Paradoxes (\cite{Aharonov-2005}) in their analysis of the Aharonov-{}Bohm effect: 
 $ \Lambda =-{\displaystyle \underset{}{\overset{\tau }{\int }}\text{d}{\tau }^{\prime }\Phi \left({\tau }^{\prime }\right)}$.}:

\begin{equation}
\label{equation-scatter-elect-scatter-elect-tg}\hyperlabel{equation-scatter-elect-scatter-elect-tg}%
{\Lambda }_{\tau }\left({t}_{\tau },x\right)={\displaystyle \underset{0}{\overset{{t}_{\tau }}{\int }}\text{d}{{t}^{\prime }}_{\tau }{\Phi }_{\tau }\left({{t}^{\prime }}_{\tau },\overrightarrow{x}\right)}
\end{equation}

If the potential is constant the time gauge is 
 $ \Phi \left(\overrightarrow{x}\right){t}_{\tau }$~ as in the constant potential case.

In the Schrödinger equation we again eliminate the quadratic 
 $ \Phi $~ and we again pull down a term linear in the potential:

\begin{equation}
\label{equation-scatter-elect-scatter-elect-6}\hyperlabel{equation-scatter-elect-scatter-elect-6}%
e{\Phi }_{\tau }\left({t}_{\tau },\overrightarrow{x}\right)
\end{equation}

We write this potential term as:

\begin{equation}
\label{equation-scatter-elect-scatter-elect-tauterm1}\hyperlabel{equation-scatter-elect-scatter-elect-tauterm1}%
{\Phi }_{\tau }\left({t}_{\tau },\overrightarrow{x}\right)={\Phi }_{\tau }\left(0,\overrightarrow{x}\right)+{\displaystyle \underset{0}{\overset{{t}_{\tau }}{\int }}\text{d}{{t}^{\prime }}_{\tau }\frac{\partial }{\partial {{t}^{\prime }}_{\tau }}{\Phi }_{\tau }\left({{t}^{\prime }}_{\tau },\overrightarrow{x}\right)}
\end{equation}

The cross potential is the same as the cross potential for the time independent case except that the time-{}smoothed electric field replaces the longitudinal electric field:

\begin{equation}
\label{equation-scatter-elect-scatter-elect-time_pot}\hyperlabel{equation-scatter-elect-scatter-elect-time_pot}%
{V}_{\tau }\left({t}_{\tau },\overrightarrow{x}\right)={V}_{\tau }^{\left(elec\right)\left(1\right)}{t}_{\tau }+{V}_{\tau }^{\left(elec\right)\left(2\right)}{t}_{\tau }^{2}
\end{equation}

With the first and second order terms:

\begin{equation}
\label{equation-scatter-elect-scatter-elect-first}\hyperlabel{equation-scatter-elect-scatter-elect-first}%
{V}_{\tau }^{\left(elec\right)\left(1\right)}\equiv e\frac{\overrightarrow{p}\cdot \langle {\overrightarrow{E}}_{\tau }^{\left(long\right)}\left({t}_{\tau },\overrightarrow{x}\right)\rangle +\langle {\overrightarrow{E}}_{\tau }^{\left(long\right)}\left({t}_{\tau },\overrightarrow{x}\right)\rangle \cdot \overrightarrow{p}}{2m}
\end{equation}

\begin{equation}
\label{equation-scatter-elect-scatter-elect-second}\hyperlabel{equation-scatter-elect-scatter-elect-second}%
{V}_{\tau }^{\left(elec\right)\left(2\right)}\equiv \frac{{\langle {\overrightarrow{E}}_{\tau }^{\left(long\right)}\left({t}_{\tau },\overrightarrow{x}\right)\rangle }^{2}}{2m}
\end{equation}

And with the time-{}smoothed electric field defined as:

\begin{equation}
\label{equation-scatter-elect-scatter-elect-smooth}\hyperlabel{equation-scatter-elect-scatter-elect-smooth}%
\langle {\overrightarrow{E}}_{\tau }^{\left(long\right)}\left({t}_{\tau },\overrightarrow{x}\right)\rangle \equiv -\frac{1}{{t}_{\tau }}{\displaystyle \underset{0}{\overset{{t}_{\tau }}{\int }}\text{d}{{t}^{\prime }}_{\tau }\nabla {\Phi }_{\tau }\left({{t}^{\prime }}_{\tau },\overrightarrow{x}\right)}
\end{equation}

The time smoothed electric field includes only the longitudinal component, the part of the electric field derived from 
 $ \Phi $.

We have one new term; the derivative with respect to 
 $ \tau $~ on the left of the Schrödinger equation pulls down a correction:

\begin{equation}
\label{equation-scatter-elect-scatter-elect-tauterm}\hyperlabel{equation-scatter-elect-scatter-elect-tauterm}%
i\frac{d}{d\tau }\mathrm{exp}\left(-ie{\displaystyle \underset{0}{\overset{{t}_{\tau }}{\int }}\text{d}{{t}^{\prime }}_{\tau }{\Phi }_{\tau }\left({{t}^{\prime }}_{\tau },\overrightarrow{x}\right)}\right)\Rightarrow e{\displaystyle \underset{0}{\overset{{t}_{\tau }}{\int }}\text{d}{{t}^{\prime }}_{\tau }\frac{\partial }{\partial \tau }{\Phi }_{\tau }\left({{t}^{\prime }}_{\tau },\overrightarrow{x}\right)}
\end{equation}

We move the new term over to the right hand side of the equation to get the total electric potential term as:

\begin{equation}
\label{equation-scatter-elect-scatter-elect-tauterm2}\hyperlabel{equation-scatter-elect-scatter-elect-tauterm2}%
e{\Phi }_{\tau }\left(0,\overrightarrow{x}\right)+e{\displaystyle \underset{0}{\overset{{t}_{\tau }}{\int }}\text{d}{{t}^{\prime }}_{\tau }\left(\frac{\partial }{\partial {{t}^{\prime }}_{\tau }}{\Phi }_{\tau }\left({{t}^{\prime }}_{\tau },\overrightarrow{x}\right)-\frac{\partial }{\partial \tau }{\Phi }_{\tau }\left({{t}^{\prime }}_{\tau },\overrightarrow{x}\right)\right)}
\end{equation}

Since:

\begin{equation}
\label{equation-scatter-elect-scatter-elect-tauterm3}\hyperlabel{equation-scatter-elect-scatter-elect-tauterm3}%
\frac{\partial }{\partial {t}_{\tau }}{\Phi }_{\tau }\left({t}_{\tau },\overrightarrow{x}\right)=\frac{\partial }{\partial \tau }{\Phi }_{\tau }\left({t}_{\tau },\overrightarrow{x}\right)
\end{equation}

We are left with:

\begin{equation}
\label{equation-scatter-elect-scatter-elect-tauterm4}\hyperlabel{equation-scatter-elect-scatter-elect-tauterm4}%
e{\Phi }_{\tau }\left(0,\overrightarrow{x}\right)
\end{equation}

This gives the Schrödinger equation:

\begin{equation}
\label{equation-scatter-elect-scatter-elect-seqnfin}\hyperlabel{equation-scatter-elect-scatter-elect-seqnfin}%
H={\stackrel{\frown }{H}}^{\left(free\right)}+{\overline{H}}_{\tau }^{\left(elec\right)}+{V}_{\tau }^{\left(elec\right)\left(1\right)}{t}_{\tau }+{V}_{\tau }^{\left(elec\right)\left(2\right)}{t}_{\tau }^{2}
\end{equation}

With the usual standard quantum theory Hamiltonian:

\begin{equation}
\label{equation-scatter-elect-scatter-elect-hamelec}\hyperlabel{equation-scatter-elect-scatter-elect-hamelec}%
{\overline{H}}_{\tau }^{\left(elec\right)}\equiv \frac{{\overrightarrow{p}}^{2}}{2m}+e{\Phi }_{\tau }\left(\overrightarrow{x}\right)
\end{equation}

This is exact.

It is composed of the free Hamiltonian for quantum time, the usual standard quantum theory term for a potential, and a cross potential of order relative time to first and higher
    powers.

It is the same as the result for the time independent case, with the time-{}smoothed electric field replacing the electric field.

\subsubsection{General Fields}\label{scatter-disc}\hyperlabel{scatter-disc}%

When 
 $ \Phi $~ is zero the time gauge is zero. Therefore the work in 
 \hyperlink{scatter-magx}{Time Independent Magnetic Field}~ and 
 \hyperlink{scatter-magt}{Time Dependent Magnetic Field}~ can be seen as already using the time gauge. The general Schrödinger equation is:

\begin{equation}
\label{equation-scatter-disc-scatter-gen-0}\hyperlabel{equation-scatter-disc-scatter-gen-0}%
H={\stackrel{\frown }{H}}^{\left(free\right)}+{\overline{H}}^{\left(elec+mag\right)}+{V}_{\tau }^{\left(mag\right)\left(1\right)}{t}_{\tau }+{V}_{\tau }^{\left(mag\right)\left(2\right)}{t}_{\tau }^{2}+{V}_{\tau }^{\left(elec\right)\left(1\right)}{t}_{\tau }+{V}_{\tau }^{\left(elec\right)\left(2\right)}{t}_{\tau }^{2}
\end{equation}

With:

\begin{equation}
\label{equation-scatter-disc-scatter-gen-1}\hyperlabel{equation-scatter-disc-scatter-gen-1}%
{\overline{H}}^{\left(elec+mag\right)}\equiv \frac{{\left(\overrightarrow{p}-e{\overrightarrow{A}}_{\tau }\left(\overrightarrow{x}\right)\right)}^{2}}{2m}+e{\Phi }_{\tau }\left(\overrightarrow{x}\right)
\end{equation}

Or spelled out:

\begin{equation}
\label{equation-scatter-disc-scatter-gen-3}\hyperlabel{equation-scatter-disc-scatter-gen-3}%
H=\left(\begin{array}{l}E-\frac{{E}^{2}}{2m}+\frac{{\left(\overrightarrow{p}-e{\overrightarrow{A}}_{\tau }\left(\overrightarrow{x}\right)-e\langle \frac{\partial \overrightarrow{A}}{\partial {t}_{\tau }}\rangle {t}_{\tau }\right)}^{2}}{2m}+e{\Phi }_{\tau }\left(\overrightarrow{x}\right)\\ +e\frac{\overrightarrow{p}\cdot \langle {\overrightarrow{E}}_{\tau }^{\left(long\right)}\left({t}_{\tau },\overrightarrow{x}\right)\rangle +\langle {\overrightarrow{E}}_{\tau }^{\left(long\right)}\left({t}_{\tau },\overrightarrow{x}\right)\rangle \cdot \overrightarrow{p}}{2m}{t}_{\tau }+\frac{{\langle {\overrightarrow{E}}_{\tau }^{\left(long\right)}\left({t}_{\tau },\overrightarrow{x}\right)\rangle }^{2}}{2m}{t}_{\tau }^{2}\end{array}\right)
\end{equation}

This is exact.

It is not, however, frame independent. This is unavoidable. While temporal quantization is frame independent, standard quantum theory is not. Therefore any connection from temporal
    quantization to standard quantum theory must point to a specific frame
 \footnote{That we can only define standard quantum theory for a specific frame is of course the problem we set out to address in this work.}.

With that said, it is clear from this Hamiltonian \textendash{} and regardless of the choice of frame for standard quantum theory \textendash{} with temporal quantization we will see an additional effective
    potential proportional to the first and higher powers of the relative time.

We will look at some specific applications below in 
 \hyperlink{xt-fields}{Time-{}varying Magnetic and Electric Fields}.

\subsection{Bound States}\label{bound}\hyperlabel{bound}%

\subsubsection{Overview}\label{bound-intro}\hyperlabel{bound-intro}%
\begin{quote}

If you are out to describe the truth, leave elegance to the tailor.

\hspace*\fill---~A. Einstein\end{quote}

We would like to give the reduction from temporal quantization to standard quantum theory for bound states, in a way that is independent of the specifics of the potential.

Unfortunately the time gauge is singular for the central ($ 1/{r}^{n}$) potentials we are interested in. This is not fatal \textendash{} Kleinert (\cite{Kleinert-2004}) found a way to handle such integrals for the hydrogen atom \textendash{} but it does make the time gauge trickier to work with.

A complete treatment would require we look at the underlying physics, the exchange of virtual photons, and the construction and summing of ladder diagrams. However to do this we would have
    to first extend temporal quantization to the multi-{}particle case. And that is a separate project, not part of this "testable chunk".

Therefore we will employ an ad hoc methodology with our goals being to:
\begin{enumerate}

\item{}Show that temporal quantization is not inconsistent with standard quantum theory,

\item{}Define the stationary temporal quantization wave functions corresponding to the familiar standard quantum theory bound states,

\item{}Show how the non-{}stationary temporal quantization states will propagate with laboratory time,

\item{}And estimate the dispersion in time of the temporal quantization states.

\end{enumerate}

An arbitrary solution of the temporal quantization Hamiltonian can be written as a sum over the product space of plane waves in time and standard quantum theory solutions in space:

\begin{equation}
\label{equation-bound-intro-bound-zero-ham-0}\hyperlabel{equation-bound-intro-bound-zero-ham-0}%
{\psi }_{\tau }\left(t,\overrightarrow{x}\right)={\displaystyle \sum _{E,\overline{n}}{c}_{E,\overline{n}}\left(\tau \right){\chi }_{E}\left(t\right){\xi }_{\overline{n}}\left(\overrightarrow{x}\right)}
\end{equation}

Where 
 $ \chi $~ are the plane wave solutions:

\begin{equation}
\label{equation-bound-intro-bound-zero-chi}\hyperlabel{equation-bound-intro-bound-zero-chi}%
{\chi }_{E}\left(t\right)=\frac{1}{\sqrt{2\pi }}\mathrm{exp}\left(-iEt\right)
\end{equation}

And the 
 $ {\xi }_{\overline{n}}\left(\overrightarrow{x}\right)$~ are the solutions of
 \footnote{
We are using 
 $ \overline{n}$~ to stand for whatever set of quantum numbers are being used to identify the standard quantum theory bound state. For instance, for the hydrogen atom these might be the
        principal quantum number 
 \emph{n}, the angular momentum quantum number 
 \emph{l}, and the magnetic quantum number 
 \emph{m}: 
 $ \overline{n}\rightarrow \left\lbrace n,l,m\right\rbrace $.
}:

\begin{equation}
\label{equation-bound-intro-bound-intro-2}\hyperlabel{equation-bound-intro-bound-intro-2}%
{\overline{H}}^{\left(sqt\right)}\left(\overrightarrow{x}\right){\xi }^{\left(\overline{n}\right)}\left(\overrightarrow{x}\right)={\overline{E}}_{\overline{n}}{\xi }_{\overline{n}}\left(\overrightarrow{x}\right)
\end{equation}

We take the stationary state condition as the defining condition:

\begin{equation}
\label{equation-bound-intro-bound-intro-0}\hyperlabel{equation-bound-intro-bound-intro-0}%
H\left(t,\overrightarrow{x}\right){\psi }^{\left(n\right)}\left(t,\overrightarrow{x}\right)=0
\end{equation}

This implies the stationary states are massively degenerate. 
 \emph{All}~ have laboratory energy zero:

\begin{equation}
\label{equation-bound-intro-bound-intro-1}\hyperlabel{equation-bound-intro-bound-intro-1}%
{\mathscr{E}}^{\left(n\right)}=0
\end{equation}

We show that for each bound state in standard quantum theory there is a corresponding stationary state in temporal quantization
 \footnote{
Where 
 \emph{n}~ represents the set of quantum numbers 
 \emph{E}~ and the set 
 $ \overline{n}$.
}:

\begin{equation}
\label{equation-bound-intro-bound-intro-3}\hyperlabel{equation-bound-intro-bound-intro-3}%
{\psi }_{n}\left(t,\overrightarrow{x}\right)\leftrightarrow {\chi }_{{E}_{\overline{n}}}\left(t\right){\xi }_{\overline{n}}\left(\overrightarrow{x}\right)
\end{equation}

Where the quantum energy 
 $ {E}_{\overline{n}}$~ is determined by 
 $ \overline{n}$.

Each stationary state will be surrounded by a cloud of "nearly-{}stationary states", states with an energy that is not quite the energy of the corresponding stationary state. We look at the
    evolution of such in 
 \hyperlink{bound-offaxis}{Evolution of General Wave Function}.

We make an order-{}of-{}magnitude estimate of the width of these clouds in time in 
 \hyperlink{bound-disp}{Estimate of Uncertainty in Time}.

\subsubsection{Stationary States}\label{bound-zero}\hyperlabel{bound-zero}%

\paragraph*{Hamiltonian}

\noindent

We start with a general Hamiltonian:

\begin{equation}
\label{equation-bound-zero-bound-zero-ham_tq}\hyperlabel{equation-bound-zero-bound-zero-ham_tq}%
H\left(t,\overrightarrow{x}\right)=-\frac{1}{2m}{\left(E-e\Phi \left(\overrightarrow{x}\right)\right)}^{2}+\frac{1}{2m}{\left(\overrightarrow{p}-e\overrightarrow{A}\right)}^{2}+\frac{m}{2}
\end{equation}

This has eigenfunctions:

\begin{equation}
\label{equation-bound-zero-bound-zero-eigen_func}\hyperlabel{equation-bound-zero-bound-zero-eigen_func}%
H\left(t,\overrightarrow{x}\right){\psi }_{\tau }\left(t,\overrightarrow{x}\right)={\mathscr{E}}_{n}{\psi }_{\tau }\left(t,\overrightarrow{x}\right)
\end{equation}

The stationary states are those with eigenvalue zero. We expect a countably infinite number of these.

To avoid inessential complications we assume the vector potential is zero, so the Hamiltonian is now:

\begin{equation}
\label{equation-bound-zero-bound-zero-ham_bound}\hyperlabel{equation-bound-zero-bound-zero-ham_bound}%
{H}^{\left(bound\right)}\left(t,\overrightarrow{x}\right)=-\frac{1}{2m}{\left(E-e\Phi \left(\overrightarrow{x}\right)\right)}^{2}+\frac{1}{2m}{\overrightarrow{p}}^{2}+\frac{m}{2}
\end{equation}

The stationary condition is:

\begin{equation}
\label{equation-bound-zero-bound-zero-stationary}\hyperlabel{equation-bound-zero-bound-zero-stationary}%
{H}^{\left(bound\right)}\left(t,\overrightarrow{x}\right)\psi \left(t,\overrightarrow{x}\right)=0
\end{equation}

We take as the standard quantum theory Hamiltonian:

\begin{equation}
\label{equation-bound-zero-bound-zero-ham_bound3}\hyperlabel{equation-bound-zero-bound-zero-ham_bound3}%
\overline{H}\equiv \sqrt{{m}^{2}+{\overrightarrow{p}}^{2}}-m+e\Phi \left(\overrightarrow{x}\right)
\end{equation}

We will assume we are able to solve this and that the solutions are indexed by 
 $ \overline{n}$.

\paragraph*{Locating the Stationary States}

\noindent

We would like to pick out the stationary states in a way that is independent of the specifics of the potential. To do this we start by looking at individual terms in the sum, simple product
    states of a plane wave in quantum time and the standard quantum theory state in space:

\begin{equation}
\label{equation-bound-zero-bound-zero-phi}\hyperlabel{equation-bound-zero-bound-zero-phi}%
{\chi }_{E}\left(t\right){\xi }_{\overline{n}}\left(\overrightarrow{x}\right)
\end{equation}

We are looking for a value of 
 \emph{E}~ that corresponds to a specific set 
 $ \overline{n}$. We know that if a state is stationary, it will obey the condition:

\begin{equation}
\label{equation-bound-zero-bound-zero-ham-stationary}\hyperlabel{equation-bound-zero-bound-zero-ham-stationary}%
\langle E,\overline{n}\vert {H}^{\left(bound\right)}\vert E,\overline{n}\rangle =0
\end{equation}

We can write the standard quantum theory Hamiltonian as:

\begin{equation}
\label{equation-bound-zero-bound-zero-ham-3}\hyperlabel{equation-bound-zero-bound-zero-ham-3}%
\overline{H}=\mathcal{K}+e\Phi \left(\overrightarrow{x}\right)
\end{equation}

Where we use script 
 $ \mathcal{K}$~ for the kinetic energy:

\begin{equation}
\label{equation-bound-zero-bound-zero-ham-4b}\hyperlabel{equation-bound-zero-bound-zero-ham-4b}%
\mathcal{K}\equiv \sqrt{{m}^{2}+{\overrightarrow{p}}^{2}}-m=\frac{{\overrightarrow{p}}^{2}}{2m}-\frac{{\overrightarrow{p}}^{4}}{8{m}^{3}}+\frac{{\overrightarrow{p}}^{6}}{16{m}^{3}}-\frac{5{\overrightarrow{p}}^{8}}{128{m}^{5}}\cdots 
\end{equation}

The main complication is the presence of the 
 $ {\Phi }^{2}$~ term in the temporal quantization 
 \emph{H}:

\begin{equation}
\label{equation-bound-zero-bound-zero-ham-5b}\hyperlabel{equation-bound-zero-bound-zero-ham-5b}%
H=-\frac{1}{2m}\left({\left(E-e\Phi \right)}^{2}-{\overrightarrow{p}}^{2}-{m}^{2}\right)
\end{equation}

As noted above in 
 \hyperlink{scatter-elect}{Time Dependent Electric Field}, the time gauge is likely to be singular, so is not a good choice for getting rid of this complication.

Our goal is a Hamiltonian linear in both 
 $ \Phi $~ and 
 $ \mathcal{K}$.

We rewrite the Hamiltonian as:

\begin{equation}
\label{equation-bound-zero-bound-zero-ham-5}\hyperlabel{equation-bound-zero-bound-zero-ham-5}%
H=\frac{-{\left(E-\overline{H}+\mathcal{K}\right)}^{2}+{\left(m+\mathcal{K}\right)}^{2}}{2m}
\end{equation}

Using:

\begin{equation}
\label{equation-bound-zero-bound-zero-ham-6}\hyperlabel{equation-bound-zero-bound-zero-ham-6}%
e\Phi =\overline{H}-\mathcal{K}
\end{equation}

And:

\begin{equation}
\label{equation-bound-zero-bound-zero-ham-7}\hyperlabel{equation-bound-zero-bound-zero-ham-7}%
{m}^{2}+{\overrightarrow{p}}^{2}={\left(m+\mathcal{K}\right)}^{2}
\end{equation}

The 
 $ {\mathcal{K}}^{2}$~ terms cancel, giving:

\begin{equation}
\label{equation-bound-zero-bound-zero-ham-linear}\hyperlabel{equation-bound-zero-bound-zero-ham-linear}%
H=\frac{-{E}^{2}+2E\overline{H}-2E\mathcal{K}-{\overline{H}}^{2}+2\mathcal{K}\overline{H}+\left\lbrack \overline{H},\mathcal{K}\right\rbrack +2m\mathcal{K}+{m}^{2}}{2m}
\end{equation}

For a single 
 $ {\xi }_{\overline{n}}\left(\overrightarrow{x}\right)$, we have:

\begin{equation}
\label{equation-bound-zero-bound-zero-ham-9}\hyperlabel{equation-bound-zero-bound-zero-ham-9}%
\overline{H}\rightarrow {\overline{\mathscr{E}}}_{\overline{n}}^{\left(bind\right)}
\end{equation}

The stationary states are given by the solutions for 
 \emph{E}~ of:

\begin{equation}
\label{equation-bound-zero-bound-zero-ham-quadratic}\hyperlabel{equation-bound-zero-bound-zero-ham-quadratic}%
-{E}^{2}+2E{\overline{\mathscr{E}}}_{\overline{n}}^{\left(bind\right)}-2E\mathcal{K}-{\overline{\mathscr{E}}}_{\overline{n}}^{\left(bind\right)2}+2\mathcal{K}{\overline{\mathscr{E}}}_{\overline{n}}^{\left(bind\right)}+2m\mathcal{K}+{m}^{2}=0
\end{equation}

There are two solutions, a positive energy one:

\begin{equation}
\label{equation-bound-zero-bound-zero-ham-Enbar}\hyperlabel{equation-bound-zero-bound-zero-ham-Enbar}%
{\overline{E}}_{\overline{n}}^{(+)}=m+{\overline{\mathscr{E}}}_{\overline{n}}^{\left(bind\right)}
\end{equation}

And a negative energy one:

\begin{equation}
\label{equation-bound-zero-bound-zero-ham-c2}\hyperlabel{equation-bound-zero-bound-zero-ham-c2}%
{\overline{E}}_{\overline{n}}^{(-)}=-m+{\overline{\mathscr{E}}}_{\overline{n}}^{\left(bind\right)}-2\mathcal{K}
\end{equation}

Presumably this latter corresponds to an anti-{}particle. We will focus on the positive energy solution:

\begin{equation}
\label{equation-bound-zero-bound-zero-ham-qE}\hyperlabel{equation-bound-zero-bound-zero-ham-qE}%
{\overline{E}}_{\overline{n}}={\overline{E}}_{\overline{n}}^{(+)}
\end{equation}

So the time part, to lowest order is:

\begin{equation}
\label{equation-bound-zero-bound-zero-ham-chi_nbar}\hyperlabel{equation-bound-zero-bound-zero-ham-chi_nbar}%
{\chi }_{\overline{n}}\left(t\right)\equiv \frac{1}{\sqrt{2\pi }}\mathrm{exp}\left(-i{\overline{E}}_{\overline{n}}t\right)=\frac{1}{\sqrt{2\pi }}\mathrm{exp}\left(-imt-i{\overline{\mathscr{E}}}_{\overline{n}}^{\left(bind\right)}t\right)
\end{equation}

We see that with the correct choice of quantum energy, the chosen product wave function has a zero expectation of the Hamiltonian.

Therefore, for each bound state in standard quantum theory, we have a stationary state in temporal quantization.

If the standard quantum theory bound states are not degenerate, the temporal quantization states are not. And if the standard quantum theory bound states are degenerate, then the temporal
    quantization states will have the same quantum energy and be degenerate as well.

\paragraph*{Stationary States in Relative Time}

\noindent

As with the free case, we can rewrite the plane wave's time in terms of the relative time:

\begin{equation}
\label{equation-bound-zero-bound-zero-reltime-chi}\hyperlabel{equation-bound-zero-bound-zero-reltime-chi}%
{\chi }_{\overline{n}}\left(t\right)=\frac{1}{\sqrt{2\pi }}\mathrm{exp}\left(-i{\overline{E}}_{\overline{n}}{t}_{\tau }\right)\mathrm{exp}\left(-i{\overline{E}}_{\overline{n}}\tau \right)
\end{equation}

So that the plane wave becomes a product of relative time and laboratory time parts.

The full wave function is now:

\begin{equation}
\label{equation-bound-zero-bound-zero-reltime-wf}\hyperlabel{equation-bound-zero-bound-zero-reltime-wf}%
{\psi }_{\tau }^{\left(E,\overline{n}\right)}\left({t}_{\tau },\overrightarrow{x}\right)=\frac{1}{\sqrt{2\pi }}\mathrm{exp}\left(-i{\overline{E}}_{\overline{n}}{t}_{\tau }\right){\xi }_{\overline{n}}\left(\overrightarrow{x}\right)\mathrm{exp}\left(-i{\overline{E}}_{\overline{n}}\tau \right)
\end{equation}

The laboratory energy is the sum of the mass and the bound state energy. As the mass part will cancel out in all transitions, only the bound state energy will matter.

\subsubsection{Evolution of General Wave Function}\label{bound-offaxis}\hyperlabel{bound-offaxis}%

We have argued that the bound states may be identified as those that are stationary with respect to laboratory time. Given non-{}zero uncertainty in quantum time, the bound/stationary states
    will be accompanied by a cloud of offshell states, a kind of temporal fuzz. How does this cloud evolve in time? What of states that are "almost stationary"?

Taking the stationary states as a starting point, we construct an arbitrary state as:

\begin{equation}
\label{equation-bound-offaxis-bound-offaxis-series}\hyperlabel{equation-bound-offaxis-bound-offaxis-series}%
{\psi }_{\tau }\left(t,\overrightarrow{x}\right)={\displaystyle \sum _{n}{c}_{\tau }^{\left(n\right)}\frac{1}{\sqrt{2\pi }}\mathrm{exp}\left(-i{E}_{n}t\right){\xi }_{\overline{n}}\left(\overrightarrow{x}\right)\mathrm{exp}\left(-i{\mathscr{E}}_{n}\tau \right)}
\end{equation}

With the sum taken over both stationary and non-{}stationary states.

This is completely general; any laboratory time dependence not captured in the exponentials:

\begin{equation}
\label{equation-bound-offaxis-bound-offaxis-lab}\hyperlabel{equation-bound-offaxis-bound-offaxis-lab}%
\mathrm{exp}\left(-i{\mathscr{E}}_{n}\tau \right)
\end{equation}

Will be tracked in the coefficients:

\begin{equation}
\label{equation-bound-offaxis-bound-offaxis-coef}\hyperlabel{equation-bound-offaxis-bound-offaxis-coef}%
{c}_{\tau }^{\left(n\right)}
\end{equation}

The quantum energy of each state can be broken out into the onshell part and the time part:

\begin{equation}
\label{equation-bound-offaxis-bound-offaxis-qE}\hyperlabel{equation-bound-offaxis-bound-offaxis-qE}%
{E}_{n}={\overline{E}}_{\overline{n}}+{\stackrel{\frown }{E}}_{n}
\end{equation}

The laboratory energy can also be broken out into the onshell part and the time part:

\begin{equation}
\label{equation-bound-offaxis-bound-offaxis-labE}\hyperlabel{equation-bound-offaxis-bound-offaxis-labE}%
{\mathscr{E}}_{n}={\overline{E}}_{\overline{n}}+{\stackrel{\frown }{\mathscr{E}}}_{n}
\end{equation}

The onshell parts of both are the same, composed of mass plus binding energy.

To estimate the laboratory energy associated with each plane wave, we put it in a Hamiltonian sandwich, taking advantage of the fact that the laboratory energy of stationary states is zero
    (by definition).

\begin{equation}
\label{equation-bound-offaxis-bound-offaxis-4}\hyperlabel{equation-bound-offaxis-bound-offaxis-4}%
{\mathscr{E}}_{n}=\langle E,\overline{n}\vert H\vert E,\overline{n}\rangle =-\frac{m+\mathcal{K}}{m}{\stackrel{\frown }{E}}_{n}-\frac{{\stackrel{\frown }{E}}_{n}^{2}}{2m}
\end{equation}

We can write the coefficient of the linear term as an expectation of 
 $ \gamma $, now treated as an operator:

\begin{equation}
\label{equation-bound-offaxis-bound-offaxis-4b}\hyperlabel{equation-bound-offaxis-bound-offaxis-4b}%
\frac{m+\mathcal{K}}{m}=\frac{\sqrt{{m}^{2}+{\overrightarrow{p}}^{2}}}{m}=\langle {\gamma }_{\overline{n}}\rangle 
\end{equation}

Giving:

\begin{equation}
\label{equation-bound-offaxis-bound-offaxis-4c}\hyperlabel{equation-bound-offaxis-bound-offaxis-4c}%
{\mathscr{E}}_{n}=-\langle {\gamma }_{\overline{n}}\rangle {\stackrel{\frown }{E}}_{n}-\frac{{\stackrel{\frown }{E}}_{n}^{2}}{2m}
\end{equation}

So the time part of the laboratory energy is given as a function of the time part of the quantum energy by
 \footnote{This is essentially the same result as for free waves (\hyperlink{free-fourt}{Time/Space Representation}): 
 $ {\stackrel{\frown }{\mathscr{E}}}_{p}=-{\gamma }_{\overrightarrow{p}}{\stackrel{\frown }{E}}_{p}-\frac{{\stackrel{\frown }{E}}_{p}^{2}}{2m}$.}:

\begin{equation}
\label{equation-bound-offaxis-bound-offaxis-parts}\hyperlabel{equation-bound-offaxis-bound-offaxis-parts}%
{\stackrel{\frown }{\mathscr{E}}}_{n}=-\langle {\gamma }_{\overline{n}}\rangle {\stackrel{\frown }{E}}_{n}-\frac{{\stackrel{\frown }{E}}_{n}^{2}}{2m}
\end{equation}

If we shift to relative time we can rewrite the time part of the wave function:

\begin{equation}
\label{equation-bound-offaxis-bound-offaxis-factor}\hyperlabel{equation-bound-offaxis-bound-offaxis-factor}%
\mathrm{exp}\left(-i{E}_{n}t-i{\mathscr{E}}_{n}\tau \right)\rightarrow \mathrm{exp}\left(-i{E}_{n}{t}_{\tau }-i{\mathscr{E}}_{n}^{\left(rel\right)}\tau \right)
\end{equation}

With the relative time energy:

\begin{equation}
\label{equation-bound-offaxis-bound-offaxis-relE}\hyperlabel{equation-bound-offaxis-bound-offaxis-relE}%
{\mathscr{E}}_{n}^{\left(rel\right)}\equiv {\mathscr{E}}_{n}+{E}_{n}={\overline{E}}_{\overline{n}}-\left(\langle {\gamma }_{\overline{n}}\rangle -1\right){\stackrel{\frown }{E}}_{n}-\frac{{\stackrel{\frown }{E}}_{n}^{2}}{2m}
\end{equation}

In either block time or relative time, the time/offshell part of the laboratory energy is quadratic in the time/offshell part of the quantum energy, so that there is a significant penalty
    for going offshell.

\subsubsection{Estimate of Uncertainty in Time}\label{bound-disp}\hyperlabel{bound-disp}%
\begin{figure}[H]

\begin{center}
\imgexists{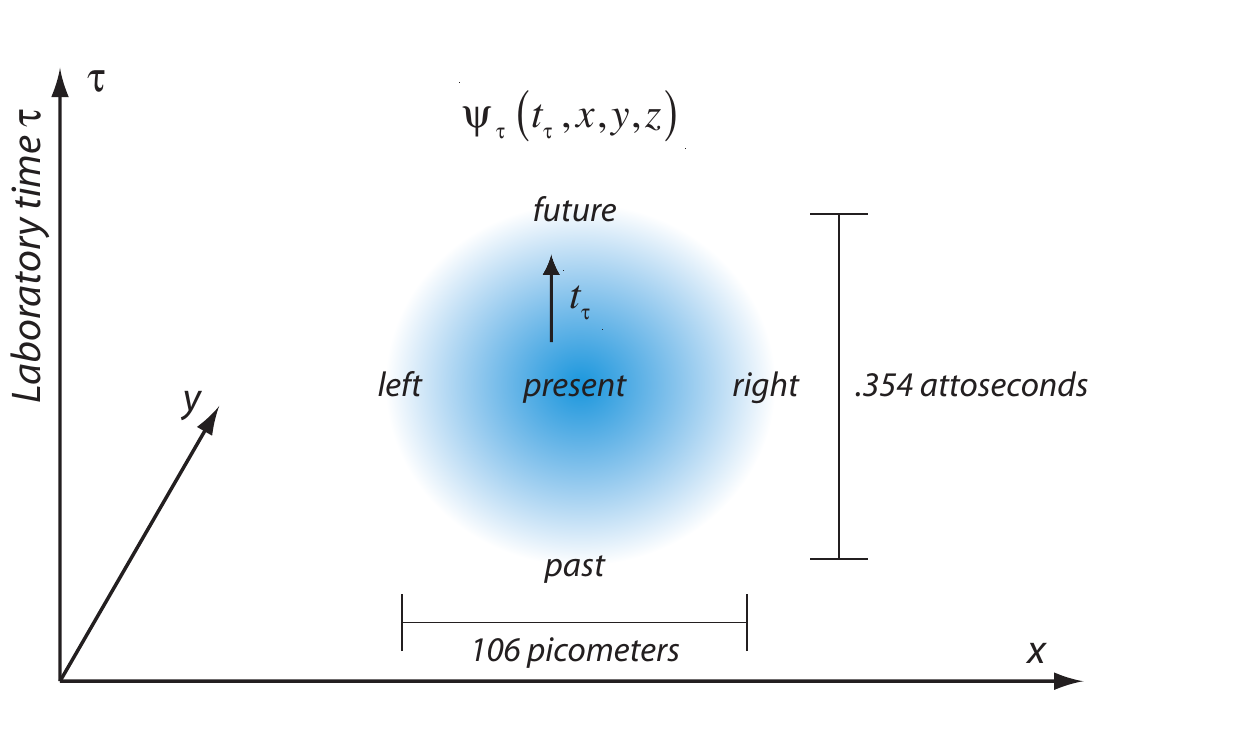}{{\imgevalsize{images/bound-disp-0.pdf}{\includegraphics[width=\imgwidth,height=\imgheight,keepaspectratio=true]{images/bound-disp-0.pdf}}\quad
}
}{}
\end{center}
\caption{Extension of a Wave Function in Time and Space}
\label{figure-images-bound-disp-0}\hyperlabel{figure-images-bound-disp-0}%
\end{figure}
\begin{quote}

It takes about 150 attoseconds for an electron to circle the nucleus of an atom. An attosecond is 10-{}18 seconds long, or, expressed in another way, an attosecond is related to a second
        as a second is related to the age of the universe.

\hspace*\fill---~Johan Mauritsson
 \cite{Mauritsson-2008b}\end{quote}

We now estimate the dispersion in quantum time of the wave functions.

We recall that the Coulomb potential is mediated by the exchange of virtual photons between electron and nucleus. These make it energetically favorable for oppositely charged particles to
    stay closer, subject to the restrictions of the Heisenberg uncertainty principle and the Pauli exclusion principle.

It is reasonable to suppose that if the paths of nucleus and electron get too far apart in time \textendash{} or too close \textendash{} the joint path will be disfavored for the same reasons.

By our principle of the maximum possible symmetry between time and space, our best order of magnitude estimate of the width of an atomic state in time is its width in space, divided by the
    speed of light. In natural units:

\begin{equation}
\label{equation-bound-disp-bound-disp-9}\hyperlabel{equation-bound-disp-bound-disp-9}%
\frac{{\sigma }_{\overline{n}}^{2}}{2}\sim {\displaystyle \int \text{d}\overrightarrow{r}{\left\vert {\xi }_{\overline{n}}\left(\overrightarrow{r}\right)\right\vert }^{2}{\overrightarrow{r}}^{2}}
\end{equation}

For an argon atom, as used in 
 \cite{Lindner-2005}, the radius of the atom is 106 picometers. So our order of magnitude estimate of the width in time is less than an attosecond:

\begin{equation}
\label{equation-bound-disp-bound-disp-8}\hyperlabel{equation-bound-disp-bound-disp-8}%
\frac{106\cdot {10}^{-12}m}{3.00\cdot {10}^{8}m/s}=.354\cdot {10}^{-18}s
\end{equation}

This is exceedingly small
 \footnote{In practice we might want to work with Rydberg atoms, with very high principal quantum number, as these can be much wider.}.

As noted, to compute the dispersion from first principles we would have to calculate the bound states. To do this we would have to extend temporal quantization to the multi-{}particle case,
    then compute the ladder diagrams corresponding to virtual photon exchange between electron and nucleus. The poles in the resulting kernel give the bound state energies; the residues at the poles
    the wave functions.

However extending temporal quantization to the multi-{}particle case is a significant project in its own right, not part of this "testable chunk".


\chapter{Experimental Tests}\label{xt}\hyperlabel{xt}%

\section{Overview}\label{xt-intro}\hyperlabel{xt-intro}%
\begin{quote}

I will listen to any hypothesis but on one condition\textendash{}that you show me a method by which it can be tested.

\hspace*\fill---~August William Von Hoffman
 \cite{Mackay-1991}\end{quote}

In the previous chapter we were looking at limits in which temporal quantization reduces to standard quantum theory. In this we are looking for differences, for ways to test for temporal
    quantization.

Our analysis is simplified by the absence of free parameters in temporal quantization; the properties of quantum time are forced by the requirement of manifest covariance. It is complicated
    by the fact we do not know what the initial wave function looks like.

In general we can force the issue by running the particle or beam through a chopper before making use of it in the experiment proper. If the chopper is open for a time 
 $ \Delta {\tau }_{chopper}$, and the particle has a roughly equal chance of getting through at any time, then it is reasonable to assume that the wave function has dispersion in time of order 
 $ \Delta {\tau }_{chopper}$:

\begin{equation}
\label{equation-xt-intro-xt-initial-1}\hyperlabel{equation-xt-intro-xt-initial-1}%
\langle {t}^{2}\rangle -{\langle t\rangle }^{2}\sim \Delta {\tau }_{chopper}^{2}
\end{equation}

If we think in terms of a classic experiment \textendash{} beam aimed at target \textendash{} then in addition to working with a particle with known or estimated dispersion in time, we will need to work with a
    target which in some way varies in time. Otherwise any effects are likely to be averaged away.

There are three obvious approaches. We can:
\begin{enumerate}

\item{}Start with an existing foundational experiment in quantum mechanics
 \footnote{See the reviews of experimental tests of quantum mechanics in 
 \cite{Lamoreaux-1992}\cite{Ghose-1999}\cite{Auletta-2000}.}, then interchange time and a space dimension. For instance, the double slit experiment becomes the double slit in time experiment if we
            replace the slits with choppers.

\item{}Run a particle with known dispersion in time through an electromagnetic field which is varying rapidly in time.

\item{}Take an experiment which implicitly takes a holistic view of time \textendash{} the quantum eraser (\cite{Scully-1982}), the Aharonov-{}Bohm experiment (\cite{Aharonov-1959}), tests of the Aharonov-{}Bergmann-{}Lebowitz (\cite{Aharonov-1964}) time-{}symmetric measurement \textendash{} and re-{}examine from the perspective of quantum time.

\end{enumerate}

We will look at examples of each of these approaches in turn, considering:
\begin{enumerate}

\item{}  \hyperlink{xt-gates}{Slits in Time},

\item{}  \hyperlink{xt-fields}{Time-{}varying Magnetic and Electric Fields},

\item{}And the 
 \hyperlink{xt-ab}{Aharonov-{}Bohm Experiment}.

\end{enumerate}

As noted, we primarily interested in "proof-{}of-{}concept" here, so we will only look at the lowest nontrivial corrections resulting from quantum time. We will discuss various ways to make the
    treatments exact 
 \hyperlink{xt-disc}{at the end of this chapter}.

\section{Slits in Time}\label{xt-gates}\hyperlabel{xt-gates}%

\subsection{Overview}\label{xt-gintro}\hyperlabel{xt-gintro}%
\begin{quote}

Here no one else can enter, since this gate was only for you. I now close it.

\hspace*\fill---~K's Gatekeeper
 \cite{Kafka-1925}\end{quote}
\begin{quote}

At the narrow passage, there is no brother, no friend.

\hspace*\fill---~Arab Proverb\end{quote}

\paragraph*{Experiments}

\noindent

We will look at:
\begin{enumerate}

\item{}The 
 \hyperlink{xt-free}{Free Case},

\item{}The 
 \hyperlink{xt-single}{Single Slit Experiment},

\item{}The 
 \hyperlink{xt-double}{Double Slit Experiment},

\item{}And Lindner's 
 \hyperlink{xt-lind}{Attosecond Double Slit in Time}~ (\cite{Lindner-2005}).

\end{enumerate}

\paragraph*{Approach}

\noindent

We take the particles as going left to right in the 
 \emph{x}~ direction; the slits as choppers in the time direction. We start the particles off at 
 \emph{x}~ position zero. The distance from the start to the gate is 
 $ {L}_{G}$, the distance from the start to the detector is 
 $ {L}_{D}$.

We model the particles as a set of rays in momentum space
 \footnote{This is an example of the breakout of the initial wave function into Gaussian test functions proposed in 
 \hyperlink{covar}{Covariant Definition of Laboratory Time}.}, each with an associated wave function in quantum time:

\begin{equation}
\label{equation-xt-gintro-xt-gates-3}\hyperlabel{equation-xt-gintro-xt-gates-3}%
\psi \left(t,p\right)=\chi \left(t,p\right)\widehat{\xi }\left(p\right)
\end{equation}

To do the corresponding analysis for the standard quantum theory case we drop the wave function in time. Essentially we are using the momentum rays as carriers.

To simplify the analysis we assume the laboratory time and corresponding momentum are related in a deterministic way by
 \footnote{Since 
 \emph{p}~ and 
 $ \delta \tau $~ are equivalent we can parameterize the space part/carrier ray with either.}:

\begin{equation}
\label{equation-xt-gintro-xt-gates-ptr}\hyperlabel{equation-xt-gintro-xt-gates-ptr}%
p=\frac{mL}{\tau }
\end{equation}

Where 
 $ \tau $~ and 
 \emph{L}~ are the laboratory time and the distance. We will frequently expand this around the averages, keeping only terms of first order:

\begin{equation}
\label{equation-xt-gintro-xt-gates-first}\hyperlabel{equation-xt-gintro-xt-gates-first}%
\overline{p}+\delta p=\frac{mL}{\overline{\tau }+\delta \tau }\rightarrow \delta p\approx -\overline{p}\frac{\delta \tau }{\overline{\tau }},\delta \tau \approx -\overline{\tau }\frac{\delta p}{\overline{p}}
\end{equation}

To further simplify we will work with Gaussian gates, rather than the more traditional square. This was suggested by Feynman and Hibbs (\cite{Feynman-1965d}). These are easier to work with and a better fit to Lindner's 
 \hyperlink{xt-lind}{Attosecond Double Slit in Time}, discussed below. In principle, thanks to Morlet wavelet decomposition, they are fully general.

We work in the non-{}relativistic regime. We use the non-{}relativistic decomposition in 
 \hyperlink{free-hybrid}{Time/Momentum Representation}. We assume 
 $ \gamma =\frac{E}{m}\approx 1$.

We will assume we are able to estimate the initial dispersion in time, and that this dispersion does not vary significantly as a function of the momentum.

\subsection{Free Case}\label{xt-free}\hyperlabel{xt-free}%
\begin{figure}[H]

\begin{center}
\imgexists{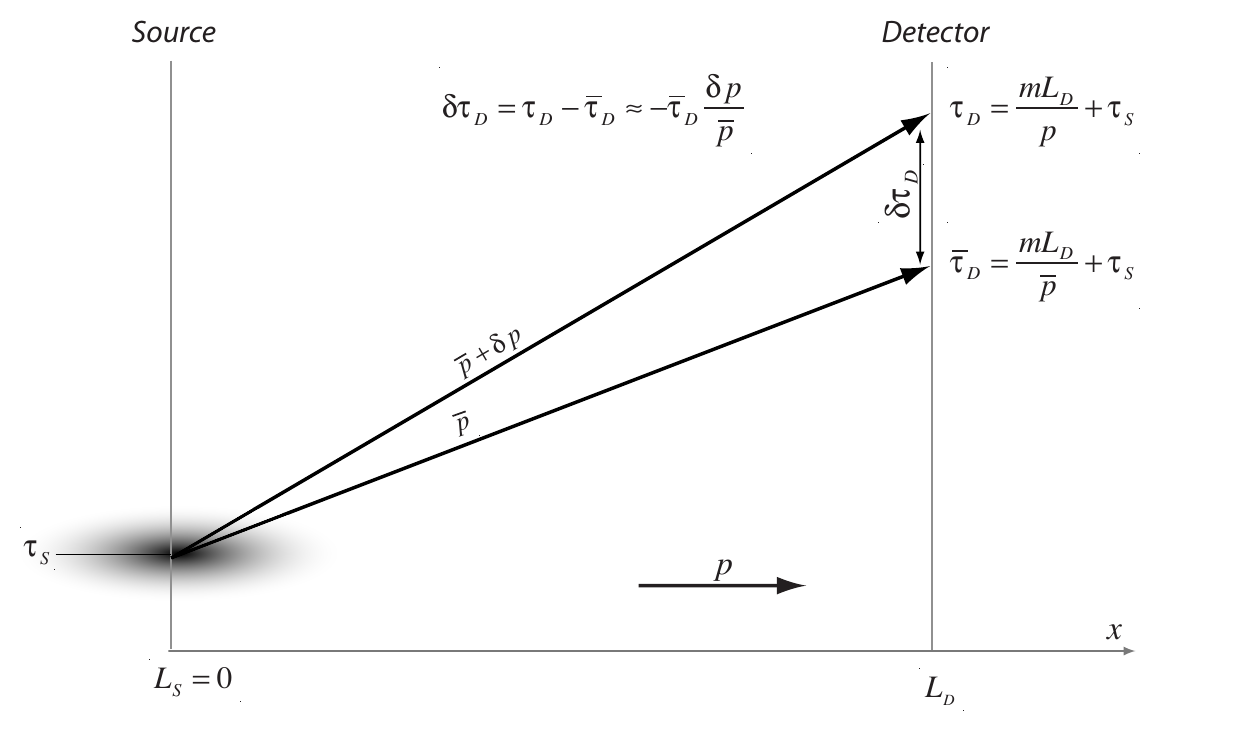}{{\imgevalsize{images/xt-free-0.pdf}{\includegraphics[width=\imgwidth,height=\imgheight,keepaspectratio=true]{images/xt-free-0.pdf}}\quad
}
}{}
\end{center}
\caption{Free Case}
\label{figure-images-xt-free-0}\hyperlabel{figure-images-xt-free-0}%
\end{figure}

We start by looking at the free case, to provide a reference point and to check the normalization.

\paragraph*{Standard Quantum Theory}

\noindent

Per the analysis in 
 \hyperlink{free-hybrid}{Time/Momentum Representation}~ we take as the initial wave function in momentum:

\begin{equation}
\label{equation-xt-free-xt-free-carrier-0}\hyperlabel{equation-xt-free-xt-free-carrier-0}%
{\widehat{\xi }}_{S}\left(p\right)=\sqrt[4]{\frac{1}{\pi {\widehat{\sigma }}_{1}^{2}}}\mathrm{exp}\left(-\frac{\delta {p}^{2}}{2{\widehat{\sigma }}_{1}^{2}}\right)
\end{equation}

And get that the wave function at the detector is:

\begin{equation}
\label{equation-xt-free-xt-free-carrier-2b}\hyperlabel{equation-xt-free-xt-free-carrier-2b}%
{\widehat{\xi }}_{D}\left(p\right)=\sqrt[4]{\frac{1}{\pi {\widehat{\sigma }}_{1}^{2}}}\mathrm{exp}\left(-\frac{\delta {p}^{2}}{2{\widehat{\sigma }}_{1}^{2}}-i\frac{{p}^{2}}{2m}{\tau }_{DS}\right)
\end{equation}

With probability density:

\begin{equation}
\label{equation-xt-free-xt-free-carrier-3}\hyperlabel{equation-xt-free-xt-free-carrier-3}%
{\widehat{\overline{\rho }}}_{D}\left(\delta p\right)=\sqrt{\frac{1}{\pi {\widehat{\sigma }}_{1}^{2}}}\mathrm{exp}\left(-\frac{\delta {p}^{2}}{{\widehat{\sigma }}_{1}^{2}}\right)
\end{equation}

Actual measurements are of clicks per unit time, not per unit momentum.

In going from momentum to laboratory time the probability density scales as the Jacobian:

\begin{equation}
\label{equation-xt-free-xt-free-carrier-8}\hyperlabel{equation-xt-free-xt-free-carrier-8}%
\text{d}p\rightarrow \frac{\overline{p}}{{\overline{\tau }}_{D}}\text{d}\tau \Rightarrow \widehat{\overline{\rho }}\left(\delta p\right)\rightarrow \frac{\overline{p}}{{\overline{\tau }}_{D}}\widehat{\overline{\rho }}\left(-\overline{p}\frac{\delta {\tau }_{D}}{{\overline{\tau }}_{D}}\right)=\overline{\rho }\left(\delta {\tau }_{D}\right)
\end{equation}

So the wave function scales as the square root of the Jacobian:

\begin{equation}
\label{equation-xt-free-xt-free-carrier-9}\hyperlabel{equation-xt-free-xt-free-carrier-9}%
{\widehat{\xi }}_{D}\left(\delta {p}_{D}\right)\rightarrow \sqrt{\frac{\overline{p}}{{\overline{\tau }}_{D}}}{\xi }_{D}\left(\delta {\tau }_{D}\right)
\end{equation}

The wave function as a function of 
 $ \delta {\tau }_{D}$~ is then (dropping a term cubic in 
 $ \delta {\tau }_{D}$):

\begin{equation}
\label{equation-xt-free-xt-free-carrier-c}\hyperlabel{equation-xt-free-xt-free-carrier-c}%
{\xi }_{D}\left(\delta {\tau }_{D}\right)=\sqrt[4]{\frac{1}{\pi {\overline{\sigma }}_{D}^{2}}}\mathrm{exp}\left(-\frac{\delta {\tau }_{D}^{2}}{2{\overline{\sigma }}_{D}^{2}}-i\frac{{\overline{p}}^{2}}{2m}{\overline{\tau }}_{D}+i\frac{{\overline{p}}^{2}}{2m}\delta {\tau }_{D}+i\frac{{\overline{p}}^{2}}{2m}\frac{\delta {\tau }_{D}^{2}}{{\overline{\tau }}_{D}}\right)
\end{equation}

With the dispersion in laboratory time:

\begin{equation}
\label{equation-xt-free-xt-free-carrier-f}\hyperlabel{equation-xt-free-xt-free-carrier-f}%
{\overline{\sigma }}_{D}^{2}=\frac{{\widehat{\sigma }}_{1}^{2}}{{\overline{p}}^{2}}{\overline{\tau }}_{D}^{2}
\end{equation}

This is the angular width of the beam, 
 $ \frac{{\widehat{\sigma }}_{1}}{\overline{p}}$, scaled by the laboratory time.

The probability density as a function of 
 $ \delta {\tau }_{D}$~ is:

\begin{equation}
\label{equation-xt-free-xt-free-carrier-e}\hyperlabel{equation-xt-free-xt-free-carrier-e}%
{\overline{\rho }}_{D}\left(\delta {\tau }_{D}\right)=\sqrt{\frac{1}{\pi {\overline{\sigma }}_{D}^{2}}}\mathrm{exp}\left(-\frac{\delta {\tau }_{D}^{2}}{{\overline{\sigma }}_{D}^{2}}\right)
\end{equation}

The normalization is already correct; no adjustment is required.

\paragraph*{Temporal Quantization}

\noindent

We start at laboratory time 
 $ {\tau }_{\text{S}}$. We have the corresponding wave function in quantum time, non-{}relativistic, again from the analysis in 
 \hyperlink{free-hybrid}{Time/Momentum Representation}:

\begin{equation}
\label{equation-xt-free-xt-free-00}\hyperlabel{equation-xt-free-xt-free-00}%
{\chi }_{S}\left({t}_{S},p\right)=\sqrt[4]{\frac{1}{\pi {\sigma }_{0}^{2}}}\mathrm{exp}\left(-i{\overline{E}}_{p}{t}_{S}-\frac{\left({t}_{S}-{\overline{t}}_{S}\right)}{2{\sigma }_{0}^{2}}\right)
\end{equation}

The relative time at the start is defined as 
 $ {t}_{S}\equiv t-{\tau }_{S}$. We assume the average relative time at the start is zero: 
 $ {\overline{t}}_{S}=0$.

We assume that the initial energy is onshell:

\begin{equation}
\label{equation-xt-free-xt-free-1}\hyperlabel{equation-xt-free-xt-free-1}%
{\overline{E}}_{p}\equiv \sqrt{{m}^{2}+{p}^{2}}\approx m+\frac{{p}^{2}}{2m}
\end{equation}

And therefore at the detector:

\begin{equation}
\label{equation-xt-free-xt-free-0}\hyperlabel{equation-xt-free-xt-free-0}%
{\chi }_{D}\left({t}_{D},p\right)=\sqrt[4]{\frac{1}{\pi {\sigma }_{0}^{2}{f}_{DS}^{\left(0\right)}}}\mathrm{exp}\left(-i{\overline{E}}_{p}{t}_{D}-\frac{{\left({t}_{D}-{\overline{t}}_{D}\right)}^{2}}{2{\sigma }_{0}^{2}{f}_{DS}^{\left(0\right)}}-im{\tau }_{DS}\right)
\end{equation}

With:

\begin{equation}
\label{equation-xt-free-xt-free-2}\hyperlabel{equation-xt-free-xt-free-2}%
{\overline{t}}_{D}={\overline{t}}_{S}+\left(\frac{{\overline{E}}_{p}}{m}-1\right){\tau }_{DS}\approx 0
\end{equation}

Further we assume that the variation in starting time 
 $ {\tau }_{\text{S}}$~ is small compared to the time to the detector 
 $ {\tau }_{\text{D}}$~ so:

\begin{equation}
\label{equation-xt-free-xt-free-3b}\hyperlabel{equation-xt-free-xt-free-3b}%
{f}_{DS}^{\left(0\right)}=1-i\frac{{\tau }_{DS}}{m{\sigma }_{0}^{2}}\approx 1-i\frac{{\tau }_{D}}{m{\sigma }_{0}^{2}}
\end{equation}

The wave function in time is then:

\begin{equation}
\label{equation-xt-free-xt-free-5}\hyperlabel{equation-xt-free-xt-free-5}%
{\chi }_{D}\left({t}_{D},p\right)=\sqrt[4]{\frac{1}{\pi {\sigma }_{0}^{2}{f}_{D}^{\left(0\right)}}}\mathrm{exp}\left(-i{\overline{E}}_{p}{t}_{D}-\frac{{t}_{D}^{2}}{2{\sigma }_{0}^{2}{f}_{D}^{\left(0\right)}}-im{\tau }_{DS}\right)
\end{equation}

With the corresponding probability density in time for a specific ray 
 \emph{p}:

\begin{equation}
\label{equation-xt-free-xt-free-time-8}\hyperlabel{equation-xt-free-xt-free-time-8}%
{\stackrel{\frown }{\rho }}_{D}\left({t}_{D},p\right)=\sqrt{\frac{1}{\pi {\stackrel{\frown }{\sigma }}_{D}^{2}}}\mathrm{exp}\left(-\frac{{t}_{D}^{2}}{{\stackrel{\frown }{\sigma }}_{D}^{2}}\right),{\stackrel{\frown }{\sigma }}_{D}^{2}\equiv {\sigma }_{0}^{2}{\left\vert {f}_{D}^{\left(0\right)}\right\vert }^{2}={\sigma }_{0}^{2}\left(1+\frac{{\overline{\tau }}_{D}^{2}}{{m}^{2}{\sigma }_{0}^{4}}\right)
\end{equation}

The full wave function, in time and momentum, is:

\begin{equation}
\label{equation-xt-free-xt-free-6}\hyperlabel{equation-xt-free-xt-free-6}%
{\psi }_{D}\left({t}_{D},p\right)=\sqrt[4]{\frac{1}{\pi {\sigma }_{0}^{2}{f}_{D}^{\left(0\right)}}}\mathrm{exp}\left(-i{\overline{E}}_{p}{t}_{D}-\frac{{t}_{D}^{2}}{2{\sigma }_{0}^{2}{f}_{D}^{\left(0\right)}}-im{\tau }_{DS}\right)\sqrt[4]{\frac{1}{\pi {\widehat{\sigma }}_{1}^{2}}}\mathrm{exp}\left(-\frac{\delta {p}^{2}}{2{\widehat{\sigma }}_{1}^{2}}-i\frac{{p}^{2}}{2m}{\tau }_{DS}\right)
\end{equation}

And the corresponding probability density is:

\begin{equation}
\label{equation-xt-free-xt-free-time-9}\hyperlabel{equation-xt-free-xt-free-time-9}%
{\rho }_{D}\left({t}_{D},\delta p\right)={\stackrel{\frown }{\rho }}_{D}\left({t}_{D},\delta p\right){\overline{\rho }}_{D}\left(\delta p\right)
\end{equation}

Each wave function in time is centered on a different 
 \emph{p}~ and therefore a different time at the detector 
 $ {\tau }_{\text{D}}$. To combine the wave functions in time, we first expand the laboratory time as a function of the momentum 
 \emph{p}, keeping only the first order correction in 
 $ \delta p$:

\begin{equation}
\label{equation-xt-free-xt-free-time-b}\hyperlabel{equation-xt-free-xt-free-time-b}%
{t}_{D}=t-{\tau }_{D}=t-{\overline{\tau }}_{D}-\delta {\tau }_{D}\approx t-{\overline{\tau }}_{D}+{\overline{\tau }}_{D}\frac{\delta p}{\overline{p}}-\cdots 
\end{equation}

We rewrite the probability density for the time part:

\begin{equation}
\label{equation-xt-free-xt-free-time-c}\hyperlabel{equation-xt-free-xt-free-time-c}%
{\rho }_{D}\left(t,\delta p\right)=\sqrt{\frac{1}{\pi {\stackrel{\frown }{\sigma }}_{D}^{2}}}\mathrm{exp}\left(-\frac{{\left(t-{\overline{\tau }}_{D}+{\overline{\tau }}_{D}\frac{\delta p}{\overline{p}}\right)}^{2}}{{\stackrel{\frown }{\sigma }}_{D}^{2}}\right)\sqrt{\frac{1}{\pi {\widehat{\sigma }}_{1}^{2}}}\mathrm{exp}\left(-\frac{\delta {p}^{2}}{{\widehat{\sigma }}_{1}^{2}}\right)
\end{equation}

\paragraph*{Probability Density as a Function of Time}

\noindent

To get the probability density as a function of time we integrate over the momentum 
 \emph{p}:

\begin{equation}
\label{equation-xt-free-xt-free-time-f}\hyperlabel{equation-xt-free-xt-free-time-f}%
{\rho }_{D}\left(t\right)={\displaystyle \int \text{d}p\sqrt{\frac{1}{\pi {\stackrel{\frown }{\sigma }}_{D}^{2}}}\mathrm{exp}\left(-\frac{{\left(t-{\overline{\tau }}_{D}+{\overline{\tau }}_{D}\frac{\delta p}{\overline{p}}\right)}^{2}}{{\stackrel{\frown }{\sigma }}_{D}^{2}}\right)\sqrt{\frac{1}{\pi {\widehat{\sigma }}_{1}^{2}}}\mathrm{exp}\left(-\frac{\delta {p}^{2}}{{\widehat{\sigma }}_{1}^{2}}\right)}
\end{equation}

Which gives:

\begin{equation}
\label{equation-xt-free-xt-free-time-g}\hyperlabel{equation-xt-free-xt-free-time-g}%
{\rho }_{D}\left(t\right)=\frac{1}{\sqrt{\pi {\sigma }_{D}^{2}}}\mathrm{exp}\left(-\frac{{\left(t-{\overline{\tau }}_{D}\right)}^{2}}{{\sigma }_{D}^{2}}\right)
\end{equation}

Where the total dispersion in time breaks out cleanly into the sum of 
 $ t$~ and 
 \emph{p}~ parts:

\begin{equation}
\label{equation-xt-free-xt-free-time-h}\hyperlabel{equation-xt-free-xt-free-time-h}%
{\sigma }_{D}^{2}={\stackrel{\frown }{\sigma }}_{D}^{2}+{\overline{\sigma }}_{D}^{2}
\end{equation}

Or:

\begin{equation}
\label{equation-xt-free-xt-free-time-i}\hyperlabel{equation-xt-free-xt-free-time-i}%
{\sigma }_{D}^{2}={\sigma }_{0}^{2}+\left(\frac{1}{{m}^{2}{\sigma }_{0}^{2}}+\frac{{\widehat{\sigma }}_{1}^{2}}{{\overline{p}}^{2}}\right){\overline{\tau }}_{D}^{2}>{\overline{\sigma }}_{D}^{2}=\frac{{\widehat{\sigma }}_{1}^{2}}{{\overline{p}}^{2}}{\overline{\tau }}_{D}^{2}
\end{equation}

By comparison to the standard quantum theory's dispersion in laboratory time, the dispersion in laboratory time in temporal quantization:
\begin{enumerate}

\item{}Starts off non-{}zero (from the initial dispersion of the wave function in time),

\item{}Grows more rapidly with laboratory time (as one would expect from a theory of fuzzy time).

\end{enumerate}

\subsection{Single Slit Experiment}\label{xt-single}\hyperlabel{xt-single}%
\begin{figure}[H]

\begin{center}
\imgexists{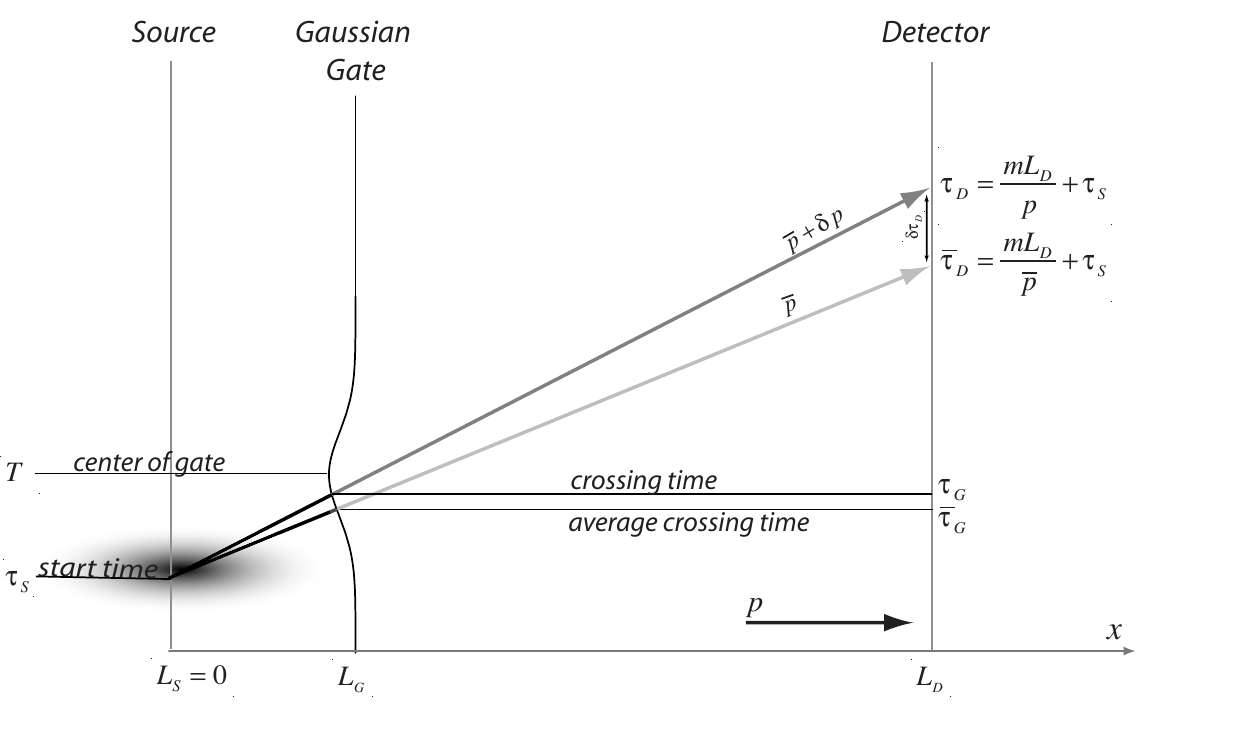}{{\imgevalsize{images/xt-single-1.pdf}{\includegraphics[width=\imgwidth,height=\imgheight,keepaspectratio=true]{images/xt-single-1.pdf}}\quad
}
}{}
\end{center}
\caption{Single Slit With Gaussian Gate}
\label{figure-images-xt-single-1}\hyperlabel{figure-images-xt-single-1}%
\end{figure}

With the results for the free particle in hand, we look at what happens when the free Gaussian test function encounters a gate.

The single gate experiment is:
\begin{enumerate}

\item{}Interesting in its own right,

\item{}A fundamental building block in the analysis of the double slit experiment,

\item{}And a useful preparatory step in most tests of quantum time.

\end{enumerate}

\subsubsection{Gaussian Gates}\label{xt-sgate}\hyperlabel{xt-sgate}%

\paragraph*{In Laboratory Time}

\noindent

We work with Gaussian gates (see 
 \cite{Feynman-1965d}). We take the gate as centered on time 
 \emph{T}~ with width 
 $ {\Sigma }_{G}$. As a function of laboratory time 
 $ {\tau }_{\text{G}}$~ the gate is given by:

\begin{equation}
\label{equation-xt-sgate-xt-sgate-00}\hyperlabel{equation-xt-sgate-xt-sgate-00}%
{\overline{G}}_{G}^{\left(T\right)}=\mathrm{exp}\left(-\frac{{\left({\tau }_{G}-T\right)}^{2}}{2{\Sigma }_{G}^{2}}\right)=\mathrm{exp}\left(-\frac{{T}_{G}^{2}}{2{\Sigma }_{G}^{2}}\right)
\end{equation}

Given 
 \emph{p}, the time of arrival at the gate is given by 
 $ {\tau }_{\text{G}}$:

\begin{equation}
\label{equation-xt-sgate-xt-sgate-0}\hyperlabel{equation-xt-sgate-xt-sgate-0}%
{\tau }_{G}=\frac{m{L}_{G}}{p}+{\tau }_{S},{\overline{\tau }}_{G}=\frac{m{L}_{G}}{\overline{p}}+{\tau }_{S}
\end{equation}

We can write the 
 \emph{T}~ associated with the gate in relative time:

\begin{equation}
\label{equation-xt-sgate-xt-sgate-3}\hyperlabel{equation-xt-sgate-xt-sgate-3}%
{T}_{G}\equiv T-{\tau }_{G},{T}_{\overline{G}}\equiv T-{\overline{\tau }}_{G}
\end{equation}

  $ {T}_{\overline{G}}=0$~ means the beam is dead center on the gate.

\paragraph*{In Momentum Space}

\noindent

To solve for the single slit case in standard quantum theory we will need the gate as a function of the momentum 
 \emph{p}. To get this, we expand 
 $ {T}_{G}$~ in terms of the momentum:

\begin{equation}
\label{equation-xt-sgate-xt-sgate-1}\hyperlabel{equation-xt-sgate-xt-sgate-1}%
{T}_{G}={T}_{\overline{G}}+{\overline{\tau }}_{G}\frac{\delta p}{\overline{p}}
\end{equation}

We use this to rewrite the gate function in momentum space:

\begin{equation}
\label{equation-xt-sgate-xt-sgate-7}\hyperlabel{equation-xt-sgate-xt-sgate-7}%
{\widehat{\overline{G}}}^{\left(P\right)}\left(\delta p\right)=\mathrm{exp}\left(-\frac{{\left(\delta p-P\right)}^{2}}{2{\widehat{\Sigma }}_{G}^{2}}\right)
\end{equation}

With the "effective momentum" of the gate:

\begin{equation}
\label{equation-xt-sgate-xt-sgate-9}\hyperlabel{equation-xt-sgate-xt-sgate-9}%
P\equiv -\frac{{T}_{\overline{G}}}{{\overline{\tau }}_{G}}\overline{p}
\end{equation}

And the dispersion of the gate in momentum space:

\begin{equation}
\label{equation-xt-sgate-xt-sgate-8}\hyperlabel{equation-xt-sgate-xt-sgate-8}%
{\widehat{\Sigma }}_{G}^{2}\equiv \frac{{\Sigma }_{G}^{2}}{{\overline{\tau }}_{G}^{2}}{\overline{p}}^{2}
\end{equation}

\paragraph*{In Quantum Time}

\noindent

In block time the gate is:

\begin{equation}
\label{equation-xt-sgate-xt-sgate-e}\hyperlabel{equation-xt-sgate-xt-sgate-e}%
{G}^{\left(T\right)}\left(t\right)=\mathrm{exp}\left(-\frac{{\left(t-T\right)}^{2}}{2{\Sigma }_{G}^{2}}\right)
\end{equation}

In relative time the gate is:

\begin{equation}
\label{equation-xt-sgate-xt-sgate-f}\hyperlabel{equation-xt-sgate-xt-sgate-f}%
{G}^{\left({T}_{G}\right)}\left({t}_{G}\right)=\mathrm{exp}\left(-\frac{{\left({t}_{G}-{T}_{G}\right)}^{2}}{2{\Sigma }_{G}^{2}}\right)
\end{equation}

\subsubsection{Standard Quantum Theory}\label{xt-ssqt}\hyperlabel{xt-ssqt}%

We first solve for the probability distribution in time at the detector in standard quantum theory. To get the wave function at the detector we use a kernel to get the wave function up to
    the detector, multiply by the gate function, then apply another kernel to get from gate to detector.

\paragraph*{In Momentum Space}

\noindent

The wave function at the detector in momentum space is given by:

\begin{equation}
\label{equation-xt-ssqt-xt-ssqt-deriv-0}\hyperlabel{equation-xt-ssqt-xt-ssqt-deriv-0}%
{\widehat{\xi }}_{D}\left({p}_{D}\right)={\displaystyle \int \text{d}{p}_{G}{\widehat{\overline{K}}}_{DG}\left({p}_{D};{p}_{G}\right){\widehat{\overline{G}}}^{\left(P\right)}\left({p}_{G}\right){\displaystyle \int \text{d}{p}_{S}{\widehat{\overline{K}}}_{G}\left({p}_{G};{p}_{S}\right){\widehat{\xi }}_{S}\left({p}_{S}\right)}}
\end{equation}

The kernels are just 
 $ \delta $~ functions with a bit of phase, so the integrals are trivial:

\begin{equation}
\label{equation-xt-ssqt-xt-ssqt-deriv-3}\hyperlabel{equation-xt-ssqt-xt-ssqt-deriv-3}%
{\widehat{\xi }}_{D}\left(\delta p\right)=\sqrt[4]{\frac{1}{\pi {\widehat{\sigma }}_{1}^{2}}}\mathrm{exp}\left(-\frac{{P}^{2}}{2{\widehat{\Sigma }}_{G}^{2}}+\frac{P}{{\widehat{\Sigma }}_{G}^{2}}\delta p-\frac{\delta {p}^{2}}{2{\widehat{\Sigma }}_{G}^{2}}-\frac{\delta {p}^{2}}{2{\widehat{\sigma }}_{1}^{2}}\right)\mathrm{exp}\left(-i\frac{{p}^{2}}{2m}{\tau }_{DS}\right)
\end{equation}

We rewrite as a Gaussian test function times a normalization factor
 \footnote{Essentially we are completing the square.}:

\begin{equation}
\label{equation-xt-ssqt-xt-ssqt-deriv-4}\hyperlabel{equation-xt-ssqt-xt-ssqt-deriv-4}%
{\widehat{\xi }}_{D}\left(\delta p\right)={\widehat{\overline{N}}}^{\left(sqt\right)}\sqrt[4]{\frac{1}{\pi {\widehat{\overline{\sigma }}}_{G}^{\left(sqt\right)2}}}\mathrm{exp}\left(-\frac{{\left(\delta p-\delta {p}^{\left(sqt\right)}\right)}^{2}}{2{\widehat{\overline{\sigma }}}_{G}^{\left(sqt\right)2}}-i\frac{{p}^{2}}{2m}{\tau }_{DS}\right)
\end{equation}

With dispersion, offset, and normalization:

\begin{equation}
\label{equation-xt-ssqt-xt-ssqt-deriv-5}\hyperlabel{equation-xt-ssqt-xt-ssqt-deriv-5}%
{\widehat{\overline{\sigma }}}_{G}^{\left(sqt\right)2}\equiv \frac{{\widehat{\Sigma }}_{G}^{2}{\widehat{\sigma }}_{1}^{2}}{{\widehat{\Sigma }}_{G}^{2}+{\widehat{\sigma }}_{1}^{2}}
\end{equation}

\begin{equation}
\label{equation-xt-ssqt-xt-ssqt-deriv-6}\hyperlabel{equation-xt-ssqt-xt-ssqt-deriv-6}%
\delta {p}^{\left(sqt\right)}\equiv \frac{P}{{\widehat{\Sigma }}_{G}^{2}}{\widehat{\overline{\sigma }}}_{G}^{\left(sqt\right)2}
\end{equation}

\begin{equation}
\label{equation-xt-ssqt-xt-ssqt-deriv-7}\hyperlabel{equation-xt-ssqt-xt-ssqt-deriv-7}%
{\widehat{\overline{N}}}^{\left(sqt\right)}\equiv \sqrt[4]{\frac{{\widehat{\overline{\sigma }}}_{G}^{\left(sqt\right)2}}{{\widehat{\sigma }}_{1}^{2}}}\mathrm{exp}\left(-\frac{{P}^{2}}{2{\widehat{\Sigma }}_{G}^{2}}+\frac{\delta {p}^{\left(sqt\right)2}}{2{\widehat{\overline{\sigma }}}_{G}^{\left(sqt\right)2}}\right)=\sqrt[4]{\frac{{\widehat{\Sigma }}_{G}^{2}}{{\widehat{\Sigma }}_{G}^{2}+{\widehat{\sigma }}_{1}^{2}}}\mathrm{exp}\left(-\frac{1}{2}\frac{{P}^{2}}{{\widehat{\Sigma }}_{G}^{2}+{\widehat{\sigma }}_{1}^{2}}\right)
\end{equation}

And probability density:

\begin{equation}
\label{equation-xt-ssqt-xt-ssqt-deriv-8}\hyperlabel{equation-xt-ssqt-xt-ssqt-deriv-8}%
{\widehat{\overline{\rho }}}_{D}\left(\delta p\right)={\overline{N}}^{\left(sqt\right)2}\sqrt{\frac{1}{\pi {\widehat{\overline{\sigma }}}_{G}^{\left(sqt\right)2}}}\mathrm{exp}\left(-\frac{{\left(\delta p-\delta {p}^{\left(sqt\right)}\right)}^{2}}{{\widehat{\overline{\sigma }}}_{G}^{\left(sqt\right)2}}\right)
\end{equation}

\paragraph*{In Laboratory Time}

\noindent

We change coordinates from 
 $ \delta p$~ to 
 $ \delta {\tau }_{D}$~ using 
 $ \delta p\approx -\overline{p}\frac{\delta \tau }{\overline{\tau }}$:

\begin{equation}
\label{equation-xt-ssqt-xt-single-tau-2}\hyperlabel{equation-xt-ssqt-xt-single-tau-2}%
{\xi }_{D}\left(\delta {\tau }_{D}\right)={\overline{N}}^{\left(sqt\right)}\sqrt[4]{\frac{1}{\pi {\overline{\sigma }}_{D}^{\left(sqt\right)2}}}\mathrm{exp}\left(-\frac{{\left(\delta {\tau }_{D}-\delta {\tau }_{D}^{\left(sqt\right)}\right)}^{2}}{2{\overline{\sigma }}_{D}^{\left(sqt\right)2}}\right)\mathrm{exp}\left(-i\frac{{\overline{p}}^{2}}{2m}{\overline{\tau }}_{D}+i\frac{{\overline{p}}^{2}}{2m}\delta {\tau }_{D}+i\frac{{\overline{p}}^{2}}{2m}\frac{\delta {\tau }_{D}^{2}}{{\overline{\tau }}_{D}}\right)
\end{equation}

With dispersion, offset, and normalization:

\begin{equation}
\label{equation-xt-ssqt-xt-single-tau-3}\hyperlabel{equation-xt-ssqt-xt-single-tau-3}%
{\overline{\sigma }}_{D}^{\left(sqt\right)2}=\frac{{\Sigma }_{G}^{2}{\overline{\sigma }}_{G}^{2}}{{\Sigma }_{G}^{2}+{\overline{\sigma }}_{G}^{2}}\frac{{\overline{\tau }}_{D}^{2}}{{\overline{\tau }}_{G}^{2}},{\overline{\sigma }}_{G}^{2}=\frac{{\widehat{\sigma }}_{1}^{2}}{{\overline{p}}^{2}}{\overline{\tau }}_{G}^{2}
\end{equation}

\begin{equation}
\label{equation-xt-ssqt-xt-single-tau-4}\hyperlabel{equation-xt-ssqt-xt-single-tau-4}%
\delta {\tau }_{D}^{\left(sqt\right)}=-\frac{{\overline{\sigma }}_{G}^{2}}{{\Sigma }_{G}^{2}+{\overline{\sigma }}_{G}^{2}}\frac{{T}_{\overline{G}}}{{\overline{\tau }}_{G}}{\overline{\tau }}_{D}
\end{equation}

\begin{equation}
\label{equation-xt-ssqt-xt-single-tau-5}\hyperlabel{equation-xt-ssqt-xt-single-tau-5}%
{\overline{N}}^{\left(sqt\right)}=\sqrt[4]{\frac{{\Sigma }_{G}^{2}}{{\Sigma }_{G}^{2}+{\overline{\sigma }}_{G}^{2}}}\mathrm{exp}\left(-\frac{1}{2}\frac{{T}_{\overline{G}}^{2}}{{\Sigma }_{G}^{2}+{\overline{\sigma }}_{G}^{2}}\right)
\end{equation}

The dispersion is dominated by the narrower of the gate and the beam. The offset has the opposite sense of 
 $ {T}_{\overline{G}}\equiv T-{\tau }_{\overline{G}}$, so nudges the beam closer to the center of the gate. The square of the normalization constant gives the probability of going through the gate.

The probability density is:

\begin{equation}
\label{equation-xt-ssqt-xt-single-tau-9}\hyperlabel{equation-xt-ssqt-xt-single-tau-9}%
{\overline{\rho }}_{D}\left(\delta {\tau }_{D}\right)={\overline{N}}^{\left(sqt\right)2}\sqrt{\frac{1}{\pi {\overline{\sigma }}_{D}^{\left(sqt\right)2}}}\mathrm{exp}\left(-\frac{{\left(\delta {\tau }_{D}-\delta {\tau }_{D}^{\left(sqt\right)}\right)}^{2}}{{\overline{\sigma }}_{D}^{\left(sqt\right)2}}\right)
\end{equation}

\paragraph*{Fast and Slow Gates}

\noindent

As the gate shuts, the effective dispersion in laboratory time scales as the width of the gate:

\begin{equation}
\label{equation-xt-ssqt-xt-single-sqt-4}\hyperlabel{equation-xt-ssqt-xt-single-sqt-4}%
\underset{{\Sigma }_{G}\rightarrow 0}{\mathrm{lim}}{\overline{\sigma }}_{D}^{\left(sqt\right)2}\rightarrow \frac{{\Sigma }_{G}^{2}}{{\overline{\tau }}_{G}^{2}}{\overline{\tau }}_{D}^{2}
\end{equation}

As the gate opens, the offset goes to zero, the effective dispersion in laboratory time goes to the free dispersion, and the probability distribution in 
 $ \delta {\tau }_{D}$~ goes to the free probability distribution:

\begin{equation}
\label{equation-xt-ssqt-xt-single-sqt-5}\hyperlabel{equation-xt-ssqt-xt-single-sqt-5}%
\underset{{\Sigma }_{G}\rightarrow \infty }{\mathrm{lim}}{\overline{\sigma }}_{D}^{\left(sqt\right)2}={\overline{\sigma }}_{D}^{2},\underset{{\Sigma }_{G}\rightarrow \infty }{\mathrm{lim}}{\overline{\rho }}_{D}\left(\delta {\tau }_{D}\right)=\sqrt{\frac{1}{\pi {\overline{\sigma }}_{D}^{2}}}\mathrm{exp}\left(-\frac{\delta {\tau }_{D}^{2}}{{\overline{\sigma }}_{D}^{2}}\right)
\end{equation}

\subsubsection{Temporal Quantization}\label{xt-stq}\hyperlabel{xt-stq}%

Now we compute the probability density in time at the detector in temporal quantization. The final wave function in time for one ray 
 \emph{p}~ is given by the product of the wave function at the gate, the gate itself, and the kernel from gate to detector, integrated over the quantum time and momentum at the
    gate:

\begin{equation}
\label{equation-xt-stq-xt-soft-time-00}\hyperlabel{equation-xt-stq-xt-soft-time-00}%
{\psi }_{D}\left({t}_{D},{p}_{D}\right)={\displaystyle \int \text{d}{t}_{G}\text{d}{p}_{G}{K}_{DG}\left({t}_{D},{p}_{D};{t}_{G},{p}_{G}\right){G}^{\left({T}_{G}\right)}\left({t}_{G}\right){\displaystyle \int \text{d}{t}_{S}\text{d}{p}_{S}{K}_{G}\left({t}_{G},{p}_{G};{t}_{S},{p}_{S}\right){\psi }_{S}\left({t}_{S},{p}_{S}\right)}}
\end{equation}

\paragraph*{Probability Density in Time and Momentum}

\noindent

We break out the time part:

\begin{equation}
\label{equation-xt-stq-xt-soft-time-0}\hyperlabel{equation-xt-stq-xt-soft-time-0}%
{\chi }_{D}\left({t}_{D},p\right)={\displaystyle \int \text{d}{t}_{G}{\stackrel{\frown }{K}}_{DG}\left({t}_{D};{t}_{G}\right){G}^{\left({T}_{G}\right)}\left({t}_{G}\right){\displaystyle \int \text{d}{t}_{S}{\stackrel{\frown }{K}}_{G}\left({t}_{G};{t}_{S}\right){\chi }_{S}\left({t}_{S},p\right)}}
\end{equation}

Explicitly:

\begin{equation}
\label{equation-xt-stq-xt-stq-check3-end}\hyperlabel{equation-xt-stq-xt-stq-check3-end}%
{\chi }_{D}\left({t}_{D},p\right)={\displaystyle \int \text{d}{t}_{G}\sqrt{\frac{im}{2\pi {\tau }_{DG}}}\left(\begin{array}{l}\mathrm{exp}\left(-im\frac{{\left({t}_{D}-{t}_{G}\right)}^{2}}{2{\tau }_{DG}}-im\left({t}_{D}-{t}_{G}\right)-im{\tau }_{DG}\right)\mathrm{exp}\left(-\frac{{\left({t}_{G}-{T}_{G}\right)}^{2}}{2{\Sigma }_{G}^{2}}\right)\\ \times \sqrt[4]{\frac{1}{\pi {\sigma }_{0}^{2}{f}_{G}^{\left(0\right)2}}}\mathrm{exp}\left(-i{\overline{E}}_{p}{t}_{G}-\frac{{\left({t}_{G}-\left(\gamma -1\right){\tau }_{GS}\right)}^{2}}{2{\sigma }_{0}^{2}{f}_{G}^{\left(0\right)}}-im{\tau }_{GS}\right)\end{array}\right)}
\end{equation}

We solve this by ansatz; we compare this integral to that for a free wave function, then set the parameters for the free wave function to match those for the gated integral. Since we know
    the free result, we are then done.

Consider a free wave function with initial dispersion 
 $ {\sigma }_{X}^{2}$~ and starting point 
 $ {\overline{t}}_{X}$. We can get the wave function at the detector by integrating the kernel from gate to detector times the wave function at the gate.

\begin{equation}
\label{equation-xt-stq-xt-stq-check3-start}\hyperlabel{equation-xt-stq-xt-stq-check3-start}%
{\chi }_{D}^{\left(X\right)}\left({t}_{D},p\right)={\displaystyle \int \text{d}{t}_{G}\sqrt{\frac{im}{2\pi {\tau }_{DG}}}\left(\begin{array}{l}\mathrm{exp}\left(-im\frac{{\left({t}_{D}-{t}_{G}\right)}^{2}}{2{\tau }_{DG}}-im\left({t}_{D}-{t}_{G}\right)-im{\tau }_{DG}\right)\\ \times \sqrt[4]{\frac{1}{\pi {\sigma }_{X}^{2}{f}_{G}^{\left(X\right)\left(0\right)}}}\mathrm{exp}\left(-i{\overline{E}}_{p}{t}_{G}-\frac{{\left({t}_{G}-\left({\overline{t}}_{X}+\left(\gamma -1\right){\tau }_{GS}\right)\right)}^{2}}{2{\sigma }_{X}^{2}{f}_{G}^{\left(X\right)\left(0\right)}}-im{\tau }_{GS}\right)\end{array}\right)}
\end{equation}

Which gives (\hyperlink{free-fourt}{Time/Space Representation}):

\begin{equation}
\label{equation-xt-stq-xt-stq-check3-answer}\hyperlabel{equation-xt-stq-xt-stq-check3-answer}%
{\chi }_{D}^{\left(X\right)}\left({t}_{D},p\right)=\sqrt[4]{\frac{1}{\pi {\sigma }_{X}^{2}{f}_{DS}^{\left(X\right)\left(0\right)}}}\mathrm{exp}\left(-i{\overline{E}}_{p}{t}_{D}-\frac{{\left({t}_{D}-\left({\overline{t}}_{X}+\left(\gamma -1\right){\tau }_{DS}\right)\right)}^{2}}{2{\sigma }_{X}^{2}{f}_{DS}^{\left(X\right)\left(0\right)}}-im{\tau }_{DS}\right)
\end{equation}

We make the assumptions that the system is nonrelativistic, that there is a short run from source to gate ("near gate"), and that there is a long run from gate to detector ("far
    detector"):

\begin{equation}
\label{equation-xt-stq-xt-stq-check3-assump}\hyperlabel{equation-xt-stq-xt-stq-check3-assump}%
\begin{array}{l}\gamma =1\\ {\tau }_{G}\rightarrow 0\Rightarrow {f}_{G}^{\left(0\right)},{f}_{G}^{\left(X\right)\left(0\right)}=1-i\frac{{\tau }_{G}}{m{\sigma }_{0,X}^{2}}\rightarrow 1\\ {\tau }_{D}\gg {\tau }_{G}\rightarrow {f}_{DS}^{\left(X\right)\left(0\right)}=1-i\frac{{\tau }_{DG}}{m{\sigma }_{X}^{2}}\rightarrow {f}_{D}^{\left(X\right)\left(0\right)}=1-i\frac{{\tau }_{D}}{m{\sigma }_{X}^{2}}\end{array}
\end{equation}

With these assumptions the gated case is:

\begin{equation}
\label{equation-xt-stq-xt-stq-check4-end1}\hyperlabel{equation-xt-stq-xt-stq-check4-end1}%
{\chi }_{D}\left({t}_{D},p\right)={\displaystyle \int \text{d}{t}_{G}\sqrt{\frac{im}{2\pi {\tau }_{DG}}}\left(\begin{array}{l}\mathrm{exp}\left(-im\frac{{\left({t}_{D}-{t}_{G}\right)}^{2}}{2{\tau }_{DG}}-im\left({t}_{D}-{t}_{G}\right)-im{\tau }_{DG}\right)\mathrm{exp}\left(-\frac{{\left({t}_{G}-{T}_{G}\right)}^{2}}{2{\Sigma }_{G}^{2}}\right)\\ \times \sqrt[4]{\frac{1}{\pi {\sigma }_{0}^{2}}}\mathrm{exp}\left(-i{\overline{E}}_{p}{t}_{G}-\frac{{t}_{G}^{2}}{2{\sigma }_{0}^{2}}-im{\tau }_{GS}\right)\end{array}\right)}
\end{equation}

And the free case is:

\begin{equation}
\label{equation-xt-stq-xt-stq-check4-start1}\hyperlabel{equation-xt-stq-xt-stq-check4-start1}%
{\chi }_{D}^{\left(X\right)}\left({t}_{D},p\right)={\displaystyle \int \text{d}{t}_{G}\sqrt{\frac{im}{2\pi {\tau }_{DG}}}\left(\begin{array}{l}\mathrm{exp}\left(-im\frac{{\left({t}_{D}-{t}_{G}\right)}^{2}}{2{\tau }_{DG}}-im\left({t}_{D}-{t}_{G}\right)-im{\tau }_{DG}\right)\\ \times \sqrt[4]{\frac{1}{\pi {\sigma }_{X}^{2}}}\mathrm{exp}\left(-i{\overline{E}}_{p}{t}_{G}-\frac{{\left({t}_{G}-{\overline{t}}_{X}\right)}^{2}}{2{\sigma }_{X}^{2}}-im{\tau }_{GS}\right)\end{array}\right)}
\end{equation}

With result:

\begin{equation}
\label{equation-xt-stq-xt-stq-check4-answer1}\hyperlabel{equation-xt-stq-xt-stq-check4-answer1}%
{\chi }_{D}^{\left(X\right)}\left({t}_{D},p\right)=\sqrt[4]{\frac{1}{\pi {\sigma }_{X}^{2}{f}_{D}^{\left(X\right)\left(0\right)}}}\mathrm{exp}\left(-i{\overline{E}}_{p}{t}_{D}-\frac{{\left({t}_{D}-{\overline{t}}_{X}\right)}^{2}}{2{\sigma }_{X}^{2}{f}_{D}^{\left(X\right)\left(0\right)}}-im{\tau }_{DS}\right)
\end{equation}

To match the term quadratic in 
 $ {t}_{G}$~ we set:

\begin{equation}
\label{equation-xt-stq-xt-stq-check4-quad}\hyperlabel{equation-xt-stq-xt-stq-check4-quad}%
\frac{1}{{\sigma }_{X}^{2}}=\frac{1}{{\Sigma }_{G}^{2}}+\frac{1}{{\sigma }_{0}^{2}}\rightarrow {\sigma }_{X}^{2}=\frac{{\Sigma }_{G}^{2}{\sigma }_{0}^{2}}{{\Sigma }_{G}^{2}+{\sigma }_{0}^{2}}
\end{equation}

To match the term linear in 
 $ {t}_{G}$~ we set:

\begin{equation}
\label{equation-xt-stq-xt-stq-check4-linear}\hyperlabel{equation-xt-stq-xt-stq-check4-linear}%
\frac{{\overline{t}}_{X}}{{\sigma }_{X}^{2}}{t}_{G}=\frac{{T}_{G}}{{\Sigma }_{G}^{2}}{t}_{G}\rightarrow {\overline{t}}_{X}=\frac{{T}_{G}{\sigma }_{0}^{2}}{{\Sigma }_{G}^{2}+{\sigma }_{0}^{2}}
\end{equation}

And to match the normalization we define:

\begin{equation}
\label{equation-xt-stq-xt-stq-check4-norm}\hyperlabel{equation-xt-stq-xt-stq-check4-norm}%
\stackrel{\frown }{N}=\sqrt[4]{\frac{{\sigma }_{X}^{2}}{{\sigma }_{0}^{2}}}\mathrm{exp}\left(-\frac{{T}_{G}^{2}}{2{\Sigma }_{G}^{2}}+\frac{{\overline{t}}_{X}^{2}}{2{\sigma }_{X}^{2}}\right)=\sqrt[4]{\frac{{\sigma }_{X}^{2}}{{\sigma }_{0}^{2}}}\mathrm{exp}\left(-\frac{1}{2}\frac{{T}_{G}^{2}}{{\Sigma }_{G}^{2}+{\sigma }_{0}^{2}}\right)
\end{equation}

With this normalization the free and gated cases match giving:

\begin{equation}
\label{equation-xt-stq-xt-stq-check4-chi}\hyperlabel{equation-xt-stq-xt-stq-check4-chi}%
{\chi }_{D}\left({t}_{D},p\right)=\stackrel{\frown }{N}\sqrt[4]{\frac{1}{\pi {\sigma }_{X}^{2}{f}_{D}^{\left(X\right)\left(0\right)}}}\mathrm{exp}\left(-i{\overline{E}}_{p}{t}_{D}-\frac{{\left({t}_{D}-{\overline{t}}_{X}\right)}^{2}}{2{\sigma }_{X}^{2}{f}_{D}^{\left(X\right)\left(0\right)}}-im{\tau }_{DS}\right)
\end{equation}

With the time part of the probability density:

\begin{equation}
\label{equation-xt-stq-xt-stq-check4-rho}\hyperlabel{equation-xt-stq-xt-stq-check4-rho}%
{\stackrel{\frown }{\rho }}_{D}\left({t}_{D},p\right)={\stackrel{\frown }{N}}^{2}\sqrt{\frac{1}{\pi {\stackrel{\frown }{\sigma }}_{D}^{2}}}\mathrm{exp}\left(-\frac{{\left({t}_{D}-{\overline{t}}_{X}\right)}^{2}}{{\stackrel{\frown }{\sigma }}_{D}^{2}}\right)
\end{equation}

And the dispersion in time:

\begin{equation}
\label{equation-xt-stq-xt-stq-check4-sigtd}\hyperlabel{equation-xt-stq-xt-stq-check4-sigtd}%
{\stackrel{\frown }{\sigma }}_{D}^{2}\equiv {\sigma }_{X}^{2}{\left\vert {f}_{D}^{\left(X\right)\left(0\right)}\right\vert }^{2}={\sigma }_{X}^{2}\left\vert 1+{\left(\frac{{\tau }_{D}}{m{\sigma }_{X}^{2}}\right)}^{2}\right\vert 
\end{equation}

The probability density in time and momentum is:

\begin{equation}
\label{equation-xt-stq-xt-stq-check4-rho2}\hyperlabel{equation-xt-stq-xt-stq-check4-rho2}%
{\rho }_{D}\left({t}_{D},p\right)={\stackrel{\frown }{N}}^{2}\sqrt{\frac{1}{\pi {\stackrel{\frown }{\sigma }}_{D}^{2}}}\mathrm{exp}\left(-\frac{{\left({t}_{D}-{\overline{t}}_{X}\right)}^{2}}{{\stackrel{\frown }{\sigma }}_{D}^{2}}\right)\sqrt{\frac{1}{\pi {\widehat{\sigma }}_{1}^{2}}}\mathrm{exp}\left(-\frac{\delta {p}^{2}}{{\widehat{\sigma }}_{1}^{2}}\right)
\end{equation}

\paragraph*{Gateless Gate}

\noindent

We verify that in the limit as the width of the gate, 
 $ {\Sigma }_{G}$, goes to infinity we recover the free case. We have:

\begin{equation}
\label{equation-xt-stq-xt-stq-check5-free1}\hyperlabel{equation-xt-stq-xt-stq-check5-free1}%
\underset{{\Sigma }_{G}\rightarrow \infty }{\mathrm{lim}}{\sigma }_{X}^{2}\rightarrow {\sigma }_{0}^{2},{t}_{X}\rightarrow 0,\stackrel{\frown }{N}\rightarrow 1,\rho \left({t}_{D},\delta p\right)={\rho }_{D}^{\left(free\right)}\left({t}_{D},\delta p\right)
\end{equation}

\paragraph*{Probability Density in Time Only}

\noindent

To get the probability density in time only we integrate over the momentum 
 \emph{p}:

\begin{equation}
\label{equation-xt-stq-xt-stq-check7-int0}\hyperlabel{equation-xt-stq-xt-stq-check7-int0}%
{\rho }_{D}\left({t}_{D}\right)={\displaystyle \int \text{d}\delta p{\rho }_{D}\left({t}_{D},\delta p\right)}
\end{equation}

Explicitly:

\begin{equation}
\label{equation-xt-stq-xt-stq-check7-explicit}\hyperlabel{equation-xt-stq-xt-stq-check7-explicit}%
{\displaystyle \int \text{d}\delta p\sqrt{\frac{{\sigma }_{X}^{2}}{{\sigma }_{0}^{2}}}\mathrm{exp}\left(-\frac{{\left({T}_{\overline{G}}+{\overline{\tau }}_{G}\frac{\delta p}{\overline{p}}\right)}^{2}}{{\Sigma }_{G}^{2}+{\sigma }_{0}^{2}}\right)\sqrt{\frac{1}{\pi {\stackrel{\frown }{\sigma }}_{D}^{2}}}\mathrm{exp}\left(-\frac{{\left(t-{\overline{\tau }}_{D}+{\overline{\tau }}_{D}\frac{\delta p}{\overline{p}}\right)}^{2}}{{\stackrel{\frown }{\sigma }}_{D}^{2}}\right)\sqrt{\frac{1}{\pi {\widehat{\sigma }}_{1}^{2}}}\mathrm{exp}\left(-\frac{\delta {p}^{2}}{{\widehat{\sigma }}_{1}^{2}}\right)}
\end{equation}

We translate the dispersion in time, 
 $ {\sigma }_{0}$, of the wave function to momentum space:

\begin{equation}
\label{equation-xt-stq-xt-stq-check7-trick}\hyperlabel{equation-xt-stq-xt-stq-check7-trick}%
{\widehat{\sigma }}_{0}^{2}\equiv \frac{{\sigma }_{0}^{2}}{{\overline{\tau }}_{G}^{2}}{\overline{p}}^{2}
\end{equation}

And we translate the norm to momentum space:

\begin{equation}
\label{equation-xt-stq-xt-stq-check7-norm}\hyperlabel{equation-xt-stq-xt-stq-check7-norm}%
\mathrm{exp}\left(-\frac{{\left({T}_{\overline{G}}+{\overline{\tau }}_{G}\frac{\delta p}{\overline{p}}\right)}^{2}}{{\Sigma }_{G}^{2}+{\sigma }_{0}^{2}}\right)=\mathrm{exp}\left(-\frac{{\left(P-\delta p\right)}^{2}}{{\widehat{\Sigma }}_{G}^{2}+{\widehat{\sigma }}_{0}^{2}}\right)
\end{equation}

We define the momentum part of the probability density:

\begin{equation}
\label{equation-xt-stq-xt-stq-check7-rescale}\hyperlabel{equation-xt-stq-xt-stq-check7-rescale}%
{\widehat{\overline{\rho }}}_{D}^{\left(tq\right)}\left(\delta p\right)\equiv \mathrm{exp}\left(-\frac{{\left(P-\delta p\right)}^{2}}{{\widehat{\Sigma }}_{G}^{2}+{\widehat{\sigma }}_{0}^{2}}\right)\sqrt{\frac{1}{\pi {\widehat{\sigma }}_{1}^{2}}}\mathrm{exp}\left(-\frac{\delta {p}^{2}}{{\widehat{\sigma }}_{1}^{2}}\right)
\end{equation}

Now the integral is:

\begin{equation}
\label{equation-xt-stq-xt-stq-check7-int2}\hyperlabel{equation-xt-stq-xt-stq-check7-int2}%
{\rho }_{D}\left({t}_{D}\right)={\displaystyle \int \text{d}\delta p\sqrt{\frac{{\sigma }_{X}^{2}}{{\sigma }_{0}^{2}}}\sqrt{\frac{1}{\pi {\stackrel{\frown }{\sigma }}_{D}^{2}}}\mathrm{exp}\left(-\frac{{\left(t-{\overline{\tau }}_{D}+{\overline{\tau }}_{D}\frac{\delta p}{\overline{p}}\right)}^{2}}{{\stackrel{\frown }{\sigma }}_{D}^{2}}\right){\widehat{\overline{\rho }}}_{D}^{\left(tq\right)}\left(\delta p\right)}
\end{equation}

We rewrite the momentum part of the probability density as:

\begin{equation}
\label{equation-xt-stq-xt-stq-check7-newxi}\hyperlabel{equation-xt-stq-xt-stq-check7-newxi}%
{\widehat{\overline{\rho }}}_{D}^{\left(tq\right)}\left(\delta p\right)={\overline{N}}^{\left(tq\right)2}\sqrt{\frac{1}{\pi {\widehat{\overline{\sigma }}}_{G}^{\left(tq\right)2}}}\mathrm{exp}\left(-\frac{{\left(\delta p-\delta {p}^{\left(tq\right)}\right)}^{2}}{{\widehat{\overline{\sigma }}}_{G}^{\left(tq\right)2}}\right)
\end{equation}

With dispersion, offset, and normalization:

\begin{equation}
\label{equation-xt-stq-xt-stq-check7-others}\hyperlabel{equation-xt-stq-xt-stq-check7-others}%
\begin{array}{l}{\widehat{\overline{\sigma }}}_{G}^{\left(tq\right)2}\equiv \frac{\left({\widehat{\Sigma }}_{G}^{2}+{\widehat{\sigma }}_{0}^{2}\right){\widehat{\sigma }}_{1}^{2}}{{\widehat{\Sigma }}_{G}^{2}+{\widehat{\sigma }}_{0}^{2}+{\widehat{\sigma }}_{1}^{2}}\\ \delta {p}^{\left(tq\right)}\equiv \frac{P}{{\widehat{\Sigma }}_{G}^{2}+{\widehat{\sigma }}_{0}^{2}}{\widehat{\overline{\sigma }}}_{G}^{\left(tq\right)2}\\ {\widehat{\overline{N}}}^{\left(tq\right)}\equiv \sqrt[4]{\frac{{\widehat{\overline{\sigma }}}_{G}^{\left(tq\right)2}}{{\widehat{\sigma }}_{1}^{2}}}\mathrm{exp}\left(-\frac{1}{2}\frac{{P}^{2}}{{\widehat{\Sigma }}_{G}^{2}+{\widehat{\sigma }}_{0}^{2}}+\frac{1}{2}\frac{\delta {p}^{\left(tq\right)2}}{{\widehat{\overline{\sigma }}}_{G}^{\left(tq\right)2}}\right)=\sqrt[4]{\frac{{\widehat{\Sigma }}_{G}^{2}+{\widehat{\sigma }}_{0}^{2}}{{\widehat{\Sigma }}_{G}^{2}+{\widehat{\sigma }}_{0}^{2}+{\widehat{\sigma }}_{1}^{2}}}\mathrm{exp}\left(-\frac{1}{2}\frac{{P}^{2}}{{\widehat{\Sigma }}_{G}^{2}+{\widehat{\sigma }}_{0}^{2}+{\widehat{\sigma }}_{1}^{2}}\right)\end{array}
\end{equation}

The temporal quantization probability density is the standard quantum theory probability density with the rescaling:

\begin{equation}
\label{equation-xt-stq-xt-stq-check7-trans}\hyperlabel{equation-xt-stq-xt-stq-check7-trans}%
{\widehat{\Sigma }}_{G}^{2}\rightarrow {\widehat{\Sigma }}_{G}^{2}+{\widehat{\sigma }}_{0}^{2}\Rightarrow {\widehat{\overline{\rho }}}_{D}^{\left(sqt\right)}\left(\delta p\right)\rightarrow {\widehat{\overline{\rho }}}_{D}^{\left(tq\right)}\left(\delta p\right)
\end{equation}

We have already done this integral a few times. Again we shift variables from 
 \emph{p}~ to 
 $ \delta {\tau }_{D}$:

\begin{equation}
\label{equation-xt-stq-xt-stq-check8-0}\hyperlabel{equation-xt-stq-xt-stq-check8-0}%
{\rho }_{D}\left({t}_{D}\right)={\displaystyle \int \text{d}\delta {\tau }_{D}\sqrt{\frac{{\sigma }_{X}^{2}}{{\sigma }_{0}^{2}}}\sqrt{\frac{1}{\pi {\stackrel{\frown }{\sigma }}_{D}^{2}}}\mathrm{exp}\left(-\frac{{\left(t-{\overline{\tau }}_{D}-\delta {\tau }_{D}\right)}^{2}}{{\stackrel{\frown }{\sigma }}_{D}^{2}}\right){\overline{\rho }}_{D}^{\left(tq\right)}\left(\delta {\tau }_{D}\right)}
\end{equation}

With the space part of the probability density:

\begin{equation}
\label{equation-xt-stq-xt-stq-check8-1}\hyperlabel{equation-xt-stq-xt-stq-check8-1}%
{\overline{\rho }}_{D}^{\left(tq\right)}\left(\delta {\tau }_{D}\right)={\overline{N}}^{\left(tq\right)2}\sqrt{\frac{1}{\pi {\overline{\sigma }}_{D}^{\left(tq\right)2}}}\mathrm{exp}\left(-\frac{{\left(\delta {\tau }_{D}-\delta {\tau }_{D}^{\left(tq\right)}\right)}^{2}}{{\overline{\sigma }}_{D}^{\left(tq\right)2}}\right)
\end{equation}

With dispersion, offset, and normalization:

\begin{equation}
\label{equation-xt-stq-xt-stq-check8-2}\hyperlabel{equation-xt-stq-xt-stq-check8-2}%
{\overline{\sigma }}_{D}^{\left(tq\right)2}=\frac{\left({\Sigma }_{G}^{2}+{\sigma }_{0}^{2}\right){\overline{\sigma }}_{G}^{2}}{{\Sigma }_{G}^{2}+{\sigma }_{0}^{2}+{\overline{\sigma }}_{G}^{2}}\frac{{\overline{\tau }}_{D}^{2}}{{\overline{\tau }}_{G}^{2}}
\end{equation}

\begin{equation}
\label{equation-xt-stq-xt-stq-check8-3}\hyperlabel{equation-xt-stq-xt-stq-check8-3}%
\delta {\tau }_{D}^{\left(tq\right)}=-\frac{{\overline{\sigma }}_{G}^{2}}{{\Sigma }_{G}^{2}+{\sigma }_{0}^{2}+{\overline{\sigma }}_{G}^{2}}\frac{{T}_{\overline{G}}}{{\overline{\tau }}_{G}}{\overline{\tau }}_{D}
\end{equation}

\begin{equation}
\label{equation-xt-stq-xt-stq-check8-4}\hyperlabel{equation-xt-stq-xt-stq-check8-4}%
{\overline{N}}^{\left(tq\right)}=\sqrt[4]{\frac{{\Sigma }_{G}^{2}}{{\Sigma }_{G}^{2}+{\sigma }_{0}^{2}+{\overline{\sigma }}_{G}^{2}}}\mathrm{exp}\left(-\frac{1}{2}\frac{{T}_{\overline{G}}^{2}}{{\Sigma }_{G}^{2}+{\sigma }_{0}^{2}+{\overline{\sigma }}_{G}^{2}}\right)
\end{equation}

This is the standard quantum theory result with the replacement 
 $ {\Sigma }_{G}^{2}\rightarrow {\Sigma }_{G}^{2}+{\sigma }_{0}^{2}$.

The final result for the probability density in time is:

\begin{equation}
\label{equation-xt-stq-xt-stq-check8-5}\hyperlabel{equation-xt-stq-xt-stq-check8-5}%
{\rho }_{D}\left({t}_{D}\right)=\sqrt{\frac{{\Sigma }_{G}^{2}}{{\Sigma }_{G}^{2}+{\sigma }_{0}^{2}}}\sqrt{\frac{1}{\pi {\sigma }_{D}^{2}}}\mathrm{exp}\left(-\frac{{\left(t-{\overline{\tau }}_{D}-\delta {\tau }_{D}^{\left(tq\right)}\right)}^{2}}{{\sigma }_{D}^{2}}\right)
\end{equation}

With the dispersion the sum of the time and space parts, as before:

\begin{equation}
\label{equation-xt-stq-xt-stq-check8-6}\hyperlabel{equation-xt-stq-xt-stq-check8-6}%
{\sigma }_{D}^{2}={\stackrel{\frown }{\sigma }}_{D}^{2}+{\overline{\sigma }}_{D}^{\left(tq\right)2}
\end{equation}

The effect of temporal quantization is to increase the dispersion in time. The total dispersion in time is increased by the dispersion of the time part:

\begin{equation}
\label{equation-xt-stq-xt-stq-check8-7}\hyperlabel{equation-xt-stq-xt-stq-check8-7}%
{\stackrel{\frown }{\sigma }}_{D}^{2}={{\sigma }_{D}^{\left(free\right)2}\vert }_{{\sigma }_{0}^{2}\rightarrow {\sigma }_{X}^{2}\equiv \frac{{\Sigma }_{G}^{2}{\sigma }_{0}^{2}}{{\Sigma }_{G}^{2}+{\sigma }_{0}^{2}}}
\end{equation}

And by the increased dispersion in the space part.

\begin{equation}
\label{equation-xt-stq-xt-stq-check8-8}\hyperlabel{equation-xt-stq-xt-stq-check8-8}%
{\overline{\sigma }}_{D}^{\left(tq\right)2}={{\overline{\sigma }}_{D}^{\left(sqt\right)2}\vert }_{{\Sigma }_{G}^{2}\rightarrow {\Sigma }_{G}^{2}+{\sigma }_{0}^{2}}
\end{equation}

In spite of the careless enthusiasm of our approximations, these results seem reasonable enough: temporal quantization induces fuzziness in time and additional fuzziness in space.

\subsection{Double Slit Experiment}\label{xt-double}\hyperlabel{xt-double}%
\begin{quote}

You'll take the high road

And I'll take the low road

And I'll be in Scotland afore ye

\hspace*\fill---~Loch Lomond\end{quote}

We will take the gates as:

\begin{equation}
\label{equation-xt-double-xt-double-4}\hyperlabel{equation-xt-double-xt-double-4}%
{G}^{\left(1,2\right)}\left(t\right)=\mathrm{exp}\left(-\frac{{\left(t-{T}_{1,2}\right)}^{2}}{2{\Sigma }_{G}^{2}}\right),{T}_{1,2}=T\mp \Delta T
\end{equation}

We will make the same "near gate, far detector" assumptions of the previous section. In keeping with that, we will take the distance between the gates small relative to the distance to the
    detector 
 $ \Delta T\ll {\tau }_{D}$.

\subsubsection{Relative Lack of Interference in the Simplest Case}\label{xt-dnull}\hyperlabel{xt-dnull}%
\begin{figure}[H]

\begin{center}
\imgexists{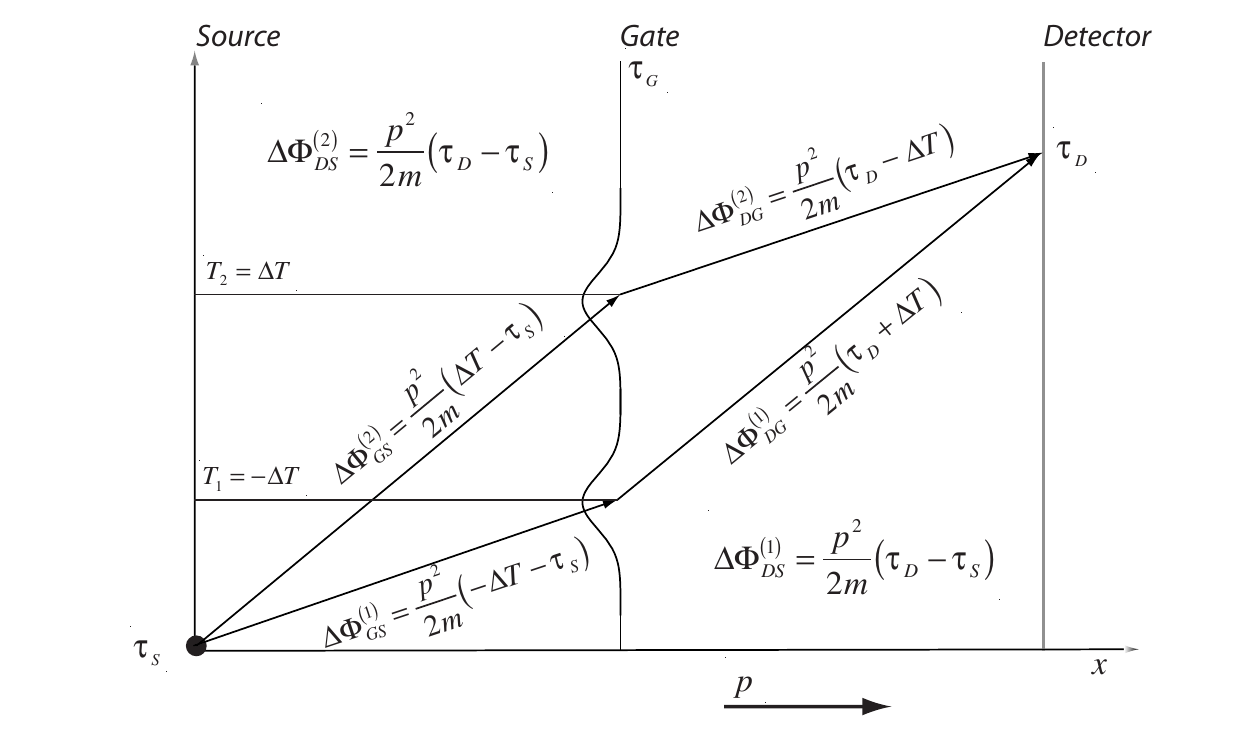}{{\imgevalsize{images/xt-double-4.pdf}{\includegraphics[width=\imgwidth,height=\imgheight,keepaspectratio=true]{images/xt-double-4.pdf}}\quad
}
}{}
\end{center}
\caption{Double Slit With a Single Source}
\label{figure-images-xt-double-4}\hyperlabel{figure-images-xt-double-4}%
\end{figure}

The classic double slit experiment posits a single source that then shines through two slits.

The phase comes from the kinetic energy times the time. From source to gate we have:

\begin{equation}
\label{equation-xt-dnull-xt-double-6}\hyperlabel{equation-xt-dnull-xt-double-6}%
\Delta {\Phi }_{GS}=\frac{{p}^{2}}{2m}\left({T}_{1,2}-{\tau }_{S}\right)
\end{equation}

And from gate to detector:

\begin{equation}
\label{equation-xt-dnull-xt-double-7}\hyperlabel{equation-xt-dnull-xt-double-7}%
\Delta {\Phi }_{DG}=\frac{{p}^{2}}{2m}\left({\tau }_{D}-{T}_{1,2}\right)
\end{equation}

So, for a specific ray 
 \emph{p}, the total contribution of the kinetic energy to the phase is the same for either gate:

\begin{equation}
\label{equation-xt-dnull-xt-double-8}\hyperlabel{equation-xt-dnull-xt-double-8}%
\Delta {\Phi }_{DS}=\Delta {\Phi }_{DG}+\Delta {\Phi }_{GS}=\frac{{p}^{2}}{2m}{\tau }_{DS}
\end{equation}

Since we are associating each time 
 $ {\tau }_{\text{D}}$~ at the detector with one ray 
 \emph{p}, the kinetic energy contribution to the phase will be the same whichever path we take
 \footnote{We will see the same logic, with kinetic energy replaced by mass, when we look at the time part of the wave function in 
 \hyperlink{xt-dtq}{Double Slit in Temporal Quantization}.}.

This does not eliminate all interference at the detector. Rays with different 
 \emph{p}~ will interfere one with another. But it does mean that any interference pattern will be weak.

We therefore look at a more interesting case which is, in addition, a better fit to Lindner's 
 \hyperlink{xt-lind}{Attosecond Double Slit in Time}. We model the double slit experiment in terms of two correlated sources. We look at the simplest possible approach: two free sources.
    We start the sources at 
 $ {\tau }_{S}^{\left(1,2\right)}=\mp \Delta T$~ with identical distributions in momentum and quantum time, but with relative phase:

\begin{equation}
\label{equation-xt-dnull-xt-double-a}\hyperlabel{equation-xt-dnull-xt-double-a}%
{\psi }_{G}^{\left(1,2\right)}\sim \mathrm{exp}\left(-i{\phi }_{1,2}\right),{\phi }_{1,2}={\phi }_{0}\mp \Delta \phi 
\end{equation}

\subsubsection{Double Slit in Standard Quantum Theory}\label{xt-dsqt}\hyperlabel{xt-dsqt}%
\begin{figure}[H]

\begin{center}
\imgexists{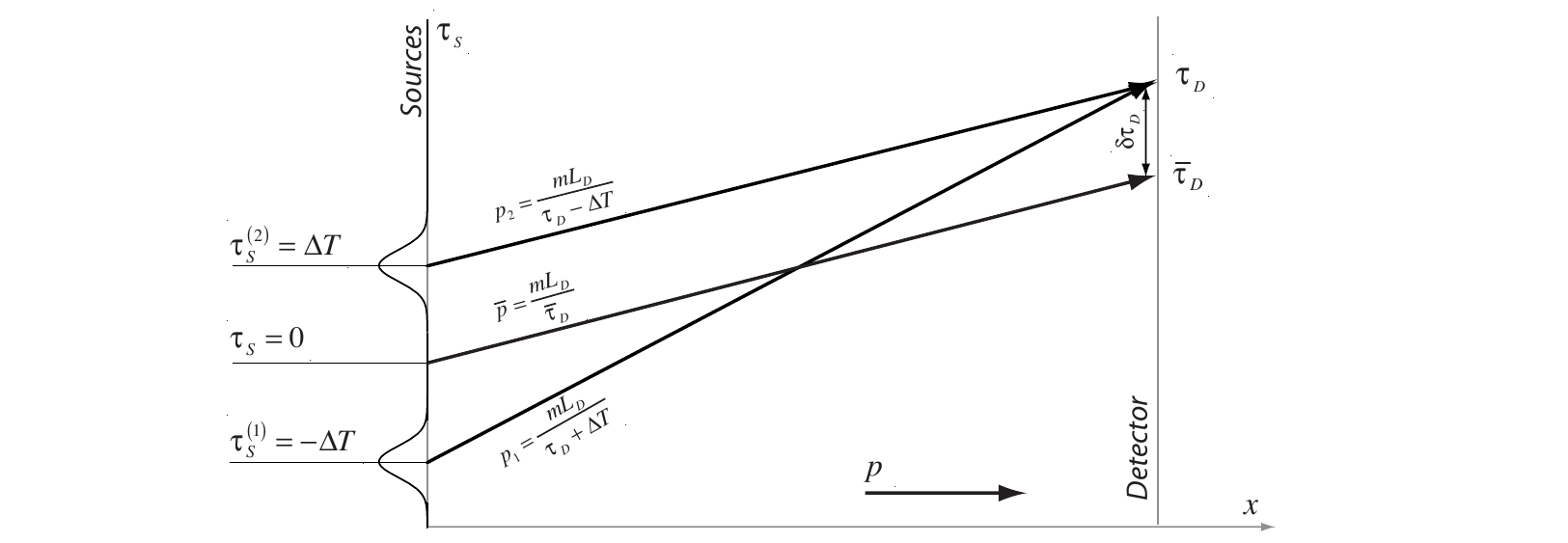}{{\imgevalsize{images/xt-dtq-1.pdf}{\includegraphics[width=\imgwidth,height=\imgheight,keepaspectratio=true]{images/xt-dtq-1.pdf}}\quad
}
}{}
\end{center}
\caption{Double Slit With Two Correlated Sources}
\label{figure-images-xt-dtq-1}\hyperlabel{figure-images-xt-dtq-1}%
\end{figure}

We start with the wave function from 
 \hyperlink{xt-free}{the free case}:

\begin{equation}
\label{equation-xt-dsqt-xt-dsrcs-sqt-xi12}\hyperlabel{equation-xt-dsqt-xt-dsrcs-sqt-xi12}%
{\xi }_{D}^{\left(1,2\right)}\left(\delta {\tau }_{D}\right)=\sqrt[4]{\frac{1}{\pi {\overline{\sigma }}_{D}^{2}}}\mathrm{exp}\left(-\frac{{\left(\delta {\tau }_{D}\pm \Delta T\right)}^{2}}{2{\overline{\sigma }}_{D}^{2}}-i\frac{{\overline{p}}^{2}}{2m}{\overline{\tau }}_{D}+i\frac{{\overline{p}}^{2}}{2m}\left(\delta {\tau }_{D}\pm \Delta T\right)+i\frac{{\overline{p}}^{2}}{2m}\frac{{\left(\delta {\tau }_{D}\pm \Delta T\right)}^{2}}{{\overline{\tau }}_{D}}-i{\phi }_{0}\pm i\Delta \phi \right)
\end{equation}

The full wave function is the sum of the terms from each gate:

\begin{equation}
\label{equation-xt-dsqt-xt-dsrcs-sqt-8}\hyperlabel{equation-xt-dsqt-xt-dsrcs-sqt-8}%
{\xi }_{D}\left(\delta {\tau }_{D}\right)={\xi }_{D}^{\left(1\right)}\left(\delta {\tau }_{D}\right)+{\xi }_{D}^{\left(2\right)}\left(\delta {\tau }_{D}\right)
\end{equation}

The probability density includes a normalization constant:

\begin{equation}
\label{equation-xt-dsqt-xt-dsrcs-sqt-9}\hyperlabel{equation-xt-dsqt-xt-dsrcs-sqt-9}%
{\overline{\rho }}_{D}\left(\delta {\tau }_{D}\right)={\overline{N}}^{2}{\left\vert {\xi }_{D}^{\left(1\right)}\left(\delta {\tau }_{D}\right)+{\xi }_{D}^{\left(2\right)}\left(\delta {\tau }_{D}\right)\right\vert }^{2}
\end{equation}

Which is:

\begin{equation}
\label{equation-xt-dsqt-xt-dsrcs-sqt-a}\hyperlabel{equation-xt-dsqt-xt-dsrcs-sqt-a}%
{\overline{N}}^{2}=\frac{1}{{\displaystyle \int \text{d}{\tau }_{D}{\left\vert {\xi }_{D}^{\left(1\right)}\left(\delta {\tau }_{D}\right)+{\xi }_{D}^{\left(2\right)}\left(\delta {\tau }_{D}\right)\right\vert }^{2}}}
\end{equation}

The probability density is given by three terms:

\begin{equation}
\label{equation-xt-dsqt-xt-dsrcs-sqt-b}\hyperlabel{equation-xt-dsqt-xt-dsrcs-sqt-b}%
{\overline{\rho }}_{D}\left(\delta {\tau }_{D}\right)={\overline{N}}^{2}\left({\overline{\rho }}_{D}^{\left(1\right)}\left(\delta {\tau }_{D}\right)+2{\overline{\rho }}_{D}^{\left(1\otimes 2\right)}\left(\delta {\tau }_{D}\right)+{\overline{\rho }}_{D}^{\left(2\right)}\left(\delta {\tau }_{D}\right)\right)
\end{equation}

There are two outer humps, corresponding to each gate considered singly:

\begin{equation}
\label{equation-xt-dsqt-xt-dsrcs-sqt-c}\hyperlabel{equation-xt-dsqt-xt-dsrcs-sqt-c}%
{\overline{\rho }}_{D}^{\left(1,2\right)}\left(\delta {\tau }_{D}\right)={\overline{\rho }}_{D}^{\left(free\right)}\left(\delta {\tau }_{D}\pm \Delta T\right)
\end{equation}

With:

\begin{equation}
\label{equation-xt-dsqt-xt-dsrcs-sqt-d}\hyperlabel{equation-xt-dsqt-xt-dsrcs-sqt-d}%
{\overline{\rho }}_{D}^{\left(free\right)}\left(\delta {\tau }_{D}\right)=\sqrt{\frac{1}{\pi {\overline{\sigma }}_{D}^{2}}}\mathrm{exp}\left(-\frac{\delta {\tau }_{D}^{2}}{{\overline{\sigma }}_{D}^{2}}\right)
\end{equation}

And a cross term in the middle:

\begin{equation}
\label{equation-xt-dsqt-xt-dsrcs-sqt-f}\hyperlabel{equation-xt-dsqt-xt-dsrcs-sqt-f}%
{\overline{\rho }}_{D}^{\left(1\otimes 2\right)}\left(\delta {\tau }_{D}\right)=\mathrm{exp}\left(-\frac{\Delta {T}^{2}}{{\overline{\sigma }}_{D}^{2}}\right)\mathrm{cos}\left(\overline{\phi }+\overline{f}\delta {\tau }_{D}\right){\overline{\rho }}_{D}^{\left(free\right)}\left(\delta {\tau }_{D}\right)
\end{equation}

With frequency and angular offset:

\begin{equation}
\label{equation-xt-dsqt-xt-dsrcs-sqt-g}\hyperlabel{equation-xt-dsqt-xt-dsrcs-sqt-g}%
\overline{f}\equiv 4\frac{{\overline{p}}^{2}}{2m}\frac{\Delta T}{{\overline{\tau }}_{D}}
\end{equation}

\begin{equation}
\label{equation-xt-dsqt-xt-dsrcs-sqt-h}\hyperlabel{equation-xt-dsqt-xt-dsrcs-sqt-h}%
\overline{\phi }\equiv 2\frac{{\overline{p}}^{2}}{2m}\Delta T+2\Delta \phi 
\end{equation}

Normalization:

\begin{equation}
\label{equation-xt-dsqt-xt-dsrcs-sqt-i}\hyperlabel{equation-xt-dsqt-xt-dsrcs-sqt-i}%
{\displaystyle \int \text{d}\delta {\tau }_{D}\left({\overline{\rho }}_{D}^{\left(1\right)}\left(\delta {\tau }_{D}\right)+2{\overline{\rho }}_{D}^{\left(1\otimes 2\right)}\left(\delta {\tau }_{D}\right)+{\overline{\rho }}_{D}^{\left(2\right)}\left(\delta {\tau }_{D}\right)\right)}=2+2\mathrm{exp}\left(-\frac{\Delta {T}^{2}}{{\overline{\sigma }}_{D}^{2}}-\frac{{\overline{f}}^{2}}{4}{\overline{\sigma }}_{D}^{2}\right)\mathrm{cos}\left(\overline{\phi }\right)
\end{equation}

Normalized probability density:

\begin{equation}
\label{equation-xt-dsqt-xt-dsrcs-sqt-k}\hyperlabel{equation-xt-dsqt-xt-dsrcs-sqt-k}%
{\overline{\rho }}_{D}\left(\delta {\tau }_{D}\right)=\sqrt{\frac{1}{\pi {\overline{\sigma }}_{D}^{2}}}\frac{\mathrm{exp}\left(-\frac{{\left(\delta {\tau }_{D}+\Delta T\right)}^{2}}{{\overline{\sigma }}_{D}^{2}}\right)+2\mathrm{exp}\left(-\frac{\delta {\tau }_{D}^{2}+\Delta {T}^{2}}{{\overline{\sigma }}_{D}^{2}}\right)\mathrm{cos}\left(\overline{\phi }+\overline{f}\delta {\tau }_{D}\right)+\mathrm{exp}\left(-\frac{{\left(\delta {\tau }_{D}-\Delta T\right)}^{2}}{{\overline{\sigma }}_{D}^{2}}\right)}{2+2\mathrm{exp}\left(-\frac{\Delta {T}^{2}}{{\overline{\sigma }}_{D}^{2}}-\frac{{\overline{f}}^{2}}{4}{\overline{\sigma }}_{D}^{2}\right)\mathrm{cos}\left(\overline{\phi }\right)}
\end{equation}

When the initial sources are well-{}separated we see only the two outer humps. When they are closer together we see an interference pattern in the center.

\subsubsection{Double Slit in Temporal Quantization}\label{xt-dtq}\hyperlabel{xt-dtq}%

We apply the same approach, using two free sources separated in time, to analyze the temporal quantization case. The two sources give:

\begin{equation}
\label{equation-xt-dtq-xt-dsrcs-tq-1}\hyperlabel{equation-xt-dtq-xt-dsrcs-tq-1}%
{\psi }_{D}^{\left(1,2\right)}\left(t,\delta p\right)={\chi }_{D}^{\left(1,2\right)}\left(t,\delta p\right){\xi }_{D}^{\left(1,2\right)}\left(\delta p\right)
\end{equation}

With time parts:

\begin{equation}
\label{equation-xt-dtq-xt-dsrcs-tq-2}\hyperlabel{equation-xt-dtq-xt-dsrcs-tq-2}%
{\chi }_{D}^{\left(1,2\right)}\left(t,\delta p\right)=\sqrt[4]{\frac{1}{\pi {\sigma }_{0}^{2}{f}_{D}^{\left(0\right)2}}}\mathrm{exp}\left(-i{\overline{E}}_{1,2}{t}_{D}-\frac{{t}_{D}^{2}}{2{\sigma }_{0}^{2}{f}_{D}^{\left(0\right)}}-im\left({\tau }_{D}\pm \Delta T\right)\right)
\end{equation}

And space parts:

\begin{equation}
\label{equation-xt-dtq-xt-dsrcs-tq-3}\hyperlabel{equation-xt-dtq-xt-dsrcs-tq-3}%
{\widehat{\xi }}_{D}^{\left(1,2\right)}\left(\delta p\right)=\sqrt[4]{\frac{1}{\pi {\widehat{\sigma }}_{1}^{2}}}\mathrm{exp}\left(-\frac{\delta {p}^{2}}{2{\widehat{\sigma }}_{1}^{2}}\right)\mathrm{exp}\left(-i\frac{{p}^{2}}{2m}\left({\tau }_{D}\pm \Delta T\right)-i{\phi }_{1,2}\right)
\end{equation}

\paragraph*{Probability Density in Time and Space}

\noindent

If we imagine a pencil beam coming from each of our two gates, aimed at a specific point, it will take sharply different momenta to arrive at the detector at the same 
 $ {\tau }_{\text{D}}$:

\begin{equation}
\label{equation-xt-dtq-xt-dsrcs-point-0}\hyperlabel{equation-xt-dtq-xt-dsrcs-point-0}%
{\tau }_{D}=\frac{m{L}_{D}}{{p}_{1}}-\Delta T=\frac{m{L}_{D}}{{p}_{2}}+\Delta T
\end{equation}

Implying (working as usual only to first order):

\begin{equation}
\label{equation-xt-dtq-xt-dsrcs-point-1}\hyperlabel{equation-xt-dtq-xt-dsrcs-point-1}%
{p}_{1,2}=\frac{m{L}_{D}}{{\tau }_{D}\pm \Delta T}=\frac{m{L}_{D}}{{\overline{\tau }}_{D}+\delta {\tau }_{D}\pm \Delta T}=\overline{p}\left(1-\frac{\delta {\tau }_{D}\pm \Delta T}{{\overline{\tau }}_{D}}\right)\Rightarrow \delta {p}_{1,2}=-\frac{\delta {\tau }_{D}\pm \Delta T}{{\overline{\tau }}_{D}}\overline{p}
\end{equation}

These shifts give us, not surprisingly, the previous standard quantum theory wave functions:

\begin{equation}
\label{equation-xt-dtq-xt-dsrcs-sqt-xi12}\hyperlabel{equation-xt-dtq-xt-dsrcs-sqt-xi12}%
{\xi }_{D}^{\left(1,2\right)}\left(\delta {\tau }_{D}\right)=\sqrt[4]{\frac{1}{\pi {\overline{\sigma }}_{D}^{2}}}\mathrm{exp}\left(-\frac{{\left(\delta {\tau }_{D}\pm \Delta T\right)}^{2}}{2{\overline{\sigma }}_{D}^{2}}-i\frac{{\overline{p}}^{2}}{2m}{\overline{\tau }}_{D}+i\frac{{\overline{p}}^{2}}{2m}\left(\delta {\tau }_{D}\pm \Delta T\right)+i\frac{{\overline{p}}^{2}}{2m}\frac{{\left(\delta {\tau }_{D}\pm \Delta T\right)}^{2}}{{\overline{\tau }}_{D}}-i{\phi }_{0}\pm i\Delta \phi \right)
\end{equation}

In the time part of the wave function we have still to consider the factor 
 $ \mathrm{exp}\left(\mp im\Delta T\right)$.

We make the same argument as was made earlier with respect to the non-{}interference in the single source case (\hyperlink{xt-dnull}{Relative Lack of Interference in the Simplest Case}). The part of the wave function going through the first gate gets a total phase:

\begin{equation}
\label{equation-xt-dtq-xt-dsrcs-point-4}\hyperlabel{equation-xt-dtq-xt-dsrcs-point-4}%
-m\left(-\Delta T-{\tau }_{S}\right)-m\left({\tau }_{D}+\Delta T\right)
\end{equation}

The second:

\begin{equation}
\label{equation-xt-dtq-xt-dsrcs-point-5}\hyperlabel{equation-xt-dtq-xt-dsrcs-point-5}%
-m\left(\Delta T-{\tau }_{S}\right)-m\left({\tau }_{D}-\Delta T\right)
\end{equation}

These are equal so irrelevant. Formally, we can absorb these phases into the angle:

\begin{equation}
\label{equation-xt-dtq-xt-dsrcs-point-6}\hyperlabel{equation-xt-dtq-xt-dsrcs-point-6}%
\begin{array}{l}{\phi }_{1}\rightarrow {\phi }_{1}+m\left(-\Delta T-{\tau }_{S}\right)\\ {\phi }_{2}\rightarrow {\phi }_{2}+m\left(\Delta T-{\tau }_{S}\right)\end{array}
\end{equation}

The probability density in time and space is now:

\begin{equation}
\label{equation-xt-dtq-xt-dsrcs-pdens-0}\hyperlabel{equation-xt-dtq-xt-dsrcs-pdens-0}%
{\rho }_{D}\left({t}_{D},\delta {\tau }_{D}\right)={\left\vert {\psi }_{D}^{\left(1\right)}\left({t}_{D},\delta {\tau }_{D}\right)+{\psi }_{D}^{\left(2\right)}\left({t}_{D},\delta {\tau }_{D}\right)\right\vert }^{2}
\end{equation}

There are three terms as before:

\begin{equation}
\label{equation-xt-dtq-xt-dsrcs-pdens-1}\hyperlabel{equation-xt-dtq-xt-dsrcs-pdens-1}%
{\rho }_{D}\left({t}_{D},\delta {\tau }_{D}\right)={\rho }_{D}^{\left(1\right)}\left({t}_{D},\delta {\tau }_{D}\right)+2{\rho }_{D}^{\left(1\otimes 2\right)}\left({t}_{D},\delta {\tau }_{D}\right)+{\rho }_{D}^{\left(2\right)}\left({t}_{D},\delta {\tau }_{D}\right)
\end{equation}

The outer humps are:

\begin{equation}
\label{equation-xt-dtq-xt-dsrcs-pdens-2}\hyperlabel{equation-xt-dtq-xt-dsrcs-pdens-2}%
{\rho }_{D}^{\left(1,2\right)}\left({t}_{D},\delta {\tau }_{D}\right)={\stackrel{\frown }{\rho }}_{D}^{\left(free\right)}\left({t}_{D},\delta {\tau }_{D}\right){\overline{\rho }}_{D}^{\left(free\right)}\left(\delta \tau \pm \Delta T\right)
\end{equation}

With:

\begin{equation}
\label{equation-xt-dtq-xt-dsrcs-pdens-3}\hyperlabel{equation-xt-dtq-xt-dsrcs-pdens-3}%
{\stackrel{\frown }{\rho }}_{D}^{\left(free\right)}\left({t}_{D},\delta {\tau }_{D}\right)=\sqrt{\frac{1}{\pi {\sigma }_{0}^{2}{\left\vert {f}_{D}^{\left(0\right)}\right\vert }^{2}}}\mathrm{exp}\left(-\frac{{t}_{D}^{2}}{{\sigma }_{0}^{2}{\left\vert {f}_{D}^{\left(0\right)}\right\vert }^{2}}\right)
\end{equation}

The cross term is now:

\begin{equation}
\label{equation-xt-dtq-xt-dsrcs-pdens-5}\hyperlabel{equation-xt-dtq-xt-dsrcs-pdens-5}%
{\rho }_{D}^{\left(1\otimes 2\right)}\left({t}_{D},\delta {\tau }_{D}\right)=\frac{1}{2}{\stackrel{\frown }{\rho }}_{D}^{\left(free\right)}\left({t}_{D},\delta {\tau }_{D}\right)\left({\xi }_{D}^{\left(1\right)\ast }\left(\delta {\tau }_{D}+\Delta T\right){\xi }_{D}^{\left(2\right)}\left(\delta {\tau }_{D}-\Delta T\right)+{\xi }_{D}^{\left(1\right)}\left(\delta {\tau }_{D}+\Delta T\right){\xi }_{D}^{\left(2\right)\ast }\left(\delta {\tau }_{D}-\Delta T\right)\right)
\end{equation}

Which is the free quantum time wave function times the momentum space cross term.

Therefore we can factor out the probability density in time from the total probability density
 \footnote{
If we had generalized the dispersion in time to include dependence on momentum 
 $ {\sigma }_{0}^{2}\rightarrow {\sigma }_{0}^{2}\left(p\right)$~ or had taken 
 $ \gamma >1$~ we would not have been able to factor out the time part.
}:

\begin{equation}
\label{equation-xt-dtq-xt-dsrcs-pdens-7}\hyperlabel{equation-xt-dtq-xt-dsrcs-pdens-7}%
{\rho }_{D}\left(t,\delta {\tau }_{D}\right)={\stackrel{\frown }{\rho }}_{D}^{\left(free\right)}\left({t}_{D},\delta {\tau }_{D}\right){\overline{\rho }}_{D}\left(\delta {\tau }_{D}\right)
\end{equation}

\paragraph*{Probability Density in Time Only}

\noindent

The probability density in time only is given by:

\begin{equation}
\label{equation-xt-dtq-xt-dsrcs-final-1}\hyperlabel{equation-xt-dtq-xt-dsrcs-final-1}%
{\rho }_{D}\left({t}_{\overline{D}}\right)={\displaystyle \int \text{d}\delta {\tau }_{D}{\stackrel{\frown }{\rho }}_{D}\left({t}_{D},\delta {\tau }_{D}\right){\overline{\rho }}_{D}\left(\delta {\tau }_{D}\right)}
\end{equation}

Explicitly:

\begin{equation}
\label{equation-xt-dtq-xt-dsrcs-final-1b}\hyperlabel{equation-xt-dtq-xt-dsrcs-final-1b}%
{\rho }_{D}\left({t}_{\overline{D}}\right)={\displaystyle \int \text{d}\delta {\tau }_{D}\sqrt{\frac{1}{\pi {\stackrel{\frown }{\sigma }}_{D}^{2}}}\mathrm{exp}\left(-\frac{{\left({t}_{\overline{D}}-\delta {\tau }_{D}\right)}^{2}}{{\stackrel{\frown }{\sigma }}_{D}^{2}}\right)\sqrt{\frac{1}{\pi {\overline{\sigma }}_{D}^{2}}}\left(\begin{array}{l}\mathrm{exp}\left(-\frac{{\left(\delta {\tau }_{D}+\Delta T\right)}^{2}}{{\overline{\sigma }}_{D}^{2}}\right)\\ +2\mathrm{exp}\left(-\frac{\delta {\tau }_{D}^{2}+\Delta {T}^{2}}{{\overline{\sigma }}_{D}^{2}}\right)\mathrm{cos}\left(\overline{\phi }+\overline{f}\delta {\tau }_{D}\right)\\ +\mathrm{exp}\left(-\frac{{\left(\delta {\tau }_{D}-\Delta T\right)}^{2}}{{\overline{\sigma }}_{D}^{2}}\right)\end{array}\right)}
\end{equation}

Giving:

\begin{equation}
\label{equation-xt-dtq-xt-dsrcs-final-2}\hyperlabel{equation-xt-dtq-xt-dsrcs-final-2}%
{\rho }_{D}\left({t}_{\overline{D}}\right)=\left(\begin{array}{l}\sqrt{\frac{1}{\pi {\sigma }_{D}^{2}}}\mathrm{exp}\left(-\frac{{\left({t}_{\overline{D}}+\Delta T\right)}^{2}}{{\sigma }_{D}^{2}}\right)\\ +2\mathrm{exp}\left(-\frac{\Delta {T}^{2}}{{\overline{\sigma }}_{D}^{2}}-\frac{{\stackrel{\frown }{\sigma }}_{D}^{2}{\overline{\sigma }}_{D}^{2}}{4{\sigma }_{D}^{2}}{\overline{f}}^{2}\right)\mathrm{cos}\left(\overline{\phi }+\frac{{\overline{\sigma }}_{D}^{2}}{{\stackrel{\frown }{\sigma }}_{D}^{2}+{\overline{\sigma }}_{D}^{2}}\overline{f}{t}_{\overline{D}}\right)\sqrt{\frac{1}{\pi {\sigma }_{D}^{2}}}\mathrm{exp}\left(-\frac{{t}_{\overline{D}}^{2}}{{\sigma }_{D}^{2}}\right)\\ +\sqrt{\frac{1}{\pi {\sigma }_{D}^{2}}}\mathrm{exp}\left(-\frac{{\left({t}_{\overline{D}}-\Delta T\right)}^{2}}{{\sigma }_{D}^{2}}\right)\end{array}\right)
\end{equation}

The effects of temporal quantization are three:
\begin{enumerate}

\item{}All three humps are widened by the familiar factor:

\begin{equation}
\label{equation-xt-dtq-xt-dsrcs-final-3}\hyperlabel{equation-xt-dtq-xt-dsrcs-final-3}%
{\overline{\sigma }}_{D}^{2}\rightarrow {\sigma }_{D}^{2}={\stackrel{\frown }{\sigma }}_{D}^{2}+{\overline{\sigma }}_{D}^{2}
\end{equation}

\item{}The frequency is reduced:

\begin{equation}
\label{equation-xt-dtq-xt-dsrcs-final-4}\hyperlabel{equation-xt-dtq-xt-dsrcs-final-4}%
\overline{f}\rightarrow \frac{{\overline{\sigma }}_{D}^{2}}{{\stackrel{\frown }{\sigma }}_{D}^{2}+{\overline{\sigma }}_{D}^{2}}\overline{f}
\end{equation}

\item{}The central hump is suppressed:

\begin{equation}
\label{equation-xt-dtq-xt-dsrcs-final-5}\hyperlabel{equation-xt-dtq-xt-dsrcs-final-5}%
\mathrm{exp}\left(-\frac{\Delta {T}^{2}}{{\overline{\sigma }}_{D}^{2}}\right)\rightarrow \mathrm{exp}\left(-\frac{\Delta {T}^{2}}{{\overline{\sigma }}_{D}^{2}}-\frac{{\stackrel{\frown }{\sigma }}_{D}^{2}{\overline{\sigma }}_{D}^{2}}{4{\sigma }_{D}^{2}}{\overline{f}}^{2}\right)
\end{equation}

\end{enumerate}

Essentially each individual crest is widened while the oscillatory "comb" is stretched out and flattened at the same time.

\subsection{Attosecond Double Slit in Time}\label{xt-lind}\hyperlabel{xt-lind}%

\subsubsection{Overview}\label{xt-linintro}\hyperlabel{xt-linintro}%
\begin{figure}[H]

\begin{center}
\imgexists{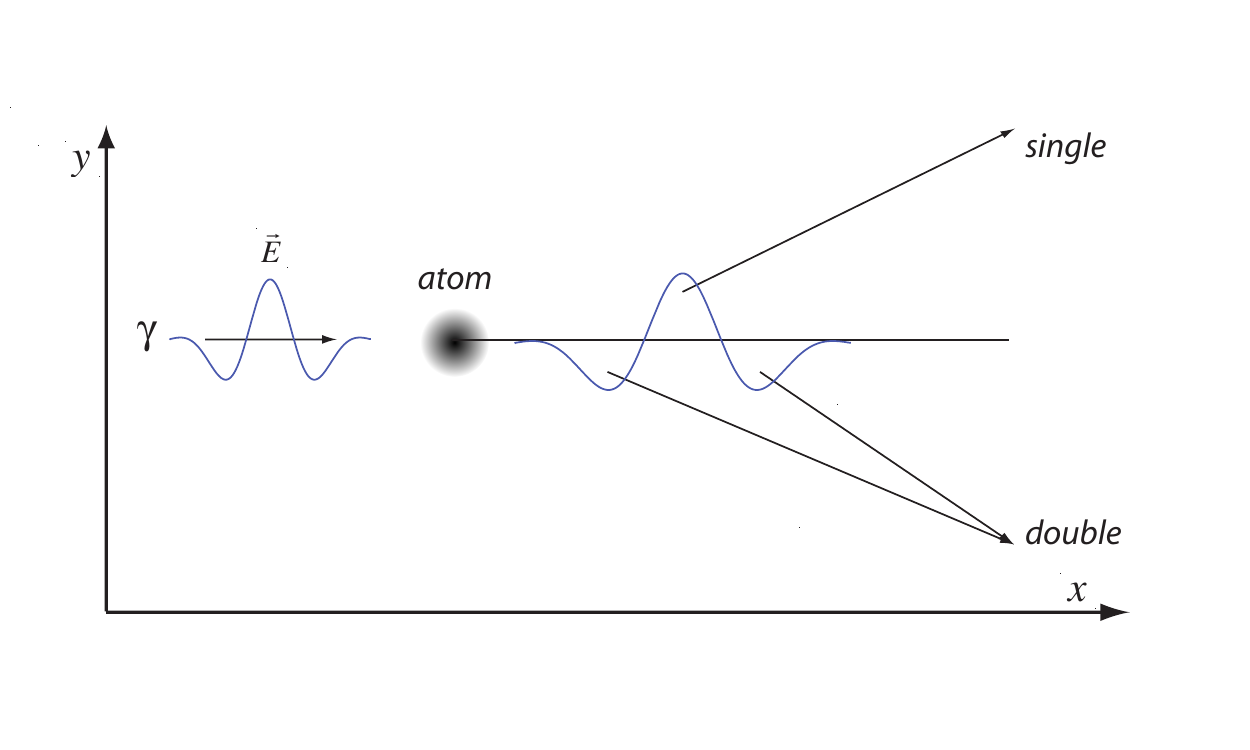}{{\imgevalsize{images/xt-lind-4.pdf}{\includegraphics[width=\imgwidth,height=\imgheight,keepaspectratio=true]{images/xt-lind-4.pdf}}\quad
}
}{}
\end{center}
\caption{Electric Field as Source}
\label{figure-images-xt-lind-4}\hyperlabel{figure-images-xt-lind-4}%
\end{figure}

In Lindner et al's 2005 experiment Attosecond Double Slit in Time 
 \cite{Lindner-2005}, a strong short electric pulse ionizes a bound electron. Like water shaken off by a dog, the electron can come off to right or left.

The pulses are extremely short, with one hump on one side, two on the other. If the electron is shaken off by the one humped side, we have a single slit experiment. If the electron is
    shaken off by the two humped side, we have a double slit experiment: the wave functions generated by the two humps will interfere with each other.

The experiment itself is one of those that put the "non" into "nontrivial"; the associated calculations are difficult as well. We will therefore look only at an extremely simplified model
    of the Lindner experiment.

\subsubsection{Model Experiment}\label{xt-linmodel}\hyperlabel{xt-linmodel}%

\paragraph*{Electric Field}

\noindent

We take a photon going from left to right in 
 \emph{x}~ direction, with an electric field in 
 \emph{y}, centered at time zero on the atom, located at position zero. We write the electric field as a Gaussian:

\begin{equation}
\label{equation-xt-linmodel-xt-lind-field-0}\hyperlabel{equation-xt-linmodel-xt-lind-field-0}%
{\overrightarrow{{\rm E}}}_{\tau }\left(\overrightarrow{x}\right)={{\rm E}}_{0}\mathrm{cos}\left(\omega \tau -kx+\phi \right)\mathrm{exp}\left(-\frac{{\left(\omega \tau -kx\right)}^{2}}{{\sigma }^{2}}\right)\widehat{y}
\end{equation}

For a physically acceptable wave form, we have to subtract out the 
 \emph{DC}~ component. This, as it happens, is the defining condition of a Morlet wavelet (\cite{Kaiser-1994}), so we can write the photon electric field as the real part of a Morlet wavelet
 \footnote{
We have used Morlet wavelets to 
 \hyperlink{tq-fpi-waves}{define the initial wave functions}, 
 \hyperlink{tq-fpi-conv}{ensure convergence of the path integral}, 
 \hyperlink{covar}{define a covariant laboratory time}, 
 \hyperlink{semi-calc}{ensure convergence of the fluctuation factor in the semi-{}classical approximation}, 
 \hyperlink{xt-gintro}{argue that our Gaussian gates are fully general}, and now define the electric field. Perhaps we should follow Humpty Dumpty's advice: "'When I make a word do a
        lot of work like that,' said Humpty Dumpty, 'I always pay it extra.'" Unfortunately Alice did not find from him what he paid them with.
}:

\begin{equation}
\label{equation-xt-linmodel-xt-lind-field-1}\hyperlabel{equation-xt-linmodel-xt-lind-field-1}%
{\overrightarrow{{\rm E}}}_{\tau }\left(\overrightarrow{x}\right)=\mathrm{Re}\left({{\rm E}}_{0}\mathrm{exp}\left(i\phi \right)\frac{\sqrt{2}}{\sqrt{\sigma }}\left({e}^{-i\left(\frac{\omega \tau -kx}{\sigma }\right)}-{e}^{-\frac{1}{4}}\right)\mathrm{exp}\left(-{\left(\frac{\omega \tau -kx}{\sigma }\right)}^{2}\right)\right)\widehat{y}
\end{equation}

By correct tuning of 
 $ \phi $~ and the other parameters, we can get one hump, two humps, or as many humps as the Loch Ness monster, depending on our requirements. For now we will assume (with Lindner)
    one or two.

Ionization only happens at the peak of the field, letting us approximate the peaks as point sources:

\begin{equation}
\label{equation-xt-linmodel-xt-lind-field-2}\hyperlabel{equation-xt-linmodel-xt-lind-field-2}%
\begin{array}{l}{\overrightarrow{{\rm E}}}_{\tau }^{\left(single\right)}\sim \overrightarrow{E}\delta \left(\tau \right)\\ {\overrightarrow{E}}_{\tau }^{\left(double\right)}\sim \overrightarrow{E}\left(\delta \left(\tau +\Delta T\right)+\delta \left(\tau -\Delta T\right)\right)\end{array}
\end{equation}

We ignore the off-{}axis part of momentum: the 
 $ {p}_{x}$~ and 
 $ {p}_{z}$~ components.

\paragraph*{Initial Wave Function}

\noindent

The initial wave function of the atomic electron is a bound state 
 \emph{a}:

\begin{equation}
\label{equation-xt-linmodel-xt-linmodel-bound-3}\hyperlabel{equation-xt-linmodel-xt-linmodel-bound-3}%
{\psi }_{\tau }^{\left(a\right)}\left({t}_{\tau },\overrightarrow{p}\right)={\chi }_{a}\left({t}_{\tau }\right){\widehat{\xi }}_{\overline{a}}\left(\overrightarrow{p}\right)\mathrm{exp}\left(-i{\overline{E}}_{\overline{a}}\tau \right)
\end{equation}

With wave function in time:

\begin{equation}
\label{equation-xt-linmodel-xt-linmodel-bound-4}\hyperlabel{equation-xt-linmodel-xt-linmodel-bound-4}%
{\chi }_{a}\left(t\right)=\sqrt[4]{\frac{1}{\pi {\sigma }_{\overline{a}}^{2}}}\mathrm{exp}\left(-i{\overline{E}}_{\overline{a}}{t}_{\tau }-\frac{{t}_{\tau }^{2}}{2{\sigma }_{\overline{a}}^{2}}\right)
\end{equation}

With quantum energy 
 $ {\overline{E}}_{\overline{a}}=m+{\overline{\mathscr{E}}}_{\overline{a}}$~ (where 
 $ {\overline{\mathscr{E}}}_{\overline{a}}$~ is the binding energy of the electron) and with estimated dispersion in time (\hyperlink{bound-disp}{Estimate of Uncertainty in Time}):

\begin{equation}
\label{equation-xt-linmodel-xt-linmodel-bound-6}\hyperlabel{equation-xt-linmodel-xt-linmodel-bound-6}%
{\sigma }_{\overline{a}}^{2}\sim 2{\displaystyle \int \text{d}\overrightarrow{r}{\left\vert {\xi }_{\overline{a}}\left(\overrightarrow{r}\right)\right\vert }^{2}{\overrightarrow{r}}^{2}}
\end{equation}

In energy space (\hyperlink{free-fourE}{Energy/Momentum Representation}):

\begin{equation}
\label{equation-xt-linmodel-xt-linmodel-bound-7}\hyperlabel{equation-xt-linmodel-xt-linmodel-bound-7}%
{\widehat{\chi }}_{a}\left(E\right)=\sqrt[4]{\frac{1}{\pi {\sigma }_{\overline{a}}^{2}}}\mathrm{exp}\left(-i\left(E-{\overline{E}}_{\overline{a}}\right)\langle {t}_{\tau }\rangle -\frac{{\left(E-{\overline{E}}_{\overline{a}}\right)}^{2}}{2{\widehat{\sigma }}_{\overline{a}}^{2}}\right)\approx \sqrt[4]{\frac{1}{\pi {\sigma }_{\overline{a}}^{2}}}\mathrm{exp}\left(-\frac{{\left(E-{\overline{E}}_{\overline{a}}\right)}^{2}}{2{\widehat{\sigma }}_{\overline{a}}^{2}}\right)
\end{equation}

With dispersion estimated as 
 $ {\widehat{\sigma }}_{\overline{a}}^{2}=1/{\sigma }_{\overline{a}}^{2}$~ and the average relative time estimated as zero: 
 $ \langle {t}_{\tau }\rangle \approx 0$.

\paragraph*{Final Wave Function of the Ionized Electron}

\noindent

The transition matrix from the initial state 
 \emph{a}~ to the final state 
 \emph{p}~ is:

\begin{equation}
\label{equation-xt-linmodel-xt-linmodel-final-0}\hyperlabel{equation-xt-linmodel-xt-linmodel-final-0}%
\langle p\vert {\widehat{V}}_{\tau }\left(\omega \right)\vert a\rangle 
\end{equation}

We assume we can write this as a product of time and space parts:

\begin{equation}
\label{equation-xt-linmodel-xt-linmodel-final-1}\hyperlabel{equation-xt-linmodel-xt-linmodel-final-1}%
\langle p\vert {\widehat{V}}_{\tau }\vert a\rangle =\langle {E}_{p}\vert {\widehat{\stackrel{\frown }{V}}}_{\tau }\left(\omega \right)\vert {E}_{a}\rangle \langle \overrightarrow{p}\vert {\widehat{\overline{V}}}_{\tau }\left(\omega \right)\vert \overline{a}\rangle \delta \left(\tau -T\right)
\end{equation}

And we take the time/energy part as the simplest possible matrix element, a 
 $ \delta $~ function in energy:

\begin{equation}
\label{equation-xt-linmodel-xt-linmodel-final-2}\hyperlabel{equation-xt-linmodel-xt-linmodel-final-2}%
\langle {E}_{p}\vert {\widehat{\stackrel{\frown }{V}}}_{\tau }\left(\omega \right)\vert {E}_{a}\rangle =\delta \left({E}_{p}-{E}_{a}-\omega \right)
\end{equation}

The wave function at the detector is given in first order perturbation by:

\begin{equation}
\label{equation-xt-linmodel-xt-linmodel-final-3}\hyperlabel{equation-xt-linmodel-xt-linmodel-final-3}%
{\widehat{\psi }}_{\tau }\left(E,p\right)=-i{\displaystyle \underset{-\infty }{\overset{\infty }{\int }}\text{d}{\tau }^{\prime }{\displaystyle \int \text{d}{E}^{\prime }\text{d}{p}^{\prime }\text{d}{E}^{{''}}\text{d}{p}^{{''}}{\widehat{K}}_{\tau {\tau }^{\prime }}\left(E,p;{E}^{\prime },{p}^{\prime }\right){\widehat{V}}_{{\tau }^{\prime }}\left({E}^{\prime },{p}^{\prime };{E}^{{''}},{p}^{{''}}\right){\widehat{\psi }}_{{\tau }^{\prime }}^{\left(a\right)}\left({E}^{{''}},{p}^{{''}}\right)}}
\end{equation}

The space part of it, just after ionization:

\begin{equation}
\label{equation-xt-linmodel-xt-linmodel-final-4}\hyperlabel{equation-xt-linmodel-xt-linmodel-final-4}%
{\widehat{\xi }}_{T}\left({p}^{\prime }\right)=-i{\displaystyle \int \text{d}{p}^{{''}}{\widehat{\overline{V}}}_{T}\left({p}^{\prime },{p}^{{''}}\right){\widehat{\xi }}_{T}^{\left(\overline{a}\right)}\left({p}^{{''}}\right)}
\end{equation}

The space part at the detector:

\begin{equation}
\label{equation-xt-linmodel-xt-linmodel-final-5}\hyperlabel{equation-xt-linmodel-xt-linmodel-final-5}%
{\widehat{\xi }}_{\tau }\left(p\right)={\widehat{\xi }}_{T}\left(p\right)\mathrm{exp}\left(-i\frac{{p}^{2}}{2m}\left(\tau -T\right)\right)
\end{equation}

We know the average energy from conservation of energy:

\begin{equation}
\label{equation-xt-linmodel-xt-linmodel-final-6}\hyperlabel{equation-xt-linmodel-xt-linmodel-final-6}%
\frac{{\overline{p}}^{2}}{2m}=\omega +{\mathscr{E}}_{\overline{a}}
\end{equation}

The energy parts of the kernel and transition matrix are just 
 $ \delta $~ functions, so they merely push the energy part of the bound state wave function out to the detector:

\begin{equation}
\label{equation-xt-linmodel-xt-linmodel-final-8}\hyperlabel{equation-xt-linmodel-xt-linmodel-final-8}%
{\widehat{\chi }}_{\tau }\left(E,p\right)=\sqrt[4]{\frac{1}{\pi {\widehat{\sigma }}_{\overline{a}}^{2}}}\mathrm{exp}\left(-\frac{{\left(E-{\overline{E}}_{p}\right)}^{2}}{2{\widehat{\sigma }}_{\overline{a}}^{2}}\right)
\end{equation}

With the average energy being:

\begin{equation}
\label{equation-xt-linmodel-xt-linmodel-final-9}\hyperlabel{equation-xt-linmodel-xt-linmodel-final-9}%
{\overline{E}}_{p}=m+\frac{{p}^{2}}{2m}
\end{equation}

So the final electron wave function in time is:

\begin{equation}
\label{equation-xt-linmodel-xt-linmodel-final-a}\hyperlabel{equation-xt-linmodel-xt-linmodel-final-a}%
{\chi }_{\tau }\left({t}_{\tau }\right)=\sqrt[4]{\frac{1}{\pi {\sigma }_{\overline{a}}^{2}}}\mathrm{exp}\left(-i{\overline{E}}_{p}{t}_{\tau }-\frac{{t}_{\tau }^{2}}{2{\sigma }_{\overline{a}}^{2}}\right)
\end{equation}

We therefore have what we need to use the results of the previous sections for single and double slit.

\subsubsection{Single Slit}\label{xt-linsingle}\hyperlabel{xt-linsingle}%

For the single source case, we have one prediction: the width of the hump will be widened by an amount dependent on the width of the bound state in time. If the standard quantum theory
    dispersion is:

\begin{equation}
\label{equation-xt-linsingle-xt-linsingle-0}\hyperlabel{equation-xt-linsingle-xt-linsingle-0}%
{\overline{\sigma }}_{D}^{2}
\end{equation}

Then the temporal quantization prediction is:

\begin{equation}
\label{equation-xt-linsingle-xt-linsingle-1}\hyperlabel{equation-xt-linsingle-xt-linsingle-1}%
{\overline{\sigma }}_{D}^{2}\rightarrow {\sigma }_{\overline{a}}^{2}+{\overline{\sigma }}_{D}^{2}
\end{equation}

\subsubsection{Double Slit}\label{xt-lindouble}\hyperlabel{xt-lindouble}%

For the double source case, we predict the same widening for each crest as in the single source case
 \footnote{
As expected the factor of 
 $ \mathrm{exp}\left(-im\Delta T\right)$~ cancels out. The path from the first hump accumulates this factor after it leaves the atom; the path from the second hump accumulates this factor before it
        leaves.
}.

Further we predict that the "comb" associated with the central peak with be spread out, with the frequency decreasing as:

\begin{equation}
\label{equation-xt-lindouble-xt-lindouble-1}\hyperlabel{equation-xt-lindouble-xt-lindouble-1}%
\overline{f}\rightarrow \frac{{\overline{\sigma }}_{D}^{2}}{{\sigma }_{\overline{a}}^{2}+{\overline{\sigma }}_{D}^{2}}\overline{f}
\end{equation}

And the central peak flattening as:

\begin{equation}
\label{equation-xt-lindouble-xt-lindouble-3}\hyperlabel{equation-xt-lindouble-xt-lindouble-3}%
\mathrm{exp}\left(-\frac{\Delta {T}^{2}}{{\overline{\sigma }}_{D}^{2}}\right)\rightarrow \mathrm{exp}\left(-\frac{\Delta {T}^{2}}{{\overline{\sigma }}_{D}^{2}}\right)\mathrm{exp}\left(-\frac{1}{4}\frac{{\sigma }_{\overline{a}}^{2}{\overline{\sigma }}_{D}^{2}}{{\sigma }_{\overline{a}}^{2}+{\overline{\sigma }}_{D}^{2}}{\overline{f}}^{2}\right)
\end{equation}

Both effects are independent of the original size; it should not matter how we computed the original frequency or height; the relative prediction is the important one.

\subsection{Discussion}\label{xt-gdisc}\hyperlabel{xt-gdisc}%

If temporal quantization is true, there will be additional dispersion in time.

In the single slit experiment we see reduced dispersion in standard quantum theory, increased in temporal quantization. In standard quantum theory, a particle going through a gate in time
    will be clipped by the gate, reducing its dispersion in time. In temporal quantization, the particle will be clipped but it will also be diffracted as well, increasing its dispersion in time
    (relative to the standard quantum theory result). This can only happen with temporal quantization and is, in principle, a clear signal.

In the double slit experiment we see similar effects: in temporal quantization the individual peaks will be widened and the "comb" of interference peaks widened as well.

Lindner's double slit in time experiment acts like an escalator, lifting the fuzziness in time associated with a bound state out to the detector.

From 
 \hyperlink{bound-disp}{Estimate of Uncertainty in Time}~ above, we estimate this fuzziness as of order a third of an attosecond. This is small and would probably be difficult to
    pick out from other sources of dispersion in time.

It might be possible to increase the initial dispersion in time by working with Rydberg atoms, given their greater dispersion in space. To make more than order of magnitude estimates, we
    have to work out the implications of temporal quantization for the multi-{}particle case.

\section{Time-{}varying Magnetic and Electric Fields}\label{xt-fields}\hyperlabel{xt-fields}%

\subsection{Overview}\label{xt-fintro}\hyperlabel{xt-fintro}%

We will look at what happens to a particle when it goes through a time-{}varying electromagnetic field.

The effects of quantum time will depend on the extent to which the particle is extended in time; we will look at corrections from terms linear and quadratic in 
 $ {t}_{\tau }$.

We will assume we have forced the values of 
 $ \langle {t}_{\tau }\rangle $~ and 
 $ \langle {t}_{\tau }^{2}\rangle $~ using a chopper, as above (\hyperlink{xt-gates}{Slits in Time}).

The derivative of the expectation of an operator 
 \emph{O}~ with respect to 
 $ \tau $~ is given (\hyperlink{tq-op}{Operators in Time}) by:

\begin{equation}
\label{equation-xt-fintro-tq-op-proof-first}\hyperlabel{equation-xt-fintro-tq-op-proof-first}%
\frac{d\langle O\rangle }{d\tau }=-i\left\lbrack O,H\right\rbrack +\frac{\partial O}{\partial \tau }
\end{equation}

We assume that we can write the Hamiltonian as:

\begin{equation}
\label{equation-xt-fintro-xt-fields-1}\hyperlabel{equation-xt-fintro-xt-fields-1}%
H={\stackrel{\frown }{H}}^{\left(free\right)}+\overline{H}+{V}_{\tau }
\end{equation}

To lowest order the change in 
 \emph{O}~ resulting from temporal quantization will be given by:

\begin{equation}
\label{equation-xt-fintro-xt-fields-2}\hyperlabel{equation-xt-fintro-xt-fields-2}%
\frac{d}{d\tau }\delta \langle O\rangle =-i\langle \left\lbrack O,{V}_{\tau }\right\rbrack \rangle 
\end{equation}

With a cumulative effect of temporal quantization:

\begin{equation}
\label{equation-xt-fintro-xt-fields-3}\hyperlabel{equation-xt-fintro-xt-fields-3}%
\delta {\langle O\rangle }_{total}=-i{\displaystyle \underset{0}{\overset{\tau }{\int }}\text{d}{\tau }^{\prime }\langle \left\lbrack O,{V}_{\tau }\right\rbrack \rangle }
\end{equation}

If we write the potential in terms of powers of the relative time:

\begin{equation}
\label{equation-xt-fintro-xt-fields-4}\hyperlabel{equation-xt-fintro-xt-fields-4}%
{V}_{\tau }={V}_{\tau }^{\left(1\right)}{t}_{\tau }+{V}_{\tau }^{\left(2\right)}{t}_{\tau }^{2}+O\left\lbrack {t}_{\tau }^{3}\right\rbrack 
\end{equation}

Then the first order effect of quantum time on an observable will be given by:

\begin{equation}
\label{equation-xt-fintro-xt-fields-5}\hyperlabel{equation-xt-fintro-xt-fields-5}%
\delta {\langle O\rangle }_{total}\approx -i{\displaystyle \underset{0}{\overset{\tau }{\int }}\text{d}{\tau }^{\prime }\left({V}_{\tau }^{\left(1\right)}{t}_{\tau }+{V}_{\tau }^{\left(2\right)}{t}_{\tau }^{2}\right)}
\end{equation}

For example, we apply this to the evolution of the expectation of the relative time:

\begin{equation}
\label{equation-xt-fintro-xt-fields-6}\hyperlabel{equation-xt-fintro-xt-fields-6}%
\frac{d}{d\tau }\langle {t}_{\tau }\rangle =-i\left\lbrack {t}_{\tau },H\right\rbrack =-i\left\lbrack {t}_{\tau },{\stackrel{\frown }{H}}^{\left(free\right)}\right\rbrack =\frac{E}{m}-1=\gamma -1
\end{equation}

Getting same the result seen in 
 \hyperlink{free-fourt}{Time/Space Representation}:

\begin{equation}
\label{equation-xt-fintro-xt-fields-7}\hyperlabel{equation-xt-fintro-xt-fields-7}%
\langle {t}_{\tau }\rangle =\langle {t}_{0}\rangle +\left(\gamma -1\right)\tau 
\end{equation}

We expect increased dispersion in time from temporal quantization in 
 \emph{all}~ cases \textendash{} time-{}varying and constant fields alike. We will get the clearest signal if we look at time-{}varying fields. We look at the cases:
\begin{enumerate}

\item{}  \hyperlink{xt-magt}{Time Dependent Magnetic Field},

\item{}And 
 \hyperlink{xt-elect}{Time Dependent Electric Field}.

\end{enumerate}

\subsection{Time Dependent Magnetic Field}\label{xt-magt}\hyperlabel{xt-magt}%

We assume a time-{}varying magnetic field pointing in the 
 \emph{z}~ direction. We write the magnetic field in a power expansion in time, keeping only terms up through second order:

\begin{equation}
\label{equation-xt-magt-xt-magt-calc-0}\hyperlabel{equation-xt-magt-xt-magt-calc-0}%
\overrightarrow{B}\left(t\right)=B\left(t\right)\widehat{z}=\left({B}_{0}+{B}_{1}t+\frac{{B}_{2}}{2}{t}^{2}\right)\widehat{z}
\end{equation}

Or in relative time:

\begin{equation}
\label{equation-xt-magt-xt-magt-calc-1}\hyperlabel{equation-xt-magt-xt-magt-calc-1}%
{\overrightarrow{B}}_{\tau }\left({t}_{\tau }\right)=\left({B}_{0}+{B}_{1}\left(\tau +{t}_{\tau }\right)+\frac{{B}_{2}}{2}{\left(\tau +{t}_{\tau }\right)}^{2}\right)\widehat{z}=\left({B}_{\tau }^{\left(0\right)}+{B}_{\tau }^{\left(1\right)}{t}_{\tau }+\frac{{B}_{\tau }^{\left(2\right)}}{2}{t}_{\tau }^{2}\right)\widehat{z}={B}_{\tau }\left({t}_{\tau }\right)\widehat{z}
\end{equation}

With:

\begin{equation}
\label{equation-xt-magt-xt-magt-calc-2}\hyperlabel{equation-xt-magt-xt-magt-calc-2}%
{B}_{\tau }^{\left(0\right)}={B}_{0}+{B}_{1}\tau +{B}_{2}{\tau }^{2}={B}_{\tau }\left(0\right),{B}_{\tau }^{\left(1\right)}={B}_{1}+{B}_{2}\tau ,{B}_{\tau }^{\left(2\right)}={B}_{2}
\end{equation}

We write the magnetic field as the curl of the vector potential:

\begin{equation}
\label{equation-xt-magt-xt-magt-calc-3}\hyperlabel{equation-xt-magt-xt-magt-calc-3}%
{\overrightarrow{A}}_{\tau }\left({t}_{\tau },\overrightarrow{x}\right)=\frac{{B}_{\tau }\left({t}_{\tau }\right)}{2}\left(-y,x,0\right)={\overrightarrow{A}}_{\tau }^{\left(0\right)}+{\overrightarrow{A}}_{\tau }^{\left(1\right)}{t}_{\tau }+{\overrightarrow{A}}_{\tau }^{\left(2\right)}{t}_{\tau }^{2}
\end{equation}

With:

\begin{equation}
\label{equation-xt-magt-xt-magt-calc-4}\hyperlabel{equation-xt-magt-xt-magt-calc-4}%
{\overrightarrow{A}}_{\tau }^{\left(n\right)}\left(\overrightarrow{x}\right)=\frac{{B}_{\tau }^{\left(n\right)}}{2}\left(-y,x,0\right)
\end{equation}

The time-{}varying magnetic field will induce an electric field:

\begin{equation}
\label{equation-xt-magt-xt-magt-calc-5}\hyperlabel{equation-xt-magt-xt-magt-calc-5}%
\overrightarrow{E}=-\frac{\partial \overrightarrow{A}}{\partial t}=-\frac{1}{2}\frac{\partial {B}_{\tau }\left(t\right)}{\partial t}\left(-y,x,0\right)
\end{equation}

As noted in 
 \hyperlink{scatter-magt}{Time Dependent Magnetic Field}, this is already accounted for in the Hamiltonian.

In this case, rather than explicitly expand the cross potential in powers of the relative time, it is easier to rewrite the full Hamiltonian as:

\begin{equation}
\label{equation-xt-magt-xt-magt-9}\hyperlabel{equation-xt-magt-xt-magt-9}%
H=\left(E-\frac{{E}^{2}}{2m}+\frac{m}{2}\right)+\left(\frac{{\overrightarrow{p}}^{2}}{2m}-e\frac{\overrightarrow{p}\cdot {\overrightarrow{A}}_{\tau }\left({t}_{\tau },\overrightarrow{x}\right)}{m}+{e}^{2}\frac{{\overrightarrow{A}}_{\tau }{\left({t}_{\tau },\overrightarrow{x}\right)}^{2}}{2m}\right)
\end{equation}

We get the equations of motion by taking the commutators (\hyperlink{tq-op}{Operators in Time}), getting the Euler-{}Lagrange equations for the space coordinates:

\begin{equation}
\label{equation-xt-magt-xt-magt-coord-3}\hyperlabel{equation-xt-magt-xt-magt-coord-3}%
\frac{{d}^{2}\overrightarrow{x}}{d{\tau }^{2}}=-\left\lbrack \left\lbrack \overrightarrow{x},H\right\rbrack ,H\right\rbrack =\left({\omega }_{\tau }\left({t}_{\tau }\right)\dot{y},-{\omega }_{\tau }\left({t}_{\tau }\right)\dot{x},0\right)
\end{equation}

With time dependent Larmor frequency:

\begin{equation}
\label{equation-xt-magt-xt-magt-coord-4}\hyperlabel{equation-xt-magt-xt-magt-coord-4}%
{\omega }_{\tau }\left({t}_{\tau }\right)\equiv \frac{e}{m}{B}_{\tau }\left({t}_{\tau }\right)
\end{equation}

Which we can break up into standard quantum theory and temporal quantization parts:

\begin{equation}
\label{equation-xt-magt-xt-magt-coord-5}\hyperlabel{equation-xt-magt-xt-magt-coord-5}%
{\omega }_{\tau }\left({t}_{\tau }\right)={\overline{\omega }}_{\tau }+{\stackrel{\frown }{\omega }}_{\tau }\left({t}_{\tau }\right)=\frac{e}{m}{B}_{\tau }^{\left(0\right)}+\frac{e}{m}\left({B}_{\tau }\left({t}_{\tau }\right)-{B}_{\tau }^{\left(0\right)}\right)
\end{equation}

The practical effect is to change the effective Larmor frequency, adding terms depending on 
 $ \langle {t}_{\tau }\rangle $~ and 
 $ \langle {t}_{\tau }^{2}\rangle $:

\begin{equation}
\label{equation-xt-magt-xt-magt-coord-9}\hyperlabel{equation-xt-magt-xt-magt-coord-9}%
\begin{array}{l}{\overline{\omega }}_{\tau }\rightarrow {\omega }_{\tau }={\overline{\omega }}_{\tau }+{\stackrel{\frown }{\omega }}_{\tau },\\ {\overline{\omega }}_{\tau }=\frac{e}{m}{B}_{\tau }^{\left(0\right)},\\ {\stackrel{\frown }{\omega }}_{\tau }=\frac{e}{m}\left({B}_{\tau }^{\left(1\right)}\langle {t}_{\tau }\rangle +\frac{1}{2}{B}_{\tau }^{\left(2\right)}\frac{{\sigma }_{0}^{2}}{2}\right)=\frac{e}{m}\left({B}_{1}+\frac{1}{2}{B}_{2}\tau \right)\langle {t}_{\tau }\rangle +\frac{1}{4}\frac{e}{m}{B}_{2}{\sigma }_{0}^{2}\end{array}
\end{equation}

We can integrate over 
 $ \tau $~ to get the cumulative effect.

We can easily extend the approach here to include more complex behavior of the magnetic field (i.e. sinusoidal), more interesting wave functions, and higher orders of perturbative
    corrections, depending on our requirements.

The general result is that if our particle is sufficiently spread out in time, and going through a magnetic field that varies rapidly enough in time, the effective magnetic field will
    include components that result from sampling past and future.

\subsection{Time Dependent Electric Field}\label{xt-elect}\hyperlabel{xt-elect}%
\begin{quote}

The present life of man, O king, seems to me, in comparison of that time which is unknown to us, like to the swift flight of a sparrow through the room wherein you sit at supper in
        winter, with your commanders and ministers, and a good fire in the midst, whilst the storms of rain and snow prevail abroad; the sparrow, I say, flying in at one door, and immediately out at
        another, whilst he is within, is safe from the wintry storm; but after a short space of fair weather, he immediately vanishes out of your sight, into the dark winter from which he had
        emerged.

\hspace*\fill---~The Venerable Bede
 \cite{Bede-1910}\end{quote}

We assume a time-{}varying electric field in the 
 \emph{x}~ direction. For instance, we might have a capacitor with a hole in it, a particle going through the hole, and the voltage on the capacitor changing even while the
    particle is in flight. We keep the terms through 
 $ {t}^{2}$. We assume no space dependence in the electric field. We take the potential:

\begin{equation}
\label{equation-xt-elect-xt-elect-0}\hyperlabel{equation-xt-elect-xt-elect-0}%
\Phi \left(t,x\right)=E\left(t\right)x=-\left({E}_{0}+{E}_{1}t+\frac{1}{2}{E}_{2}{t}^{2}\right)x
\end{equation}

The potential produces a longitudinal electric field:

\begin{equation}
\label{equation-xt-elect-xt-elect-1}\hyperlabel{equation-xt-elect-xt-elect-1}%
{\overrightarrow{E}}^{\left(long\right)}\left(t\right)=-\nabla \Phi =E\left(t\right)\widehat{x}
\end{equation}

The time derivative of the electric field is the displacement current, which induces a magnetic field:

\begin{equation}
\label{equation-xt-elect-xt-elect-2}\hyperlabel{equation-xt-elect-xt-elect-2}%
\nabla \times \overrightarrow{B}=\frac{\partial {\overrightarrow{E}}^{\left(long\right)}}{\partial t},\overrightarrow{B}=\frac{1}{2}\frac{\partial E}{\partial t}\left(0,-z,y\right)
\end{equation}

We write the magnetic field in terms of a vector potential 
 $ \overrightarrow{A}$:

\begin{equation}
\label{equation-xt-elect-xt-elect-4}\hyperlabel{equation-xt-elect-xt-elect-4}%
\overrightarrow{B}=\nabla \times \overrightarrow{A},\overrightarrow{A}=-\frac{1}{4}\frac{\partial E}{\partial t}\left({y}^{2}+{z}^{2},0,0\right)
\end{equation}

The time derivative of the vector potential induces a transverse electric field:

\begin{equation}
\label{equation-xt-elect-xt-elect-6}\hyperlabel{equation-xt-elect-xt-elect-6}%
{\overrightarrow{E}}^{\left(trans\right)}\equiv -\frac{\partial \overrightarrow{A}}{\partial t}=\frac{1}{4}\frac{{\partial }^{2}E}{\partial {t}^{2}}\left({y}^{2}+{z}^{2},0,0\right)
\end{equation}

In principle, this should induce a further correction to the magnetic field. However because we are keeping only through the second order in time, this is zero:

\begin{equation}
\label{equation-xt-elect-xt-elect-7}\hyperlabel{equation-xt-elect-xt-elect-7}%
\nabla \times {\overrightarrow{B}}^{\prime }=\frac{\partial {\overrightarrow{E}}^{\left(trans\right)}}{\partial t}=\frac{1}{4}\frac{{\partial }^{3}E}{\partial {t}^{3}}\left({y}^{2}+{z}^{2},0,0\right)=0
\end{equation}

The longitudinal electric field can be expanded in terms of the relative time:

\begin{equation}
\label{equation-xt-elect-xt-elect-8}\hyperlabel{equation-xt-elect-xt-elect-8}%
{\overrightarrow{E}}_{\tau }^{\left(long\right)}\left({t}_{\tau }\right)=\left({E}_{0}+{E}_{1}\left(\tau +{t}_{\tau }\right)+\frac{1}{2}{E}_{2}{\left(\tau +{t}_{\tau }\right)}^{2}\right)\widehat{x}=\left({E}_{\tau }^{\left(0\right)}+{E}_{\tau }^{\left(1\right)}{t}_{\tau }+\frac{1}{2}{E}_{\tau }^{\left(2\right)}{t}_{\tau }^{2}\right)\widehat{x}={E}_{\tau }\left({t}_{\tau }\right)\widehat{x}
\end{equation}

With:

\begin{equation}
\label{equation-xt-elect-xt-elect-9}\hyperlabel{equation-xt-elect-xt-elect-9}%
{E}_{\tau }^{\left(0\right)}={E}_{0}+{E}_{1}\tau +\frac{1}{2}{E}_{2}{\tau }^{2},{E}_{\tau }^{\left(1\right)}={E}_{1}+{E}_{2}\tau ,{E}_{\tau }^{\left(2\right)}={E}_{2}
\end{equation}

And we can expand the vector potential in powers of the relative time as well:

\begin{equation}
\label{equation-xt-elect-xt-elect-a}\hyperlabel{equation-xt-elect-xt-elect-a}%
{\overrightarrow{A}}_{\tau }=\left(-\frac{1}{4}\left({E}_{\tau }^{\left(1\right)}+{E}_{\tau }^{\left(2\right)}{t}_{\tau }\right)\left({y}^{2}+{z}^{2}\right),0,0\right)
\end{equation}

The standard quantum theory parts of the electric and vector potentials ($ {t}_{\tau }=0$) are:

\begin{equation}
\label{equation-xt-elect-xt-elect-b}\hyperlabel{equation-xt-elect-xt-elect-b}%
{\Phi }_{\tau }\left(\overrightarrow{x}\right)=-{E}_{\tau }^{\left(0\right)}x,{\overrightarrow{A}}_{\tau }\left(\overrightarrow{x}\right)=\left(-\frac{1}{4}{E}_{\tau }^{\left(1\right)}\left({y}^{2}+{z}^{2}\right),0,0\right)
\end{equation}

Time smoothed longitudinal electric field:

\begin{equation}
\label{equation-xt-elect-xt-elect-c}\hyperlabel{equation-xt-elect-xt-elect-c}%
\langle {\overrightarrow{E}}_{\tau }^{\left(long\right)}\left({t}_{\tau },\overrightarrow{x}\right)\rangle =\left({E}_{\tau }^{\left(0\right)}+\frac{1}{2}{E}_{\tau }^{\left(1\right)}{t}_{\tau }+\frac{1}{6}{E}_{\tau }^{\left(2\right)}{t}_{\tau }^{2}\right)\widehat{x}
\end{equation}

And time smoothed vector potential:

\begin{equation}
\label{equation-xt-elect-xt-elect-d}\hyperlabel{equation-xt-elect-xt-elect-d}%
\langle \frac{\partial {\overrightarrow{A}}_{\tau }\left({t}_{\tau },\overrightarrow{x}\right)}{\partial {t}_{\tau }}\rangle =-\frac{1}{4}{E}_{\tau }^{\left(2\right)}\left({y}^{2}+{z}^{2}\right)\widehat{x}
\end{equation}

The full temporal quantization Hamiltonian is:

\begin{equation}
\label{equation-xt-elect-xt-elect-final-0}\hyperlabel{equation-xt-elect-xt-elect-final-0}%
H={\stackrel{\frown }{H}}^{\left(free\right)}+{\overline{H}}^{\left(elec+mag\right)}+{V}_{\tau }
\end{equation}

Standard quantum theory part:

\begin{equation}
\label{equation-xt-elect-xt-elect-final-2}\hyperlabel{equation-xt-elect-xt-elect-final-2}%
{\overline{H}}^{\left(elec+mag\right)}=\frac{{\left(\overrightarrow{p}+\frac{e}{4}{E}_{\tau }^{\left(1\right)}\left({y}^{2}+{z}^{2}\right)\right)}^{2}}{2m}-e{E}_{\tau }^{\left(0\right)}x
\end{equation}

Cross potential (dropping terms higher than quadratic in relative time):

\begin{equation}
\label{equation-xt-elect-xt-elect-final-3}\hyperlabel{equation-xt-elect-xt-elect-final-3}%
{V}_{\tau }=\left(\begin{array}{l}e\frac{\left({p}_{x}+\frac{1}{4}e{E}_{\tau }^{\left(1\right)}\left({y}^{2}+{z}^{2}\right)\right)\left({y}^{2}+{z}^{2}\right){E}_{\tau }^{\left(2\right)}}{4m}{t}_{\tau }+{e}^{2}\frac{{E}_{\tau }^{\left(2\right)2}{\left({y}^{2}+{z}^{2}\right)}^{2}}{32m}{t}_{\tau }^{2}\\ +e\frac{\left({E}_{\tau }^{\left(0\right)}+\frac{1}{2}{E}_{\tau }^{\left(1\right)}{t}_{\tau }\right)}{m}{p}_{x}{t}_{\tau }+{e}^{2}\frac{{E}_{\tau }^{\left(0\right)2}}{2m}{t}_{\tau }^{2}\end{array}\right)
\end{equation}

Rate of change of discrepancy in 
 \emph{x}~ is:

\begin{equation}
\label{equation-xt-elect-xt-elect-final-4}\hyperlabel{equation-xt-elect-xt-elect-final-4}%
\frac{\partial }{\partial \tau }\langle \delta x\rangle =-i\left\lbrack x,{V}_{\tau }\right\rbrack =e\frac{{E}_{\tau }^{\left(2\right)}}{4m}\left({y}^{2}+{z}^{2}\right){t}_{\tau }+e\frac{{E}_{\tau }^{\left(0\right)}}{m}{t}_{\tau }+e\frac{{E}_{\tau }^{\left(1\right)}}{2m}{t}_{\tau }^{2}
\end{equation}

So the rate of change of the discrepancy in 
 \emph{x}~ is approximately:

\begin{equation}
\label{equation-xt-elect-xt-elect-final-6}\hyperlabel{equation-xt-elect-xt-elect-final-6}%
\frac{\partial }{\partial \tau }\langle \delta x\rangle \approx e\frac{{E}_{\tau }^{\left(2\right)}}{4m}\langle {y}^{2}+{z}^{2}\rangle \langle {t}_{\tau }\rangle +e\frac{{E}_{\tau }^{\left(0\right)}}{m}\langle {t}_{\tau }\rangle +e\frac{{E}_{\tau }^{\left(1\right)}}{2m}\langle {t}_{\tau }^{2}\rangle 
\end{equation}

In general, terms linear in the relative time are likely to average to zero, so the final result is:

\begin{equation}
\label{equation-xt-elect-xt-elect-final-7}\hyperlabel{equation-xt-elect-xt-elect-final-7}%
\frac{\partial }{\partial \tau }\langle \delta x\rangle \approx e\frac{{E}_{\tau }^{\left(1\right)}}{2m}\langle {t}_{\tau }^{2}\rangle 
\end{equation}

\section{Aharonov-{}Bohm Experiment}\label{xt-ab}\hyperlabel{xt-ab}%
\begin{figure}[H]

\begin{center}
\imgexists{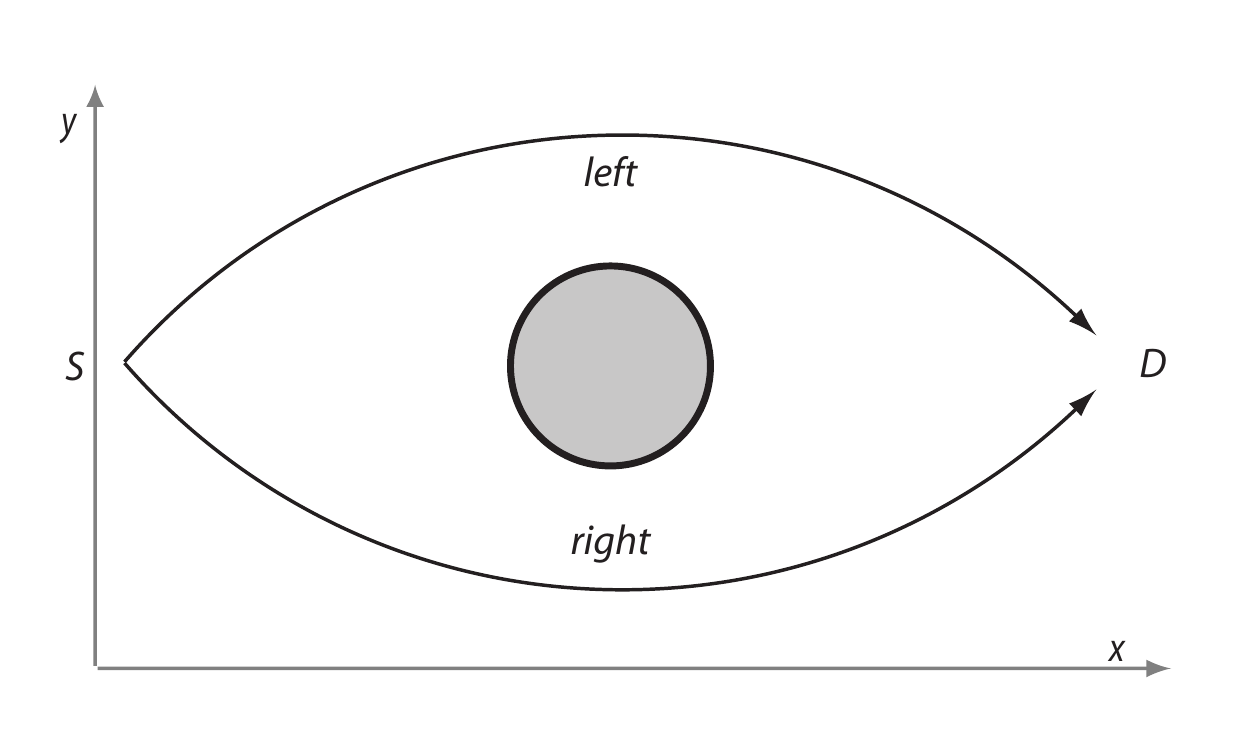}{{\imgevalsize{images/xt-ab-0.pdf}{\includegraphics[width=\imgwidth,height=\imgheight,keepaspectratio=true]{images/xt-ab-0.pdf}}\quad
}
}{}
\end{center}
\caption{Aharonov-{}Bohm Experiment in Space}
\label{figure-images-xt-ab-0}\hyperlabel{figure-images-xt-ab-0}%
\end{figure}

\subsection{The Aharonov-{}Bohm Experiment in Space}
In the Aharonov-{}Bohm experiment (\cite{Aharonov-1959}) a particle is sent through a splitter so its wave function is divided in two. One part of the wave function is routed around the left of an ideal solenoid,
        the other around the right. Outside an ideal solenoid the magnetic field is zero, so the two parts of the wave function see no magnetic field.

They do see a vector potential however. In semi-{}classical approximation the lowest order phase shift due to the vector potential is given by:

\begin{equation}
\label{equation-xt-ab-xt-ab-0}\hyperlabel{equation-xt-ab-xt-ab-0}%
ie{\displaystyle \underset{{\tau }^{\prime }}{\overset{{\tau }^{{''}}}{\int }}\text{d}\tau \dot{\overrightarrow{x}}\cdot \overrightarrow{A}\left(\overrightarrow{x}\left(\tau \right)\right)}
\end{equation}

It is a function of the magnetic field in the solenoid, but different for the left and right paths.

By changing the magnetic field in the solenoid we can induce a change in the relative phase shift experienced along each path. This in turn includes changes in the interference pattern
        at the detector.

These have been detected (\cite{Osakabe-1986}). Extraordinary.

If the vector potential still has an effect when the magnetic field is zero then the vector potential appears more fundamental than the magnetic field, contrary to all classical
        thinking.

\subsection{The Aharonov-{}Bohm Experiment in Time}
As noted, one way to build experimental tests of temporal quantization is to look at "flipped" versions of tests of standard quantum theory, versions with time and a space dimension
        interchanged.

Flipping time and a space dimension also interchanges the electric and magnetic fields.

Therefore we look a variation on the Aharonov-{}Bohm experiment where we vary the electric potential in time rather than vary the magnetic potential in space.
\begin{figure}[H]

\begin{center}
\imgexists{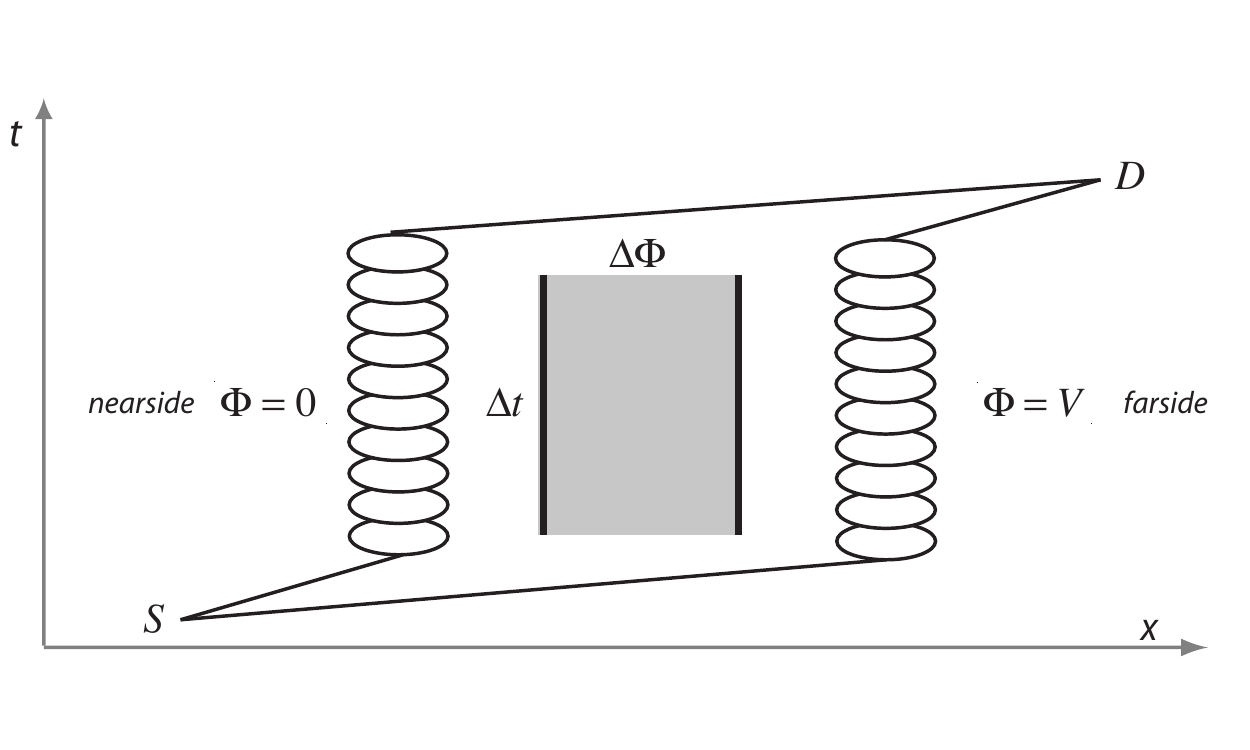}{{\imgevalsize{images/xt-ab-1.pdf}{\includegraphics[width=\imgwidth,height=\imgheight,keepaspectratio=true]{images/xt-ab-1.pdf}}\quad
}
}{}
\end{center}
\caption{Aharonov-{}Bohm Experiment in Time}
\label{figure-images-xt-ab-1}\hyperlabel{figure-images-xt-ab-1}%
\end{figure}

The setup consists of a splitter, a capacitor we can turn on and off, and a pair of delay loops, one in front of the capacitor, one behind it.

The delay loop has to "park" an electron or other charged particle for a specified (laboratory) time without loss of coherence.

While the capacitor is off we route a particle through the splitter. We send one part of the wave function through a hole in the capacitor, reserving the other part. The near and far
        sides of the wave function are then fed into delay loops. Then the capacitor is turned on. There is no field outside the capacitor (it is an ideal capacitor) so neither part of the wave
        function sees a field. However, there is a potential difference between the two sides, so one part sees a potential of 
 \emph{V}~ relative to the other.

Therefore one part of the wave function experiences a phase change given by the integral (per above 
 \hyperlink{semi-const}{Constant Potentials}):

\begin{equation}
\label{equation-xt-ab-xt-ab-1}\hyperlabel{equation-xt-ab-xt-ab-1}%
ie{\displaystyle \underset{{\tau }^{\prime }}{\overset{{\tau }^{{''}}}{\int }}\text{d}\tau \dot{t}\Phi \left(x\left(\tau \right)\right)}
\end{equation}

Now the capacitor is turned off and the nearside part of the wave function sent through the hole in the capacitor to be combined with the farside part.

They will interfere destructively or constructively depending on the relative phase change. The relative phase change can be tuned by changing the voltage on the capacitor and the
        amount of time the capacitor is turned on.

If the potential is 
 \emph{V}~ on the near side and zero on the far side, then the phase shift is:

\begin{equation}
\label{equation-xt-ab-xt-ab-2}\hyperlabel{equation-xt-ab-xt-ab-2}%
\Delta \phi \sim -\frac{E}{m}eV\Delta \tau =-\gamma eV\Delta \tau 
\end{equation}

On the farside it is zero.

The interference pattern may therefore be controlled by changing a field after and before the particle passes through the field, with the particle never seeing the field, only the
        potential 
 $ \Phi $.

\paragraph*{Discussion}

\noindent

As it happens, we see essentially the same effect in standard quantum theory. The non-{}relativistic Lagrangian gives:

\begin{equation}
\label{equation-xt-ab-xt-ab-3}\hyperlabel{equation-xt-ab-xt-ab-3}%
\Delta \overline{\phi }\approx -eV\Delta \tau 
\end{equation}

In the non-{}relativistic case, the two phase shifts are the same. Therefore the Aharonov-{}Bohm experiment in time effect is present in standard quantum theory and temporal quantization
        both.

The original paper by Aharonov and Bohm looked first at the integral over:

\begin{equation}
\label{equation-xt-ab-xt-ab-0b}\hyperlabel{equation-xt-ab-xt-ab-0b}%
-ie{\displaystyle \underset{{\tau }^{\prime }}{\overset{{\tau }^{{''}}}{\int }}\text{d}\tau E\left(\overrightarrow{x}\left(\tau \right)\right)-\dot{\overrightarrow{x}}\cdot \overrightarrow{A}\left(\overrightarrow{x}\left(\tau \right)\right)}
\end{equation}

Which includes this contribution from the electric field. Contributions to the Aharonov-{}Bohm experiment from the electric part have been seen 
 \cite{vanOudenaarden-1998}.

The novelty is that we are modulating the electric field in time, not just in space, to produce the effect.

Since Aharonov and Rohrlich (\cite{Aharonov-2005}) have shown the effect is needed to ensure the self-{}consistency of quantum mechanics in the case of the vector potential, we expect it is needed to ensure
        the self-{}consistency of quantum mechanics in the case of the electric potential as well.

The Aharonov-{}Bohm experiment in time does not discriminate decisively between temporal quantization and standard quantum theory. But it does suggest that temporal quantization offers a
        useful way to develop new experiments in time.

\section{Discussion}\label{xt-disc}\hyperlabel{xt-disc}%
\begin{quote}

A theory is an experiment's way of creating new experiments.

\hspace*\fill---~Anonymous\end{quote}

\paragraph*{Exact Solutions}

\noindent

We have used some fairly rough approximations here.

There is an indirect benefit to use of approximate methods: it avoids excessive dependence on the specifics of temporal quantization, making the predictions if cruder more robust.

With that said, the expressions for the kernel and the Schrödinger equation are exact and in some cases can be solved exactly. Approaches:
\begin{enumerate}

\item{}We can solve the Schrödinger equation directly, posit reasonable initial boundary conditions (i.e. a plane wave to left of the gate, emptiness to its right) and work along the lines
            developed by Moshinsky (\cite{Moshinsky-1951},
 \cite{Moshinsky-1952}).

\item{}We can solve for the path integral kernel by using a discrete grid for space and time and discrete steps in laboratory time, then sum over the paths using generating functions and
            other techniques from discrete analysis, along the lines developed in Feller (\cite{Feller-1968},
 \cite{Feller-1971}) and Graham, Knuth, and Patashnik (\cite{Graham-1994}).

\item{}We can model the gates and detector as absorbing walls and apply the methods developed in Marchewka and Schuss (\cite{Marchewka-1998},
 \cite{Marchewka-1999b}).

\end{enumerate}

\paragraph*{Further Experiments}

\noindent

Temporal quantization is an experiment factory.

We have developed just a few experiments here, enough to establish the basic rules, but obviously only a small subset of what is possible.

To date there have been relatively few experiments aimed directly at time: in addition to Lindner's, we have diffraction in time (\cite{Godoy-2002}) and the delayed choice quantum eraser (\cite{Scully-1982},
 \cite{Scully-1991b},
 \cite{Kim-2000b}) as the most obvious.

Further, the most obvious effect of quantum time is to produce increased dispersion in time. It is possible re-{}mining existing datasets may provide evidence for or against quantum
    time.

As noted, most foundational experiments can serve as a starting point to develop a test of quantum time; there are potentially hundreds of tests.

We might look at:
\begin{enumerate}

\item{}The 
 \hyperlink{xt-ab}{Aharonov-{}Bohm Experiment}~ (in space) with time-{}varying solenoid.

\item{}Tunneling in time: a quantum cat nervous about the occasional appearances of hydrocyanic acid in its chamber might understandably choose to tunnel past the times when the acid is
            present (\cite{Schroedinger-1935}).

\item{}Gates in energy rather than time.

\item{}Particle-{}particle scattering experiments looking for the effects of dispersion in time.

\item{}Particle-{}particle scattering experiments taking advantage of the point that if the particle's wave function is anti-{}symmetric in time it can have the "wrong" symmetry in space and
            still satisfy its symmetry requirements.

\item{}Experiments where we mix rather than flip time and space, as relativistic scattering experiments analyzed from multiple frames.

\item{}Variations on the Michelson-{}Morley experiment (\cite{Michelson-1887}). See for instance 
 \cite{Cahill-2003}.

\item{}Apparent violations of dispersion relations. While temporal quantization satisfies unitarity (see 
 \hyperlink{tq-seqn-unitarity}{Unitarity}~ above) it does so in four dimensions, not three. This is the sort of thing that will not normally be seen unless explicitly looked
            for (but see 
 \cite{Bennett-1987},
 \cite{Bennett-1987b},
 \cite{Bennett-1987c}, disputed in 
 \cite{Valentini-1988}).

\item{}Variations on the experimental designs being used by Hestenes and Catillion to look for Zitterbewegung, used to look for fluctuations in time as well.

\item{}Variations on electro-{}magnetically induced transparency (EIT), slow light, the quantum Zeno effect, the quantum eraser, and so on.

\end{enumerate}

To analyze some of these we would have to extend the ideas in this work to the multi-{}particle case.


\chapter{Discussion}\label{disc}\hyperlabel{disc}%
\begin{quote}

The only way of discovering the limits of the possible is to venture a little way past them into the impossible.

\hspace*\fill---~Arthur C. Clarke's Second Law
 \cite{Clarke-1984}\end{quote}
\begin{quote}

Theories need to be pushed as far as they can go before deciding they no longer apply in a given domain.

\hspace*\fill---~Victor Stenger
 \cite{Stenger-2000}\end{quote}

\paragraph*{Analysis}

\noindent

Our goal has been to quantize time using the rules that we use to quantize space, then see what breaks.

We started with laboratory time 
 $ \tau $, time as defined by clocks, laser beams, and graduate students. We posited quantum time 
 $ t$, time quantized the same way we quantize space. Laboratory and quantum time represent two ways of looking at same thing: laboratory time or clock time is time as parameter,
    quantum time is time as observable.

To define what this means in operational terms, we used path integrals (\hyperlink{tq-fpi}{Feynman Path Integrals}). We generalized the paths to include motion in time, but made no other changes to the path integrals themselves.

We used Morlet wavelets to analyze the initial wave functions, demonstrate the convergence of the integrals (\hyperlink{tq-fpi-conv}{Convergence}), and define a covariant form of the laboratory time (\hyperlink{covar}{Covariant Definition of Laboratory Time}).

The work product of the analysis is the expression for the path integral:

\begin{equation}
\label{equation-disc-tq-fpi-final-full_kernel}\hyperlabel{equation-disc-tq-fpi-final-full_kernel}%
{K}_{\tau }\left({x}^{{''}};{x}^{\prime }\right)=\underset{N\rightarrow \infty }{\mathrm{lim}}{\displaystyle \int \mathcal{D}x\mathrm{exp}\left(-i{\displaystyle \sum _{j=1}^{N+1}\left(m\frac{{\left({x}_{j}-{x}_{j-1}\right)}^{2}}{2\epsilon }+e\left({x}_{j}-{x}_{j-1}\right)\frac{A\left({x}_{j}\right)+A\left({x}_{j-1}\right)}{2}+\frac{m}{2}\epsilon \right)}\right)}
\end{equation}

With measure:

\begin{equation}
\label{equation-disc-tq-fpi-final-measure}\hyperlabel{equation-disc-tq-fpi-final-measure}%
\mathcal{D}x\equiv {\left(-\frac{i{m}^{2}}{4{\pi }^{2}{\epsilon }^{2}}\right)}^{N+1}{\displaystyle \prod _{n=1}^{n=N}{\text{d}}^{\text{4}}{x}_{n}}
\end{equation}

We used the short time limit of the path integral to derive the Schrödinger equation:

\begin{equation}
\label{equation-disc-tq-seqn-deriv-seqn}\hyperlabel{equation-disc-tq-seqn-deriv-seqn}%
i\frac{d{\psi }_{\tau }\left(x\right)}{d\tau }=-\frac{1}{2m}\left({\left(p-eA\right)}^{2}-{m}^{2}\right){\psi }_{\tau }\left(x\right)
\end{equation}

Or:

\begin{equation}
\label{equation-disc-tq-seqn-coord-seqn}\hyperlabel{equation-disc-tq-seqn-coord-seqn}%
i\frac{d{\psi }_{\tau }}{d\tau }\left(t,\overrightarrow{x}\right)=\frac{1}{2m}\left({\left({\partial }_{t}+ie\Phi \left(t,\overrightarrow{x}\right)\right)}^{2}-{\left(\overrightarrow{\nabla }-ie\overrightarrow{A}\left(t,\overrightarrow{x}\right)\right)}^{2}+{m}^{2}\right){\psi }_{\tau }\left(t,\overrightarrow{x}\right)
\end{equation}

We used the Schrödinger equation to demonstrate unitarity (\hyperlink{tq-seqn-unitarity}{Unitarity}) ~ and analyze the effect of gauge transformations (\hyperlink{tq-seqn-gauge}{Gauge Transformations for the Schrödinger Equation}).

We derived an operator formalism from the Schrödinger equation (\hyperlink{tq-op}{Operators in Time}), canonical path integrals from the operator formalism (\hyperlink{tq-pq}{Canonical Path Integrals}), and the original Feynman path integrals from the canonical path integrals (\hyperlink{tq-pq-circle}{Closing the Circle}) \textendash{} thus closing the circle, and establishing the consistency of all four approaches with each other.

We worked out an invariant definition of the laboratory time (\hyperlink{covar}{Covariant Definition of Laboratory Time}) by breaking the wave function up into its component wavelets and finding an invariant definition of the laboratory time for
    each.

\paragraph*{Slow, Large, and Long Time Limits}

\noindent

We then showed we recover standard quantum theory from temporal quantization in various limits.
\begin{enumerate}

\item{}In the non-{}relativistic limit (\hyperlink{nonrel}{Non-{}relativistic Limit}), we showed we can separate the system into time and space parts, with the space part being the usual standard quantum theory part. In
            this limit the effect of temporal quantization is to create a kind of temporal fuzz around the baseline standard quantum theory solutions.

\item{}In the semi-{}classical limit (Planck's constant goes to zero), we again got temporal quantization as standard quantum theory plus some fuzz in time (\hyperlink{semi}{Semi-{}classical Limit}). The expectations are the same for both, but in temporal quantization we see additional dispersion in time. Temporal quantization is to
            standard quantum theory (with respect to time) as standard quantum theory is to classical mechanics.

\item{}In the long time limit (\hyperlink{stat}{Long Time Limit}), we got standard quantum theory as the time average of temporal quantization. Over longer periods of time (which might not be that long,
            anything more than a few hundred attoseconds might do) the stationary states, those with no variation with respect to laboratory time, will dominate. These will be the ones with
            laboratory energy 
 $ \mathscr{E}$~ zero:

\begin{equation}
\label{equation-disc-disc-0}\hyperlabel{equation-disc-disc-0}%
i\frac{d}{d\tau }{\psi }_{\tau }\left(t,\overrightarrow{x}\right)=0\Rightarrow H{\psi }_{\tau }\left(t,\overrightarrow{x}\right)=0\Rightarrow \mathscr{E}=0
\end{equation}

To take advantage of this limit we looked at the cases of non-{}singular potentials and of bound states:
\begin{enumerate}

\item{}For non-{}singular potentials (\hyperlink{scatter}{Non-{}singular Potentials}), we reformulated the Schrödinger equation using relative time 
 $ \langle {t}_{\tau }\rangle $~ and the time gauge:

\begin{equation}
\label{equation-disc-scatter-elect-tg}\hyperlabel{equation-disc-scatter-elect-tg}%
{\Lambda }_{\tau }\left({t}_{\tau },x\right)={\displaystyle \underset{0}{\overset{{t}_{\tau }}{\int }}\text{d}{{t}^{\prime }}_{\tau }{\Phi }_{\tau }\left({{t}^{\prime }}_{\tau },\overrightarrow{x}\right)}
\end{equation}

To expand the Schrödinger equation in powers of the relative time, the difference between quantum and laboratory time.

\item{}For bound states (\hyperlink{bound}{Bound States}), we used an analysis of the diagonal matrix elements to show that the stationary state condition picks out the usual Bohr orbitals: for
                    each standard quantum theory bound state 
 $ {\xi }_{\overline{n}}\left(\overrightarrow{x}\right)$~ there is an associated temporal quantization bound state:

\begin{equation}
\label{equation-disc-bound-zero-labenergy-wf}\hyperlabel{equation-disc-bound-zero-labenergy-wf}%
{\psi }_{\tau }^{\left(E,\overline{n}\right)}\left({t}_{\tau },\overrightarrow{x}\right)=\frac{1}{\sqrt{2\pi }}\mathrm{exp}\left(-i{E}_{\overline{n}}{t}_{\tau }\right){\xi }_{\overline{n}}\left(\overrightarrow{x}\right)\mathrm{exp}\left(-i{E}_{\overline{n}}\tau \right)
\end{equation}

We estimated the dispersion in time (\hyperlink{bound-disp}{Estimate of Uncertainty in Time}) as the same order as the width of the atomic orbitals in space, taking it to under an attosecond in most
                    cases.

\end{enumerate}

\end{enumerate}

\paragraph*{Experimental Tests}

\noindent

Having established that temporal quantization looks like standard quantum theory in the slow, large, and long time limits, we turned to asking when would we expect to see effects of fuzzy
    time? What experimental tests are possible?

We noted that most experiments have two halves, a scatterer and a scatteree. Both have to vary in time or the effects of temporal quantization are likely to average out. This helps explain
    why quantum time might not have been seen by accident. The most common effect of temporal quantization is additional dispersion in time, easily written off as additional experimental
    error.

To get a test of quantum time we need to work with time-{}varying inputs (not steady beams) and work with apparatus that varies in time (choppers, time-{}varying fields, capacitors that turn on
    and off, and the like).

We can manufacture tests of quantum time by:
\begin{enumerate}

\item{}Starting with an existing foundational experiment in quantum mechanics then interchanging time and a space dimension.

\item{}Running a particle with known dispersion in time through an electromagnetic field which is varying rapidly in time.

\item{}Taking an experiment which implicitly takes a holistic view of time \textendash{} quantum eraser, Aharonov-{}Bohm experiment, tests of Aharonov-{}Bergmann-{}Lebowitz time-{}symmetric measurements \textendash{} and
            reexamining it from the perspective of quantum time.

\end{enumerate}

We looked at
\begin{enumerate}

\item{}Single and double slit experiments (\hyperlink{xt-gates}{Slits in Time});

\item{}Time-{}varying magnetic fields (\hyperlink{xt-magt}{Time Dependent Magnetic Field}) and electric fields (\hyperlink{xt-elect}{Time Dependent Electric Field});

\item{}And a variation on the Aharonov-{}Bohm experiment (\hyperlink{xt-ab}{Aharonov-{}Bohm Experiment}).

\end{enumerate}

This is obviously only a small subset of the possible experiments.

\paragraph*{Requirements}

\noindent

We therefore argue that temporal quantization satisfies the 
 \hyperlink{int-intro}{requirements}:
\begin{enumerate}

\item{}Well-{}defined \textendash{} with both path integral and Schrödinger equation forms,

\item{}Manifestly covariant \textendash{} by construction,

\item{}Consistent with known experimental results \textendash{} in the slow, large, and long time limits,

\item{}Testable \textendash{} in a large number of ways,

\item{}And reasonably simple.

\end{enumerate}

The fifth requirement, reasonably simple, is the only one we have not explicitly defended to this point. Arguments in support include:
\begin{enumerate}

\item{}The only change we made to quantize in time was to add motion in time to the usual paths in path integrals. This is the least change we could have made.

\item{}The demonstration of the convergence of the path integrals is simpler than usual: there are no factors of 
 $ i\epsilon $, no artificial Wick rotation, and the like.

\item{}Unitarity is immediate.

\item{}By construction, quantum time and the three space coordinates appear in the theory in a fully covariant way 
 \footnote{Recall Weinberg's observation about the difficulties of demonstrating unitarity and covariance simultaneously.}. There is no special role for the quantum time
            coordinate.

\item{}The uncertainty principle between quantum time and quantum energy stands on the same basis as the uncertainty principle between space and momentum.

\end{enumerate}

\paragraph*{Merits}

\noindent

The principal merit of temporal quantization is that it lets us look at the relationship between time and the quantum in an experimentally testable way.

As noted in the introduction, the possible results are:
\begin{enumerate}

\item{}The behavior of time in quantum mechanics is fully covariant; all quantum effects seen along the space dimensions are seen along the time dimension.

\item{}We see quantum mechanical effects along the time direction, but they are not fully covariant; the effects along the time direction are less or greater or different than those seen
            in space.

\item{}We see no quantum mechanical effects along the time dimension.

\end{enumerate}

In the second and third cases, we might look for associated failures of Lorentz invariance.

In addition, temporal quantization provides a natural starting point for investigation in other areas where time is important, i.e. quantum gravity, quantum computing, and quantum
    communication.

\paragraph*{Further Questions}

\noindent

Areas for further investigation include:
\begin{enumerate}

\item{}Exact solutions for various simple cases.

\item{}More experimental tests.

\item{}Further analysis of the semi-{}classical approximation in temporal quantization.

\item{}Further clarification of the relationship between the classical, standard quantum theory, and temporal quantization pictures.

\item{}Extension of temporal quantization to multiple particles.

\item{}Extension of temporal quantization to statistical mechanics.

\item{}Implications for the time/energy uncertainty principle.

\item{}Implications of temporal quantization for the measurement problem.

\end{enumerate}
\appendix
\chapter{Free Particles}
\label{apps}\hyperlabel{apps}%

\section{Overview}\label{free-intro}\hyperlabel{free-intro}%
\begin{quote}

There is something fascinating about science. One gets such wholesale returns of conjecture out of such a trifling investment of fact.

\hspace*\fill---~Mark Twain
 \cite{Twain-1883}\end{quote}

We assemble some useful properties of the free particle wave functions and kernels here. We look at:
\begin{enumerate}

\item{}The 
 \hyperlink{free-fourt}{Time/Space Representation},

\item{}The 
 \hyperlink{free-fourE}{Energy/Momentum Representation},

\item{}And the 
 \hyperlink{free-hybrid}{Time/Momentum Representation}.

\end{enumerate}

\section{Time/Space Representation}\label{free-fourt}\hyperlabel{free-fourt}%

We look at the block time case then the relative time case.

\subsection{In Block Time}
We look at the kernel, then apply it to plane waves and Gaussian test functions.

\paragraph*{Kernel}

\noindent

The free kernel may be written as a product of time and space parts:

\begin{equation}
\label{equation-free-fourt-free-kernel-kprodt}\hyperlabel{equation-free-fourt-free-kernel-kprodt}%
{K}_{\tau }\left({x}^{{''}};{x}^{\prime }\right)={\stackrel{\frown }{K}}_{\tau }\left({t}^{{''}};{t}^{\prime }\right){\overline{K}}_{\tau }\left(\overrightarrow{{x}^{{''}}};\overrightarrow{{x}^{\prime }}\right)
\end{equation}

Time part:

\begin{equation}
\label{equation-free-fourt-free-onet-kernel}\hyperlabel{equation-free-fourt-free-onet-kernel}%
{\stackrel{\frown }{K}}_{\tau }\left({t}^{{''}};{t}^{\prime }\right)=\sqrt{\frac{im}{2\pi \tau }}\mathrm{exp}\left(-im\frac{{\left({t}^{{''}}-{t}^{\prime }\right)}^{2}}{2\tau }-im\frac{\tau }{2}\right)
\end{equation}

Space part:

\begin{equation}
\label{equation-free-fourt-free-fourt-kernel3}\hyperlabel{equation-free-fourt-free-fourt-kernel3}%
{\overline{K}}_{\tau }\left(\overrightarrow{{x}^{{''}}};\overrightarrow{{x}^{\prime }}\right)={\sqrt{\frac{m}{2\pi i\tau }}}^{3}\mathrm{exp}\left(\frac{im}{2\tau }{\left(\overrightarrow{{x}^{{''}}}-\overrightarrow{{x}^{\prime }}\right)}^{2}\right)
\end{equation}

\paragraph*{Plane Waves}

\noindent

We take the initial plane wave as:

\begin{equation}
\label{equation-free-fourt-free-planet-0}\hyperlabel{equation-free-fourt-free-planet-0}%
{\phi }_{0}\left(x\right)=\frac{1}{4{\pi }^{2}}\mathrm{exp}\left(-ipx\right)
\end{equation}

By applying the kernel we get the dependence on laboratory time:

\begin{equation}
\label{equation-free-fourt-free-planet-3}\hyperlabel{equation-free-fourt-free-planet-3}%
{\phi }_{\tau }\left(x\right)=\frac{1}{4{\pi }^{2}}\mathrm{exp}\left(-ipx-i{\mathscr{E}}_{p}\tau \right)
\end{equation}

With the laboratory energy 
 $ {\mathscr{E}}_{p}$:

\begin{equation}
\label{equation-free-fourt-free-planet-2}\hyperlabel{equation-free-fourt-free-planet-2}%
{\mathscr{E}}_{p}\equiv -\frac{{E}^{2}-{\overrightarrow{p}}^{2}-{m}^{2}}{2m}
\end{equation}

The 
 $ {\phi }_{\tau }\left(x\right)$~ are eigenstates of the free Hamiltonian with eigenvalues 
 $ {\mathscr{E}}_{p}$:

\begin{equation}
\label{equation-free-fourt-free-planet-1}\hyperlabel{equation-free-fourt-free-planet-1}%
{H}^{\left(free\right)}{\phi }_{\tau }\left(x\right)={\mathscr{E}}_{p}{\phi }_{\tau }\left(x\right),{H}^{\left(free\right)}=\frac{1}{2m}\left({\partial }_{t}^{2}-{\nabla }^{2}+{m}^{2}\right)
\end{equation}

The stationary states are onshell:

\begin{equation}
\label{equation-free-fourt-free-planet-3d}\hyperlabel{equation-free-fourt-free-planet-3d}%
i\frac{d}{d\tau }{\phi }_{\tau }\left(x\right)=0\rightarrow {\mathscr{E}}_{p}=0\rightarrow E={\overline{E}}_{\overrightarrow{p}}
\end{equation}

With the onshell energy defined as:

\begin{equation}
\label{equation-free-fourt-free-planet-3c}\hyperlabel{equation-free-fourt-free-planet-3c}%
{\overline{E}}_{\overrightarrow{p}}\equiv \sqrt{{\overrightarrow{p}}^{2}+{m}^{2}}
\end{equation}

Therefore the stationary states are given by:

\begin{equation}
\label{equation-free-fourt-free-planet-3b}\hyperlabel{equation-free-fourt-free-planet-3b}%
{\phi }_{\tau }^{\left(onshell\right)}\left(x\right)=\frac{1}{4{\pi }^{2}}\mathrm{exp}\left(-i{\overline{E}}_{\overrightarrow{p}}t+i\overrightarrow{p}\cdot \overrightarrow{x}\right)
\end{equation}

\paragraph*{Gaussian Test Functions}

\noindent

We take the initial Gaussian test function as:

\begin{equation}
\label{equation-free-fourt-free-fourt-phi0}\hyperlabel{equation-free-fourt-free-fourt-phi0}%
{\psi }_{0}\left(x\right)=\sqrt[4]{\frac{1}{{\pi }^{4}\mathrm{det}\left({\Sigma }_{0}\right)}}\mathrm{exp}\left(-i{p}_{\mu }^{\left(0\right)}{x}^{\mu }-\frac{1}{2{\Sigma }_{0}^{\mu \nu }}\left({x}^{\mu }-{\overline{x}}_{0}^{\mu }\right)\left({x}^{n}-{\overline{x}}_{0}^{\nu }\right)\right)
\end{equation}

With the dispersion matrix 
 $ \Sigma $~ diagonal:

\begin{equation}
\label{equation-free-fourt-free-fourt-sigma0}\hyperlabel{equation-free-fourt-free-fourt-sigma0}%
{\Sigma }_{0}^{\mu \nu }=\left(\begin{array}{cccc}{\sigma }_{0}^{2}& 0& 0& 0\\ 0& {\sigma }_{1}^{2}& 0& 0\\ 0& 0& {\sigma }_{2}^{2}& 0\\ 0& 0& 0& {\sigma }_{3}^{2}\end{array}\right)
\end{equation}

By applying the kernel we get the dependence on laboratory time:

\begin{equation}
\label{equation-free-fourt-free-fourt-phitau}\hyperlabel{equation-free-fourt-free-fourt-phitau}%
{\psi }_{\tau }\left(x\right)=\sqrt[4]{\frac{\mathrm{det}\left({\Sigma }_{0}^{\mu \nu }\right)}{{\pi }^{4}}}\sqrt{\frac{1}{\mathrm{det}\left({\Sigma }_{\tau }^{\mu \nu }\right)}}\mathrm{exp}\left(-i{p}_{0}^{\mu }{x}_{\mu }-\frac{1}{2{\Sigma }_{\tau }^{\mu \nu }}\left({x}^{\mu }-{\overline{x}}_{\tau }^{\mu }\right)\left({x}^{\nu }-{\overline{x}}_{\tau }^{\nu }\right)+i\frac{{p}_{0}^{2}-{m}^{2}}{2m}\tau \right)
\end{equation}

The expectations of coordinates evolve with laboratory time:

\begin{equation}
\label{equation-free-fourt-free-fourt-average}\hyperlabel{equation-free-fourt-free-fourt-average}%
{\overline{x}}_{\tau }^{\mu }\equiv {\langle {x}^{\mu }\rangle }_{\tau }={\overline{x}}_{0}^{\mu }+\frac{{p}_{0}^{\mu }}{m}\tau 
\end{equation}

As does the dispersion matrix:

\begin{equation}
\label{equation-free-fourt-free-fourt-sigmatau}\hyperlabel{equation-free-fourt-free-fourt-sigmatau}%
{\Sigma }_{\tau }^{\mu \nu }=\left(\begin{array}{cccc}{\sigma }_{0}^{2}-i\frac{\tau }{m}& 0& 0& 0\\ 0& {\sigma }_{1}^{2}+i\frac{\tau }{m}& 0& 0\\ 0& 0& {\sigma }_{2}^{2}+i\frac{\tau }{m}& 0\\ 0& 0& 0& {\sigma }_{3}^{2}+i\frac{\tau }{m}\end{array}\right)=\left(\begin{array}{cccc}{\sigma }_{0}^{2}{f}_{\tau }^{\left(0\right)}& 0& 0& 0\\ 0& {\sigma }_{1}^{2}{f}_{\tau }^{\left(1\right)}& 0& 0\\ 0& 0& {\sigma }_{2}^{2}{f}_{\tau }^{\left(2\right)}& 0\\ 0& 0& 0& {\sigma }_{3}^{2}{f}_{\tau }^{\left(3\right)}\end{array}\right)
\end{equation}

  $ \psi $~ is not an eigenstate of the free Hamiltonian but it does satisfy the free Schrödinger equation:

\begin{equation}
\label{equation-free-fourt-free-seqn-0}\hyperlabel{equation-free-fourt-free-seqn-0}%
i\frac{\partial {\psi }_{\tau }\left(x\right)}{\partial \tau }=\frac{1}{2m}\left(\frac{{\partial }^{2}}{\partial {t}^{2}}-{\nabla }^{2}+{m}^{2}\right){\psi }_{\tau }\left(x\right)
\end{equation}

With probability density:

\begin{equation}
\label{equation-free-fourt-free-fourt-rho}\hyperlabel{equation-free-fourt-free-fourt-rho}%
{\rho }_{\tau }\left(x\right)=\sqrt{\frac{1}{{\pi }^{4}{\displaystyle \prod _{\mu =0}^{3}{\sigma }_{\mu }^{2}\left(1+\frac{{\tau }^{2}}{{m}^{2}{\sigma }_{\mu }^{4}}\right)}}}\mathrm{exp}\left(-{\displaystyle \sum _{\mu =0}^{3}\frac{{\left({x}^{\mu }-\left({\overline{x}}_{0}^{\mu }+\frac{{p}_{0}^{\mu }}{m}\tau \right)\right)}^{2}}{{\sigma }_{\mu }^{2}\left(1+\frac{{\tau }^{2}}{{m}^{2}{\sigma }_{\mu }^{4}}\right)}}\right)
\end{equation}

And uncertainties:

\begin{equation}
\label{equation-free-fourt-free-fourt-dispersions}\hyperlabel{equation-free-fourt-free-fourt-dispersions}%
\langle {\left({x}^{\mu }-{\overline{x}}_{\tau }^{\mu }\right)}^{2}\rangle =\frac{{\sigma }_{\mu }^{2}}{2}\left\vert 1+\frac{{\tau }^{2}}{{m}^{2}{\sigma }_{\mu }^{4}}\right\vert 
\end{equation}

  $ \psi $~ has a natural decomposition into time and space parts:

\begin{equation}
\label{equation-free-fourt-free-fourt-decomp}\hyperlabel{equation-free-fourt-free-fourt-decomp}%
{\psi }_{\tau }\left(t,x\right)={\chi }_{\tau }\left(t\right){\xi }_{\tau }\left(\overrightarrow{x}\right)
\end{equation}

Time part:

\begin{equation}
\label{equation-free-fourt-free-onet-chitq}\hyperlabel{equation-free-fourt-free-onet-chitq}%
{\chi }_{\tau }\left(t\right)=\sqrt[4]{\frac{1}{\pi {\sigma }_{0}^{2}}}\sqrt{\frac{1}{{f}_{\tau }^{\left(0\right)}}}\mathrm{exp}\left(-i{E}_{0}t-\frac{1}{2{\sigma }_{0}^{2}{f}_{\tau }^{\left(0\right)}}{\left(t-{\overline{t}}_{\tau }\right)}^{2}+i\frac{{E}_{0}^{2}-{m}^{2}}{2m}\tau \right)
\end{equation}

The space part is the standard quantum theory wave function:

\begin{equation}
\label{equation-free-fourt-free-threet-xitau}\hyperlabel{equation-free-fourt-free-threet-xitau}%
{\xi }_{\tau }\left(\overrightarrow{x}\right)=\sqrt[4]{\frac{1}{{\pi }^{3}\mathrm{det}\left({\overleftrightarrow{\Sigma }}_{\tau }\right)}}\mathrm{exp}\left(i{\overrightarrow{p}}_{0}\cdot \overrightarrow{x}-\left(\overrightarrow{x}-{\overline{x}}_{\tau }\right)\cdot \frac{1}{2{\overleftrightarrow{\Sigma }}_{\tau }}\cdot \left(\overrightarrow{x}-{\overline{x}}_{\tau }\right)-i\frac{{\overrightarrow{p}}_{0}^{2}}{2m}\tau \right)
\end{equation}

\subsection{In Relative Time}
Recall that the relative time is defined as the difference between the absolute quantum time and the laboratory time 
 $ {t}_{\tau }\equiv t-\tau $. We look at the relative time kernel, then apply it to plane waves and Gaussian test functions.

\paragraph*{Kernel}

\noindent

Only the time part of the kernel is affected by a switch from block time to relative time:

\begin{equation}
\label{equation-free-fourt-free-reltimet-kernel}\hyperlabel{equation-free-fourt-free-reltimet-kernel}%
{\stackrel{\frown }{K}}_{\tau }^{\left(rel\right)}\left({{t}^{{''}}}_{\tau };{{t}^{\prime }}_{0}\right)=\sqrt{\frac{im}{2\pi \tau }}\mathrm{exp}\left(-im\frac{{\left({{t}^{{''}}}_{\tau }-{{t}^{\prime }}_{0}+\tau \right)}^{2}}{2\tau }-im\frac{\tau }{2}\right)
\end{equation}

Or:

\begin{equation}
\label{equation-free-fourt-free-reltimet-kernel2}\hyperlabel{equation-free-fourt-free-reltimet-kernel2}%
{\stackrel{\frown }{K}}_{\tau }^{\left(rel\right)}\left({{t}^{{''}}}_{\tau };{{t}^{\prime }}_{0}\right)=\sqrt{\frac{im}{2\pi \tau }}\mathrm{exp}\left(-im\frac{{\left({{t}^{{''}}}_{\tau }-{{t}^{\prime }}_{0}\right)}^{2}}{2\tau }-im\left({{t}^{{''}}}_{\tau }-{{t}^{\prime }}_{0}+\tau \right)\right)
\end{equation}

\paragraph*{Plane Waves}

\noindent

The relative time plane wave is:

\begin{equation}
\label{equation-free-fourt-free-planet-7}\hyperlabel{equation-free-fourt-free-planet-7}%
{\phi }_{\tau }^{\left(rel\right)}\left(x\right)=\frac{1}{4{\pi }^{2}}\mathrm{exp}\left(-iE{t}_{\tau }+i\overrightarrow{p}\cdot \overrightarrow{x}-i\left({\mathscr{E}}_{p}+E\right)\tau \right)
\end{equation}

We break out the quantum energy 
 $ E$~ into space and time parts:

\begin{equation}
\label{equation-free-fourt-free-planet-8}\hyperlabel{equation-free-fourt-free-planet-8}%
E={\overline{E}}_{\overrightarrow{p}}+{\stackrel{\frown }{E}}_{p},{\stackrel{\frown }{E}}_{p}\equiv E-{\overline{E}}_{\overrightarrow{p}}
\end{equation}

The relative time laboratory energy is:

\begin{equation}
\label{equation-free-fourt-free-planet-9}\hyperlabel{equation-free-fourt-free-planet-9}%
{\mathscr{E}}_{p}^{\left(rel\right)}={\overline{E}}_{\overrightarrow{p}}+{\stackrel{\frown }{\mathscr{E}}}_{p}^{\left(rel\right)},{\stackrel{\frown }{\mathscr{E}}}_{p}^{\left(rel\right)}\equiv -\left({\gamma }_{\overrightarrow{p}}-1\right){\stackrel{\frown }{E}}_{p}-\frac{{\stackrel{\frown }{E}}_{p}^{2}}{2m},{\gamma }_{\overrightarrow{p}}\equiv \frac{1}{\sqrt{1-{\overrightarrow{v}}^{2}}}=\frac{{\overline{E}}_{\overrightarrow{p}}}{m}
\end{equation}

With this notation we have for a general free plane wave in relative time:

\begin{equation}
\label{equation-free-fourt-free-planet-d}\hyperlabel{equation-free-fourt-free-planet-d}%
{\phi }_{\tau }^{\left(rel\right)}\left(x\right)=\frac{1}{4{\pi }^{2}}\mathrm{exp}\left(-i\left({\overline{E}}_{\overrightarrow{p}}+{\stackrel{\frown }{E}}_{p}\right){t}_{\tau }+i\overrightarrow{p}\cdot \overrightarrow{x}-i\left({\overline{E}}_{\overrightarrow{p}}+{\stackrel{\frown }{\mathscr{E}}}_{p}^{\left(rel\right)}\right)\tau \right)
\end{equation}

These are eigenstates of the free relative time Hamiltonian, with eigenvalues 
 $ {\mathscr{E}}_{p}^{\left(rel\right)}$:

\begin{equation}
\label{equation-free-fourt-free-reltime-0b}\hyperlabel{equation-free-fourt-free-reltime-0b}%
{H}^{\left(free,rel\right)}{\phi }_{\tau }^{\left(rel\right)}\left({t}_{\tau },\overrightarrow{x}\right)={\mathscr{E}}_{p}^{\left(rel\right)}{\phi }_{\tau }^{\left(rel\right)}\left({t}_{\tau },\overrightarrow{x}\right),{H}^{\left(free,rel\right)}=\left(i{\partial }_{{t}_{\tau }}+\frac{1}{2m}{\partial }_{{t}_{\tau }}^{2}-\frac{1}{2m}{\overrightarrow{\nabla }}^{2}+\frac{m}{2}\right)
\end{equation}

The onshell states are:

\begin{equation}
\label{equation-free-fourt-free-planet-7b}\hyperlabel{equation-free-fourt-free-planet-7b}%
{\phi }_{\tau }^{\left(rel,onshell\right)}\left(x\right)=\frac{1}{4{\pi }^{2}}\mathrm{exp}\left(-i{\overline{E}}_{\overrightarrow{p}}{t}_{\tau }+i\overrightarrow{p}\cdot \overrightarrow{x}-i{\overline{E}}_{\overrightarrow{p}}\tau \right)
\end{equation}

With:

\begin{equation}
\label{equation-free-fourt-free-planet-7a}\hyperlabel{equation-free-fourt-free-planet-7a}%
i\frac{d}{d\tau }{\phi }_{\tau }^{\left(rel,onshell\right)}\left(x\right)={\overline{E}}_{\overrightarrow{p}}{\phi }_{\tau }^{\left(rel,onshell\right)}\left(x\right)
\end{equation}

In the non-{}relativistic regime there is a natural division of the laboratory energy into time (\emph{m}) and space ($ \frac{{\overrightarrow{p}}^{2}}{2m}$) parts, giving the wave function a natural division into time and space parts as well:

\begin{equation}
\label{equation-free-fourt-free-planet-7c}\hyperlabel{equation-free-fourt-free-planet-7c}%
{\overline{E}}_{\overrightarrow{p}}\approx m+\frac{{\overrightarrow{p}}^{2}}{2m}\rightarrow {\phi }_{\tau }^{\left(rel,onshell,nr\right)}\left(x\right)\approx \left(\begin{array}{l}{\stackrel{\frown }{\phi }}_{\tau }\left({t}_{\tau }\right)\equiv \frac{1}{\sqrt{2\pi }}\mathrm{exp}\left(-i{\overline{E}}_{\overrightarrow{p}}{t}_{\tau }-im\tau \right)\\ \times {\overline{\phi }}_{\tau }\left(\overrightarrow{x}\right)\equiv \frac{1}{\sqrt{8{\pi }^{3}}}\mathrm{exp}\left(i\overrightarrow{p}\cdot \overrightarrow{x}-i\frac{{\overrightarrow{p}}^{2}}{2m}\tau \right)\end{array}\right)
\end{equation}

With 
 $ {\overline{\phi }}_{\tau }\left(\overrightarrow{x}\right)$~ being the usual standard quantum theory plane wave.

\paragraph*{Gaussian Test Functions}

\noindent

The relative time wave function at 
 $ \tau =0$~ is the same as the block time wave function:

\begin{equation}
\label{equation-free-fourt-free-reltimet-000}\hyperlabel{equation-free-fourt-free-reltimet-000}%
{\psi }_{0}^{\left(rel\right)}\left({t}_{0},\overrightarrow{x}\right)={\chi }_{0}^{\left(rel\right)}\left({t}_{0}\right){\xi }_{0}\left(\overrightarrow{x}\right)
\end{equation}

By applying the kernel we get:

\begin{equation}
\label{equation-free-fourt-free-reltimet-00}\hyperlabel{equation-free-fourt-free-reltimet-00}%
{\psi }_{\tau }^{\left(rel\right)}\left({t}_{\tau },\overrightarrow{x}\right)=\left(\begin{array}{l}\sqrt[4]{\frac{1}{\pi {\sigma }_{0}^{2}}}\sqrt{\frac{1}{{f}_{\tau }^{\left(0\right)}}}\mathrm{exp}\left(-i\left({\overline{E}}_{\overrightarrow{p}}+{\stackrel{\frown }{E}}_{p}\right){t}_{\tau }-\frac{{\left({t}_{\tau }-{\overline{t}}_{\tau }^{\left(rel\right)}\right)}^{2}}{2{\sigma }_{0}^{2}{f}_{\tau }^{\left(0\right)}}-i\left({\overline{E}}_{\overrightarrow{p}}+{\stackrel{\frown }{\mathscr{E}}}_{p}^{\left(rel\right)}\right)\tau \right)\\ \times \sqrt[4]{\frac{1}{{\pi }^{3}\mathrm{det}\left({\overleftrightarrow{\Sigma }}_{\tau }\right)}}\mathrm{exp}\left(i{\overrightarrow{p}}_{0}\cdot \left(\overrightarrow{x}-{\overrightarrow{x}}_{0}\right)-\left(\overrightarrow{x}-{\overrightarrow{x}}_{0}\right)\cdot \frac{1}{2{\overleftrightarrow{\Sigma }}_{\tau }}\cdot \left(\overrightarrow{x}-{\overrightarrow{x}}_{0}\right)\right)\end{array}\right)
\end{equation}

With the average position in relative time given by:

\begin{equation}
\label{equation-free-fourt-free-reltimet-00b}\hyperlabel{equation-free-fourt-free-reltimet-00b}%
{\overline{t}}_{\tau }^{\left(rel\right)}\equiv {\overline{t}}_{0}+\left(\frac{{E}_{0}}{m}-1\right)\tau 
\end{equation}

The quantum and laboratory energy are the same as for the plane wave.

These Gaussian test functions are not eigenfunctions of the Hamiltonian but they do satisfy the free relative time Schrödinger equation:

\begin{equation}
\label{equation-free-fourt-free-reltime-3}\hyperlabel{equation-free-fourt-free-reltime-3}%
i\frac{d}{d\tau }{\psi }_{\tau }^{\left(rel\right)}\left({t}_{\tau },\overrightarrow{x}\right)=\left(i{\partial }_{{t}_{\tau }}+\frac{1}{2m}{\partial }_{{t}_{\tau }}^{2}-\frac{1}{2m}{\overrightarrow{\nabla }}^{2}+\frac{m}{2}\right){\psi }_{\tau }^{\left(rel\right)}\left({t}_{\tau },\overrightarrow{x}\right)
\end{equation}

The onshell states are those with the time parts of the quantum energy and the laboratory energy zero. As with the plane waves, in the non-{}relativistic regime we have a natural division
        of the laboratory energy into time and space parts:

\begin{equation}
\label{equation-free-fourt-free-reltimet-00c}\hyperlabel{equation-free-fourt-free-reltimet-00c}%
{\psi }_{\tau }^{\left(rel,onshell\right)}\left({t}_{\tau },\overrightarrow{x}\right)=\left(\begin{array}{c}{\chi }_{\tau }^{\left(rel,onshell\right)}\left({t}_{\tau }\right)\equiv \sqrt[4]{\frac{1}{\pi {\sigma }_{0}^{2}}}\sqrt{\frac{1}{{f}_{\tau }^{\left(0\right)}}}\mathrm{exp}\left(-i{\overline{E}}_{\overrightarrow{p}}{t}_{\tau }-\frac{{\left({t}_{\tau }-{\overline{t}}_{\tau }^{\left(rel\right)}\right)}^{2}}{2{\sigma }_{0}^{2}{f}_{\tau }^{\left(0\right)}}-im\tau \right)\\ \times {\xi }_{\tau }\left(\overrightarrow{x}\right)\equiv \sqrt[4]{\frac{1}{{\pi }^{3}\mathrm{det}\left({\overleftrightarrow{\Sigma }}_{\tau }\right)}}\mathrm{exp}\left(i{\overrightarrow{p}}_{0}\cdot \left(\overrightarrow{x}-{\overrightarrow{x}}_{0}\right)-\left(\overrightarrow{x}-{\overrightarrow{x}}_{0}\right)\cdot \frac{1}{2{\overleftrightarrow{\Sigma }}_{\tau }}\cdot \left(\overrightarrow{x}-{\overrightarrow{x}}_{0}\right)-i\frac{{\overrightarrow{p}}^{2}}{2m}\tau \right)\end{array}\right)
\end{equation}

\section{Energy/Momentum Representation}\label{free-fourE}\hyperlabel{free-fourE}%

We define the Fourier transform using opposite signs for the time/energy and space/momentum parts:

\begin{equation}
\label{equation-free-fourE-conv-trans-ft4}\hyperlabel{equation-free-fourE-conv-trans-ft4}%
\begin{array}{l}\widehat{f}\left(E,\overrightarrow{p}\right)\equiv {\displaystyle \int \frac{\text{d}t\text{d}\overrightarrow{x}}{4{\pi }^{2}}\mathrm{exp}\left(iEt-i\overrightarrow{p}\cdot \overrightarrow{x}\right)f\left(t,\overrightarrow{x}\right)}\\ f\left(t,\overrightarrow{x}\right)\equiv {\displaystyle \int \frac{\text{d}E\text{d}\overrightarrow{p}}{4{\pi }^{2}}\mathrm{exp}\left(-iEt+i\overrightarrow{p}\cdot \overrightarrow{x}\right)\widehat{f}\left(E,\overrightarrow{p}\right)}\end{array}
\end{equation}

For kernels, gates, potentials and other two sided objects we have:

\begin{equation}
\label{equation-free-fourE-conv-trans-ft5}\hyperlabel{equation-free-fourE-conv-trans-ft5}%
\begin{array}{l}\widehat{g}\left({E}^{{''}},{\overrightarrow{p}}^{{''}};{E}^{\prime },{\overrightarrow{p}}^{\prime }\right)\equiv {\displaystyle \int \frac{\text{d}{t}^{{''}}\text{d}{\overrightarrow{x}}^{{''}}\text{d}{t}^{\prime }\text{d}{\overrightarrow{x}}^{\prime }}{16{\pi }^{4}}\mathrm{exp}\left(i{E}^{{''}}{t}^{{''}}-i{\overrightarrow{p}}^{{''}}\cdot {\overrightarrow{x}}^{{''}}\right)g\left({t}^{{''}},{\overrightarrow{x}}^{{''}};{t}^{\prime },{\overrightarrow{x}}^{\prime }\right)\mathrm{exp}\left(-i{E}^{\prime }{t}^{\prime }+i{\overrightarrow{p}}^{\prime }\cdot {\overrightarrow{x}}^{\prime }\right)}\\ g\left({t}^{{''}},{\overrightarrow{x}}^{{''}};{t}^{\prime };{\overrightarrow{x}}^{\prime }\right)\equiv {\displaystyle \int \frac{\text{d}{E}^{{''}}\text{d}{\overrightarrow{p}}^{{''}}\text{d}{E}^{\prime }\text{d}{\overrightarrow{p}}^{\prime }}{16{\pi }^{4}}\mathrm{exp}\left(-i{E}^{{''}}{t}^{{''}}+i{\overrightarrow{p}}^{{''}}\cdot {\overrightarrow{x}}^{\prime }\right)\widehat{g}\left({E}^{{''}},{\overrightarrow{p}}^{{''}};{E}^{\prime },{\overrightarrow{p}}^{\prime }\right)\mathrm{exp}\left(i{E}^{\prime }{t}^{\prime }-i{\overrightarrow{p}}^{\prime }\cdot {\overrightarrow{x}}^{\prime }\right)}\end{array}
\end{equation}

We look at the block time case then the relative time case.

\subsection{In Block Time}
We look at the kernel, then apply it to plane waves and Gaussian test functions.

\paragraph*{Kernel}

\noindent

The free kernel may be written as a product of energy and momentum parts:

\begin{equation}
\label{equation-free-fourE-free-kernel-kprodE}\hyperlabel{equation-free-fourE-free-kernel-kprodE}%
{\widehat{K}}_{\tau }\left({p}^{{''}};{p}^{\prime }\right)={\widehat{\stackrel{\frown }{K}}}_{\tau }\left({E}^{{''}};{E}^{\prime }\right){\widehat{\overline{K}}}_{\tau }\left(\overrightarrow{{p}^{{''}}};\overrightarrow{{p}^{\prime }}\right)
\end{equation}

Energy part:

\begin{equation}
\label{equation-free-fourE-free-oneE-kernel}\hyperlabel{equation-free-fourE-free-oneE-kernel}%
{\widehat{\stackrel{\frown }{K}}}_{\tau }\left({E}^{{''}};{E}^{\prime }\right)=\delta \left({E}^{{''}}-{E}^{\prime }\right)\mathrm{exp}\left(i\frac{{{E}^{\prime }}^{2}-{m}^{2}}{2m}\tau \right)
\end{equation}

Momentum part:

\begin{equation}
\label{equation-free-fourE-free-fourE-kernel3}\hyperlabel{equation-free-fourE-free-fourE-kernel3}%
{\widehat{\overline{K}}}_{\tau }\left(\overrightarrow{{p}^{{''}}};\overrightarrow{{p}^{\prime }}\right)={\delta }^{\left(3\right)}\left(\overrightarrow{{p}^{{''}}}-\overrightarrow{{p}^{\prime }}\right)\mathrm{exp}\left(-i\frac{{\overrightarrow{{p}^{\prime }}}^{2}}{2m}\tau \right)
\end{equation}

\paragraph*{Plane Waves}

\noindent

At laboratory time zero the plane waves of the time/space representation turn into 
 $ \delta $~ functions:

\begin{equation}
\label{equation-free-fourE-free-planeE-0}\hyperlabel{equation-free-fourE-free-planeE-0}%
{\widehat{\phi }}_{0}\left(p\right)={\delta }^{\left(4\right)}\left(p-{p}_{0}\right)
\end{equation}

By applying the kernel we get:

\begin{equation}
\label{equation-free-fourE-free-planeE-1}\hyperlabel{equation-free-fourE-free-planeE-1}%
{\widehat{\phi }}_{\tau }\left(p\right)={\delta }^{\left(4\right)}\left(p-{p}_{0}\right)\mathrm{exp}\left(-i{\mathscr{E}}_{0}\tau \right)
\end{equation}

With laboratory energy:

\begin{equation}
\label{equation-free-fourE-free-planeE-1b}\hyperlabel{equation-free-fourE-free-planeE-1b}%
{\mathscr{E}}_{0}\equiv -\frac{{p}_{0}^{2}-{m}^{2}}{2m}
\end{equation}

The 
 $ {\widehat{\phi }}_{\tau }\left(p\right)$~ are eigenfunctions of the Hamiltonian:

\begin{equation}
\label{equation-free-fourE-free-planeE-1c}\hyperlabel{equation-free-fourE-free-planeE-1c}%
{H}^{\left(free\right)}{\widehat{\phi }}_{\tau }\left(p\right)={\mathscr{E}}_{0}{\widehat{\phi }}_{\tau }\left(p\right),{H}^{\left(free\right)}=-\frac{{p}^{2}-{m}^{2}}{2m}
\end{equation}

\paragraph*{Gaussian Test Functions}

\noindent

The initial wave function is the Fourier transform of the coordinate space initial wave function:

\begin{equation}
\label{equation-free-fourE-free-fourE-phi0}\hyperlabel{equation-free-fourE-free-fourE-phi0}%
{\widehat{\psi }}_{0}\left(p\right)=\sqrt[4]{\frac{1}{{\pi }^{4}\mathrm{det}\left({\widehat{\Sigma }}_{0}\right)}}\mathrm{exp}\left(i\left(p-{p}_{0}\right){\overline{x}}_{0}-\frac{1}{2{\widehat{\Sigma }}_{0}^{\mu \nu }}\left({p}^{\mu }-{p}_{0}^{\mu }\right)\left({p}^{v}-{p}_{0}^{v}\right)\right)
\end{equation}

With the dispersion matrix:

\begin{equation}
\label{equation-free-fourE-free-fourE-sigma0}\hyperlabel{equation-free-fourE-free-fourE-sigma0}%
{\widehat{\Sigma }}_{0}^{\mu \nu }=\left(\begin{array}{cccc}{\widehat{\sigma }}_{0}^{2}& 0& 0& 0\\ 0& {\widehat{\sigma }}_{1}^{2}& 0& 0\\ 0& 0& {\widehat{\sigma }}_{2}^{2}& 0\\ 0& 0& 0& {\widehat{\sigma }}_{3}^{2}\end{array}\right)={\left(\frac{1}{{\Sigma }_{0}}\right)}^{\mu \nu }
\end{equation}

The time/space and energy/momentum dispersions are reciprocal 
 $ {\widehat{\sigma }}_{0}^{2}=1/{\sigma }_{0}^{2},{\widehat{\sigma }}_{i}^{2}=1/{\sigma }_{i}^{2}$.

The wave function as a function of laboratory time is:

\begin{equation}
\label{equation-free-fourE-free-fourE-fulltau}\hyperlabel{equation-free-fourE-free-fourE-fulltau}%
{\widehat{\psi }}_{\tau }\left(p\right)=\sqrt[4]{\frac{1}{{\pi }^{4}\mathrm{det}\left({\widehat{\Sigma }}_{0}\right)}}\mathrm{exp}\left(i\left(p-{p}_{0}\right){\overline{x}}_{0}-\frac{1}{2{\widehat{\Sigma }}_{0}^{\mu \nu }}\left({p}^{\mu }-{p}_{0}^{\mu }\right)\left({p}^{v}-{p}_{0}^{v}\right)-i{\mathscr{E}}_{0}\tau \right)
\end{equation}

It is an eigenfunction of the Hamiltonian:

\begin{equation}
\label{equation-free-fourE-free-fourE-seqn}\hyperlabel{equation-free-fourE-free-fourE-seqn}%
{H}^{\left(free\right)}{\widehat{\psi }}_{\tau }\left(p\right)={\mathscr{E}}_{0}{\widehat{\psi }}_{\tau }\left(p\right)
\end{equation}

It has probability density:

\begin{equation}
\label{equation-free-fourE-free-fourE-rho}\hyperlabel{equation-free-fourE-free-fourE-rho}%
{\rho }_{\tau }\left(p\right)=\sqrt{\frac{1}{{\pi }^{4}{\displaystyle \prod _{\mu =0}^{3}{\widehat{\sigma }}_{\mu }^{2}}}}\mathrm{exp}\left(-\frac{1}{{\widehat{\Sigma }}_{0}^{\mu \nu }}\left({p}^{\mu }-{p}_{0}^{\mu }\right)\left({p}^{v}-{p}_{0}^{v}\right)\right)
\end{equation}

Expectations of the momenta:

\begin{equation}
\label{equation-free-fourE-free-fourE-average}\hyperlabel{equation-free-fourE-free-fourE-average}%
\langle {p}^{\mu }\rangle ={p}_{0}^{\mu }
\end{equation}

And uncertainties of the momenta:

\begin{equation}
\label{equation-free-fourE-free-fourE-dispersions}\hyperlabel{equation-free-fourE-free-fourE-dispersions}%
\langle \left({p}^{\mu }-{p}_{0}^{\mu }\right)\left({p}^{\nu }-{p}_{0}^{\nu }\right)\rangle =\frac{{\widehat{\Sigma }}_{0}^{\mu \nu }}{2}
\end{equation}

We can break out the Gaussian test function into energy and momentum parts:

\begin{equation}
\label{equation-free-fourE-free-fourE-breakout}\hyperlabel{equation-free-fourE-free-fourE-breakout}%
{\widehat{\psi }}_{\tau }\left(p\right)={\widehat{\chi }}_{0}\left(E\right){\widehat{\xi }}_{0}\left(\overrightarrow{p}\right)
\end{equation}

Energy part:

\begin{equation}
\label{equation-free-fourE-free-fourE-energy_side}\hyperlabel{equation-free-fourE-free-fourE-energy_side}%
{\widehat{\chi }}_{\tau }\left(E\right)=\sqrt[4]{\frac{1}{\pi {\widehat{\sigma }}_{0}^{2}}}\mathrm{exp}\left(i\left(E-{E}_{0}\right){t}_{0}-\frac{1}{2{\widehat{\sigma }}_{0}^{2}}{\left(E-{E}_{0}\right)}^{2}+i\frac{{E}^{2}-{m}^{2}}{2m}\tau \right)
\end{equation}

Momentum part:

\begin{equation}
\label{equation-free-fourE-free-fourE-mom_side}\hyperlabel{equation-free-fourE-free-fourE-mom_side}%
{\widehat{\xi }}_{\tau }\left(\overrightarrow{p}\right)=\sqrt[4]{\frac{1}{{\pi }^{3}\mathrm{det}\left({\widehat{\Sigma }}_{ij}^{\left(0\right)}\right)}}\mathrm{exp}\left(-i\left(\overrightarrow{p}-{\overrightarrow{p}}_{0}\right)\cdot {\overrightarrow{x}}_{0}-{\left(\overrightarrow{p}-{\overrightarrow{p}}_{0}\right)}_{i}\cdot \frac{1}{2{\widehat{\Sigma }}_{ij}^{\left(0\right)}}\cdot {\left(\overrightarrow{p}-{\overrightarrow{p}}_{0}\right)}_{j}-i\frac{{\overrightarrow{p}}^{2}}{2m}\tau \right)
\end{equation}

\subsection{In Relative Time}
We look at the relative time kernel, then apply it to plane waves and Gaussian test functions.

\paragraph*{Kernel}

\noindent

We can get the energy part of the kernel in relative time as a Fourier transform of the time kernel:

\begin{equation}
\label{equation-free-fourE-free-reltimek-def}\hyperlabel{equation-free-fourE-free-reltimek-def}%
{\widehat{\stackrel{\frown }{K}}}_{\tau }^{\left(rel\right)}\left({E}^{{''}};{E}^{\prime }\right)=\frac{1}{2\pi }{\displaystyle \int \text{d}{{t}^{{''}}}_{\tau }}\text{d}{{t}^{\prime }}_{\tau }\mathrm{exp}\left(i{E}^{{''}}{{t}^{{''}}}_{\tau }-i{E}^{\prime }{{t}^{\prime }}_{\tau }\right){\stackrel{\frown }{K}}_{\tau }^{\left(rel\right)}\left({{t}^{{''}}}_{\tau };{{t}^{\prime }}_{0}\right)
\end{equation}

Or:

\begin{equation}
\label{equation-free-fourE-free-reltimek-kernel}\hyperlabel{equation-free-fourE-free-reltimek-kernel}%
{\widehat{\stackrel{\frown }{K}}}_{\tau }^{\left(rel\right)}\left({E}^{{''}};{E}^{\prime }\right)=\mathrm{exp}\left(-i{E}^{\prime }\tau +i\frac{{{E}^{\prime }}^{2}-{m}^{2}}{2m}\tau \right)\delta \left({E}^{{''}}-{E}^{\prime }\right)
\end{equation}

With all four dimensions:

\begin{equation}
\label{equation-free-fourE-free-reltimek-kernel4}\hyperlabel{equation-free-fourE-free-reltimek-kernel4}%
{\widehat{K}}_{\tau }^{\left(rel\right)}\left({p}^{{''}};{p}^{\prime }\right)=\mathrm{exp}\left(-i{E}^{\prime }\tau +i\frac{{{p}^{\prime }}^{2}-{m}^{2}}{2m}\tau \right){\delta }^{\left(4\right)}\left({p}^{{''}}-{p}^{\prime }\right)
\end{equation}

\paragraph*{Plane Waves}

\noindent

We can get the free plane waves in the energy-{}momentum representation as the Fourier transforms of the free plane waves in the time/space representation:

\begin{equation}
\label{equation-free-fourE-free-planeE-3}\hyperlabel{equation-free-fourE-free-planeE-3}%
{\widehat{\phi }}_{\tau }^{\left(rel\right)}\left(E,\overrightarrow{p}\right)\equiv {\displaystyle \int \frac{\text{d}{t}_{\tau }\text{d}\overrightarrow{x}}{4{\pi }^{2}}\mathrm{exp}\left(iE{t}_{\tau }-i\overrightarrow{p}\cdot \overrightarrow{x}\right){\phi }_{\tau }^{\left(rel\right)}\left({t}_{\tau },\overrightarrow{x}\right)}
\end{equation}

Or:

\begin{equation}
\label{equation-free-fourE-free-planeE-4}\hyperlabel{equation-free-fourE-free-planeE-4}%
{\widehat{\phi }}_{\tau }^{\left(rel\right)}\left(E,\overrightarrow{p}\right)={\delta }^{\left(4\right)}\left(p-{p}_{0}\right)\mathrm{exp}\left(-i{\mathscr{E}}_{0}^{\left(rel\right)}\tau \right)
\end{equation}

With relative time laboratory energy:

\begin{equation}
\label{equation-free-fourE-free-planeE-4b}\hyperlabel{equation-free-fourE-free-planeE-4b}%
{\mathscr{E}}_{0}^{\left(rel\right)}={E}_{0}-\frac{{E}_{0}^{2}-{\overrightarrow{p}}^{2}-{m}^{2}}{2m}
\end{equation}

This is an eigenfunction of the relative time Hamiltonian:

\begin{equation}
\label{equation-free-fourE-free-planeE-5}\hyperlabel{equation-free-fourE-free-planeE-5}%
{H}^{\left(free,rel\right)}{\widehat{\phi }}_{\tau }^{\left(rel\right)}\left(E,\overrightarrow{p}\right)={\mathscr{E}}_{0}^{\left(rel\right)}{\widehat{\phi }}_{\tau }^{\left(rel\right)}\left(E,\overrightarrow{p}\right),{H}^{\left(free,rel\right)}=E-\frac{{p}^{2}-{m}^{2}}{2m}
\end{equation}

\paragraph*{Gaussian Test Functions}

\noindent

We can define the Gaussian test function in relative time as the Fourier transform of the relative time Gaussian test function in time and space:

\begin{equation}
\label{equation-free-fourE-free-reltimeE-0}\hyperlabel{equation-free-fourE-free-reltimeE-0}%
{\widehat{\psi }}_{\tau }^{\left(rel\right)}\left(E,\overrightarrow{p}\right)=\left({\displaystyle \int \frac{\text{d}{t}_{\tau }}{\sqrt{2\pi }}\mathrm{exp}\left(iE{t}_{\tau }\right){\chi }_{\tau }^{\left(rel\right)}\left({t}_{\tau }\right)}\right)\left({\displaystyle \int \frac{\text{d}\overrightarrow{x}}{{\sqrt{2\pi }}^{3}}\mathrm{exp}\left(-i\overrightarrow{p}\cdot \overrightarrow{x}\right){\xi }_{\tau }\left(\overrightarrow{x}\right)}\right)
\end{equation}

The momentum part is unchanged.

By applying the kernel we get the dependence on laboratory time:

\begin{equation}
\label{equation-free-fourE-free-reltimeE-8}\hyperlabel{equation-free-fourE-free-reltimeE-8}%
{\widehat{\psi }}_{\tau }^{\left(rel\right)}\left(p\right)={\widehat{\psi }}_{0}\left(p\right)\mathrm{exp}\left(-i{\mathscr{E}}_{0}^{\left(rel\right)}\tau \right)
\end{equation}

The 
 $ {\widehat{\psi }}_{\tau }^{\left(rel\right)}\left(p\right)$~ are eigenfunctions of the free relative time Hamiltonian:

\begin{equation}
\label{equation-free-fourE-free-reltimeE-c}\hyperlabel{equation-free-fourE-free-reltimeE-c}%
{H}^{\left(free,rel\right)}{\widehat{\psi }}_{\tau }^{\left(rel\right)}\left(p\right)={\mathscr{E}}_{p}^{\left(rel\right)}{\widehat{\psi }}_{\tau }^{\left(rel\right)}\left(p\right)
\end{equation}

The onshell states have laboratory energy equal to the onshell energy:

\begin{equation}
\label{equation-free-fourE-free-reltimeE-a}\hyperlabel{equation-free-fourE-free-reltimeE-a}%
{\mathscr{E}}_{p}^{\left(rel,onshell\right)}={\overline{E}}_{\overrightarrow{p}},{\widehat{\psi }}_{\tau }^{\left(rel,onshell\right)}\left(p\right)={\widehat{\psi }}_{0}\left(p\right)\mathrm{exp}\left(-i{\overline{E}}_{\overrightarrow{p}}\tau \right)
\end{equation}

\section{Time/Momentum Representation}\label{free-hybrid}\hyperlabel{free-hybrid}%

In the 
 \hyperlink{xt-gates}{analysis of slit experiments}~ it is useful to work with a hybrid representation in quantum time and momentum
 \footnote{Here we take 
 \emph{x}~ and 
 \emph{p}~ as the single 
 \emph{x}~ and 
 \emph{p}~ dimensions, rather than four-{}vectors.}. The momentum side gives the standard quantum theory part and functions as carrier. The quantum time side gives the
    temporal quantization part and functions as signal. This is a clean division of labor.

The hybrid wave function at laboratory time zero is given by the product of the time and momentum wave functions:

\begin{equation}
\label{equation-free-hybrid-free-hybrid-5b}\hyperlabel{equation-free-hybrid-free-hybrid-5b}%
{\psi }_{0}^{\left(hyb\right)}\left({t}_{0},p\right)={\chi }_{0}^{\left(rel\right)}\left({t}_{0}\right){\widehat{\xi }}_{0}\left(p\right)
\end{equation}

The relative time part of the wave function is:

\begin{equation}
\label{equation-free-hybrid-free-hybrid-0}\hyperlabel{equation-free-hybrid-free-hybrid-0}%
{\chi }_{0}^{\left(rel\right)}\left({t}_{0}\right)=\sqrt[4]{\frac{1}{\pi {\sigma }_{0}^{2}}}\mathrm{exp}\left(-i{E}_{0}{t}_{0}-\frac{{\left({t}_{0}-{\overline{t}}_{0}\right)}^{2}}{2{\sigma }_{0}^{2}}\right)
\end{equation}

We start with onshell wave functions, with energy 
 $ {E}_{0}={\overline{E}}_{p}=\sqrt{{m}^{2}+{p}^{2}}$.

The momentum part of the wave function is:

\begin{equation}
\label{equation-free-hybrid-free-hybrid-3}\hyperlabel{equation-free-hybrid-free-hybrid-3}%
{\widehat{\xi }}_{0}\left(p\right)=\sqrt[4]{\frac{1}{\pi {\widehat{\sigma }}_{1}^{2}}}\mathrm{exp}\left(-i\left(p-\overline{p}\right){\overline{x}}_{0}-\frac{{\left(p-\overline{p}\right)}^{2}}{2{\widehat{\sigma }}_{1}^{2}}\right)
\end{equation}

The hybrid kernel is given by the product of the time and momentum kernels:

\begin{equation}
\label{equation-free-hybrid-free-hybrid-6}\hyperlabel{equation-free-hybrid-free-hybrid-6}%
{K}_{\tau }^{\left(hyb\right)}\left({{t}^{{''}}}_{\tau },{p}^{{''}};{{t}^{\prime }}_{0}{p}^{\prime }\right)={\stackrel{\frown }{K}}_{\tau }^{\left(rel\right)}\left({{t}^{{''}}}_{\tau };{{t}^{\prime }}_{0}\right){\widehat{\overline{K}}}_{\tau }\left({p}^{{''}};{p}^{\prime }\right)
\end{equation}

To get the hybrid wave function as a function of laboratory time we apply the hybrid kernel to the hybrid wave function:

\begin{equation}
\label{equation-free-hybrid-free-hybrid-8}\hyperlabel{equation-free-hybrid-free-hybrid-8}%
{\psi }_{\tau }^{\left(hyb\right)}\left({t}_{\tau },p\right)={\displaystyle \int \text{d}{{t}^{\prime }}_{0}\text{d}{p}^{\prime }{K}_{\tau }^{\left(hyb\right)}\left({{t}^{{''}}}_{\tau },{p}^{{''}};{{t}^{\prime }}_{0},{p}^{\prime }\right){\psi }_{0}^{\left(hyb\right)}\left({{t}^{\prime }}_{0},{p}^{\prime }\right)}
\end{equation}

The wave function is:

\begin{equation}
\label{equation-free-hybrid-free-hybrid-a}\hyperlabel{equation-free-hybrid-free-hybrid-a}%
{\psi }_{\tau }^{\left(hyb\right)}\left({{t}^{{''}}}_{\tau },{p}^{{''}}\right)=\left(\begin{array}{l}\sqrt[4]{\frac{1}{\pi {\sigma }_{0}^{2}{f}_{\tau }^{\left(1\right)}}}\mathrm{exp}\left(-i{E}_{0}{t}_{\tau }-\frac{{\left({t}_{\tau }-{\overline{t}}_{\tau }^{\left(rel\right)}\right)}^{2}}{2{\sigma }_{0}^{2}{f}_{\tau }^{\left(1\right)}}\right)\\ \times \sqrt[4]{\frac{1}{\pi {\widehat{\sigma }}_{1}^{2}}}\mathrm{exp}\left(-i\left(p-\overline{p}\right){\overline{x}}_{0}-\frac{{\left(p-\overline{p}\right)}^{2}}{2{\widehat{\sigma }}_{1}^{2}}\right)\\ \times \mathrm{exp}\left(-i{\overline{E}}_{p}\tau \right)\end{array}\right)
\end{equation}

In the non-{}relativistic case we have 
 $ {E}_{0}\approx m+\frac{{p}^{2}}{2m}$~ so have a natural division of the laboratory energy and wave function into time and space parts:

\begin{equation}
\label{equation-free-hybrid-free-hybrid-d}\hyperlabel{equation-free-hybrid-free-hybrid-d}%
{\psi }_{\tau }^{\left(hyb\right)}\left({t}_{\tau },p\right)=\left(\begin{array}{l}{\chi }_{\tau }^{\left(rel\right)}\left({t}_{\tau },p\right)\equiv \sqrt[4]{\frac{1}{\pi {\sigma }_{0}^{2}{f}_{\tau }^{\left(1\right)}}}\mathrm{exp}\left(-i{E}_{0}{t}_{\tau }-\frac{{\left({t}_{\tau }-{\overline{t}}_{\tau }^{\left(rel\right)}\right)}^{2}}{2{\sigma }_{0}^{2}{f}_{\tau }^{\left(1\right)}}-im\tau \right)\\ \times {\widehat{\xi }}_{\tau }^{\left(sqt\right)}\left(p\right)\equiv \sqrt[4]{\frac{1}{\pi {\widehat{\sigma }}_{1}^{2}}}\mathrm{exp}\left(-i\left(p-\overline{p}\right){\overline{x}}_{0}-\frac{{\left(p-\overline{p}\right)}^{2}}{2{\widehat{\sigma }}_{1}^{2}}-i\frac{{p}^{2}}{2m}\tau \right)\end{array}\right)
\end{equation}
\chapter{Acknowledgments}
\label{acks}\hyperlabel{acks}%
\begin{quote}

Lately it occurs to me what a long strange trip it's been.

\hspace*\fill---~Robert Hunter of The Grateful Dead
 \cite{Hunter-1970}\end{quote}
\begin{quote}

We are all travellers in the wilderness of the world, and the best we can find in our travels is an honest friend.

\hspace*\fill---~Robert Louis Stephenson\end{quote}

I thank my long time friend Jonathan Smith for invaluable encouragement, guidance, and practical assistance.

I thank the anonymous reviewer who pointed out that I was using time used in multiple senses in an earlier work 
 \cite{Ashmead-2003b}.

I thank Ferne Cohen Welch for extraordinary moral and practical support.

I thank Linda Marie Kalb and Diane Dugan for their long and ongoing moral and practical support. I thank my brothers Graham and Gaylord Ashmead and my brother-{}in-{}law Steve Robinson for
    continued encouragement. I thank Oz Fontecchio, Bruce Bloom, Shelley Handin, and Lee and Diane Weinstein for listening to a perhaps baroque take on free will and determinism. I thank Arthur
    Tansky for many helpful conversations and some proofreading. I thank Chris Kalb for suggesting the title.

I thank John Cramer, Robert Forward, and Catherine Asaro for helpful conversations (and for writing some fine SF novels). I thank Connie Willis for several entertaining conversations about
    wormhole physics, closed causal loops and the like (and also for writing several fine SF stories).

I thank Stewart Personick for many constructive discussions. I thank Matt Riesen for suggesting the use of Rydberg atoms. I thank Terry the Physicist for useful thoughts on tunneling and
    for generally hammering the ideas here. I thank Andy Love for some useful experimental suggestions, especially the frame mixing idea. I thank Dave Kratz for helpful conversations. I thank Paul
    Nahin for some useful email. I thank Jay Wile for some necessary sarcasm.

I thank John Myers and others at QUIST and DARPA for useful conversations.

I thank the participants at the third Feynman festival for many good discussions, including Gary Bowson, Fred Herz, Y. S. Kim, Marilyn Noz, A. Vourdas, and others. I thank Howard Brandt for
    his suggestion of internal decoherence.

I thank the participants at The Clock and The Quantum Conference at the Perimeter Institute for many good discussions, including J. Barbour, L. Vaidman, R. Tumulka, S. Weinstein, J.
    Vaccaro, R. Penrose, H. Price, and L. Smolin.

I thank the participants at the Third International Conference on the Nature and Ontology of Spacetime for many good discussions, including V. Petkov, W. Unruh, J. Ferret, H. Brown, and O.
    Maroney.

I thank the participants at the fourth Feynman festival for many good discussions, including N. Gisin, J. Pe\v{r}ina, Y. S. Kim, L. Skála, A. Vourdas, A. Khrennikov, A Zeilinger, J. H. Samson,
    and H. Yadsan-{}Appleby.

I thank the librarians of Bryn Mawr College, Haverford College, and the University of Pennsylvania for their unflagging helpfulness. I thank Mark West and Ashleigh Thomas for help getting
    set up at the University of Pennsylvania.

I thank countless other friends and acquaintances, not otherwise acknowledged, for listening to and often contributing to the ideas here.

I acknowledge a considerable intellectual debt to Yakir Aharonov, Julian Barbour, Paul Nahin, Huw Price, L. S. Schulman, Victor J. Stenger, and Dieter Zeh.

Finally, I thank the six German students at the Cafe Destiny in Olomouc who over a round of excellent Czech beer helped push this to its final form.

And of course, none of the above are in any way responsible for any errors of commission or omission in this work.
\begin{btSect}[plain]{/Users/taqm/qt/single/qt}

\chapter{Bibliography}
\label{bibliography-single}\hyperlabel{bibliography-single}%
\btPrintCited
\end{btSect}
\end{document}